\documentclass[12pt, a4paper]{snkmodthesis}

\usepackage{fancyhdr}
\usepackage{mathptmx}
\usepackage{lscape}
\usepackage{psfrag}
\usepackage{amsmath}
\usepackage{amssymb}
\usepackage{graphics}
\usepackage{epsfig}
\input{diagrams.tex}

\diagramstyle[PostScript=dvips,flushleft]
\newarrow{to}---->
\newarrow{mto}|--->
\newarrow{inject}{hooka}---{vee}
\newarrow{surject}----{>>}

\newtheorem{definition}{Definition}[chapter]
\newtheorem{remark}{Remark}[chapter]
\newtheorem{lemma}{Lemma}[chapter]
\newtheorem{theorem}{Theorem}[chapter]
\newtheorem{corollary}{Corollary}[chapter]
\newtheorem{conjecture}{Conjecture}[chapter]

\newcommand{\Tr}{\operatorname{Tr}}
\newcommand{\Sp}{\operatorname{Sp}}

\newenvironment{proof}[1][Proof]{\begin{trivlist}
\item[\hskip \labelsep {\bfseries #1}]}{\end{trivlist}}

\newcommand{\qed}{\raisebox{2pt}{\framebox[7pt]{}}}

\begin{document}
\begin{titlepage}
\title{The Arithmetic of\\ Calabi--Yau Manifolds\\ and Mirror Symmetry}
\author{\sc Shabnam Nargis Kadir}
\college{Christ Church}
\degree{Doctor of Philosophy}
\degreedate{Trinity 2004}
\thispagestyle{empty} \maketitle
\end{titlepage}

\include{dedi}
\include{ackno}

\setlength{\parskip}{10pt}

\begin{dedication}
This thesis is dedicated to my parents.
\end{dedication}

\begin{acknowledgements}
I am greatly indebted to my supervisor, Philip Candelas, for
introducing me to and encouraging me to pursue this exciting new
area of research straddling both String Theory and Number Theory.
I would also like to thank Xenia de la Ossa for many very fruitful
and enjoyable discussions.

Many thanks to Klaus Hulek for inviting me to Universit\"{a}t
Hannover on a European Algebraic Geometry Research Training
(EAGER) fellowship during April 2004, where I had several
stimulating discussions with him, Helena Verrill, Noriko Yui, and
others. In addition, I am grateful to Noriko Yui for organizing
and allowing me to attend the Fields Institute Workshop:
`Arithmetic, Geometry and Physics around Calabi--Yau Varieties and
Mirror Symmetry' in July 2001. I would also like to thank Charles
Doran for interesting discussions, encouragement and great career
advice.

Thanks to all my friends both at Oxford and elsewhere for
interesting times during the last few years; in particular to Ben
Green, Steve Lukito and Sakura Schafer-Nameki. A special thank you
must go to David Sanders for excellent \LaTeX tips and to Misha
Gavrilovich for proofreading at the last minute.

Finally, I have a great deal of gratitude towards my parents,
Maksuda and Nurul Kadir and sisters, Shaira and Shereen, for their
continual love and support.

This research was supported by a PhD studentship from the UK
Engineering and Physical Sciences Research Council (EPSRC). I
would like to thank the Oxford Mathematical Institute and Christ
Church for help with conference expenses.

\end{acknowledgements}

\clearpage \thispagestyle{empty} \cleardoublepage
\begin{center}
\thispagestyle{empty} {\LARGE\bf The Arithmetic of}\\
\vspace{6pt}
{\LARGE\bf  Calabi--Yau Manifolds}\\
\vspace{6pt}
{\LARGE\bf  and Mirror Symmetry}\\
\vspace{20pt}
{\large \sc Shabnam Nargis Kadir}\\
\vspace{2pt}
{\large Christ Church,  University of Oxford}\\
\vspace{2pt}
{\large D.Phil. Thesis}\\
\vspace{2pt}
{\large Trinity 2004}\\
\vspace{20pt}

{\large \bf Abstract}\\\end{center} \vspace{10pt}

 We study mirror symmetric pairs of Calabi--Yau
manifolds over finite fields. In particular we compute the number
of rational points
of the manifolds as a function of the complex structure parameters. 
The data of the number of rational points of a Calabi--Yau
$X/\mathbb{F}_q$ can be encoded in a generating function known as
the congruent zeta function. The Weil Conjectures (proved in the
1970s) show that for smooth varieties, these functions take a very
interesting form in terms of the Betti numbers of the variety.
This has interesting implications for mirror symmetry, as mirror
symmetry exchanges the odd and even Betti numbers. Here the zeta
functions for a one-parameter family of K3 surfaces,
$\mathbb{P}_3[4]$, and a two-parameter family of octics in
weighted projective space,  $\mathbb{P}_4{}^{\,(1, 1, 2, 2,
2)}\,[8]$, are computed. The form of the zeta function at points
in the moduli space of complex structures where the manifold is
singular (where the Weil conjectures apart from rationality are
not applicable), is investigated. The zeta function appears to be
sensitive to monomial and non-monomial deformations of complex
structure (or equivalently on the mirror side, toric and non-toric
divisors). Various conjectures about the form of the zeta function
for mirror symmetric pairs are made in light of the results of
this calculation. Connections with $L$-functions associated to
both elliptic and Siegel modular forms are suggested.

\clearpage \thispagestyle{empty} \cleardoublepage

\listoffigures

\tableofcontents

\chapter[Introduction]{Introduction}
In this thesis we are interested in studying arithmetic properties
of Mirror symmetry. Mirror symmetry is a conjecture in string
theory according to which certain `mirror pairs' of Calabi--Yau
manifolds give rise to isomorphic physical theories. (A
Calabi--Yau manifold is a complex variety of dimension $d$ with
trivial canonical bundle and vanishing Hodge numbers $h^{i,0}$ for
$0<i<d$, e.g. a one-dimensional Calabi--Yau variety is an elliptic
curve, a two-dimensional Calabi--Yau is a K3 surface, and in
dimensions three and above there are many thousands of Calabi--Yau
manifolds).

Physicists concerned with mirror symmetry usually deal with
Calabi--Yau manifolds defined over $\mathbb{C}$, here however, in
order to study the arithmetic, we shall reduce these algebraic
Calabi--Yau varieties over discrete finite fields,
$\mathbb{F}_q,\, q = p^r$, where $p$ is prime and $r \in
\mathbb{N}$; these are the extensions of degree $r$ of the finite
field, $\mathbb{F}_p$.
The data of the number of rational points of the reduced variety, $N_{r,p}(X)=\#(X/\mathbb{F}_{p^r})$, can be encoded in a generating function known as Artin's Congruent Zeta Function, which takes the form:

\begin{equation}
Z(X/\mathbb{F}_{p^r},t)\equiv \exp\left(\sum_{r \in
   \mathbb{N}} N_{r,p}(X)\frac{t^r}{r}\right)\;.
\label{classical}
\end{equation}

The motivation for choosing the above type of generating function
is that this expression leads to rational functions in the formal
variable, $t$. This result is part of the famous Weil conjectures
(no longer conjectures for at least the last 30 years
\cite{Dwork,Deligne}). The Weil conjectures show that the Artin's
Zeta function for a smooth variety is a rational function
determined by the cohomology of the variety; in particular, that
the degree of the numerator is the sum of the odd Betti numbers of
the variety and the denominator, the sum of the even Betti
numbers. As mirror symmetry interchanges the odd and even Betti
numbers of the Calabi--Yau variety, it is a natural question to
investigate the zeta functions of pairs of mirror symmetric
families. Hence there is  speculation as to whether a `quantum
modification' to the Congruent Zeta function can be defined such
that the zeta function of mirror pairs of Calabi--Yau varieties
are inverses of each other. Notice that the above conjecture
cannot hold using the `classical definition' (\ref{classical})
above because this would mean that for a pair of manifolds,
$(X,Y)$, we would have to have $N_{r,p}(X)=-N_{r,p}(Y)$, which is
not possible.

In order to study these questions, we shall be considering
families with up to two-parameters, and use methods similar to
\cite{COV1, COV2}. In particular, we study a one-parameter family
of K3 surfaces and a two-parameter family of Calabi--Yau
threefolds, octic hypersurfaces in weighted projective space
$\mathbb{P}_4{}^{\,(1, 1, 2, 2, 2)}\,[8]$, the (non-arithmetic)
mirror symmetry for which was studied in detail in \cite{CdFKM}.
We shall be concerned with Calabi--Yau manifolds which are
hypersurfaces in toric varieties, as this provides a powerful
calculational tool, and it enables one to use the Batyrev
formulation of mirror symmetry \cite{Baty}.
In addition, higher dimensional analogues of the quintics in
$\mathbb{P}_{4}$, namely one-parameter families of
$(n-2)$-dimensional Calabi--Yau manifolds in $\mathbb{P}_{n-1}$,
will be briefly examined in Section \ref{combin}. For these the
computation is particularly simple. It is shown how the zeta
function decomposes into pieces parameterized by certain monomials
which are related to the toric data of the Calabi--Yau and, in
particular, to partitions of~$n$.

The method of computation involves use of Gauss sums, that is the
sum over an additive character times a multiplicative character.
For the purpose of counting rational points the Dwork character is
a very suitable choice for the additive character along with the
Teichm\"{u}ller character as multiplicative character. The Dwork
character was first used by Dwork \cite{Dwork} to prove the
rationality of the congruent zeta function for varieties (this is
one of the Weil conjectures). The number of rational points can be
written in terms of these Gauss sums, thus enabling computer-aided
computation of the zeta function.

The mirror of the Calabi--Yau manifolds can be found using the
Batyrev mirror construction \cite{Baty}. This is a construction in
toric geometry in which a dual pair of reflexive polytopes can be
related via toric geometry to mirror symmetric pairs of
Calabi--Yau manifolds. Using this construction and using Gauss
sums it is possible to also find the zeta function of the mirror
Calabi--Yau manifold. In \cite{ COV2}, the mirror zeta function
was found to have some factors in common with the original zeta
function, namely a contribution to the number of points associated
to the unique interior lattice point of the polyhedra. We shall
find the same phenomenon for the octic where there is a sextic,
$R_{(0,0,0,0,0)}$, that appears in both the original family and
the mirror. This phenomenon will be explained in Section
\ref{sectionlast} for the special case of Fermat hypersurfaces in
weighted projective space (these admit a Greene--Plesser
description). For the octic threefolds, on the mirror side, there
was also a contribution which may be thought of as related to a
`zero-dimensional Calabi--Yau manifold' (studied in \cite{COV1})
which was sensitive to a particular type of singularity where upon
resolution there was birationality with a one-parameter model.
This contrasts with the the fact that the zeta function of the
original family of octics has a contribution related to a
particular monomial $(4,0,3,2,1)$ (not in the Newton polyhedron
because it has degree $16$ as opposed to $8$) which is sensitive
to the presence of conifold points. This phenomenon of a special
contribution in at conifold points was also found for the quintic
threefolds in \cite{COV2}.

In \cite{COV1} it was shown that the number of rational points of
the quintic Calabi--Yau manifolds over $\mathbb{F}_p$ can be given
in terms of its periods; our calculation verified this relation
for the octic. The periods satisfy a system of differential
equations known as the Picard--Fuchs equations, with respect to
the parameters. The Picard--Fuchs equations for the quintics (and
also the octics) simplify considerably because of the group of
automorphisms, $\mathcal{A}$, of the manifold. The elements of the
polynomial ring can be classified according to their
transformation properties under the automorphisms, that is into
representations of $\mathcal{A}$, and owing to the correspondence
with the periods, the periods can be classified accordingly.

We shall be particularly interested in the behaviour of zeta
functions of singular members of such families, as for singular
varieties there is no guiding principle similar to the Weil
conjectures for the smooth case. In the computations involving the
two-parameter family of octic Calabi--Yau manifolds, it is
observed that the zeta function degenerates in a consistent way at
singularities. The nature of the degeneration depends on the type
of singularity. In this case there are two types of singularity: a
one-dimensional locus (in the base space) of conifold points and
another one-dimensional locus of points where the Calabi--Yau
manifolds are birational to a one-parameter family, whose conifold
locus has previously been studied by Rodriguez-Villegas
\cite{Villegas}.

The toric data for a Calabi--Yau manifold can be used to calculate
many important invariants, e.g. Hodge numbers. It can also be used
to calculate the number of monomial and non-monomial deformations
of complex structure \cite{Baty}, or equivalently, on the mirror
side, toric and non-toric divisors (by the monomial-divisor map
\cite{AGM}). The zeta functions appear to be sensitive to the type
of deformation/divisor.

Number theorists have been interested in the cohomological
$L$-series of Calabi--Yau varieties over $\mathbb{Q}$ or number
fields \cite{Yui}. One important question is the modularity of
Calabi--Yau varieties, i.e. are their cohomological $L$-series
completely determined by certain modular cusp forms? A brief
review of these ideas is presented in Section \ref{Hasse}.
In particular, rigid Calabi--Yau threefolds defined over the field
of rational numbers are equipped with $2$-dimensional Galois
representations, which are conjecturally equivalent to modular
forms of one variable of weight $4$ on some congruence subgroup of
$PSL(2,\mathbb{Z})$ \cite{Yui}. For not necessarily rigid
Calabi--Yau threefolds over the rationals, the Langlands Program
predicts that there should be some automorphic forms attached to
them.  Modularity was observed for the octic family of Calabi--Yau
threefolds at the special values in the moduli space where there
was birationality with the one-parameter family and
simultaneously, a conifold singularity, which reproduced the
results in \cite{Villegas}. However, other indications of
potential modularity were observed for which related cusp forms
were found using the tables of \cite{Stein}.

It is thought that higher dimensional Galois representations will
correspond to $L$-functions associated with Siegel modular forms.
We find evidence of connections to the spinor
$L$-function of genus $2$ Siegel Modular forms, and possibly to
the standard $L$-function of some genus $3$ modular forms.

The structure of this thesis is as follows: Chapter
\ref{chapLfunctions} gives a brief introduction to L-functions,
outlining what is expected for the zeta functions of mirror
symmetric manifolds by only considering the Weil Conjectures.
Chapter \ref{chapmirrorsymmetry} provides an introduction to the
Batyrev procedure and introduces the family of Calabi--Yau
manifolds we shall be considering. Chapter \ref{chappicardfuchs}
outlines the Dwork--Katz--Griffiths method for finding the
Picard--Fuchs Equations. The relation between periods and rational
points can be explained by comparing the classes of monomials
obtained using the Dwork--Katz--Griffiths method with the classes
of monomials obtained by calculating the number of rational points
in terms of Gauss sums in Chapter \ref{chapexact}. This chapter
includes an introduction to the Gauss sums employed in the
computation, and the derivation of formulae for the number of
rational points for both the original family of manifolds and its
mirror. Derivation of the formula for the number of rational
points for the family of quartic K3 surfaces is not included
because it is a direct simplification of the case for the quintic
in $\mathbb{P}_4$ studied in \cite{COV1}. In Chapter
\ref{chapzeta} we analyse the zeta functions obtained from these
derived formula, observing many instances of probable modularity.
Finally in Chapter \ref{chaplast} we review our results and
conjectures concerning the structure of the zeta functions
corresponding to Calabi--Yau manifolds. Particular interest
attaches to the form of the zeta function for values of the
parameters for which the manifold is singular. The explicit zeta
functions for $p=3,5,7,11,13,17$ for the octic Calabi--Yau
threefolds as a function of the parameters can be found in
Appendix \ref{appendix}. The results for $p=19, 23$ are also
available and support the conjectures that we make regarding the
general form of the zeta function, but are too large to display in
this thesis.


\chapter[$L$-functions]{$L$-functions}\label{chapLfunctions}
In this Chapter we review some general properties of zeta
functions; throughout the thesis we shall often refer to some of
the known results presented here. Our main interest will be the
form of the zeta function for Calabi--Yau threefolds calculated as
a function of their complex structure parameters, but we include a
review of what is known for lower-dimensional examples: elliptic
curves and K3 surfaces. Existing conjectures about modularity for
K3 manifolds and threefolds are described, as there is evidence of
modularity in the zeta functions we calculate for the quartic K3
surfaces and the octic threefolds (see Chapter \ref{chapzeta} for
full details).

\section[Artin's Congruent Zeta Function]{Artin's Congruent Zeta Function}

The arithmetic structure of Calabi--Yau varieties can be encoded
in Artin's congruent zeta function. The Weil Conjectures (finally
proven by Deligne) show that the Artin's Zeta function is a
rational function determined by the cohomology of the variety.
\begin{definition}[Artin Weil Congruent Zeta Function]
The Artin Weil Congruent Zeta Function is defined as follows:
\begin{equation}
 Z(X/\mathbb{F}_p,t):= \exp\left(\sum_{r \in
   \mathbb{N}} \#(X/\mathbb{F}_{p^r})\frac{t^r}{r}\right).
\end{equation}
\end{definition}

\section[Weil Conjectures and Mirror Symmetry]{Weil Conjectures and Mirror Symmetry}
\begin{theorem}[Weil Conjectures]
\hfill \\The three Weil conjectures are:
\begin{enumerate}
\item Functional Equation\\
The zeta function of an algebraic variety of dimension $d$
satisfies a functional equation
\begin{equation} Z\left(X/\mathbb{F}_p,\frac{1}{p^dt}\right)=
   (-1)^{\chi+\mu}p^{\frac{d\chi}{2}}t^\chi Z(X/\mathbb{F}_p,t),
\label{functional}
\end{equation}
where $\chi$ is Euler characteristic of the variety $X$ over
$\mathbb{C}$. Furthermore $\mu$ is zero when the dimension $d$ of
the variety is odd, and when $d$ is even, $\mu$ is the
multiplicity of $-p^{d/2}$ as an eigenvalue of the action induced
on the cohomology by the Frobenius automorphism:

\begin{equation}
F : X \rightarrow X, \quad x \mapsto x^p\;.
\end{equation}

\item Rationality\\
The zeta function is rational $Z(X/\mathbb{F}_p,t) \in
\mathbb{Q}(t)$. There is a factorization:

\begin{equation}
Z(X/\mathbb{F}_p,t)=
\frac{\prod_{j=1}^{d}P_{2j-1}^{(p)}(t)}{\prod_{j=0}^{d}P_{2j}^{(p)}(t)}\;,
\label{generalzeta}
\end{equation}

with each $P_j^{(p)}(t) \in \mathbb{Z}[t]$. Further $P_0^{(p)}(t)=1-t, \quad
P_{2d}^{(p)}(t)=1-p^dt$, and for each $1 \leq j \leq (2d-1), P_j^{(p)}(t)$ factors
(over $\mathbb{C}$) as

\begin{equation}
P_j^{(p)}(t)= \prod_{l=1}^{b_j} (1- \alpha_l^{(j)}(p)t)\;,
\end{equation}

where $b_j$ are the Betti numbers of the variety,
$b_j=\mathrm{dim}\,H_{\mathrm{DeRham}}^{j}(X)$. This was first
prove by Dwork \cite{Dwork}.

\item Riemann Hypothesis\\
The coefficients $\alpha_j^{(i)}(p)$ are algebraic integers which satisfy the
Riemann hypothesis:

\begin{equation}
 \|\alpha_j^{(i)}(p)\| = p^{i/2}\;.
\end{equation}

This was proved by Deligne.
\par
\end{enumerate}
\end{theorem}

 The Weil Conjectures admit explanation in terms of \'{e}tale
 cohomology \cite{FK}, \cite{M} (although we shall not be using it
 in this thesis):
Let $H_{\acute{e}t}^i(X_{\bar{\mathbb{F}}_q},\mathbb{Q}_l)$ be the
$i$-th $l$-adic \'{e}tale cohomology of
$X_{\bar{\mathbb{F}}_q}=X\otimes_{\mathbb{F}_q}\bar{\mathbb{F}}_q$,
where $\bar{\mathbb{F}}_q$ denotes an algebraic closure of
$\mathbb{F}_q$. Let $F:X_{\bar{\mathbb{F}}_q}\rightarrow
X_{\bar{\mathbb{F}}_q}$ be the $q$-th power Frobenius morphism
(note that the set of fixed points of $F$ is exactly the set of
$\mathbb{F}_q$-rational points $X(\mathbb{F}_q)$). The rationality
of the zeta function is a consequence of the Lefschetz fixed point
formula for \'{e}tale cohomology:

\begin{equation}
|X(\mathbb{F}_{q})|=\sum_{i=0}^{2n}(-1)^i\Tr(F^*;H_{\acute{e}t}^i(X_{\bar{\mathbb{F}}_q},\mathbb{Q}_l))\;;
\end{equation}

\noindent moreover we have
$P_i^{(p)}(t)=\det(1-F^*t;H_{\acute{e}t}^i(X_{\bar{\mathbb{F}}_q},\mathbb{Q}_l))$
in (\ref{generalzeta}).

\subsection[Relation to $L$-series]{Relation to $L$-Series}

Let $V_0$ be a separable algebraic variety of finite type over
$\mathbb{F}_q$. $V = V_0 \otimes \bar{\mathbb{F}_q}$, i.e. the set
of zeroes in $\mathbb{F}_q$.

The Frobenius map $F:V \rightarrow V$ raises the coordinates of
$V$ to the power $q$ (i.e. the points of $V$ fixed by $F$ are the
$\mathbb{F}_q$ rational points of $V_0$).

Let $|\text{Fix}(g)| = \#$ points on $V$ fixed by $g$ where $g \in
\mbox{End}(V)$ (i.e. $|\text{Fix}(F)|=|V_0(\mathbb{F}_q)|$; indeed
$|\text{Fix}(F^n)|=|V_0(\mathbb{F}_{q^n})|$).

Let $G$ be a finite group of automorphisms of $V_0$, and
$G\rho:\rightarrow GL(W)$ be a finite representation of $G$ in a
vector space $W$ over a field $K,\,\text{char}(K)=0$. Let $\chi$
denote the character, $ Tr(\rho)$; then the $L$-series is given
by:

\begin{equation}
L(V_0,G,\rho)= \exp\left(\sum_{n \geq
1}\frac{T^n}{n}\frac{1}{|G|}\sum_{g \in
G}\chi(g^{-1})|\text{Fix}(F^n\circ g)|\right)\;.
\end{equation}

\noindent If $G={1}$, the $L$-series reduces to the zeta function:
\begin{eqnarray*}
L(V_0,{1},\rho)&=& \exp\left(\sum_{n \geq 1}\frac{T^n}{n}|V_0(\mathbb{F}_{q^n})|\right)\\
&=&
Z(V_0/\mathbb{F}_{q},T)\;.\\
\end{eqnarray*}

\subsection[Some Examples of Congruent Zeta Functions]{Some Examples of Congruent Zeta Functions}
Here we shall look at the form of $Z(X/\mathbb{F}_p,t)$ in
(\ref{generalzeta}) when $X$ is a Calabi--Yau manifold of various
dimension.

\subsubsection[Elliptic Curves]{Elliptic Curves}
For one-dimensional Calabi--Yau manifolds, elliptic curves, the
zeta function (\ref{generalzeta}) assumes the form:
\begin{equation}
Z(E/\mathbb{F}_p,t)=\frac{P_1^{(p)}(t)}{(1-t)(1-pt)}\;,
\end{equation}
with quadratic $P_1^{(p)}(t)$.

\subsubsection[K3 Surfaces]{K3 Surfaces}
For a K3 surface the Hodge diamond is of the form:
\begin{center}
\begin{tabular}{ccccc}
 & & 1 &     &\\
 &0&   &0    &\\
1& &20& &1\\
 &0&   &0    &\\
 & & 1 &     &\\
\end{tabular}\\
\end{center}
and the zeta function (\ref{generalzeta}) assumes the form:
\begin{equation}
Z(K3/\mathbb{F}_p,t)=\frac{1}{(1-t)P_2^{(p)}(t)(1-p^2t)}\;,
\end{equation}
where $\deg P_2^{(p)}(t)=22$.

\subsubsection[Calabi--Yau Threefolds]{Calabi--Yau Threefolds}
For Calabi--Yau threefolds with finite fundamental group, i.e.
$h^{1,0}=0=h^{2,0}$, the expressions above simplify in the
following way:
\begin{equation}
Z(X/\mathbb{F}_p,t)=\frac{P_3^{(p)}(t)}{(1-t)P_2^{(p)}(t)P_4^{(p)}(t)(1-p^3t)}\;,
\end{equation}
with
\begin{equation}
\mbox{deg}(P_3^{(p)}(t)) = 2 + 2h^{2,1}, \quad
\mbox{deg}(P_2^{(p)}(t)) =\mbox{deg}(P_4^{(p)}(t))= h^{1,1}.
\end{equation}

\noindent as the Hodge diamond in this case is of the form:\\

\begin{center}
\begin{tabular}{ccccccc}
 & & & 1 &    & &\\
  &&0&   &0 &   &\\
 &0& &$h^{1,1}$& &0&\\
1~~~& &$h^{1,2}$& & $h^{2,1}$& & ~~1\\
 &0& &$h^{2,2}$& &0&\\
   &&0&   &0 &   &\\
 & & & 1 &    & &\\
\end{tabular}\\
\end{center}
where $h^{1,1}=h^{2,2}$ and $h^{2,1}=h^{1,2}$.

The Weil conjectures suggest that some residue of Mirror Symmetry
survives in the form of the zeta function for Calabi--Yau
threefolds. This is because mirror symmetry interchanges the Hodge
numbers, i.e. for a mirror symmetric pair, $\mathcal{M}$,
$\mathcal{W}$:

\begin{equation}
h_{\mathcal{M}}^{1,1}=h_{\mathcal{W}}^{2,1}, \quad
h_{\mathcal{M}}^{2,1}=h_{\mathcal{W}}^{1,1}\;. \label{interchange}
\end{equation}

\noindent Chapter \ref{chapmirrorsymmetry} provides more details
about (\ref{interchange}).

\section[Hasse--Weil $L$-functions]{Hasse--Weil $L$-functions}
\label{Hasse} The concept of the Hasse--Weil $L$-Function of an
elliptic curve can be generalized for a Calabi--Yau $d$-fold. Our
main interest is with Calabi--Yau threefolds, however, we shall
briefly review also what is known for $d=1$ and $d=2$, that is,
elliptic curves and K3 surfaces.  It can be shown that the $ith$
polynomial $P_i(t)$ is associated to the action induced by the
Frobenius morphism on the $i$th cohomology group $H^i(X)$. It is
useful to decompose the zeta function in order to find out the
arithmetic information encoded in the Frobenius action.

\begin{definition}[Local $L$-Function]
Let $P_i(t)$ be the polynomials in Artin's Zeta function over the
field $\mathbb{F}_p$. The $ith$ $L$-function of the variety $X$
over $\mathbb{F}_p$ is defined as:

\begin{equation}
L^{(i)}(X/\mathbb{F}_p,s) = \frac{1}{P_i^{(p)}(p^{-s})}\;.
\end{equation}
\end{definition}

The natural generalization of the concept of the Hasse--Weil
$L$-Function
to Calabi--Yau $d$-folds is as follows:\\

Let $X$ be a Calabi--Yau $d$-fold, and let $h^{i,0}=0$ for $1
<i<(d-1)$. Its associated Hasse--Weil $L$-function is defined as:

\begin{eqnarray}
L_{HW}(X,s) &=& \prod_{p\, good}L^{(d)}(X/\mathbb{F}_p,s)\nonumber\\
            &=& \prod_{p\, good} \frac{1}{P^{(p)}_d(p^{-s})}\nonumber\\
            &=&\prod_{p\,
            good}\frac{1}{\prod_{j=1}^{b_d}(1-\alpha_j^{(d)}(p)p^{-s})}\;.\nonumber\\
\end{eqnarray}
where the product is taken over the primes of good reduction for
the variety.

This is a generalization of the Euler product for the Riemann Zeta
Function: notice that if the variety is a point, there are no bad
primes and $P_0(t)=1-t$; we recover the Riemann zeta function:
\begin{equation}
L_{HW}(X,s)=\prod_{p}\frac{1}{(1-p^{-s})}
  =\sum_{n=1}^{\infty}\frac{1}{n^s}
  =\zeta(s)\;.
\end{equation}

\subsection[Hasse--Weil $L$-function for Elliptic Curves]{Hasse--Weil $L$-function for Elliptic Curves}

We shall quickly review some properties of the Hasse--Weil
$L$-function for an elliptic curves, for which \cite{numb} is a
good source. The main theorem concerning elliptic curves over
finite fields is due to Hasse:

\begin{theorem}[Hasse's Theorem]
Let $p$ be a prime, and let $E$ be an elliptic curve over
$\mathbb{F}_p$. Then there exists an algebraic number $\alpha_p$
such that

\begin{enumerate}
\item If $q=p^n$ then
\begin{equation}
|E(\mathbb{F}_q)|=q+1-\alpha_p^n-\bar{\alpha_p}^n\;.
\end{equation}\\
\item$\alpha_p\bar{\alpha_p}=p$, or equivalently $|\alpha_p|=\sqrt{p}$.
\end{enumerate}
\end{theorem}

\begin{corollary}
Under the same hypotheses, we have
\begin{equation}
|E(\mathbb{F}_p)|=p+1-a_p, \quad with\quad |a_p|<2\sqrt{p}\;.
\end{equation}
\label{ap}
\end{corollary}

The numbers $a_p$ are very important and are the coefficients of a
modular form of weight~$2$. This theorem gives us all the
information we need to find the congruent zeta function; hence:

\begin{corollary}
Let $E$ be an elliptic curve over $\mathbb{Q}$ and let $p$ be a
prime of good reduction (i.e. such that $E_p$ is still smooth).
Then

\begin{equation}
Z_p(E)=\frac{1-a_pt+pt^2}{(1-t)(1-pt)}\;,
\end{equation}\\
where $a_p$ is as in Corollary \ref{ap}.
\end{corollary}

Neglecting the question of bad primes, the Hasse--Weil
$L$-function for an elliptic curve is thus:
\begin{equation}
L(E,s)=\prod_{p}\frac{1}{(1-a_pp^{-s}+p^{1-2s})}\;.
\end{equation}

When bad primes are taken into consideration we obtain the following definition:

\begin{definition}
Let $E$ be an elliptic curve over $\mathbb{Q}$, and let
$y^2+a_1xy+a_3y=x^3+a_3x^2+a_4x+a_6$ be a minimal Weierstrass
equation for $E$. When $E$ has good reduction at $p$, define
$a_p=p+1-N_p$, where $N_p$ is the number of projective points of
$E$ over $\mathbb{F}_q$. If $E$ has bad reduction, define

\begin{equation}
\epsilon(p)=\begin{cases} 1, & \text{if E has split multiplicative
reduction at p};\\ -1,& \text{if E has non-split multiplicative
reduction at p};\\0, & \text{if E has additive reduction at
p.}\end{cases}
\end{equation}\\
Then we define the $L$-function of $E$ for $Re (s)>3/2$ as follows:\\
\begin{equation}
L(E,s)=\prod_{bad p}\frac{1}{1-\epsilon(p)p^{-s}}\prod_{good
p}\frac{1}{1-a_pp^{-s}+p^{1-2s}}\;.
\end{equation}
\end{definition}

By expanding the product, it is clear that $L(E,s)$ is a Dirichlet
series, i.e. a series of the form $\sum_{n\geq1}a_nn^{-s}$ (which
is, of course, the case for all zeta functions of varieties). Set

\begin{equation}
f_{\textsl{E}}(\tau)=\sum_{n\geq1}a(n)q^n, \quad \text{where as
usual}\, q=e^{2i\pi\tau}\;.
\end{equation}

\begin{theorem}
The function  $L(E,s)$ can be analytically continued to the whole
complex plane to a holomorphic function. Furthermore, there exists
a positive integer $N$, such that if we set

\begin{equation}
\Lambda(E,s)=N^{s/2}(2\pi)^{-s}\Gamma(s)L(E,s),
\end{equation}

\noindent then we have the following functional equation:
\begin{equation}
\Lambda(E,2-s)=\pm\Lambda(E,s).
\label{funct}
\end{equation}
\end{theorem}

$N$ is the conductor of $E$, and it has the form $\prod_{p\;
\text{bad}}p^{e_p}$. The product is over primes of bad reduction,
and for $p>3$, $e_p=1$ if $E$ has multiplicative reduction at $p$,
$e_p=2$ if $E$ has additive reduction. The situation is more
complicated for $p\leq3$.

It should be noted that the form of the functional equation,
(\ref{funct}),  of $L(E,s)$ is the same as the functional equation
for the Mellin transform of a modular form of weight~$2$ over the
group $\Gamma_0(N)=\left\{\left({a\atop c}{b\atop d}\right) \in
SL_2(\mathbb{Z}),\,c\equiv 0\,\mod N\right\}$.

It can be proved \cite{Wiles}, \cite{TW}, \cite{BCDT} that:

\begin{theorem}[Wiles et al.]
Let $E$ be an elliptic curve defined over $\mathbb{Q}$ with
conductor $N$, then there exists a modular cusp form, $f$, of
weight $2$ on the congruence subgroup $\Gamma_0(N)$ such that

\begin{equation}
L(E,s)=L(f,s)\;,
\end{equation}

\noindent that is if we write $f(q)=\sum_{m=1}^{\infty}a_f(m)q^m$
with $q=e^{2\pi iz}$, then $a_m=a_f(m)$ for all integers $m$,
where $z$ is the coordinate on the upper half plane on which
$\Gamma_0(N)$ acts.
\end{theorem}

\subsection[Modularity of Extremal K3 surfaces]{Modularity of Extremal K3 surfaces}

\begin{definition}[Extremal K3 surface]
A K3 surface is said to be extremal if its N\'{e}ron-Severi group,
$NS(X)$, has the maximal possible rank, namely $20$, i.e. the
Picard number of $X$ is $20$.
\end{definition}

Let $X$ be an extremal K3 surface defined over $\mathbb{Q}$; for
the sake of simplicity assume that $NS(X)$ is generated by $20$
algebraic cycles defined over $\mathbb{Q}$. The $L$-series of $X$
is determined as follow:

\begin{theorem}
Let $X$ be an extremal K3 surface defined over $\mathbb{Q}$.
Suppose that all of the $20$ algebraic cycles generating $NS(X)$
are defined over $\mathbb{Q}$. Then the $L$-series of $X$ is
given, up to finitely many Euler factors, by

\begin{equation}
L(X_{\mathbb{Q}},s)=\zeta(s-1)^{20}L(f,s)\;,
\end{equation}

\noindent where $\zeta(s)$ is the Riemann zeta-function, and
$L(f,s)$ is the $L$-series associated to a modular cusp form $f$
of weight $3$ on a congruence subgroup of $PSL_2(\mathbb{Z})$
(e.g. $\Gamma_1(N)$, or $\Gamma_0(N)$ twisted by a character). The
level $N$ depends on the discriminant of $NS(X)$, and can be
determined explicitly.
\end{theorem}

However, we shall be studying a one-parameter family of K3's with
Picard number $19$.

\subsection[Modularity Conjectures for Calabi--Yau threefolds]{Modularity Conjectures for Calabi--Yau
threefolds}\label{modularitylanglands}
 Rigid Calabi--Yau
threefolds have the property that $h^{2,1}=0$. In a certain sense,
rigid Calabi--Yau varieties are the natural generalization of
elliptic curves to higher dimensions.

\begin{conjecture}[Yui]
Every rigid Calabi--Yau three-fold over $\mathbb{Q}$ is modular,
i.e. its $L$-series coincides with the Mellin transform of a
modular cusp-form, $f$, of weight $4$ on $\Gamma_0(N)$. Here $N$
is a positive integer divisible by primes of bad reduction. More
precisely:

\begin{equation}
L(X,s)=L(f,s)\quad \mathrm{for}\, f \in S_4(\Gamma_0(N)).
\end{equation}
\end{conjecture}

At certain points in the complex structure moduli space for the
mirror family of octic Calabi--Yau threefolds that we shall
consider, the manifold is rigid. Here we observe indications of
elliptic modularity as expected by the above conjecture. Our
calculations not only verify the results of \cite{Villegas}, but
also point to connections with another modular cusp form. This is
described in detail in \ref{modularity}.

For not necessarily rigid Calabi--Yau threefolds over the
rationals, the Langlands Program predicts that there should be
some automorphic forms attached to them. In this thesis, we shall
be considering a family of threefolds which is non-rigid for all
but a few sub-cases. We observe possible connections with Siegel
modular forms.

\chapter[Toric Geometry, Mirror Symmetry and Calabi--Yau Manifolds]{Toric Geometry, Mirror Symmetry and Calabi--Yau
Manifolds} \label{chapmirrorsymmetry} In this chapter we shall
first provide a summary of Batyrev's construction for mirror pairs
of toric Calabi--Yau hypersurfaces given in \cite{Baty,CK,Voisin}
as well as numerous other sources. We shall make use of this
construction in our calculations for the zeta function for the
two-parameter mirror symmetric pair of octic threefolds. A precise
geometric description of the two-parameter family that we study is
given in Section \ref{cytwo}.

\section[Batyrev's Construction]{Batyrev's Construction}
To describe the toric variety $\mathbb{P}_{\Delta}$, let us
consider an $n$-dimensional convex integral polyhedron
$\Delta\in\mathbb{R}^n$ containing the origin $\mu_0=(0,\ldots,0)$
as an interior point. An integral polyhedron is a polyhedron with
integral vertices, and is called reflexive if its dual, defined
by:

\begin{equation}
\Delta^{\ast}=\{(x_1,\ldots,x_n)|\sum_{i=1}^{n}x_iy_i\geq
-1\;\quad \forall (y_1,\ldots,y_n) \in \Delta\}\;,
\end{equation}

\noindent is again an integral polyhedron. It is clear that if
$\Delta$ is reflexive, then $\Delta^{\ast}$ is also reflexive
since $(\Delta^{\ast})^{\ast}=\Delta$.

We associate to $\Delta$ a complete rational fan,
$\Sigma(\Delta)$, in the following way: For every $l$-dimensional
face, $\Theta_l\subset\Delta$, we define an $n$-dimensional cone
$\sigma(\Theta_l)$ by:
\begin{equation}
\sigma(\Theta_l):=\{\lambda(\acute{p}-p)|\lambda\in\mathbb{R}_{+},\;p\in\Delta,\;\acute{p}\in
\Theta_l\}.
\end{equation}
The fan, $\Sigma(\Delta)$, is then given as the collection of
$(n-l)$-dimensional dual cones $\sigma^{\ast}(\Theta)$ for
$l=0,1,\ldots,n$ for all faces of $\Delta$. The toric variety,
$\mathbb{P}_{\Delta}$, is the toric variety associated to the fan
$\Sigma(\Delta)$, i.e.
$\mathbb{P}_{\Delta}:=\mathbb{P}_{\Sigma(\Delta)}$ (See
\cite{fulton} for a nice description of how this is done).

Denote by $\mu_i\;(i=0,\ldots,s)$ the integral points of $\Delta$ and consider an affine space $\mathbb{C}^{s+1}$ with coordinates $(a_0,\ldots,a_s)$. Consider the zero locus $Z_f$ of the Laurent polynomial:

\begin{equation}
f(a,X)=\sum_{i=0}^{s}a_iX^{\mu_i},\quad
f(s,X)\in\mathbb{C}[X_1^{\pm1},\ldots,X_n^{\pm1}]\;,
\end{equation}
in the algebraic torus
$(\mathbb(C)^{\ast})^n\in\mathbb{P}_{\Delta}$, and its  closure
$Z_{f}$ in $\mathbb{P}_{\Delta}$. For conciseness, we have used
the notation $X^{\mu}:=X_1^{\mu_1}\ldots X_n^{\mu_n}$.

$f$ and $Z_f$ are called $\Delta$-regular if for all $l=1,\ldots,n$, the $f_{\Theta_l}$ and $X_i\frac{\delta}{\delta X_i}f_{\Theta_l}, \forall i= 1,\ldots,n$ are not zero simultaneously in $(\mathbb{C}^{\ast})^n$. This is equivalent to the transversality condition for the quasi-homogeneous polynomials $W_i$.

We shall now assume $\Delta$ to be reflexive. When we vary the
parameters $a_i$ under the condition of $\Delta$-regularity, we
will have a family of Calabi--Yau varieties.

In general the ambient space $\mathbb{P}_{\Delta}$ is singular,
and so in general $\bar{Z}_f$ inherits some of the singularities
of the ambient space. The $\Delta$ regularity condition ensures
that the only singularities of $\bar{Z}_f$ are the inherited ones.
$\bar{Z}_f$ can be resolved to a Calabi--Yau manifold $\hat{Z}_f$
if and only if $\mathbb{P}_{\Delta}$ only has Gorenstein
singularities, which is the case if $\Delta$ is reflexive
\cite{Baty}.

The families of Calabi--Yau manifolds $\hat{Z}_f$ will be denoted
$\mathcal{F}(\Delta)$. Of course, the above definitions also hold
for the dual polyhedron $\Delta^{\ast}$ with its integral points
$\mu_i^{\ast}(i=0,1,\ldots,s^{\ast})$.

Batyrev observed that a pair of reflexive polyhedra
$(\Delta,\Delta^{\ast})$ naturally gives rise to a pair of mirror
Calabi--Yau families
$(\mathcal{F}(\Delta),\mathcal{F}(\Delta^{\ast}))$ as the
following identities on the Hodge number hold when $n\geq 4$
($(n-1)$ is the dimension of the Calabi--Yau varieties):

\begin{eqnarray}
h^{1,1}(\hat{Z}_{f,\Delta})
&=&
h^{n-2,1}(\hat{Z}_{f,\Delta^{\ast}})\nonumber\\
&=&
l(\Delta^{\ast})-(n+1)-\sum_{\mathrm{codim}\Theta^{\ast}=1}\acute{l}(\Theta^{\ast})+\sum_{\mathrm{codim}\Theta^{\ast}=2}\acute{l}(\Theta^{\ast})\acute{l}(\Theta)\;,\nonumber\\
h^{1,1}(\hat{Z}_{f,\Delta^{\ast}}) &=&
h^{n-2,1}(\hat{Z}_{f,\Delta})\nonumber\\
&=&
l(\Delta)-(n+1)-\sum_{\mathrm{codim}\Theta=1}\acute{l}(\Theta)+\sum_{\mathrm{codim}\Theta=2}\acute{l}(\Theta)\acute{l}(\Theta^{\ast})\;.\nonumber\\
\label{Hodge}
\end{eqnarray}

Here $l(\Theta)$  and $\acute{l}(\Theta)$ are the number of
integral points on a face $\Theta$ of $\Delta$ and in its
interior, respectively. An $l$-dimensional face can be represented
by specifying its vertices $v_{i_1},\ldots,v_{i_k}$. Then the dual
face defined by $\Theta^{\ast}=\{x\in
\Delta^{\ast}|(x,v_{i_1})=\ldots=(x,v_{i_k})=-1\}$ is an
$(n-l-1)$-dimensional face of $\Delta^{\ast}$. By construction
$(\Theta^{\ast})^{\ast}=\Theta$, and we thus have a natural
pairing between $l$-dimensional faces of $\Delta$ and
$(n-l-1)$-dimensional faces of $\Delta^{\ast}$.The last sum in
(\ref{Hodge}) is over pairs of dual faces. Their contribution
cannot be associated with a monomial in the Laurent polynomial.

We shall denote by
$h^{1,1}_{\mathrm{toric}}(\hat{Z}_{f,\Delta})=h^{n-1,1}_{\mathrm{poly}}(\hat{Z}_{f,\Delta^{\ast}})$
and $h^{1,1}_{\mathrm{poly}}(\hat{Z}_{f,\Delta^{\ast}}) =
h^{n-2,1}_{\mathrm{toric}}(\hat{Z}_{f,\Delta})$ the expressions
(\ref{Hodge}) without the last terms which sum over codimension
$2$ faces. These are the dimensions of the spaces,
$H^{n-1}_{\mathrm{poly}}(\hat{Z}_{f,\Delta^{\ast}})$ and
$H^{1,1}_{\mathrm{poly}}(\hat{Z}_{f,\Delta^{\ast}})$.
$H^{n-1}_{\mathrm{poly}}(\hat{Z}_{f,\Delta^{\ast}})$ is isomorphic
to the space of first-order polynomial deformations of
$\hat{Z}_{f,\Delta^{\ast}}$ and can be generated by monomials. The
space $H^{1,1}_{\mathrm{poly}}(\hat{Z}_{f,\Delta^{\ast}})$ is the
part of the second cohomology of $\hat{Z}_{f,\Delta^{\ast}}$ which
comes from the ambient toric variety, and can be generated by
toric divisors. There is a one-to-one correspondence between
monomials and toric divisors called the \emph{monomial-divisor
map} explained in \cite{AGM}.

The formulae (\ref{Hodge}) are invariant under interchange of
$\Delta$ with $\Delta^{\ast}$ and $h^{1,1}$ with $h^{n-2,1}$, and
are thus manifestly mirror symmetric. For Calabi--Yau threefolds,
$n=4$ and we get the interchange of $h^{1,1}$ with $h^{2,1}$.

Let us now consider three dimensional Calabi--Yau hypersurfaces in
$\mathbb{P}^4(\textbf{w})$, weighted projective space. A reflexive
polyhedron can be associated to such a space
$\mathbb{P}^n(\textbf{w})$ is Gorenstein, which is the case if
$\mathrm{lcm}[w_1,\ldots,w_{n+1}]$ divides the degree $d$. In this
case we can define a simplicial, reflexive polyhedron
$\Delta(\textbf{w})$ in terms of the weights, s.t.
$\mathbb{P}_{\Delta}(\textbf{w})=\mathbb{P}(\textbf{w})$.

The Newton polyhedron can be constructed as the convex hull
(shifted by the vector $(-1,-1,-1,-1,-1)$) of the most general
polynomial $p$ of degree $d=\sum_{i=1}^{5}w_i$,
\begin{eqnarray}
\Delta=\mathrm{Conv}\left(\{v\in\mathbb{Z}^5|\sum_{i=1}^{5}v_iw_i=0,\quad
v_i \geq-1\quad \forall i\}\right), \label{mono}
\end{eqnarray}
which lies in a hyperplane in $\mathbb{R}^5$ passing through the
origin. For more details on this construction refer to
\cite{CdOK}.

\section[Cox Variables]{Cox Variables}
Global coordinates, akin to the homogenous coordinates on
projective space, can be defined for a toric variety. Hence we
shall define the homogenous coordinate ring of a toric variety. It
turns out, as we shall see later, that using these coordinates is
extremely convenient.

\begin{definition}
If $X=X_{\Sigma}$ is given by a fan $\Sigma$ in $N_{\mathbb{R}}$,
then we can introduce a variable $x_{\rho}$ for each $\rho \in
\Sigma(1)$ \textup{(}where $\Sigma(1)$ are the one-dimensional
cones\textup{)}, and consider the polynomial ring,
\begin{equation}
S=\mathbb{C}[x_{\rho}: \rho\in\Sigma(1)].
\end{equation}
A monomial in $S$ is written $x^D=\prod_{\rho}x_{\rho}^{a_p}$,
where $D=\sum_{\rho}a_{\rho}D_{\rho}$ is an effective
torus-invariant divisor on $X$ \textup{(}this uses the monomial
divisor correspondence $(\rho\leftrightarrow D_{\rho})$\textup{)},
and we say that $x^D$ has degree
\begin{equation}
\mathrm{deg}(x^D)=[D]\in A_{n-1}(X).
\end{equation}
\end{definition}
Here $A_{n-1}$ is the Chow group. The ring $S$ is graded by the
Chow group and together they give us the homogenous coordinate
ring. Given a divisor class $\alpha \in A_{n-1}(X)$, let
$S_{\alpha}$ denote the graded piece of $S$ is degree~$\alpha$.
For $\mathbb{P}^n$ it can be shown that the homogeneous coordinate
ring defined above is the usual one. Hence, for weighted
projective space we can associate a coordinate to each point in
the Newton polyhedron $\Delta$, and thus a coordinate to each
monomial $\textbf{v}$ s.t. (\ref{mono}) is satisfied.

\section[Calabi--Yau Manifolds with Two Parameters]{Calabi--Yau Manifolds with Two Parameters}
\label{cytwo} 
\begin{figure}[htbp]
\includegraphics[scale=0.5]{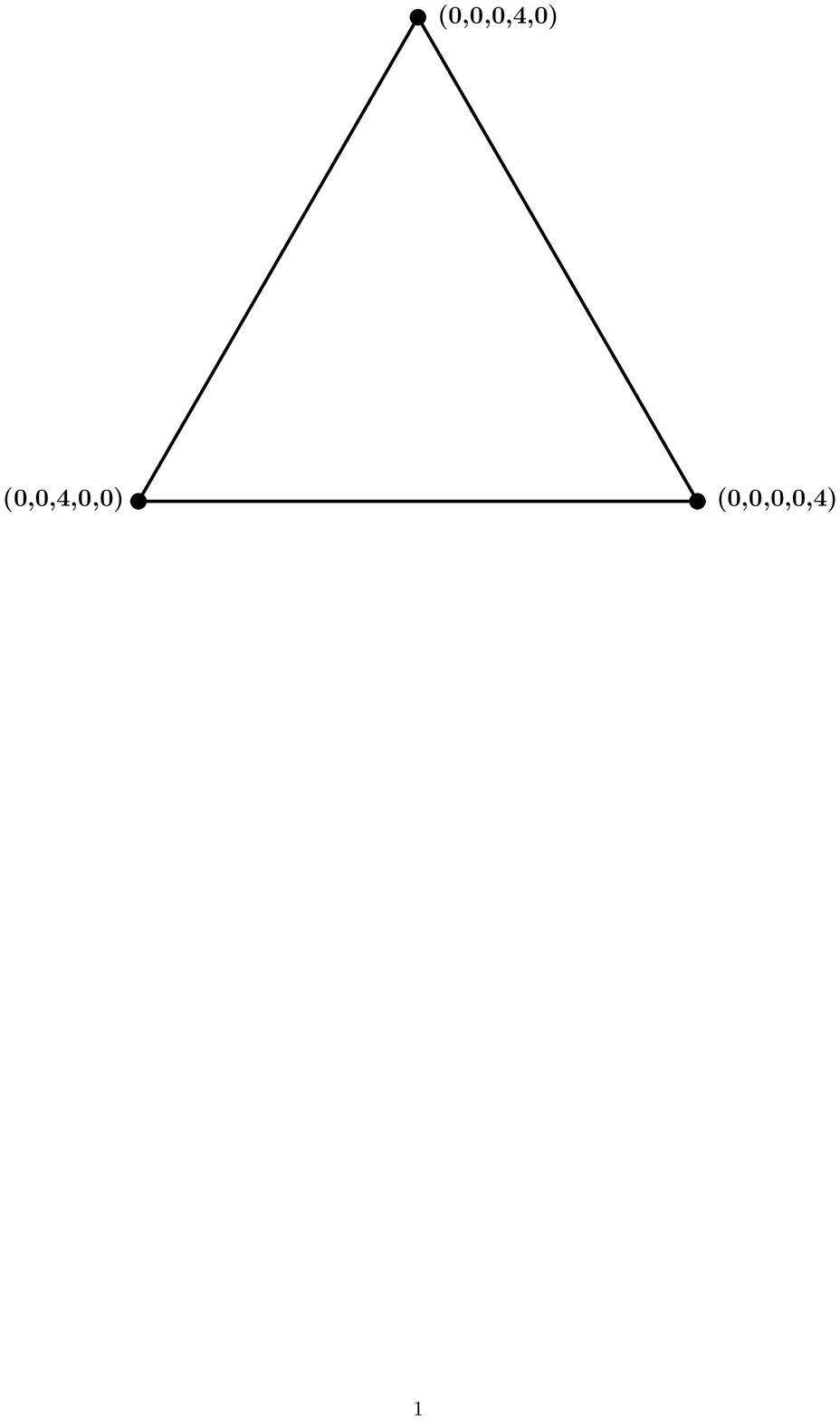}
\includegraphics[scale=0.5]{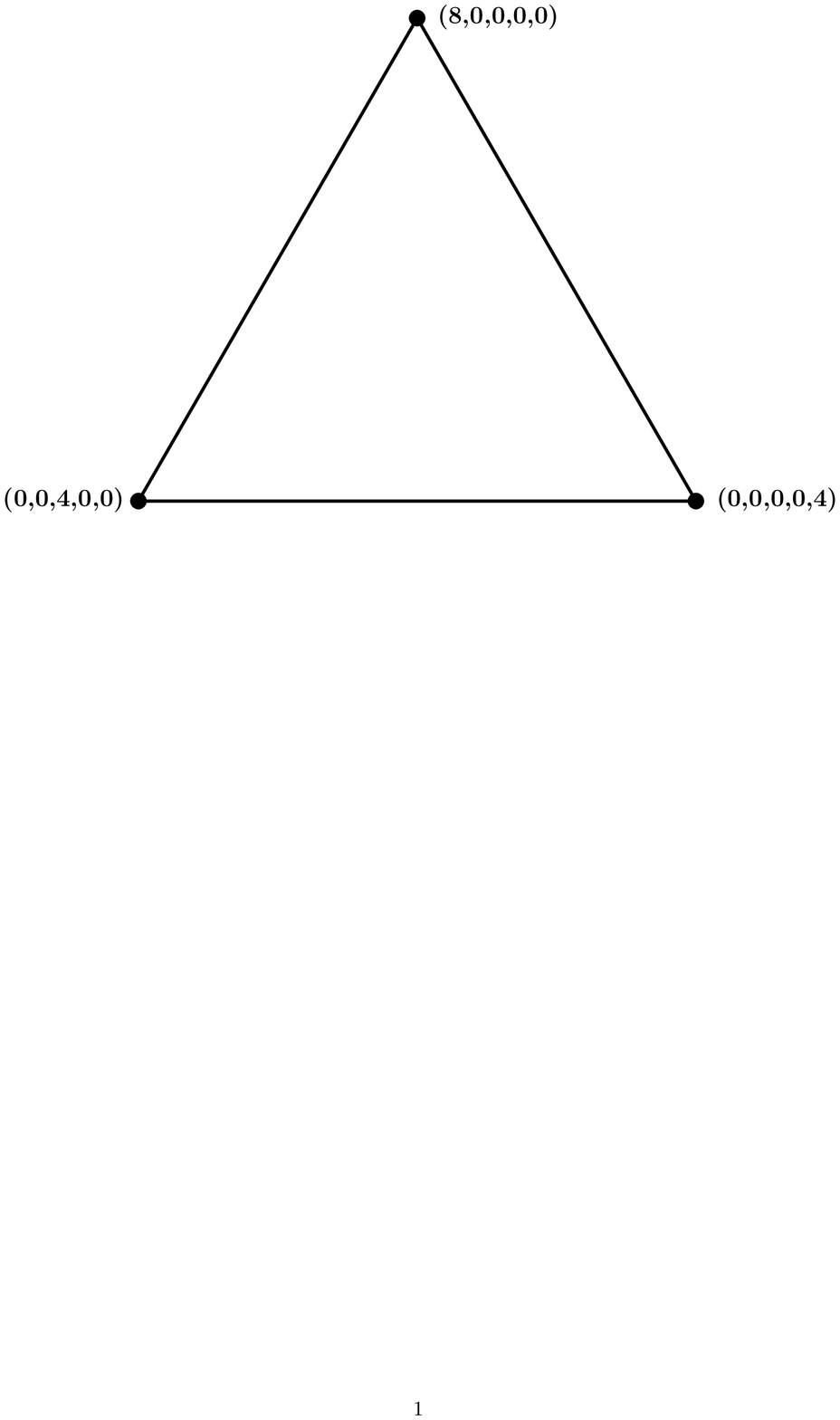}
\caption{Two-faces} \label{fig:ortwoface1}
\end{figure}

\begin{figure}[htbp]
\centering
\includegraphics[scale=1]{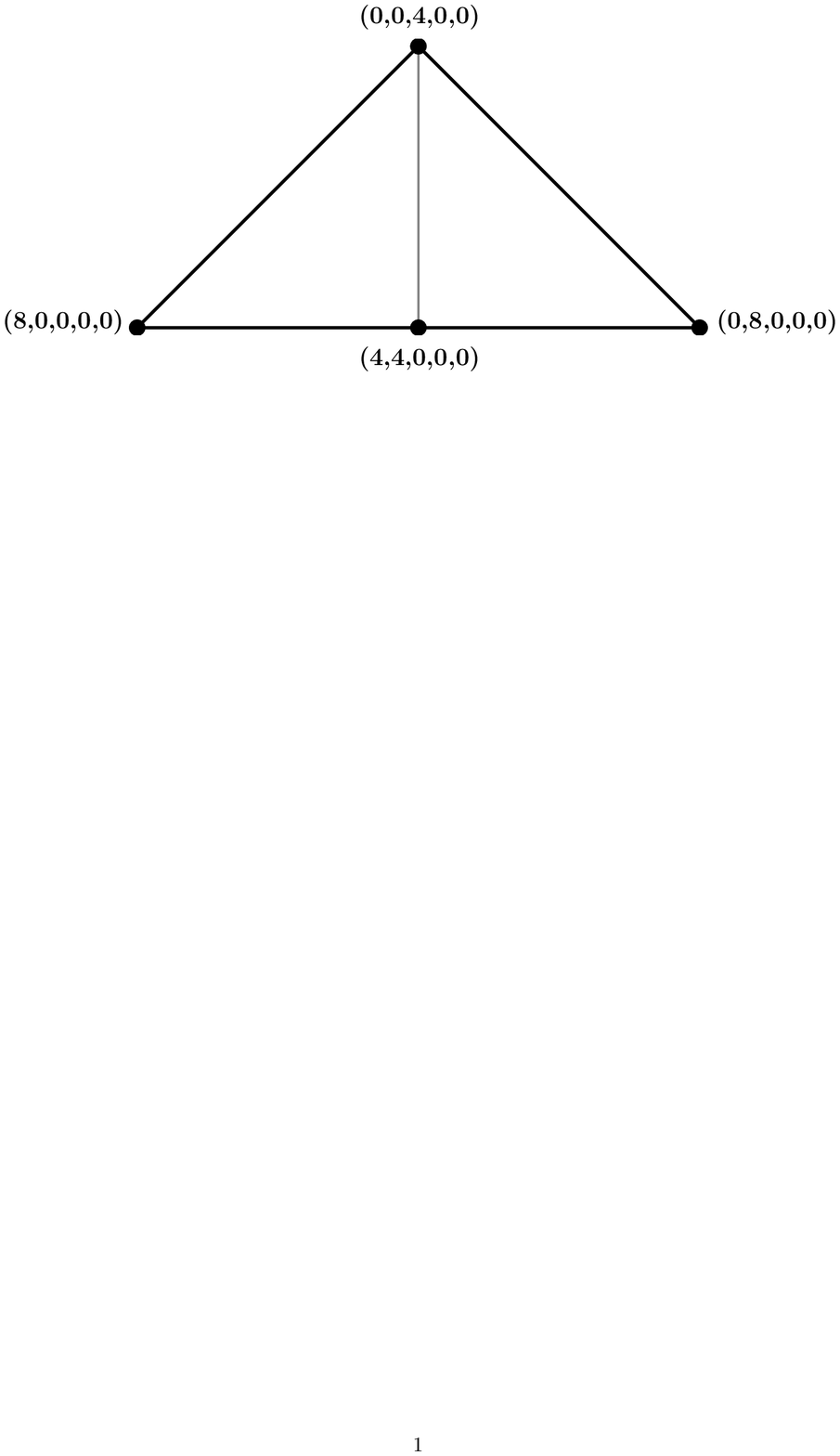}
\caption{ A two-face triangulated to resolve singularities }
\label{fig:ortwoface3}
\end{figure}

We shall consider Calabi--Yau threefolds which are
obtained by resolving singularities of degree eight (octic)
hypersurfaces in the weighted projective space
$\mathbb{P}_4^{(1,1,2,2,2)}$. A typical defining polynomial for
such a hypersurface is:
\begin{equation}
P=c_1x_1^8+c_2x_2^8+c_3x_3^4+c_4x_4^4+c_5x_5^4+\alpha
x_1^6x_3+\beta x_2^4x_3^4+\gamma x_1^7x_2\;,
\end{equation}
though in general an octic may contain many more terms. There are
$105$ degree eight monomials, but using the freedom of homogenous
coordinate redefinitions, which provide $22$ parameters, we are
left with $83$ possible monomial deformations. The Newton
polyhedron for this family has $7$ points, some two-faces of which
are illustrated in Figures \ref{fig:ortwoface1} and
\ref{fig:ortwoface3}. The mirror family has a Newton polyhedron
(the dual) containing $105$ points, two-faces of which are
illustrated in Section \ref{mirroroctic} in Figures
\ref{figtwoface1}, \ref{figtwoface2} and \ref{figtwoface3}.

We can define a 2-dimensional sub-family of the above
$83$-dimensional family, namely degree eight hypersurfaces with
two parameters of deformations defined by the equation:
\begin{equation}
P(x)=x_1^8+x_2^8+x_3^4+x_4^4+x_5^4-2\phi x_1^4x_2^4-8\psi
x_1x_2x_3x_4x_5\;.
\end{equation}

We shall be comparing the zeta functions of this $2$-dimensional
sub-family of octics, with that of the mirror family of the full
$83$-dimensional family. These are a special family because it
only contain monomials which are invariant under the group $G$ of
automorphisms by which on quotients when using the Greene--Plesser
construction.

The singularities occur along $x_1=x_2=0$, where there is a curve
$C$ of singularities of type $A_1$.  In our particular example,
the curve $C$ is described by as it only containing monomials
which are invariant under the automorphism group by
\begin{equation}
x_1=x_2=0,\quad x_3^4 + x_4^4 + x_5^4=0\;.
\end{equation}
In general it will just be a smooth quartic plane curve, which
always has genus $3$. To resolve these singularities we blow up
the locus $x_1=x_2=0$; replacing the curve of singularities by a
(exceptional divisor) $E$, which is a ruled surface over the curve
$C$. That is, each point of $C$ is blown up to a $\mathbb{P}^1$.
We point out the existence of this exceptional divisor because
later it contributes the the number of rational points in a way
not dependent on the parameters~$(\phi, \psi)$.

However, as we wish to exploit the Batyrev mathod for finding the
mirror, it is best to represent the manifolds torically:
\begin{definition}
The family of octic Calabi--Yau threefolds can be defined over the
toric variety $\frac{\mathbb{C}^6-F}{(\mathbb{C}^\ast)^2}$
\textup{(}where the r\^{o}le of $F$ is to restrict the coordinates
that are allowed by toric geometry to vanish
simultaneously\textup{)}, as a hypersurface defined by the
following polynomial:
\begin{equation}
P(\textbf{x},x_6)=x_1^8x_6^4+x_2^8x_6^4+x_3^4+x_4^4+x_5^4-8\psi
x_1x_2x_3x_4x_5x_6-2\phi x_1^4x_2^4x_6^4\;,
\end{equation}
where the $x_i$ are Cox variables associated to the monomials in
the Newton polyhedron as follows (we have shifted back the
monomials by $(1,1,1,1,1)$ for ease of manipulation):
\begin{eqnarray}
x_1&=&x_{(8,0,0,0,0,1,4)},\nonumber\\
x_2&=&x_{(0,8,0,0,0,1,4)},\nonumber\\
x_3&=&x_{(0,0,4,0,0,1,0)},\nonumber\\
x_4&=&x_{(0,0,0,4,0,1,0)},\nonumber\\
x_5&=&x_{(0,0,0,0,4,1,0)},\nonumber\\
x_6&=&x_{(4,4,0,0,0,1,4)}.\nonumber\\
\end{eqnarray}
\end{definition}
The two-faces of the triangulated Newton polyhedron which consists
of $7$ points (the dual polyhedron corresponding to the mirror
contains $105$ points) are shown below. The triangulation imposes
restrictions on the allowable simultaneous vanishing of the
coordinates.

\pagebreak
It should be noted that for this model the Hodge diamond for values of $(\phi,\psi)$ for which the variety is smooth is (which can be calculated from (\ref{Hodge})):\\

\begin{center}
\begin{tabular}{ccccccc}
 & & & 1 &    & &\\
  &&0&   &0 &   &\\
 &0& &$2$& &0&\\
1~~& &$86$& & $86$& & ~1\\
 &0& &$2$& &0&\\
   &&0&   &0 &   &\\
 & & & 1 &    & &\\
\end{tabular}\\
\end{center}
Hence we expect a zeta function of the form:

\begin{equation}
Z(X/\mathbb{F}_p,t)=\frac{R_{174}(t)}{(1-t)R_{2}(t)\acute{R}_{2}(t)(1-p^3t)}\;,
\end{equation}\label{Weil174}\\

\noindent where $R_{174}(t)$ is of degree $174$;
$R_{2}(t)$ and $\acute{R}_{2}(t)$ are both of degree $2$. We
shall see in Chapter \ref{chapexact} that $R_{2}(t)=(1-pt)^2$
and $\acute{R}_{2}(t)=(1-p^2t)^2$.

It should be noted that using Batyrev's formula for the Hodge
numbers, (\ref{Hodge}), for $n=4$ and leaving out the last terms,
we obtain $h^{2,1}_{\mathrm{poly}}=83$ (for the original family).
(Using all of (\ref{Hodge}) we obtain $h^{2,1}=86$). Thinking
about the octics as hypersurfaces in weighted projective space,
this corresponds to the fact that we can only incorporate $83$
complex structure deformations as monomial perturbations of the
homogeneous polynomial:

\begin{equation}
P_0(x)=x_1^8+x_2^8+x_3^4+x_4^4+x_5^4\;.
\end{equation}

Indeed, if one writes down the most general homogenous polynomial
of degree eight, it of course has $105$ monomials. Using the
freedom of homogenous coordinate redefinitions, which provide $22$
parameters, we are left with $83$ possible monomial deformations.

 For the mirror family,
$H^{1,1}_{\mathrm{toric}}(Y)=83$, whereas $H^{1,1}(Y)=86$, indeed
we shall see that in Chapter \ref{chapexact} in the calculation of
the number of rational points for the mirror, there is a splitting
in the denominator of the zeta function.

\subsection[K3 Fibration]{K3 Fibration}\label{k3fibration}
The octic family of Calabi--Yau threefolds is K3-fibred. A K3 can
be given by a hypersurface in $\mathbb{P}_{3}^{\textbf{k}}$, if
$\textbf{k}=(1,k_2,k_3,k_4)$ and $1+k_2+k_3+k_4=4$. A K3 fibration
can thus be formed in $\mathbb{P}_4^{(1,1,2k_2,2k_3,2k_4)}$: the
$\mathbb{P}_1$ base can be given by the ratio of the two
coordinates with weight one, and the K3 are quartic hypersurfaces
in $\mathbb{P}_{3}^{(1,k_2,k_3,k_4)}$, i.e. letting $x_2=\lambda
x_1$, $y_1=x_1^2$ and $y_{i}=x_{i+1}$ for $i=2,\ldots,4$, we get
the pencil of K3 surfaces (for fixed $\psi$ and $\phi$):

\begin{equation}
(1+\lambda^8)y_1^4+y_2^4+y_3^4+y_4^4-2\phi \lambda^4y_1^4-8\psi
\lambda y_1y_2y_3y_4\;,
\end{equation} hence, in our case $\textbf{k}=(1,1,1,1)$.

In order to see how the fibration structure affects the
arithmetic, we shall also study the one-parameter family of K3
surfaces defined by a quartic in $\mathbb{P}_3\,[4]$:

\begin{equation}
P(x)=x_1^4+x_2^4+x_3^4+x_4^4-4\psi x_1x_2x_3x_4\;.
\end{equation}

\subsection[Singularities]{Singularities}\label{singularity}
This two-parameter family of varieties was extensively studied in
\cite{CdFKM}; some essential features are summarized here.

For the purposes of examining the singularities it is easier to
think of the family as hypersurfaces in weighted projective space,
once more. Using the Greene--Plesser construction \cite{GP} (the
very first construction for finding mirror manifolds), the mirror
of $\mathbb{P}_4^{(1,1,2,2,2)}[8]$ can be found by the process of
`orbifolding', i.e. it may be identified with the family of
Calabi--Yau threefolds of the form $\{p=0\}/G$ where
\begin{eqnarray}
p(x)=x_1^8+x_2^8+x_3^4+x_4^4+x_5^4-2\phi x_1^4x_2^4-8\psi
x_1x_2x_3x_4x_5\;,
\end{eqnarray}
and where $G\cong \mathbb{Z}_4^3$, is the group with generators\\
\begin{eqnarray}
(\mathbb{Z}_4; 0,3,1,0,0),\nonumber\\
(\mathbb{Z}_4; 0,3,0,1,0),\nonumber\\
(\mathbb{Z}_4; 0,3,0,0,1).\nonumber\\
\end{eqnarray}
We have used the notation $(\mathbb{Z}_k; r_1,r_2,r_3,r_4,r_5)$
for a $\mathbb{Z}_k$ symmetry with the action:
\begin{equation}
(x_1,x_2,x_3,x_4,x_5)\rightarrow(\omega^{r_1}x_1,\omega^{r_2}x_2,\omega^{r_3}x_3,\omega^{r_4}x_4,\omega^{r_5}x_5),
\quad \mathrm{where}\;\omega^k=1.
\end{equation}
For a good description of the moduli space, it is useful to
enlarge $G$ to $\hat{G}$ consisting of elements
$g=(\alpha^{a_1},\alpha^{a_2},\alpha^{2a_3},\alpha^{2a_4},\alpha^{2a_5})$
acting as:
\begin{eqnarray}
(x_1,x_2,x_3,x_4,x_5)\mapsto
(\alpha^{a_1}x_1,\alpha^{a_2}x_2,\alpha^{2a_3}x_3,\alpha^{2a_4}x_4,\alpha^{2a_5}x_5;\alpha^{-a}\psi,\alpha^{-4a}\phi)\;,
\end{eqnarray}
where $a=a_1+a_2+2a_3+2a_4+2a_5$, and where $\alpha^{a_1}$ and
$\alpha^{a_2}$ are 8th roots of unity and
$\alpha^{2a_3},\alpha^{2a_4},\alpha^{2a_5}$ are $4$th roots of
unity.

If we quotient the family of weighted projective hypersurfaces
$\{p=0\}$ by the full group $\hat{G}$ we must quotient the
parameter space $(\phi,\psi)$ by a $\mathbb{Z}_8$ with a generator
$g_0$ acting as follows:

\begin{equation}
g_0: (\psi,\phi)\rightarrow (\alpha\psi,-\phi).
\end{equation}

The quotiented parameter space has a singularity at the origin and
can be described by three functions:

\begin{eqnarray}
\tilde{\xi}:=\psi^8,\quad\tilde{\eta}:=\psi^4\phi,\quad\tilde{\zeta}:=\phi^2,
\end{eqnarray}

\noindent subject to the relation:

\begin{eqnarray}
\tilde{\xi}\tilde{\zeta}=   \tilde{\eta}^2\;.
\end{eqnarray}

\begin{figure}[htbp]
\centering
\includegraphics[scale=1]{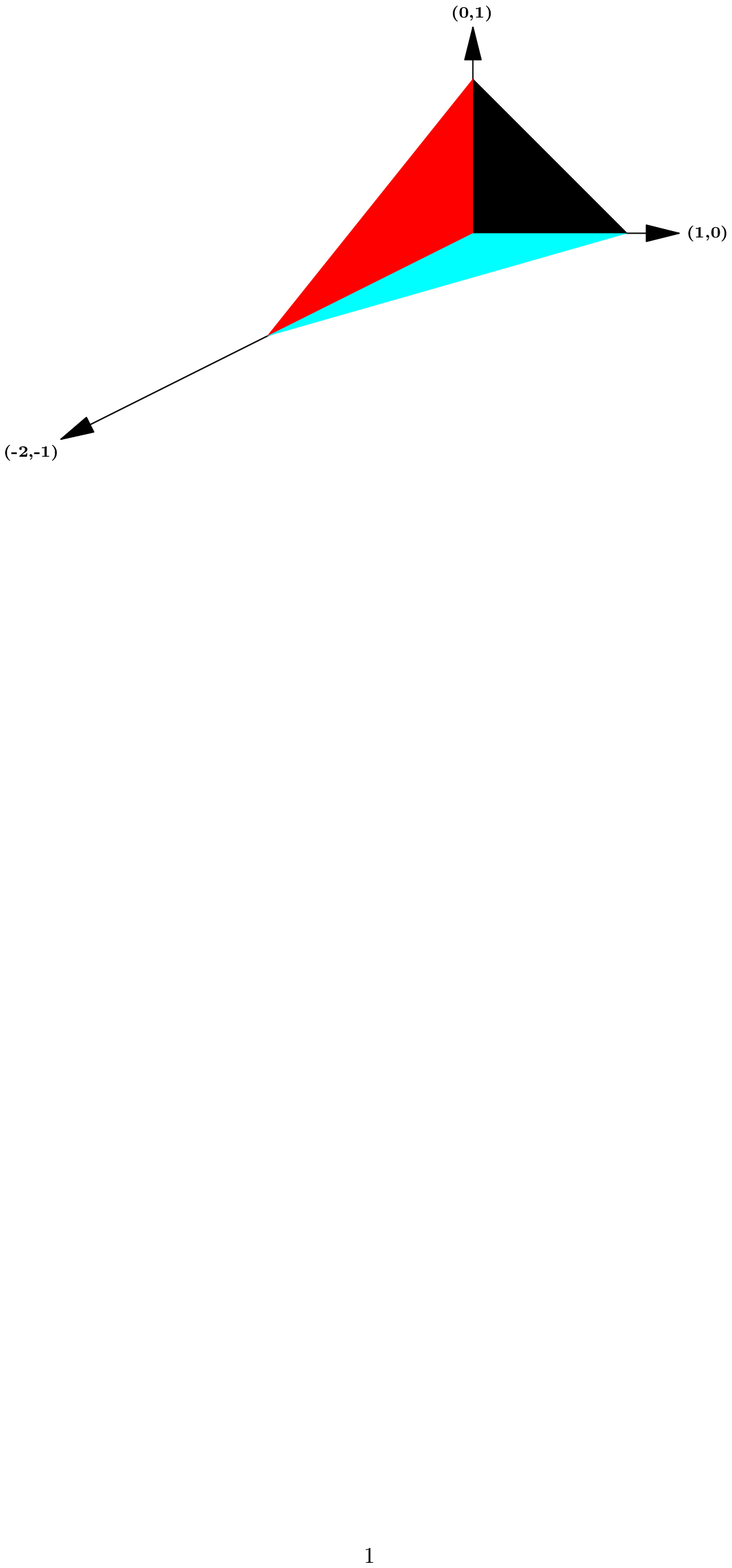} \caption{Toric Diagram for the Singular Quadric}
\label{toricsing}
\end{figure}

The above equations describe an affine quadric cone in
$\mathbb{C}^3$; this can be compactified to the projective quadric
cone in $\mathbb{P}_3$. This is isomorphic to the weighted
projective space $\mathbb{P}^{(1,1,2)}$, the toric diagram for
which is the union of the comes whose with edges that are spanned
by the vectors $(1,0$, $(0,1)$ and $(-2, -1)$, as illustrated in
Figure \ref{toricsing}. We embed $\mathbb{C}^3$ in $\mathbb{P}_3$
by sending the point of $\mathbb{C}^3$ with the coordinates
$(\tilde{\xi},\tilde{\eta},\tilde{\zeta})$ to the point in
$\mathbb{P}_3$ with homogeneous coordinates $[\xi,\eta,\zeta,1]$.

Note that the square of the generator $g_0^2$ acts trivially on
$\phi$, and so fixes the entire line $\psi =0$. This means that
the quotiented family $\{p=0\}/\hat{G}$ will have new
singularities along the $\psi=0$ locus (which becomes the
$\xi=\eta=0$ locus in the quotient).

We now locate the parameters for which the original family of
hypersurfaces is singular, and study the behaviour of these
singularities under quotienting. We obtain the following:

\begin{enumerate}
\item Along the locus $(8\psi^4+\phi)^2=\pm 1$, the original family of three-fold acquires a collection of conifold
points. For the mirror, these are identified under the G-action,
giving only one conifold point per three-fold on the quotient.

\item Along the locus $\phi^2=1$, the three-fold acquires $4$
isolated singularities, leading to only one singular point on the
quotient.

\item If we let $(\phi,\psi)$ approach infinity, we obtain a singular three-fold
\begin{equation}
2\phi x_1^4x_2^4+8\psi x_1 x_2x_3x_4x_5\;.
\end{equation}

\item $\psi=0$ - this locus requires special treatment , as it leads
to additional singularities on the quotient by $\hat{G}$.

\end{enumerate}

\begin{figure}[htbp]
\centering
\includegraphics[scale=1]{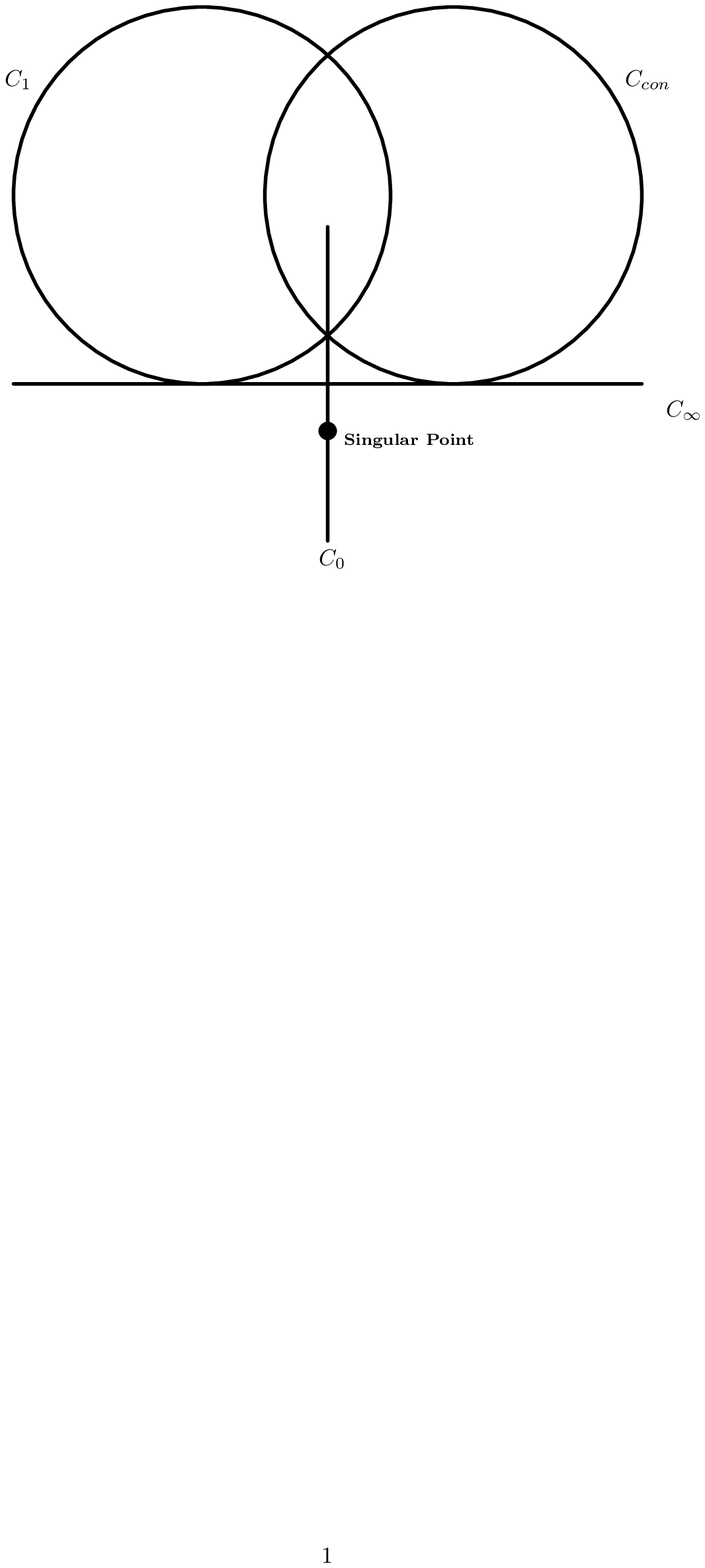} \caption{The
Compactified Moduli Space} \label{compmod}
\end{figure}

We now pass to the quotiented parameter space. We use homogeneous
coordinates$[\xi,\eta,\zeta,\tau]$ on $\mathbb{P}_3$. The
compactified quotiented parameter space can be described as the
singular quadric $Q:=\{\xi\zeta-\eta^2=0\}\subset\mathbb{P}_3$.
The singular loci can be described as follows:

\begin{enumerate}
\item $C_{\mathrm{con}}=Q\cap\{64\xi+16\eta+\zeta-\tau=0\}$,
\item $C_1=Q\cap\{\zeta-\tau=0\}$,
\item $C_{\infty}=Q\cap\{\tau =0\}$,
\item $C_0=\{\xi=\eta=0\}\subset Q$.
\end{enumerate}
The points of intersection of the above curves are (see Figure
\ref{compmod}):

\begin{itemize}
\item $[1,-8,64,0]$, the point of tangency between $C_{\mathrm{con}}$ and $C_{\infty}$,
\item $[1,0,0,0]$, the point of tangency between $C_{1}$ and $C_{\infty}$,
\item $[0,0,1,0]$, the point of tangency between $C_{0}$ and $C_{\infty}$,
\item $[0,0,1,1]$, the common point of intersection of $C_{0}$ and $C_{1}$, and $C_{\mathrm{con}}$ ,
\item $[1,-4,16,16]$, the intersection point of $C_{\mathrm{con}}$ and $C_{1}$, through which $C_0$ does not pass.
\end{itemize}


\subsubsection[The Locus $\phi^2=1$]{The Locus $\phi^2=1$}
 When we restrict $\mathbb{P}_4{}^{\,(1, 1, 2, 2,
2)}\,[8]$ to the locus $\phi^2=1$, the resulting family of
singular manifolds is birationally equivalent to the mirror family
of $\mathbb{P}_5[2,4]$. The equations defining the complete
intersection can have the form:

\begin{eqnarray*}
y_0^2+y_1^2+y_2^2+y_3^2&=&\eta y_4y_5\;,\\
y_4^4+y_5^4&=&y_0y_1y_2y_3\;.\\
\end{eqnarray*}

\noindent where $\eta$ is the one-parameter. This complete
intersection needs to be quotiented by the group $H$ of coordinate
rescalings which preserve both hypersurfaces as well as the
holomorphic three-form. $H$ can be described as follows:

\begin{equation}
\{(\alpha^{4a},\alpha^{4b},\alpha^{4c},\alpha^{4a},\alpha^e,\alpha^f)|e+f\equiv0
\mod 8, \; 4a+4b+4c+4d\equiv4e \mod 8\}\;.
\end{equation}

Now define a rational map, $\Phi:\mathbb{P}_4{}^{\,(1, 1, 2, 2,
2)}\,[8]/G\rightarrow\mathbb{P}_5/H$, as:

\begin{eqnarray}
y_0=x_1^4-x_2^4,\quad y_1=x_3^2,\quad y_2=x_4^2,\\
y_3=x_5^2,\quad y_4=x_1\sqrt{x_3x_4x_5},\quad y_5=\alpha
x_2y_4/x_1,
\end{eqnarray}
which is compatible with the actions of $G$ and $H$. The image
$X\subset\mathbb{P}_5/H$ of $\Phi$ is defined by
$y_4^4+y_5^4=y_0y_1y_2y_3$ which is $H$-invariant.

It can be checked that the rational map
$\Upsilon:X\rightarrow\Phi:\mathbb{P}_4{}^{\,(1, 1, 2, 2,
2)}\,[8]/G$ given by:

\begin{eqnarray}
x_1=y_4/\sqrt{x_3x_4x_5},\quad
x_2=\alpha^{-1}y_5/\sqrt{x_3x_4x_5},\\
x_3=\sqrt{y_1},\quad x_4=\sqrt{y_2},\quad x_5=\sqrt{y_3},
\end{eqnarray}
is the rational inverse of $\Phi$.

The Hodge diamond of the mirror of this complete intersection is
as follows \cite{CK}:

\begin{center}
\begin{tabular}{ccccccc}
 & & & 1 &    & &\\
  &&0&   &0 &   &\\
 &0& &$1$& &0&\\
1& &$89$& & $89$& & 1\\
 &0& &$1$& &0&\\
   &&0&   &0 &   &\\
 & & & 1 &    & &\\
\end{tabular}\\\label{hodge24}
\end{center}

When there is both a $\phi^2=1$ singularity and a conifold point
the (mirror) three-fold upon resolution becomes rigid, i.e.
$h^{2,1}=0$. The modularity of the zeta functions at these points
was established by Rodriguez-Villegas in \cite{Villegas}.

The Hodge numbers above can be understood by the means of an
extremal transition. This class of extremal transitions generalise
the notion of a conifold transition, for details see \cite{KMP},
where a genus $g$ curve of $A_{N-1}$ singularities was considered,
where $N\geq2$. A Calabi--Yau hypersurface will have this type of
singularity, precisely when $\Delta^{\ast}$ has an edge joining
$2$ vertices $v_0,v_N$ of $\Delta^{\ast}$ with $N-1$ equally
spaced lattice points $v_1, \ldots, v_{N-1}$ in the interior of
the edge. As explained previously, the Calabi--Yau three-fold,
$M$, is a hypersurface in the toric variety whose fan is a
suitable refinement of the fan consisting of the fan over the
faces of $\Delta^{\ast}$. In particular, there are edges
corresponding to the lattice points $v_1, \ldots, v_{N-1}$. These
correspond to toric divisors, which resolves a surface, $S$ of
$A_{N-1}$ singularities in $V$. Restricting to the hypersurface,
$M$, we see that there are $N-1$ divisors in $M$ which resolve a
curve $C$ of $A_{N-1}$ singularities. The genus, $g$, of the curve
$C$ can be found by finding the dual to the edge $D_1^{\ast}=<v_0,
v_{N}>$ which determines a $2$-dimensional face, $\Delta_2$, of
$\Delta$. The number of integral interior points of the triangle
$\Delta_2$ is equal the genus~$g$.

In our case $g=3,\,N=2$ and the relevant edge is that shown in
Figure \ref{fig:ortwoface3}.

Overall, in smoothing the singularities, the change in Hodge
numbers is given by the following formula \cite{KMP}:

\begin{eqnarray}
h^{1,1}\rightarrow h^{1,1}-(N-1)\;, \quad h^{2,1} \rightarrow
h^{2,1}+(2g-2)\begin{pmatrix}N\\2\end{pmatrix}(N-1)\;.\label{hodgechange}
\end{eqnarray}

The change above is due to the fact that the transition `kills'
$(N-1)$ independent homology cycles. Homotopically, the transition
is obtained by replacing $(2g-2){N\choose 2}$ two-spheres by
$(2g-2){N\choose 2}$ three-spheres. This results in a change in
the Euler characteristic of $-2(2g-2){N\choose 2}$, hence the
change in the $3$D-homology has to be $2(2g-2){N\choose
2}-2(N-1)$, which is equally shared between $H^{2,1}$ and
$H^{1,2}$, giving us (\ref{hodgechange}).

\chapter[Periods and Picard--Fuchs Equations]{Periods and Picard--Fuchs
Equations}\label{chappicardfuchs} The number of rational points
was shown to be related to the periods of the variety in
\cite{COV1}. In our calculation for the octic threefolds we show
that the zeta function factorizes into pieces that can be
associated with certain set of periods. In this chapter we give a
review of a method of computing the Picard--Fuchs equations, the
Dwork--Katz--Griffiths Method, and then exhibit the Picard--Fuchs
equations that we computed for the octic threefolds in Section
\ref{picfuchdiagrams}.

\section[Periods and Picard--Fuchs Equations]{Periods and Picard--Fuchs
Equations}\label{picfuchs}
 The dimension of the third
cohomology group for a Calabi--Yau is \hbox{$\dim
H^3=b_3=2(h^{2,1}+1)$}. Furthermore, the unique holomorphic three
form $\Omega$ depends only on the complex structure. If we take
derivatives with respect to the complex structure moduli, we get
elements in $H^{3,0}\bigoplus H^{2,1}\bigoplus H^{1,2} \bigoplus
H^{0,3}$. Since $b_3$ is finite, there must be linear relations
between derivatives of $\Omega$ of the form
$\mathcal{L}\Omega=d\eta$ where $\mathcal{L}$ is a differential
operator with moduli dependent coefficients. If we integrate this
equation over an element of the third homology group $H_3$, we
will get a differential equation $\mathcal{L}\Pi_{i}=0$ satisfied
by the periods of $\Omega$. They are defined as
\begin{equation}
 \Pi_i(a)=\int_{\Gamma_i}\Omega(a), \quad \Gamma_i\in H_3(X,\mathbb{Z}).
\end{equation}
where $a$ represents the complex structure moduli. These are the Picard--Fuchs Equations, and in general we obtain a set of coupled linear partial differential equations (in the case of one-parameter models (i.e. $b_3=4$) we just get a single ordinary differential equation of order~4).\\
\section[Dwork, Katz and Griffiths' Method]{Dwork, Katz and Griffiths' Method}
One procedure for determining the Picard--Fuchs equations for
hypersurfaces in $\mathbb{P}[\vec{w}]$, is the Dwork, Katz and
Griffiths' reduction method \cite{CK}. For hypersurfaces in
weighted projective space $\mathbb{P}_4^{(w_1,w_2,w_3,w_4,w_5)}$
the periods $\Pi_i(a)$ of the holomorphic three form $\Omega(a)$
can be written as:

\begin{equation}
 \Pi_i(a)=\int_{\Gamma_i}\Omega(a)=\int_{\gamma}\int_{\Gamma_i}\frac{\omega}{P(a)}, \quad \Gamma_i\in H_3(X,\mathbb{Z}),\quad \mathrm{i.e.}\; i= 1,\ldots,2(h^{2,1}+1)\;.
\end{equation}
Here
\begin{equation}
\omega=\sum_{i=1}^{5}(-1)^{i}w_{i}z_{i}dz_1\wedge\ldots\wedge \hat{dz_i}\wedge\ldots\wedge dz_5;
\end{equation}

\noindent and $\gamma$ is a small curve around the hypersurface
$P=0$ in the $4$-dimensional embedding space. The numbers, $a_i$,
are the coefficients of the perturbations of the quasi-homogeneous
polynomial~$P$.

Observe that $\frac{\partial}{\partial
z_i}\left(\frac{f(z)}{P^r}\right)\omega$ is exact if $f(z)$ is
homogeneous with degree such that the whole expression has degree
zero. This leads to the partial integration rule
$\left(\partial_i=\frac{\partial}{\partial z_i}\right)$:
\begin{equation}
\frac{f\partial_i P}{P^r}=\frac{1}{r-1}\,\frac{\partial_i
f}{P^{r-1}}. \label{intrule}
\end{equation}
In practice one chooses a basis $\{\varphi_k(z)\}$ for the
elements of the local ring $\mathcal{R}$. For hypersurfaces
\begin{equation}
\mathcal{R}=\mathbb{C}[z_1,\ldots,z_{n+1}]/(\partial_iP)\;.
\end{equation}

From the Poincar\'{e} polynomial associated to $P$ one sees that
there are $(1,\tilde{h}^{2,1},\tilde{h}^{2,1},1)$ basis elements
with degrees $(0,d,2d,3d)$ respectively. The elements of degree
$d$ are the perturbing monomials.

\noindent One then takes derivatives of the expressions:
\begin{equation}
\Pi_j^k=\int_{\Gamma_j}\frac{\varphi_k(z)}{P^{n+1}}
\quad(n=\text{deg}(\varphi_k)/d) \end{equation} with respect to
the moduli. If one produces an expression such that the numerator
in the integrand is not one of the basis elements, one relates it,
using the equations $\partial_iP=\ldots$, to the basis and uses
(\ref{intrule}). This leads to a system of first order
differential equations (known as `Gauss--Manin equations') for the
$\Pi_j^k$ which can be rewritten as a system of partial
differential equations for the period (the Picard--Fuchs
equations): \hbox{$\partial_{a_k}\Pi=M^{(k)}(a)\Pi,\;
k=1,\ldots,\tilde{h^{2,1}}$} . The Picard--Fuchs equations reflect
the structure of the local ring and expresses relations between
its elements (modulo the ideal).

Note that the method above only gives us the Picard--Fuchs
equations for which there exists a monomial perturbation. Also the
above method only works for manifolds embedded in projective
spaces. In \cite{COV1} it was shown that the number of rational
points of the quintic Calabi--Yau manifolds over $\mathbb{F}_p$
can be given in terms of the periods; our calculations verify this
relation for the octic threefolds, as we have $15$ classes of
monomials each providing its own Picard--Fuchs equations. The
following table lists the monomials (the meaning of the
$\lambda_{\textbf{v}}$ will be explained in Chapter
\ref{chapexact}):

\noindent
\begin{center}
\begin{tabular}{|c|c|c|}\hline
Monomial $\textbf{v}$ & degree & Permutations $\lambda_{\textbf{v}}$ \\
\hline\hline
(0,0,0,0,0) & 0 & 1 \\
(0,2,1,1,1) & 8 & 2  \\
(6,2,0,0,0) & 8 & 1  \\ \hline
(0,0,0,2,2) & 8 & 3  \\
(3,1,2,0,0) & 8 & 6 \\
(4,0,2,0,0) & 8 & 3 \\ \hline
(0,0,2,1,1) & 8 & 3  \\
(6,0,1,0,0) & 8 & 6  \\
(5,1,1,0,0) & 8 & 3  \\ \hline
(4,0,1,1,0) & 8 & 3  \\
(2,0,3,0,0) & 8 & 6  \\
(1,1,0,0,3) & 8 & 3 \\\hline
(0,0,3,1,0) & 8 & 6  \\
(2,0,2,1,0) & 8 & 12  \\
(7,3,2,1,0) & 16 & 6  \\ \hline
\end{tabular}
\end{center}

Later on we shall see that the zeta function for the octic
threefolds can be decomposed into pieces corresponding to certain
monomials classes which turn out to be identical to the above
monomial classes. This establishes the correspondence between
periods and the number of rational points.

Mirror symmetric pairs of threefolds $(X,Y)$, often have a certain
number of periods in common; these periods are associated to the
monomials that both reflexive polyhedra have in common.  For our
two-parameter model, $6$ periods are `shared'. The shared periods
are those associated to the monomial $(0,0,0,0,0)$ (the $G$-
invariant ones).

We now list the $15$ different Picard--Fuch Equations obtained.
Using the notation of \cite{COV1}, we associate a period to each
monomial:
\begin{equation}
\varpi_{\textbf{v}}(\phi,\psi)=\int_{\gamma_{\textbf{v}}}\Omega(\phi,\psi)=C\int_{\Gamma}\frac{x^{\textbf{v}}}{P^{w(\textbf{v})+1}}\;,
\end{equation}
where $C$ is a convenient normalization constant. In going from
the first equality to the second, we have used the residue formula
for the holomorphic $3$-form \cite{BCdOFHJG}:
\begin{equation}
\Omega:=\mathrm{Res}\left[\frac{\omega}{P}\right]
\end{equation}
In affine space $\mathbb{C}_{5}^{(w_1,w_2,w_3,w_4,w_5)}$, we may
take $\omega$ to be given by
$\omega=\prod_{i=0}^{5}\mathrm{d}x_i$. The contour $\Gamma$ is
chosen so as to reproduce the integral of the residue over the
cycle $\gamma_{\textbf{v}}$. Differentiating with respect to each
parameter we obtain (suppressing constants):
\begin{equation}
\frac{d}{d\psi}\varpi_{\textbf{v}}=\int_{\Gamma}\mathrm{d}^5x\;\frac{x^{\textbf{v+(1,1,1,1,1)}}}{P^{w(\textbf{v})+2}}8\psi(w(\textbf{v})+1)\;,
\label{differ1}
\end{equation}
\begin{equation}
\frac{d}{d\phi}\varpi_{\textbf{v}}=\int_{\Gamma}\mathrm{d}^5x\;\frac{x^{\textbf{v+(4,4,0,0,0)}}}{P^{w(\textbf{v})+2}}2\phi(w(\textbf{v})+1)\;.
\label{differ2}
\end{equation}

The Picard--Fuchs equation can be found by repeated use of the
following operations:

\noindent For $i=1,2$:
\begin{eqnarray}
\lefteqn{D_i\left(\frac{x^{\textbf{v}}}{P^{w(\textbf{v})+1}}\right)}\nonumber\\
&:=& \frac{\partial }{\partial x_i}
\left(x_i\frac{x^{\textbf{v}}}{P^{w(\textbf{v})+1}}\right)\nonumber\\
&=&
(1+v_i)\frac{x^{\textbf{v}}}{P^{w(\textbf{v})+1}}+\frac{\partial
P}{\partial
x_i}(w(\textbf{v})+1)\frac{x^{\textbf{v}}}{P^{w(\textbf{v})+2}}\nonumber\\
&=& (1+v_i)\frac{x^{\textbf{v}}}{P^{w(\textbf{v})+1}}
+(w(\textbf{v})+1)\frac{x^{\textbf{v}}}{P^{w(\textbf{v})+2}}8(x_i^8-\psi
x_1x_2x_3x_4x_5-\phi x_1^4x_2^4)\;;\label{diffid1}
\end{eqnarray}
for $i=3,4,5$:
\begin{eqnarray}
\lefteqn{D_i\left(\frac{x^{\textbf{v}}}{P^{w(\textbf{v})+1}}\right)}\nonumber\\
&:=& \frac{\partial }{\partial x_i}
\left(x_i\frac{x^{\textbf{v}}}{P^{w(\textbf{v})+1}}\right)\nonumber\\
&=&
(1+v_i)\frac{x^{\textbf{v}}}{P^{w(\textbf{v})+1}}+\frac{\partial
P}{\partial
x_i}(w(\textbf{v})+1)\frac{x^{\textbf{v}}}{P^{w(\textbf{v})+2}}\nonumber\\
&=&
(1+v_i)\frac{x^{\textbf{v}}}{P^{w(\textbf{v})+1}}+(w(\textbf{v})+1)\frac{x^{\textbf{v}}}{P^{w(\textbf{v})+2}}4(x_i^4-2\psi
x_1x_2x_3x_4x_5)\;.\label{diffid2}
\end{eqnarray}
Now
\begin{equation}
\int_{\Gamma}\mathrm{d}^5x\;D_i\left(\frac{x^{\textbf{v}}}{P^{w(\textbf{v})+1}}\right)=0\;,
\end{equation}
because
$D_i\left(\frac{x^{\textbf{v}}}{P^{w(\textbf{v})+1}}\right)$
corresponds to an exact differential form and hence
(\ref{diffid1}) and (\ref{diffid2}) establish identities between
the differential form associated to $x^{\textbf{v}}$ and the
differential forms associated to $x^{\textbf{v+(1,1,1,1,1)}}$,
$x^{\textbf{v+(4,4,0,0,0)}}$, and either
$x^{\textbf{v+(8,0,0,0,0)}}$ or $x^{\textbf{v+(0,0,4,0,0)}}$ (the
last two up to permutation). Using \ref{differ1} and \ref{differ2}
the Picard--Fuchs equation can be derived. These identities are
summarized in a straightforward way using arrows in a diagramatic
representation of each Picard--Fuchs equation.

\section[Picard--Fuchs Diagrams]{Picard--Fuchs
Diagrams}\label{picfuchdiagrams} In this section we present
diagramatic representations of the Picard--Fuchs Equations. The
notation used is as follows: the monomials are written in the
form:
\begin{equation}
\left(v_1,v_2|v_3,v_4,v_5|\frac{v_1+v_2+2(v_3+v_4+v_5)}{8},\frac{v_1+v_2}{2}\right),
\end{equation}
the last two digit denoting the degree and the values of the Cox
variable $x_{\textbf{(4,4,0,0,0)}}$, respectively. Unbroken arrows
labelled by $D_i$ relate periods associated to $x^{\textbf{v}}$ to
those associated to either (up to permutation of the added
monomial) $x^{\textbf{v+(8,0,0,0,0)}}$ or
$x^{\textbf{v+(0,0,4,0,0)}}$. Horizontal unbroken arrows denote
the correspondence between periods associated to $x^{\textbf{v}}$
and $x^{\textbf{v}+(1,1,1,1,1)}$. Curvy broken lines relate, in
the same way, $x^{\textbf{v}}$ with $x^{\textbf{v+(4,4,0,0,0)}}$.
Hence it is possible, although rather involved, to derive the
Picard--Fuchs equations from these diagrams by using
(\ref{diffid1}), and (\ref{diffid2}). The pairs of identical
monomials encased in boxes are to be identified, thus making the
diagrams `closed'. Hence these diagrams are pictorial
representations of the Picard--Fuch Equations.

The monomial $(7,3|0,2,1|2,5)$ is related to itself via the
Jacobian ideal, it is not in the ideal on the conifold locus,
$(8\psi^4+\phi)^2=1$. This is illustrated by its corresponding
equation, derived from the diagram in Section
\ref{conifoldperiod}:
\begin{eqnarray}
16[(8\psi^4+\phi)^2-1]\frac{x_1^7x_2^3x_4^2x_5}{P^3} & = &
8\psi^3[\partial_3(\frac{x_1^2x_2^6x_4}{P^2})+(8\psi^4+\phi)\partial_3(\frac{x_1^6x_2^2x_4}{P^2})]\nonumber\\
& & +4\psi^2[\partial_5(\frac{x_1^5x_2x_3^2}{P^2})+(8\psi^4+\phi)\partial_5(\frac{x_1x_2^5x_3^2}{P^2})]\nonumber\\
& &
+2\psi[\partial_4(\frac{x_2^4x_3x_5^2}{P^2})+(8\psi^4+\phi)\partial_4(\frac{x_1^4x_3x_5^2}{P^2})]\nonumber\\
& &
+[\partial_1(\frac{x_2^3x_4^2x_5}{P^2})+(8\psi^4+\phi)\partial_2(\frac{x_1^3x_4^2x_5}{P^2})]\;.\nonumber\\
\end{eqnarray}
Equivalently, the monomial $(7,3,0,2,1)$ corresponds to a
differential form which is exact except at the conifold locus,
when $(8\psi^4+\phi)^2=1$. Hence as long as
$(8\psi^4+\phi)^2\neq1$, the period associated to $(7,3,0,2,1)$ is
zero. It will be later seen that, generically, each class of
monomials contributes to the zeta function except $(7,3,0,2,1)$.
The monomial, $(7,3,0,2,1)$, only contributes when there is a
conifold singularity. This is similar to the case for the quintic
threefolds \cite{COV1}, where there was a `period' associated to
the monomial $(4,3,2,1,0)$ which was only non-zero on the conifold
locus.

\subsection[Picard--Fuchs Equation for $(0,0,0,0,0)$]{Picard--Fuchs Equation for $(0,0,0,0,0)$}
\includegraphics[scale=0.9]{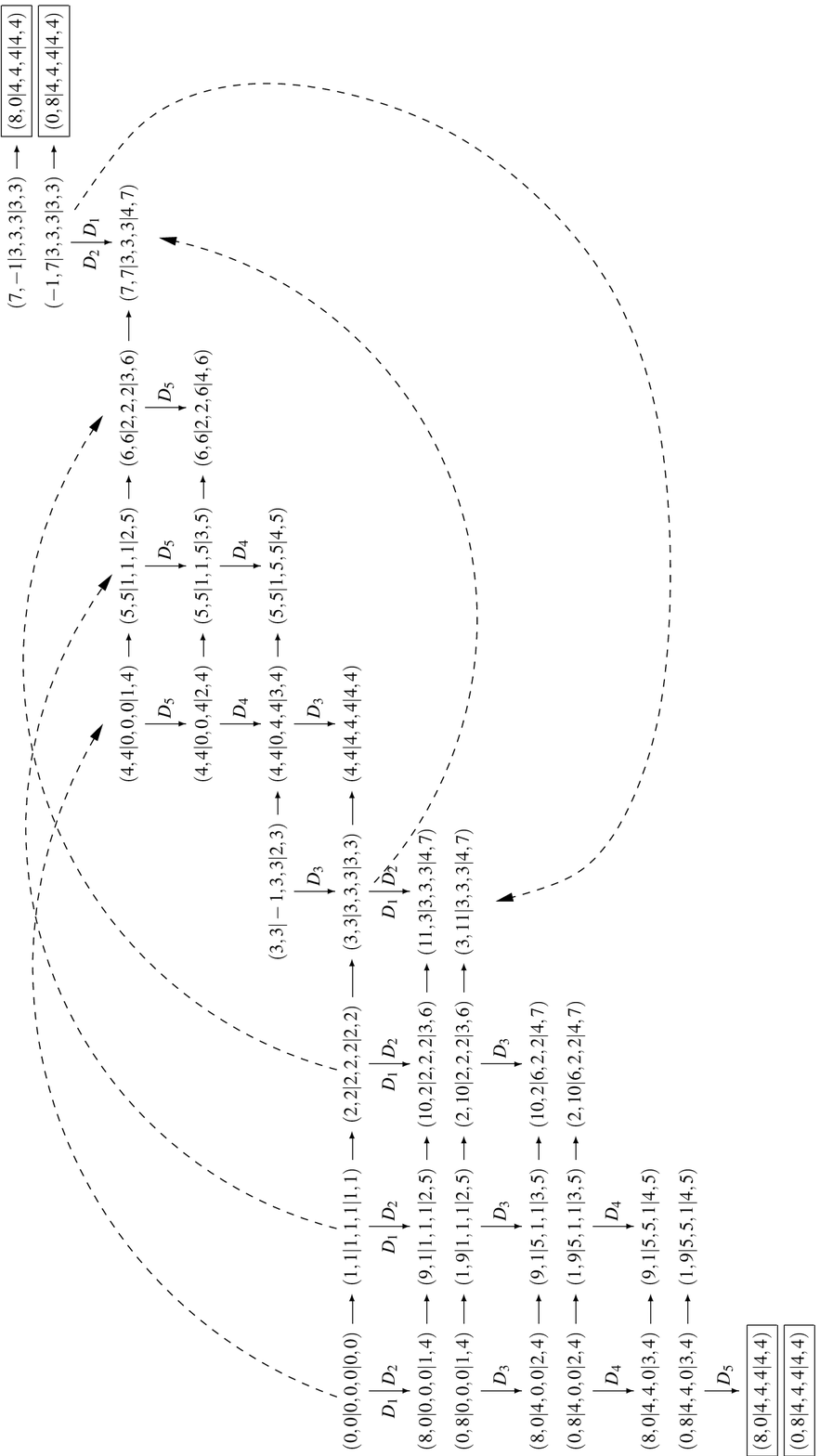}

\subsection[Picard--Fuchs Equation for $(0,2,1,1,1)$]{Picard--Fuchs Equation for $(0,2,1,1,1)$}
\includegraphics[scale=0.9]{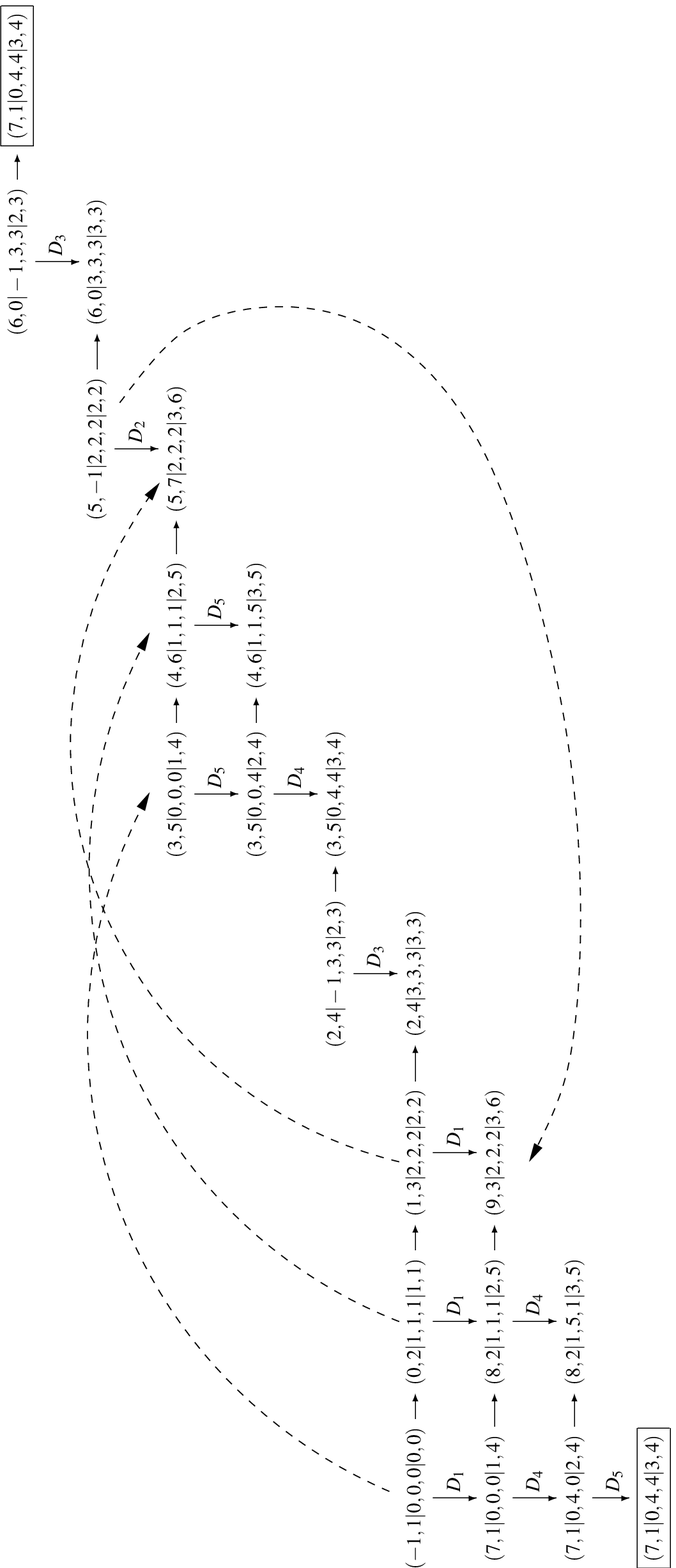}

\subsection[Picard--Fuchs Equation for $(6,2,0,0,0)$]{Picard--Fuchs Equation for $(6,2,0,0,0)$}
\includegraphics[scale=0.9]{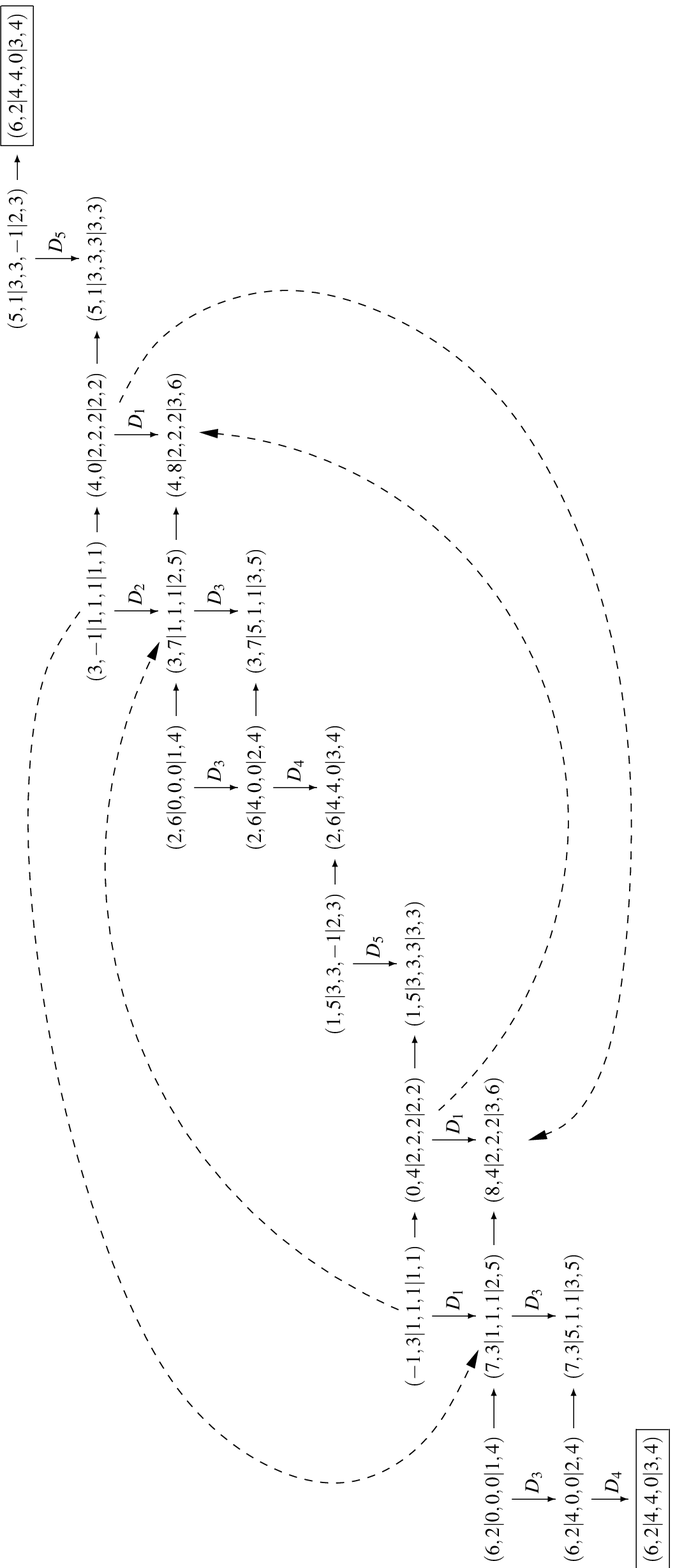}

\subsection[Picard--Fuchs Equation for $(0,0,0,2,2)$]{Picard--Fuchs Equation for $(0,0,0,2,2)$}
\includegraphics[scale=0.9]{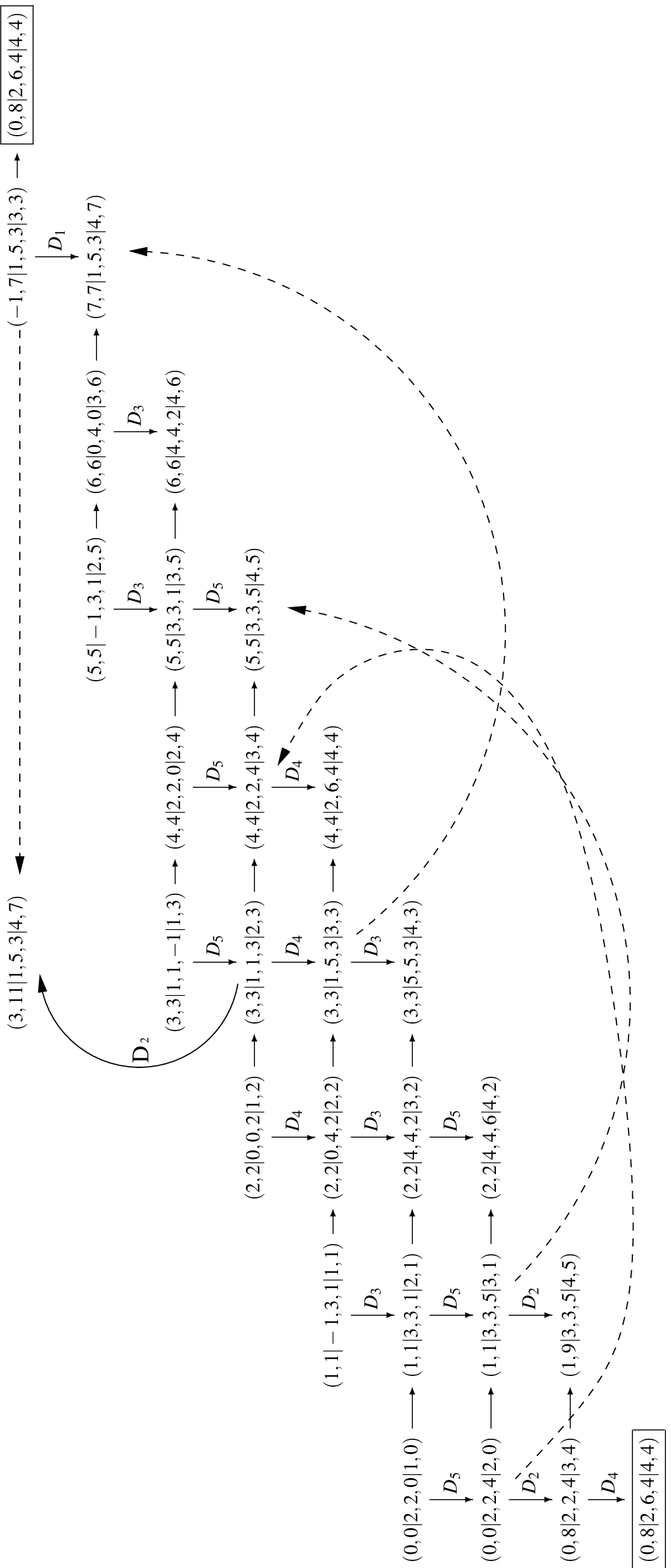}

\subsection[Picard--Fuchs Equation for $(2,0,1,3,3)$]{Picard--Fuchs Equation for $(2,0,1,3,3)$}
\includegraphics[scale=0.9]{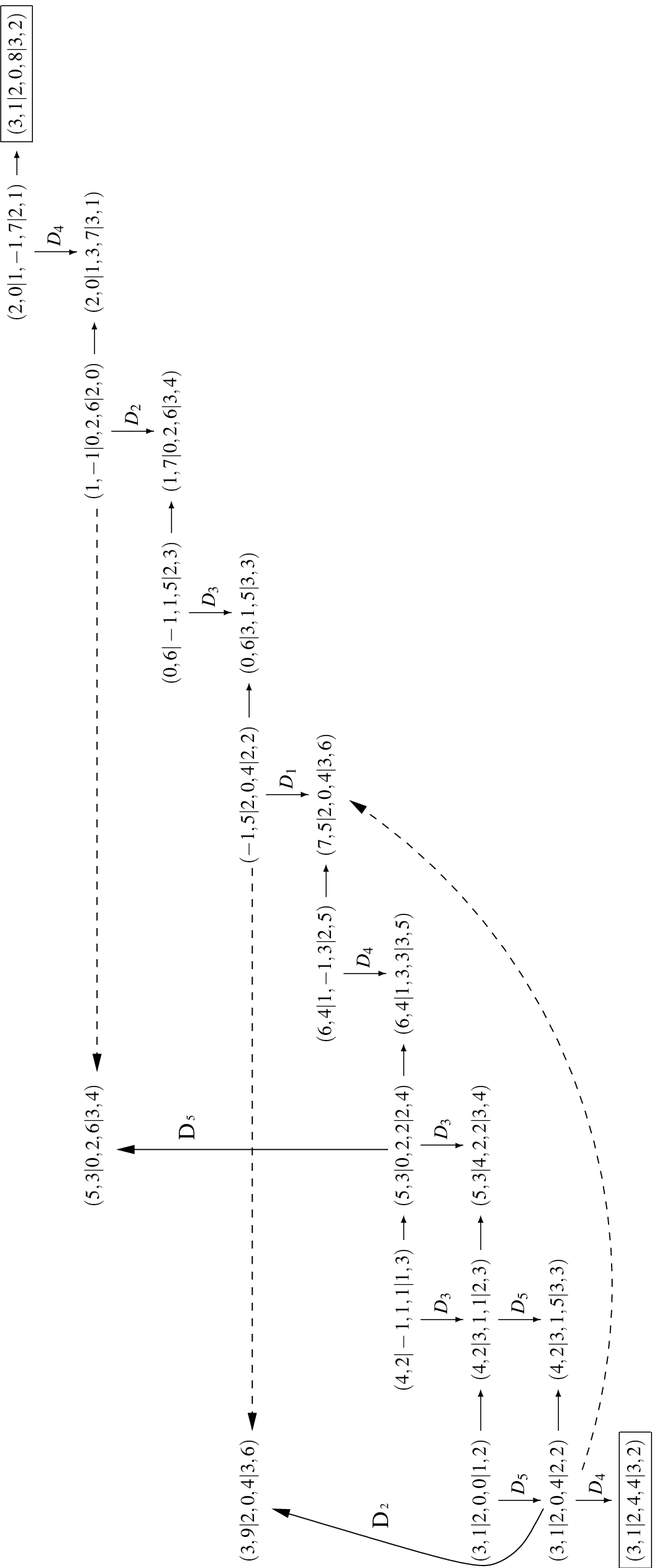}

\subsection[Picard--Fuchs Equation for $(4,0,2,0,0)$]{Picard--Fuchs Equation for $(4,0,2,0,0)$}
\includegraphics[scale=0.9]{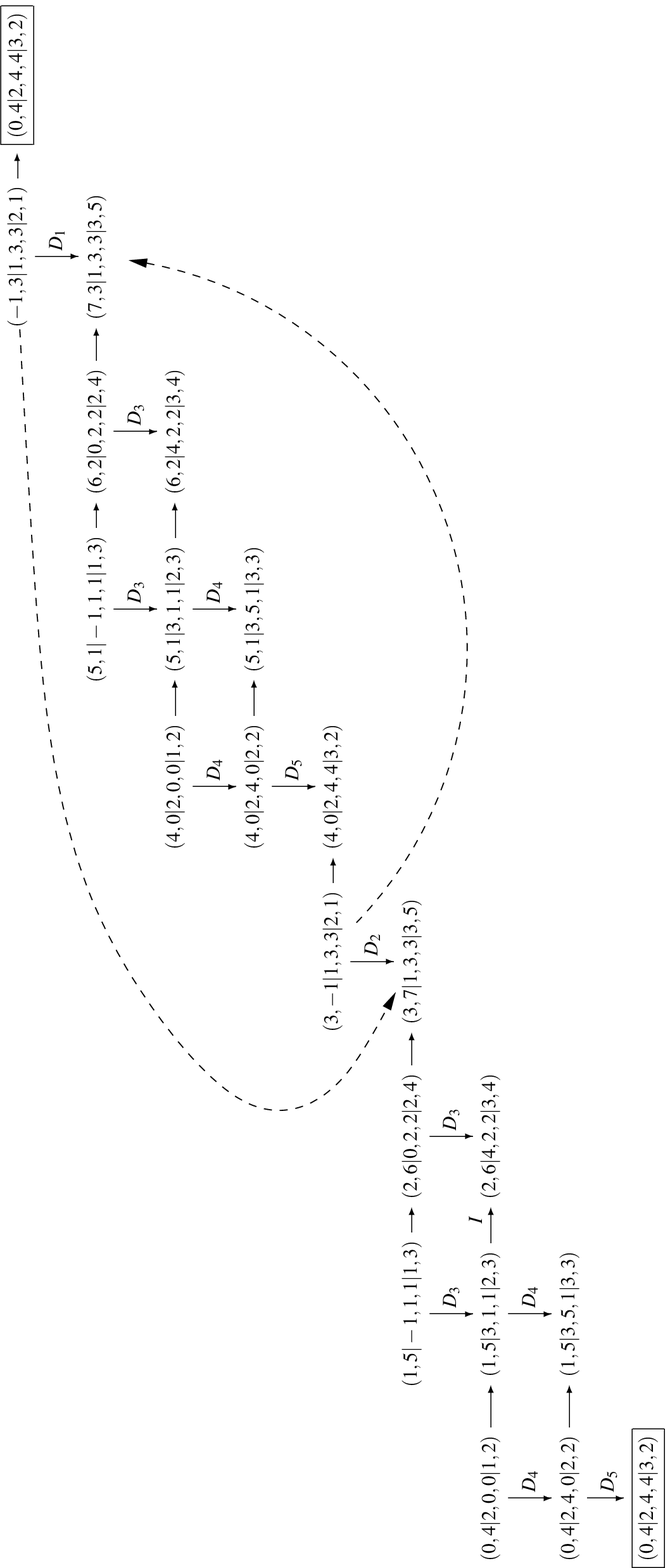}

\subsection[Picard--Fuchs Equation for $(0,0,2,1,1)$]{Picard--Fuchs Equation for $(0,0,2,1,1)$}
\includegraphics[scale=0.9]{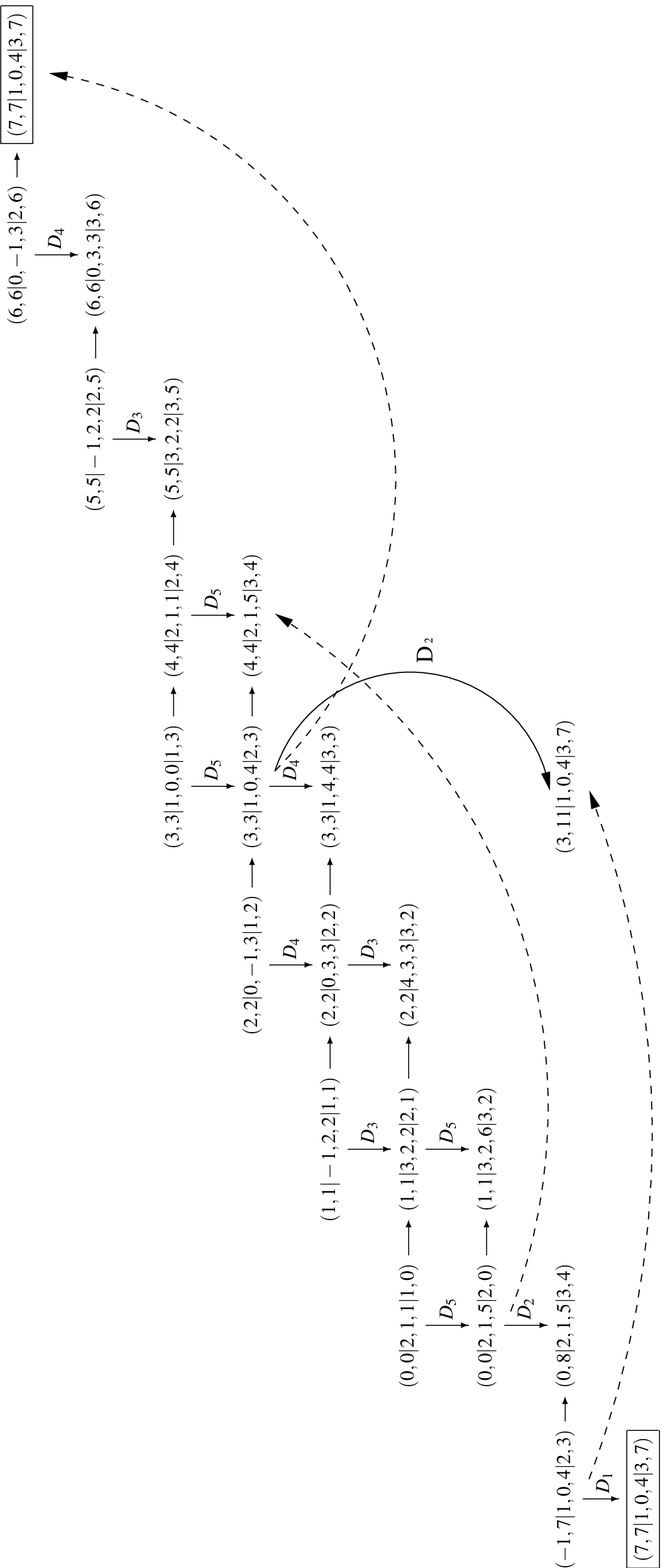}

\subsection[Picard--Fuchs Equation for $(6,0,1,0,0)$]{Picard--Fuchs Equation for $(6,0,1,0,0)$}
\includegraphics[scale=0.9]{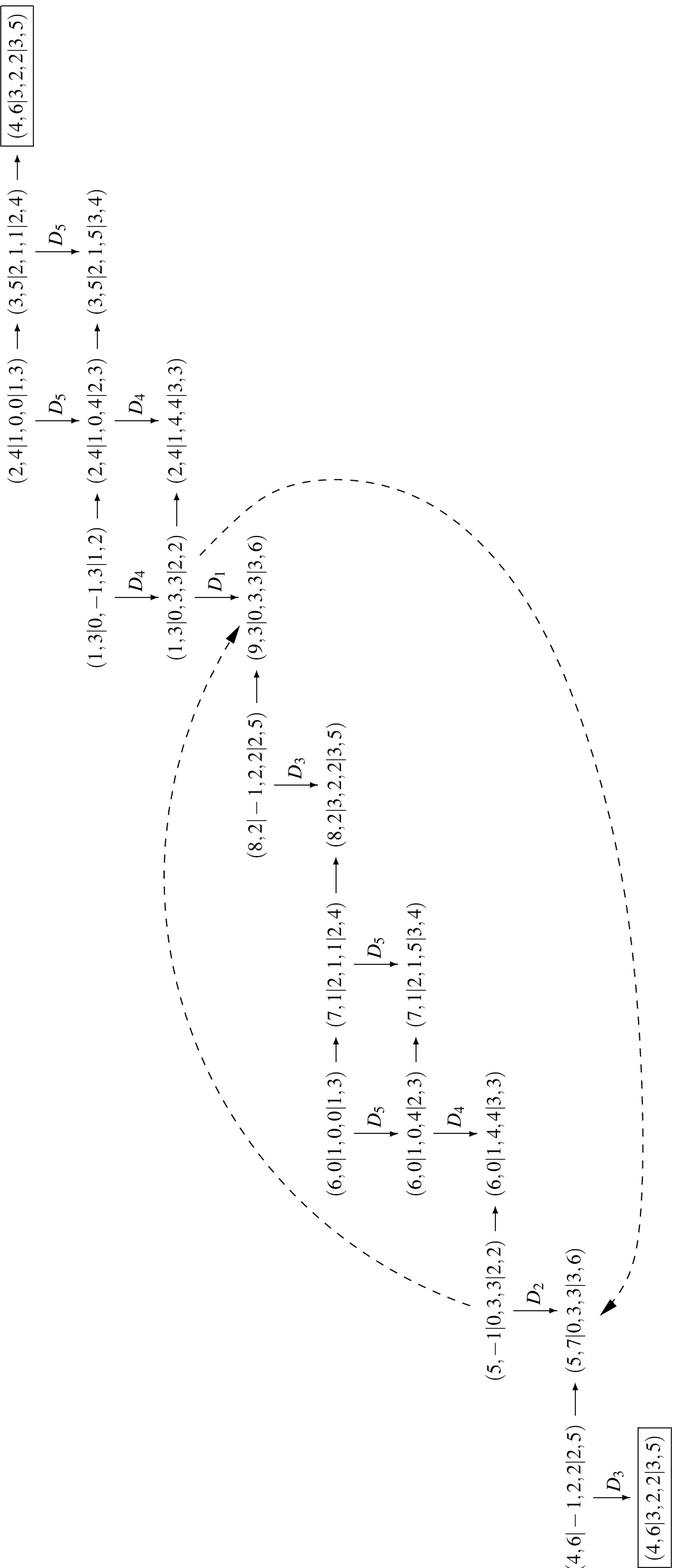}

\subsection[Picard--Fuchs Equation for $(0,4,0,3,3)$]{Picard--Fuchs Equation for $(0,4,0,3,3)$}
\includegraphics[scale=0.9]{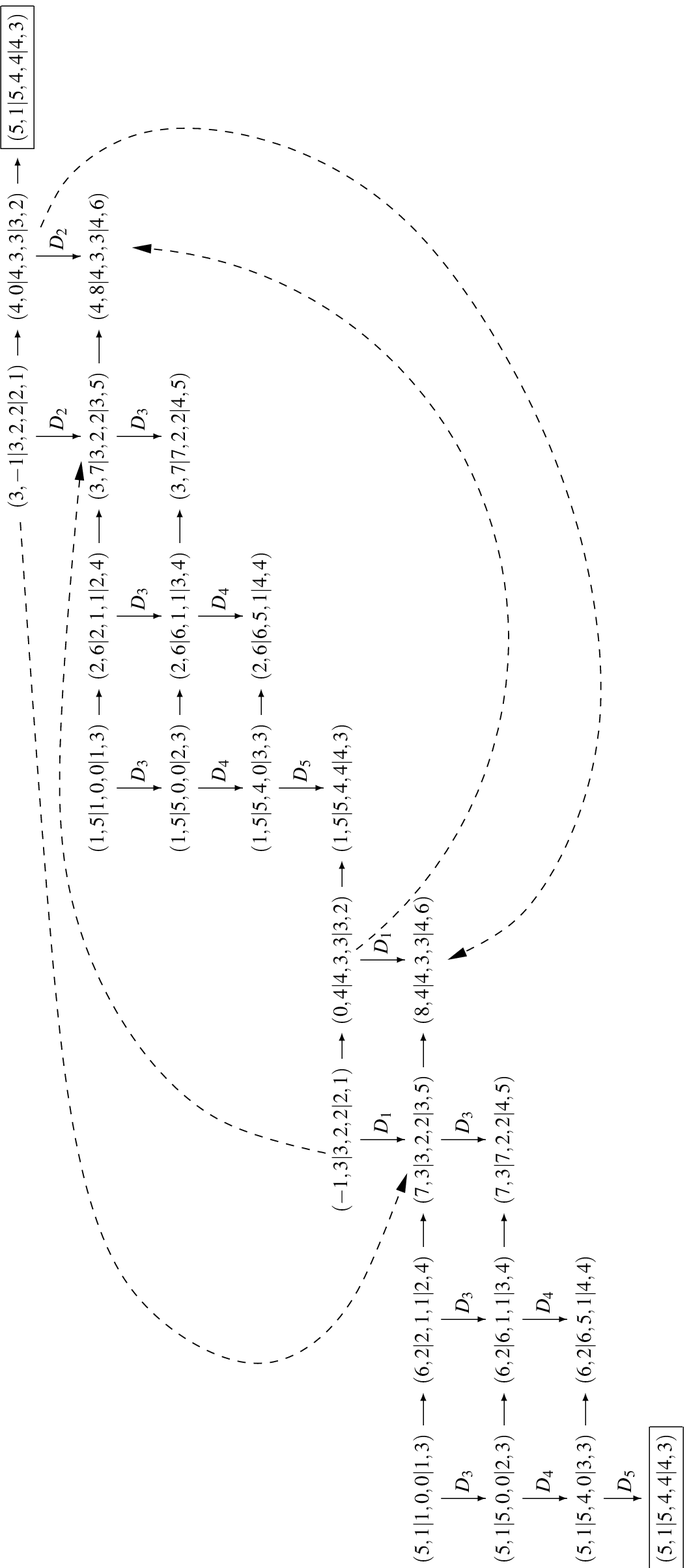}

\subsection[Picard--Fuchs Equation for $(4,0,1,1,0)$]{Picard--Fuchs Equation for $(4,0,1,1,0)$}
\includegraphics[scale=0.9]{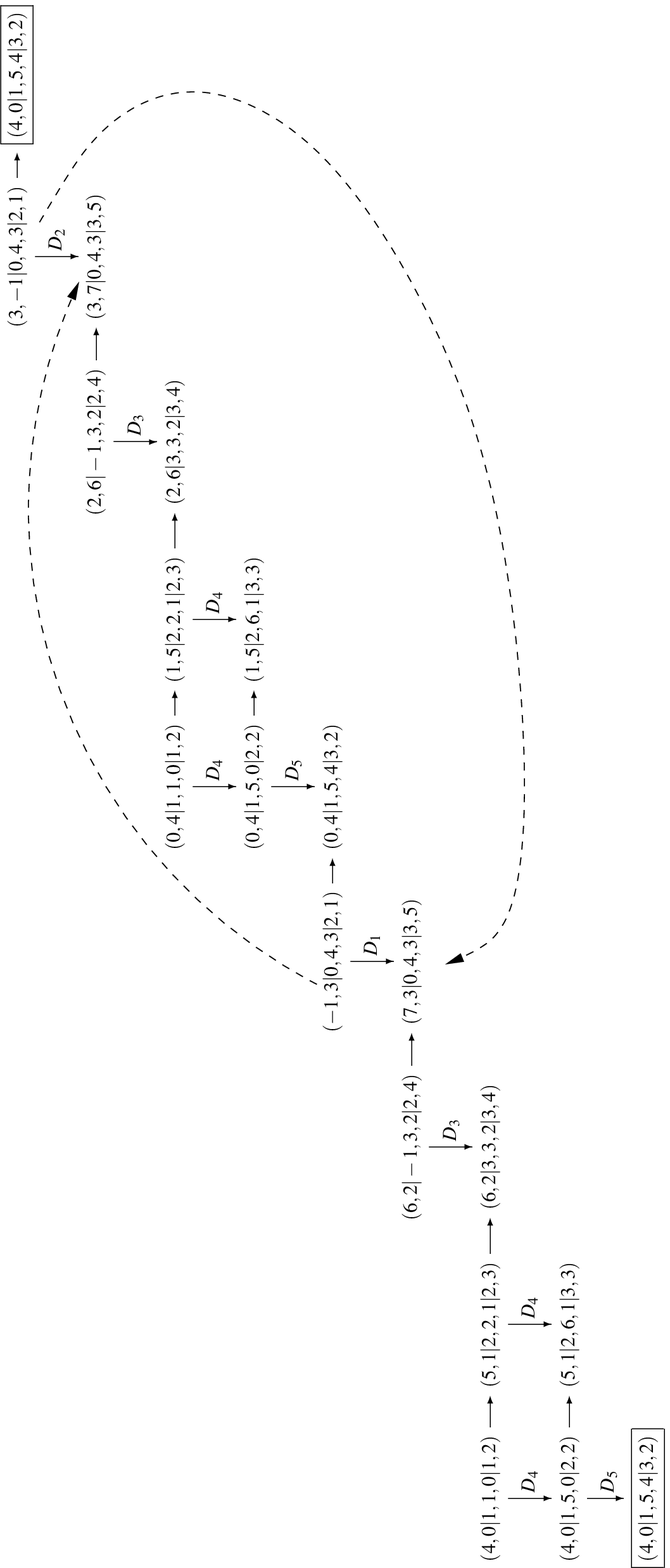}

\subsection[Picard--Fuchs Equation for $(2,0,3,0,0)$]{Picard--Fuchs Equation for $(2,0,3,0,0)$}
\includegraphics[scale=0.9]{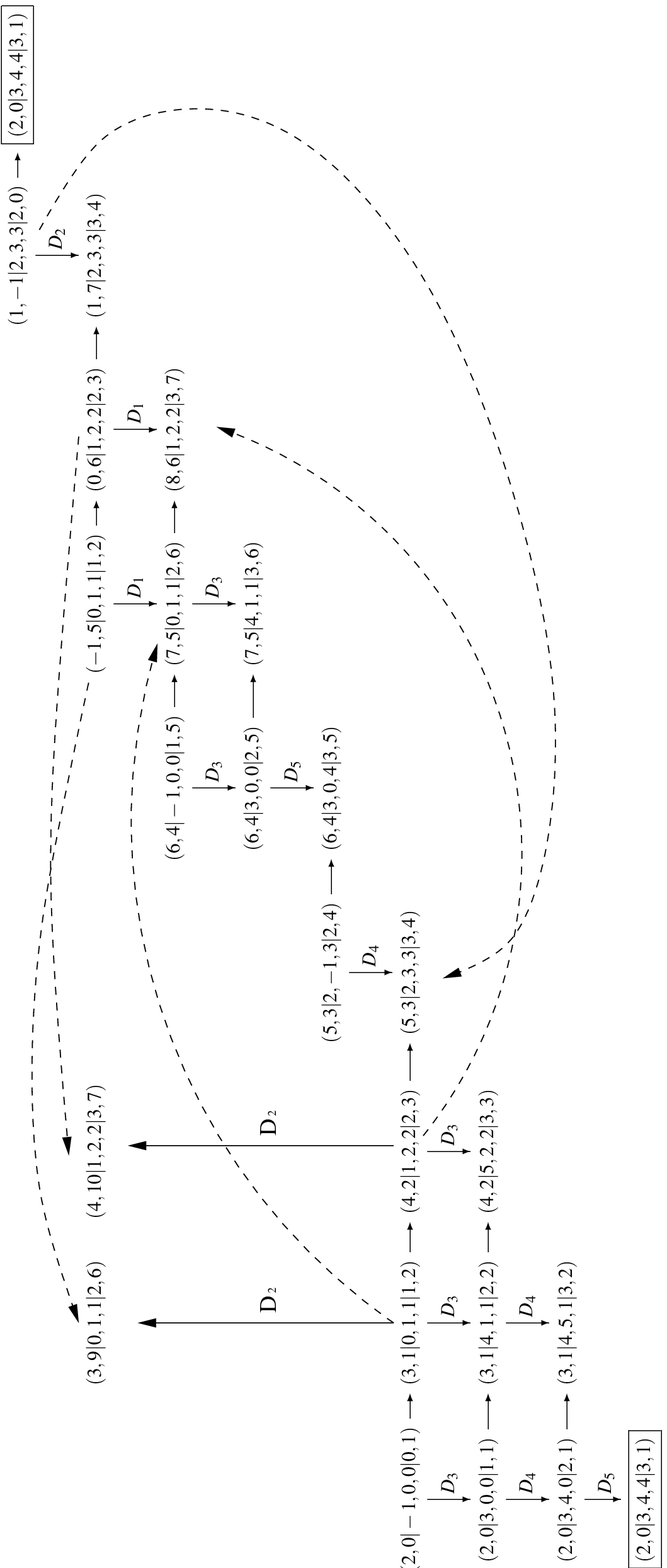}

\subsection[Picard--Fuchs Equation for $(2,2,1,1,0)$]{Picard--Fuchs Equation for $(2,2,1,1,0)$}
\includegraphics[scale=0.9]{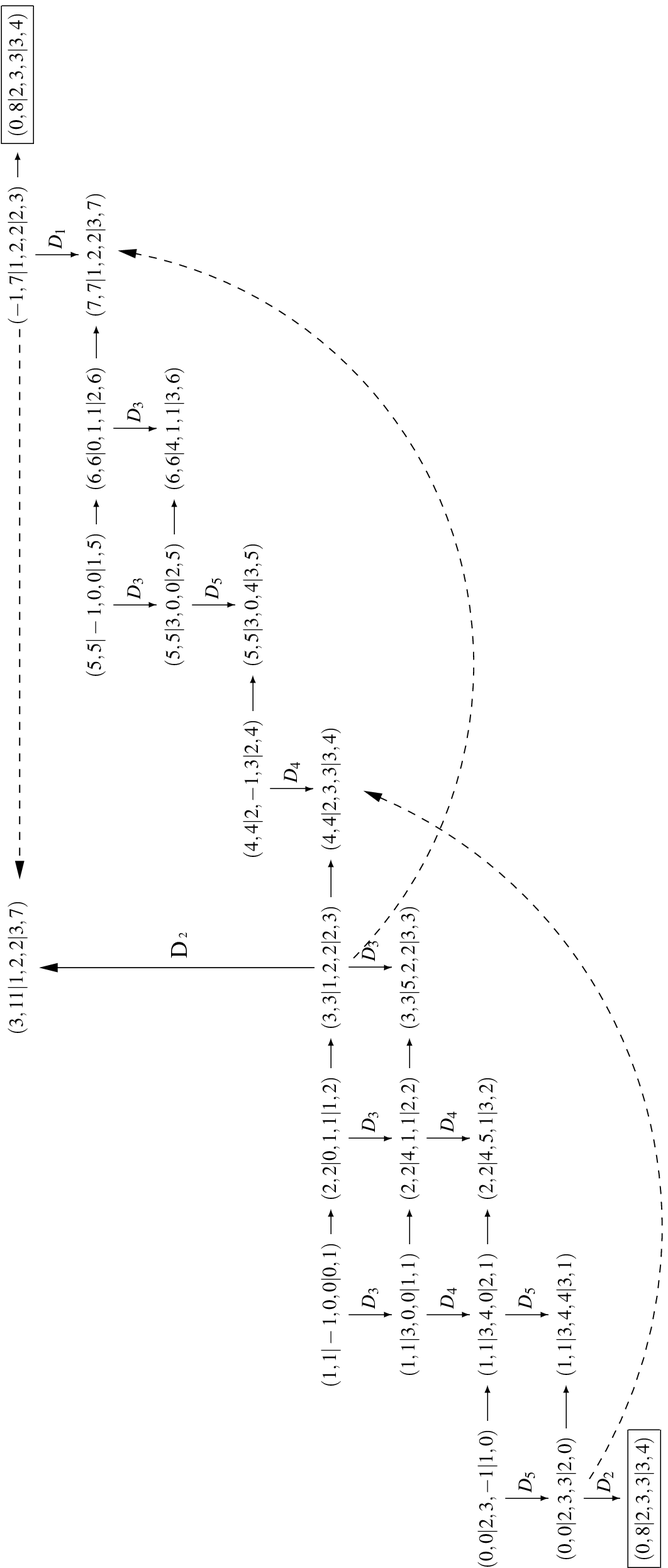}

\subsection[Picard--Fuchs Equation for $(0,0,3,1,0)$]{Picard--Fuchs Equation for $(0,0,3,1,0)$}
\includegraphics[scale=0.9]{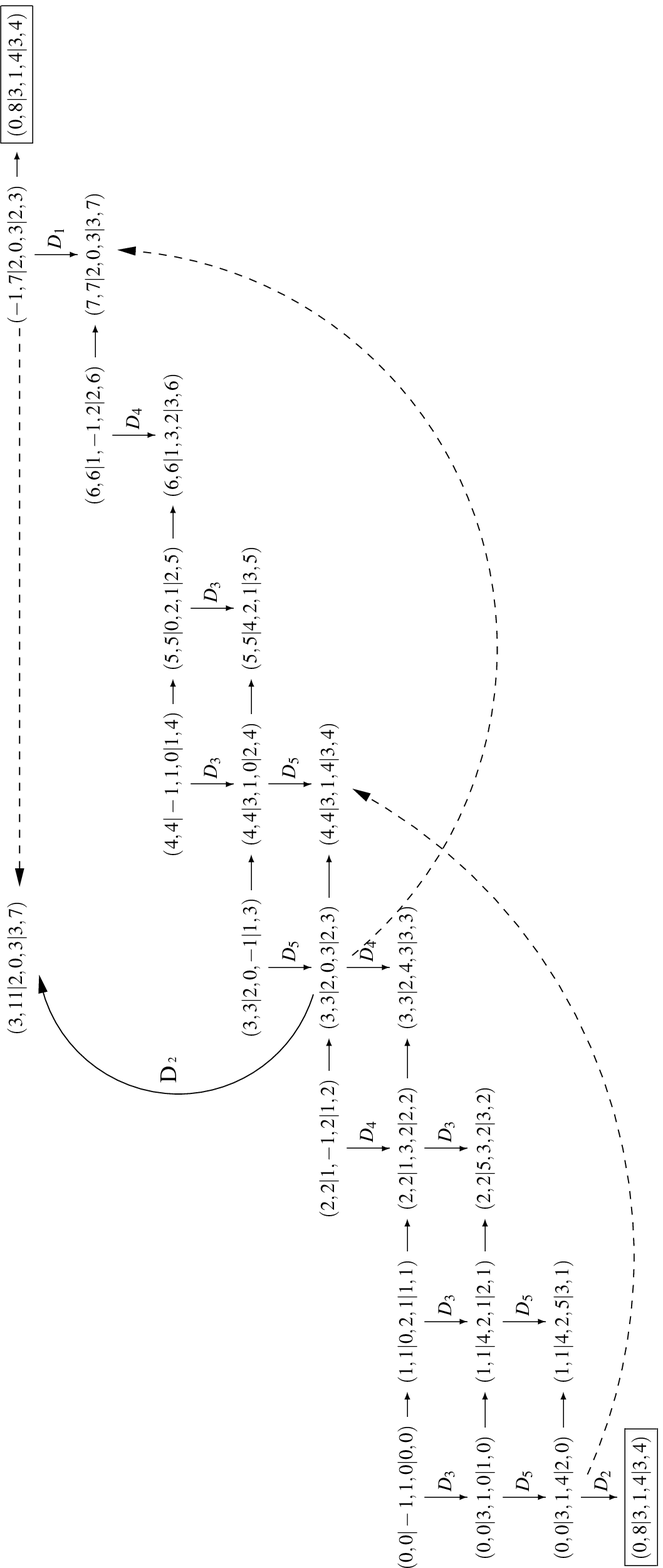}

\subsection[Picard--Fuchs Equation for $(2,0,2,1,0)$]{Picard--Fuchs Equation for $(2,0,2,1,0)$}
\includegraphics[scale=0.9]{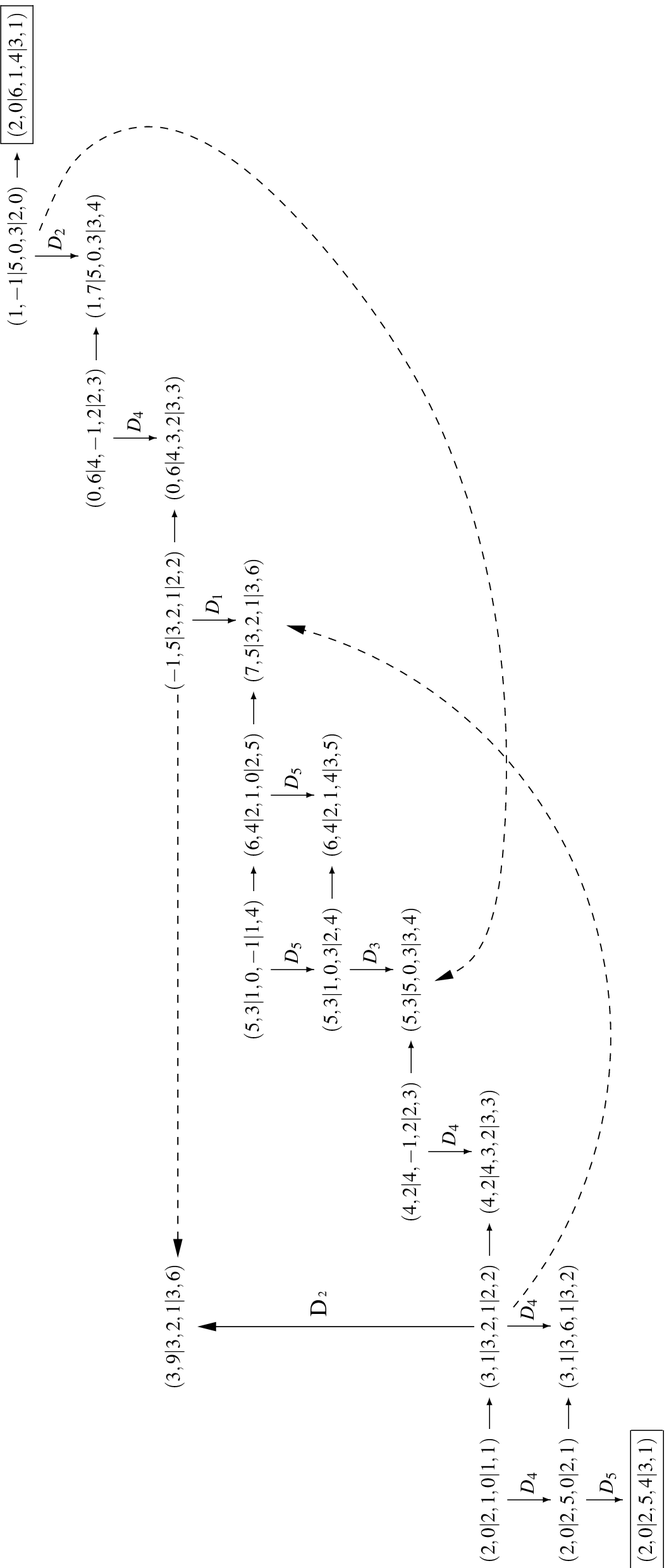}
\subsection[Picard--Fuchs Equation for $(4,0,2,3,1)$]{Picard--Fuchs Equation for
$(4,0,2,3,1)$}\label{conifoldperiod}
\includegraphics[scale=1]{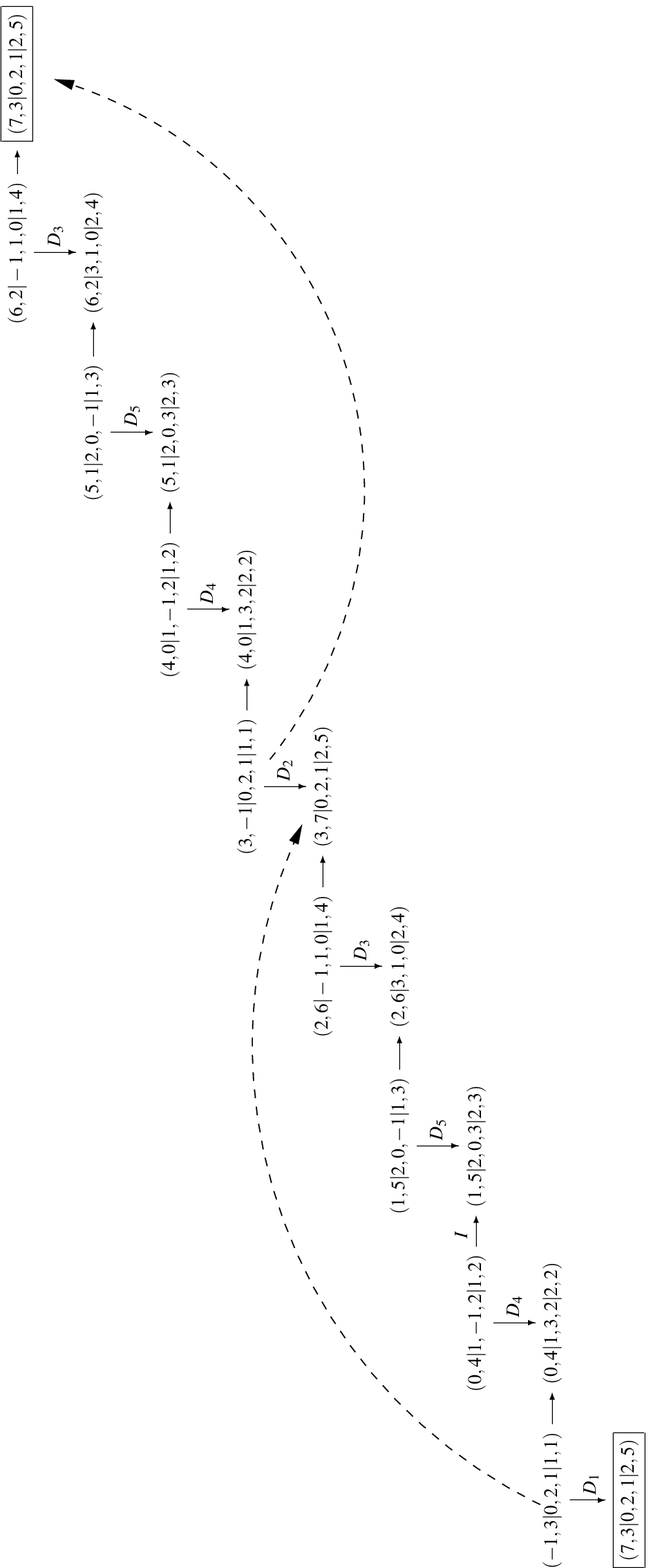}


\chapter[Exact Calculation of the Number of Rational Points]{Exact Calculation of the Number of Rational Points}
\label{chapexact}

In this chapter we describe in detail the derivation of our
results for the octic threefolds. In the first few sections, we
review some of the machinery (much of which was used by Dwork to
prove the rationality of the zeta function \cite{Dwork}) that we
use in the main computation, which begins in Section
\ref{numpoints}. The formulae we obtained for the number of
rational points are summarized in Section
\ref{sectionsummaryofresults}. Section \ref{sectionrearrangement}
briefly describes how to rearrange these expressions for easier
programming, as Mathematica code was written to compute the number
of rational points using these formulae. Section \ref{mirroroctic}
describes the computation of the number of points for the mirror
octic threefolds. Finally Section \ref{combin} gives a
combinatorial prescription on how to find the monomials that label
the sets of periods (or equivalently contributions to the number
of rational points and hence the zeta function) for Calabi--Yau
manifolds that are hypersurfaces in weighted projective space.

\section[Properties of $p$-adic Gamma Function]{Properties of Gamma
Functions}\label{sectiongammafunction} For $n \in \mathbb{Z}$, $n
\geqslant 0$ the $p$-adic gamma function is defined by
\begin{equation}
\Gamma_p(n+1) = (-1)^{n+1}\prod_{j=1}^n\,^\prime j,\quad
\Gamma_p(0)=1\;,
\end{equation}
where the prime after the product above denotes leaving out
multiples of $p$ in the product. It is known that this extends to
a continuous function on $\mathbb{Z}_p$, the ring of $p$-adic
integers.

\noindent For $X \in \mathbb{Z}_p$ it satisfies the recurrence
relation:

\begin{equation}
\frac{\Gamma_p(X+1)}{\Gamma_p(X)}=
\begin{cases} -X; \quad X\in\mathbb{Z}_p^\ast\\
-1; \quad X\in p\mathbb{Z}_p\;.
\end{cases}
\end{equation}
A useful formula is:
\begin{equation}
n!=(-1)^{n+l+1}(-p)^{\frac{n-S(n)}{p-1}}\prod_{j=0}^l\Gamma_p\left(\left[\frac{n}{p^j}\right]+1\right)\;,
\end{equation}
where for a number $n$ with a finite $p$-adic expansion
$n=\sum_{j} c_jp^j$, $S(n)=\sum_{j}c_j$ is the sum of the digits.
The reflection formula involving the $p$-adic gamma function is
\begin{equation}
\Gamma_p(X)\Gamma_p(1-X)=(-1)^{L(X)}\;,
\end{equation}
where $L(X)$ denotes the $\mod p$ reduction of the lowest digit of
$X$ to the range ${1,2,\cdot,p}$, that is $L(X)$ is the lowest
digit unless the lowest digit is $0$ in which case $L(X)=p$. It is
the analogue of the classical result,
\begin{equation}
\Gamma(\xi)\Gamma(1-\xi) = \frac{\pi}{\sin(\pi\xi)}\;.
\end{equation}
The multiplication formula for $\Gamma_p$ is:
\begin{equation}
\frac{\prod_{i=0}^{n-1}\Gamma_p\left(\frac{X+1}{n}\right)}{\Gamma_p(X)\prod_{i=0}^{n-1}\Gamma_p\left(\frac{i}{n}\right)}=
n^{1-L(X)}(n^{-(p-1)})^{X_1}; \quad \text{for} \quad p\nmid n\;,
\end{equation}
where $X_1$ is related to $X$ by
\begin{equation}
X=L(X) + pX_1\;.
\end{equation}
Recall the classical multiplication formula for the
$\Gamma$-function:
\begin{equation}
\prod_{i=0}^{n-1}\Gamma\left(\frac{\xi+i}{n}\right)=(2\pi)^{\frac{n-1}{2}}n^{\frac{1}{2}-\xi}\Gamma(\xi)\;;
\label{pi}
\end{equation}
setting $\xi=1$ gives:
\begin{equation}
\prod_{i=0}^{n-1}\Gamma\left(\frac{i}{n}\right)=(2\pi)^{\frac{n-1}{2}}n^{-\frac{1}{2}}\;.
\label{nopi}
\end{equation}
Dividing (\ref{pi}) by (\ref{nopi}) yields a classical analogue of
the $p$-adic formula:
\begin{equation}
\frac{\prod_{i=0}^{n-1}\Gamma\left(\frac{\xi+1}{n}\right)}{\Gamma(\xi)\prod_{i=0}^{n-1}\Gamma\left(\frac{i}{n}\right)}=
n^{1-\xi}\;.
\end{equation}

\section[Gauss Sums]{Gauss Sums}\label{sectiongausssumms}
\subsection[Dwork's Character]{Dwork's Character}We record the expressions for the number of points, which we shall denote $\nu(\phi,\psi)$, in terms
of Gauss sums. We use machinery originally invented by Dwork to
prove the rationality of the zeta function - part of the Weil
Conjectures

We begin with a quick review of the properties of Dwork's character \cite{COV1}. The
character is defined by
\begin{equation}\Theta(x) = F(\pi \mathrm{Teich}(x))\;,
\end{equation}\\
where $\pi$ is a number in $\mathbb{C}_p$ such that $\pi^{p-1} =
-p$, $\mathrm{Teich}$ is the Teichm\"{u}ller character, and $F$ is
the function defined by:
\begin{equation}F(X) = \exp \left(X + \frac{X^p}{p}\right) = \sum_{n=0}^\infty c_n
X^n\;.
\end{equation}
By differentiating $F(X)$ it can be seen that the $c_n$ satisfy
the following recurrence relation:
\begin{equation} nc_n = c_{n-1} + c_{n-p},\quad c_0 = 1,\quad c_n = 0
\quad \mbox{for} \quad n < 0\;.
\end{equation}
For $x,y \in \mathbb{F}_p$,
\begin{equation} \mathrm{Teich}(x + y) = \mathrm{Teich}(x) +
\mathrm{Teich}(y) + pZ\;,
\end{equation}
\noindent for some $Z \in \mathbb{Z}_p$; thus it follows that
$\Theta$ is a nontrivial additive character. i.e.
\begin{equation} \Theta(x + y) = \Theta(x)\Theta(y)\;.
\end{equation}
This construction was adapted by Dwork to give an additive
character $\Theta_s : \mathbb{F}_q \to \mathbb{C}_p$ where
$q=p^s$:
\begin{equation}  \Theta_s(x) = \Theta(\mathrm{tr}(x))\;,
\end{equation}
where $\mathrm{tr} : \mathbb{F}_q \to \mathbb{F}_p$ is the trace
map
\begin{equation} \mathrm{tr}(x) = x + x^p + x^{p^2} + \cdots + x^{p^{s-1}}\;.
\end{equation}
It follows that
\begin{equation} \mathrm{Teich}(\mathrm{tr}(x)) = \sum_{l=0}^{s-1}
\mathrm{Teich}^{p^l}(x) +pZ\;,
\end{equation}

\noindent where $Z$ is an integer of $\mathbb{C}_p$; also

\begin{equation} \Theta_s(x) = \Theta(1)^{\sum_{l=0}^{s-1}
\mathrm{Teich}^{p^l}(x)}\;.
\end{equation}
The character, $\Theta_s$, has a remarkable splitting property:
\begin{equation} \Theta_s(x) = \prod_{l=0}^{s-1}\Theta_{p^l}(x)\;.
\end{equation}

A Gauss sum is the sum of an additive character, $\Theta$, times a
multiplicative character, $\mathrm{Teich}^n(x)$. Define:

\begin{equation} G_n = \sum_{x\in \mathbb{F}_p^\ast}
\Theta(x)\mathrm{Teich}^n(x);
\label{defgauss}
\end{equation}

\noindent thus, in particular, we have $G_0 =  \sum_{x\in
\mathbb{F}_p^\ast} \Theta(x) = -1$.

\noindent The following is a standard properties of Gauss sums:
\begin{equation} G_nG_{-n} = (-1)^np \quad \mbox{for} \quad
p-1\nmid n\;.
\end{equation}
The Gross--Koblitz formula gives $G_n$ for the case $p-1 \nmid n$:
\begin{equation} G_n = p (-p)^{-\left\langle
\frac{n}{p-1}\right\rangle}\Gamma_p\left(1-\left\langle
\frac{n}{p-1}\right\rangle\right); \quad p-1 \nmid n\;,
\end{equation}
where $\langle x\rangle$ denotes the fractional part of $x$.The
following is a very useful identity, which can be derived from
(\ref{defgauss}):

\begin{equation} \Theta(x) =
\frac{1}{p-1}\sum_{m=0}^{p-2}G_{-m}\mathrm{Teich}^m(x)\;.
\end{equation}

\noindent Now since $\Theta(x)$ is a character (for a defining
polynomial of a variety, $P$):
\begin{equation}
 \sum_{y\in \mathbb{F}_p}\Theta(yP(x)) =
 \begin{cases} p & \text{if
 $P(x)$ = 0,}\\ 0 & \text{if $P(x) \not= 0$.}
 \end{cases}
 \label{counter}
\end{equation}\\

The above property is of enormous importance to us in calculating
the number of points on a variety, as we shall see in
Section(\ref{numpoints}).

\noindent Similarly for finer fields $\mathbb{F}_{p^s}$ with
$q=p^s$ points:

\begin{equation} G_{s,n} = \sum_{x\in \mathbb{F}_q^\ast}
\Theta_{s}(x)\mathrm{Teich}^n(x).
\label{sdefgauss}
\end{equation}

\noindent The Gross--Koblitz formula is now:

\begin{equation} G_{s,n} = (-1)^{s+1}q\pi^{-S(n)}\prod_{l=0}^{s-1}\Gamma_p\left(1-\left\langle \frac{p^{l}n}{q-1}\right\rangle\right); \quad
q-1 \nmid n\;. \label{grosskoblitz}
\end{equation}

\section[General Form for Calabi--Yau Manifolds as Toric Hypersurfaces]{General Form for Calabi--Yau Manifolds as Toric Hypersurfaces}
When a Calabi--Yau manifold $\mathcal{M}$, as in the case for the
quintic, can be realized as a hypersurface
 in a toric variety, then we may associate with $\mathcal{M}$ its Newton polyhedron, $\Delta$, which has the
property of being reflexive. Using the Batyrev \cite{Baty} method
one is able to find the mirror $\mathcal{W}$. The deformation
class of mirror manifolds may be associated with hypersurfaces in
the toric variety whose fan consists of the cones, with vertex at
the unique interior point of $\Delta$, over the faces of $\Delta$.
Such a toric manifold will, in general, be singular as will the
hypersurfaces since these will intersect the singular subvarieties
of the toric variety.

Let us consider a weighted projective space $\mathbb{P}^n(w)$ and
a hypersurface $X_d(w)$ with (weighted) homogeneous degree $d =
w_1 +\ldots+ w_{n+1}$. Without loss of generality, we may assume
that the weight $w$ is normalized, i.e.
$\gcd(w_1,\ldots,\hat{w}_i,\ldots,w_{n+1})=1$, for
$i=1,\ldots,n+1$. Consider the polynomial:

\begin{equation}
P(x)=\sum_{(w,m)=d}a_\textbf{m}x^\textbf{m}=
\sum_{(w,m)=d}a_{m_1,\ldots,m_{n+1}}x_1^{m_1}\ldots
x_{n+1}^{m_{n+1}},
\end{equation}
or equivalently:
\begin{equation}
P(x)=\sum_{\textbf{m}\in\Delta,\atop\textbf{m}\neq\textbf{1}}a_\textbf{m}x^\textbf{m}
- a_\textbf{1}x^\textbf{1}\;,
\end{equation}

\noindent where we have denoted the Newton Polyhedron of $P(x)$ as
$\Delta$. For generic $a_{\textbf{m}}$, the zero locus ${P(x)=0}$
defines a hypersurface $X_d(w)$.

\section[Number of Points in Terms of Gauss Sums]{Number of
Points in Terms of Gauss Sums} \label{numpoints} When a variety is
defined as the vanishing set of a polynomial $P \in
k[X_1,\ldots,X_n]$, where k is a field, a non-trivial additive
character like Dwork's character can be exploited to count points
over $k$. The generalization of (\ref{counter}) is:

\begin{equation}
\sum_{y\in k}\Theta(yP(x))=\begin{cases}0\quad \mathrm{if}\,
P(x)\neq0,\\q:=\mathrm{Card(k)}\quad\mathrm{if}\,
P(x)=0\;;\end{cases}
\end{equation}

\noindent hence:
\begin{equation}
\sum_{x_i\in k}\sum_{y\in k}\Theta(yP(x))=q\nu\;,
\end{equation}\\

\noindent where $\nu$ is the number of rational points of the
variety.

Hence, for a hypersurface given by a polynomial $P$ in a toric variety (as defined in the previous section), the number of points over ${\mathbb{F}_p}$ can be found as follows:

\begin{eqnarray*}
p\nu(a_\textbf{m}) & = & \sum_{y\in \mathbb{F}_p}\sum_{x\in\mathbb{F}_p^{N+4}}\Theta\left(y(P(x)\right)\\
&=&\sum_{y\in\mathbb{F}_p}\sum_{x\in\mathbb{F}_p^{N+4}}\Theta\left(y\left(
\sum_{\textbf{m}\in\Delta,\atop\textbf{m}\neq\textbf{1}}a_\textbf{m}x^\textbf{m}
- a_\textbf{1}x^\textbf{1}\right)\right)\\
&=& p^{N+4} + \sum_{y\in\mathbb{F}_p^\ast}\sum_{x\in\mathbb{F}_p^{N+4}}\Theta\left(y\left(\sum_{\textbf{m}\in\Delta,\atop\textbf{m}\neq\textbf{1}}a_\textbf{m}x^\textbf{m} - a_\textbf{1}x^\textbf{1}\right)\right)\\
& = & p^{N+4} + \sum_{y\in\mathbb{F}_p^\ast}\sum_{x\in\mathbb{F}_p^{N+4}}\prod_{\textbf{m}\in\Delta,\atop\textbf{m}\neq\textbf{1}}\Theta\left(y\left(a_{\textbf{m}}x^\textbf{m}\right)\right)\Theta\left(-y\left(a_{\textbf{1}}x^\textbf{1}\right)\right)\\
& = & p^{N+4} + \frac{1}{(p-1)^{|\Delta|}}\sum_{y\in\mathbb{F}_p^\ast}\sum_{x\in\mathbb{F}_p^{N+4}}\prod_{\textbf{m}\in\Delta,\atop\textbf{m}\neq\textbf{1}}\sum_{s_{\textbf{m}}=0}^{p-2}G_{-s_{\textbf{m}}}\mathrm{Teich}^{s_{\textbf{m}}}(y\left(a_{\textbf{m}}x^\textbf{m})\right)\\
& & \times
\sum_{s_{\textbf{1}}=0}^{p-2}G_{-s_{\textbf{1}}}\mathrm{Teich}^{s_{\textbf{1}}}(-y\left(a_{\textbf{1}}x^\textbf{1})\right)\;.
\end{eqnarray*}

The steps taken above shall be explained in detail in the next
subsection, where we explicitly compute the number of points for
the octic family of Calabi--Yau threefolds.

\subsection[The Octic Family of Threefolds]{The Octic Family of
Threefolds}\label{octicpoints} Recall from Section \ref{cytwo}
that the family of Octic Calabi--Yau manifolds we shall study is a
hypersurface in $\frac{\mathbb{C}^6-F}{(\mathbb{C}^\ast)^2}$,
where $F$ is the set of disallowed vanishings of the variables. It
is a family of hypersurfaces indexed by the two parameters
$(\phi,\psi)$ defined by the following polynomial:

\begin{equation}
P(\textbf{x},x_6)=x_1^8x_6^4+x_2^8x_6^4+x_3^4+x_4^4+x_5^4-8\psi
x_1x_2x_3x_4x_5x_6-2\phi x_1^4x_2^4x_6^4\;,
\end{equation}

\noindent where the $x_i$ are Cox variables related to the
monomials in the Newton polyhedron~as~follows:
\begin{eqnarray}
x_1&=&x_{(8,0,0,0,0,1,4)},\nonumber\\
x_2&=&x_{(0,8,0,0,0,1,4)},\nonumber\\
x_3&=&x_{(0,0,4,0,0,1,0)},\nonumber\\
x_4&=&x_{(0,0,0,4,0,1,0)},\nonumber\\
x_5&=&x_{(0,0,0,0,4,1,0)},\nonumber\\
x_6&=&x_{(4,4,0,0,0,1,4)}.\nonumber\\
\end{eqnarray}
We shall now apply the procedure of the previous section to this
particular set of varieties:
\begin{eqnarray}
q\nu(\psi,\phi)&=&\sum_{y,x_6\in\mathbb{F}_q}\sum_{\textbf{x}\in(\mathbb{F}_q)^5}\Theta(yP(\textbf{x},x_6))-q\nu_F\nonumber\\
&=&
\sum_{y=0,\atop x_6\neq0}\sum_{\textbf{x}\in(\mathbb{F}_q)^5}\Theta(yP(\textbf{x},x_6))+\sum_{y=0,\atop x_6=0}\sum_{\textbf{x}\in(\mathbb{F}_q)^5}\Theta(yP(\textbf{x},0))+\sum_{y\neq0,\atop x_6\neq0}\sum_{\textbf{x}\in(\mathbb{F}_q)^5}\Theta(yP(\textbf{x},x_6))\nonumber\\
& &
+\sum_{y\neq0,\atop x_6=0}\sum_{\textbf{x}\in(\mathbb{F}_q)^5}\Theta(yP(\textbf{x},0))-q\nu_F\nonumber\\
&=&
(q-1)(q^2-1)(q^3-1)+(q^2-1)(q^3-1)+\sum_{y\neq0,\atop x_6\neq0}\sum_{\textbf{x}\in(\mathbb{F}_q)^5}\Theta(yP(\textbf{x},x_6))\nonumber\\
& & -\sum_{y\neq0,\atop x_6\neq0}\sum_{x_1=x_2=0,\atop
x_3,x_4,x_5\in\mathbb{F}_q}\Theta(y(x_3^4+x_4^4+x_5^4))+\sum_{y\neq0,\atop
x_6=0}\sum_{\textbf{x}\in(\mathbb{F}_q)^5}\Theta(y(x_3^4+x_4^4+x_5^4))
\nonumber\\
&=& (q-1)^2(q+1)q(1+q+q^2)+\sum_{y\neq0,\atop x_6\neq0}\sum_{\textbf{x}\in(\mathbb{F}_q)^5}\Theta(yP(\textbf{x},x_6))\nonumber\\
& &
-\sum_{y\neq0,\atop x_6\neq0}\sum_{x_1=x_2=0,\atop x_3,x_4,x_5\in\mathbb{F}_q}\Theta(y(x_3^4+x_4^4+x_5^4))+\sum_{y\neq0,\atop x_6=0}\sum_{\textbf{x}\in(\mathbb{F}_q)^5}\Theta(y(x_3^4+x_4^4+x_5^4))\;,\nonumber\\
\end{eqnarray}
where $\nu_{F}$ is the number of points that needed to be
subtracted from the total sum due to the combinations of
coordinates which are not allowed to vanish simultaneously. The
rule given by toric geometry is that a subset of the Cox variables
can vanish simultaneously if and only if the corresponding
monomials lie in the same cone.

In this case the following vanishings are forbidden by the
triangulation (see Figure \ref{fig:ortwoface3}):
\begin{eqnarray}
y=0,&\quad x_6\neq0,&\quad x_1=x_2=0,\nonumber\\
    &               &\quad x_3=x_4=x_5=0\;;\nonumber\\
y=0,&\quad x_6=0,&\quad x_1=x_2=0,\nonumber\\
    &               &\quad x_3=x_4=x_5=0\;.\nonumber\\
\end{eqnarray}
Thus the first two terms (minus the above forbidden vanishings)
are given by:
\begin{eqnarray}
\sum_{y=0,\atop
x_6\neq0}\sum_{\textbf{x}\in(\mathbb{F}_q)^5\backslash F}\Theta(yP(\textbf{x},x_6))&=&(q-1)(q^2-1)(q^3-1)\;,\\
\sum_{y=0,\atop
x_6=0}\sum_{\textbf{x}\in(\mathbb{F}_q)^5\backslash
F}\Theta(yP(\textbf{x},0))&=&(q^2-1)(q^3-1)\;.
\end{eqnarray}
We shall use the relation $N=\nu(q-1)^2$, where
$N=N_{q=p^s}=\#(X/\mathbb{F}_{p^s})$ i.e. the number of points on
the Calabi--Yau threefold reduced over the finite field
$\mathbb{F}_{p^s}$. This relations comes from the fact that we are
looking at hypersurfaces in
$\frac{\mathbb{C}^6-F}{(\mathbb{C}^\ast)^2}$, the extra factor of
$(q-1)^2$ comes from taking the quotient by $(\mathbb{C}^\ast)^2$.
Now:
\begin{eqnarray}
\lefteqn{\left(\sum_{y\neq0,\atop
x_6=0}\sum_{(x_1,x_2)\neq(0,0)\atop
x_3,x_4,x_5\in(\mathbb{F}_q)^3}-\sum_{y\neq0,\atop
x_6\neq0}\sum_{x_1=x_2=0,\atop
x_3,x_4,x_5\in\mathbb{F}_q}\right)\Theta(y(x_3^4+x_4^4+x_5^4))}\nonumber\\
&=&((q^2-1)(q-1)-(q-1)^2)\sum_{\textbf{v}} \prod_{i=3,4,5}G_{-\frac{v_i(q-1)}{4}}\nonumber\\
&=&q(q-1)^2\sum_{\textbf{v}}\prod_{i=3,4,5}G_{-\frac{v_i(q-1)}{4}}\;.\nonumber\\
\label{exceptional}
\end{eqnarray}

Where the sum $\sum_{\textbf{v}}$ is over monomials (and their
permutations) of the form $\textbf{v}:=(v_3,v_4,v_5)$ of degree
$d=0\mod4$ with $0\leq v_i \leq 3$, namely $(2,1,1)$ and
$(3,3,2)$. This is a very simple case of use of Lemma
(\ref{lemma}), which we shall come to soon. The procedure shall be
described in detail for the more complicated term. However, note
that the contribution above, can be thought of as the contribution
from the ruled surface of $x_3^4+x_4^4+x_5^4$ over a
$\mathbb{P}^1$ described in Section \ref{cytwo} and from now on
shall be denoted $N_{\mathrm{exceptional}}$. Note that
$N_{\mathrm{exceptional}}$ is only non-zero when $4|q-1$.

\begin{eqnarray}
N(\psi,\phi)&=& 1+2q+2q^2+q^3+N_{\mathrm{exceptional}}+\frac{1}{q(q-1)^2}\sum_{y\neq0,\atop x_6\neq0}\sum_{\textbf{x}\in(\mathbb{F}_q)^5}\Theta(yP(\textbf{x},x_6))\nonumber\\
\label{denominatorfound}
\end{eqnarray}

It appears that (\ref{denominatorfound}) will give us the terms
$(1-t)(1-pt)(1-p^2t)(1-p^3t)$ in the denominator of the zeta
function. We shall see at the end of the calculation that this is
indeed the case.

\noindent The term in (\ref{denominatorfound}):
$\frac{1}{q(q-1)^2}\sum_{y\neq0,\atop
x_6\neq0}\sum_{\textbf{x}\in(\mathbb{F}_q)^5}\Theta(yP(\textbf{x},x_6))$,
shall from now on be denoted~$N_{\mathrm{mon}}$.
\begin{eqnarray}
q(q-1)^2N_{\mathrm{mon}}
&=& q(q-1)^2(N_{\mathrm{mon}}(x_i\neq 0\,\forall i)+N_{\mathrm{mon}}(x_1x_2\neq 0,x_3x_4x_5=0)\nonumber\\
& &+N_{\mathrm{mon}}(x_1x_2= 0,x_3x_4x_5=0))\nonumber\\
&\equiv&
\sum_{y\in\mathbb{F}_p^\ast}g_{\phi}(y)g_{\psi}(y)g_8(y)^2g_4(y)^3\nonumber\\
& &
+\sum_{y\in\mathbb{F}_p^\ast}g_{\phi}(y)g_8(y)^2[(g_4(y)+1)^3-g_4(y)^3]\nonumber\\
& &
+\sum_{y\in\mathbb{F}_p^\ast}[(g_8(y)+1)^2-g_8(y)^2](g_4(y)+1)^3\;,
\label{basenonzero}
\end{eqnarray}
where $g_8(y)=\Theta(yx_i^8x_6^4)$, $g_4(y)=\Theta(yx_i^4)$,
$g_{\phi}(y)=\Theta(-y2\phi x_1^4x_2^4x_6^4)$, and \linebreak
$g_{\psi}(y)=\Theta(-y8\psi x_1x_2x_3x_4x_5x_6)$. The above
expression is obtained by simply expanding out in $\Theta$ over
all the combinations of vanishing $x_i$ (as we have already
subtracted out the number of points due to the `forbidden
vanishings', $\nu_{F}$).

An important observation is that $g_8(y)+1=\sum_{s=1}^{p-2}G_{-s}\mathrm{Teich}^s(yx^8x_6^4)$. Hence whenever there are expressions involving $(g_8(y)+1)$ or $(g_4(y)+1)$, we can restrict the argument of the Gauss sum, e.g. to be non-zero. As we shall see, this shall be immensely useful in dealing with, for instance, the second and third terms in (\ref{basenonzero}).

The expression, (\ref{basenonzero}), is for the case $\phi,
\psi\neq 0$, the general case. When either $\phi$ or $\psi$ are
zero, the expressions for $N_{\mathrm{mon}}$ are as follows:

\begin{enumerate}

\item For $\phi=0$
\begin{eqnarray*}
q(q-1)^2N_{\mathrm{mon}}&=&q(q-1)^2(N_{\mathrm{mon}}(\prod_{i=1}^{5}x_i\neq 0)+N_{\mathrm{mon}}(\prod_{i=1}^{5}x_i= 0))\\
&\equiv&\sum_{y\in\mathbb{F}_p^\ast}g_{\psi}(y)g_8(y)^2g_4(y)^3\\
&&+\sum_{y\in\mathbb{F}_p^\ast}[(g_8(y)+1)^2(g_4(y)+1)^3-g_8(y)^2g_4(y)^3]\;.\\
\end{eqnarray*}

\item For $\psi=0$
\begin{eqnarray*}
q(q-1)^2N_{\mathrm{mon}}&=&q(q-1)^2(N_{\mathrm{mon}}(x_1x_2\neq 0)+N_{\mathrm{mon}}(x_1x_2= 0))\\
&\equiv&\sum_{y\in\mathbb{F}_p^\ast}g_{\phi}(y)g_8(y)^2(g_4(y)+1)^3\\
&&+\sum_{y\in\mathbb{F}_p^\ast}((g_8(y)+1)^2-g_8(y)^2)(g_4(y)+1)^3\;.\\
\end{eqnarray*}

\item For $\psi=\phi=0$
\begin{eqnarray*}
q(q-1)^2N_{\mathrm{mon}}&=&\sum_{y\in\mathbb{F}_p^\ast}(g_8(y)+1)^2(g_4(y)+1)^3\;.\\
\end{eqnarray*}
\end{enumerate}

\noindent First we shall calculate $N_{\mathrm{mon}}$ for the case
when $x_i\neq0 \,\forall i$ and $\phi\psi\neq 0$:
\begin{eqnarray}
N_{\mathrm{mon}}(x_i\neq 0)
&=&\frac{1}{q(q-1)^2}\sum_{y,x_6\in\mathbb{F}_p^\ast}\sum_{\textbf{x}\in(\mathbb{F}_p^\ast)^5}\Theta(yx_1^8x_6^4)\Theta(yx_2^8x_6^4)\Theta(yx_3^4)\Theta(yx_4^4)\nonumber\\
& &
\times\Theta(yx_5^4)\Theta(-2y\phi x_1^4x_2^4x_6^4)\Theta(-8y\psi  x_1x_2x_3x_4x_5x_6)\nonumber\\
&=&\frac{1}{q(q-1)^2}\sum_{s_{\phi},s_{\psi},s_i=0}^{q-2}\prod_{i=1}^{5}
G_{-s_i}G_{-s_{\psi}}G_{-s_{\phi}}\mathrm{Teich}^{s_{\psi}}(-8\psi)\mathrm{Teich}^{s_{\phi}}(-2\phi)\nonumber\\
&=&\frac{1}{q(q-1)^2}\sum_{s_{\phi},s_{\psi},s_i=0}^{q-2}\prod_{i=1}^{5} G_{-s_i}G_{s_{\psi}}G_{s_{\phi}}\mathrm{Teich}^{-s_{\psi}}(-8\psi)\mathrm{Teich}^{-s_{\phi}}(-2\phi)\nonumber\\
&=&
\frac{1}{q(q-1)^2}\sum_{s_{\phi},s_{\psi}}^{q-2}\sum_{\textbf{v}}\prod G_{-\left(\frac{v_i(q-1)}{8}+\frac{s_{\phi}}{2}+\frac{s_{\psi}}{8}\right)}\prod G_{-\left(\frac{v_i(q-1)}{4}+\frac{s_{\psi}}{4}\right)}\nonumber\\
& &
\times G_{s_{\psi}}G_{s_{\phi}}\mathrm{Teich}^{-s_{\psi}}(-8\psi)\mathrm{Teich}^{-s_{\phi}}(-2\phi)\;.\nonumber\\
\label{gencalc}
\end{eqnarray}

In our calculations repeated use shall be made of the following lemma:
\begin{lemma}
Let $i_1,\ldots,i_n$ be non-negative integers. Then
\begin{equation}
\sum_{(a_1,\ldots,a_n)\in
(\mathbb{F}_q)^n}\mathrm{Teich}^{i_1}(a_1)\ldots
\mathrm{Teich}^{i_n}(a_n)=0\;,
\end{equation}

\noindent unless each $i_j$ is non-zero and divisible by $q-1$.
\label{lemma}
\end{lemma}

\begin{proof}Suppose first that $n=1$. If $i=0$, then $\sum_{a \in\mathbb{F}_q}\mathrm{Teich}^0(a)=q=0$ in $\mathbb{F}_q$. Assume that $i>0$, and let $b$ denote a generator of the finite cyclic group $\mathbb{F}_q^{\ast}$. If $(q-1)\nmid i$, then

\begin{equation}
\sum_{a
\in\mathbb{F}_q}\mathrm{Teich}^i(a)=\sum_{j=0}^{q-2}(\mathrm{Teich}^j(b))^i=\frac{(\mathrm{Teich}^i(b))^{q-1}-1}{\mathrm{Teich}^1(b)-1}=0.
\end{equation}

\noindent If $i=(q-1)j$, then
\begin{equation}
\sum_{a\in\mathbb{F}_q}\mathrm{Teich}^{(q-1)j}(a)=\sum_{a\in\mathbb{F}_q^{\ast}}1^j=q-1.
\end{equation}

\noindent In general,
\begin{align}
\sum_{(a_1,\ldots,a_n)\in
(\mathbb{F}_q)^n}\mathrm{Teich}^{i_1}(a_1)\ldots
\mathrm{Teich}^{i_n}(a_n)=\left(\sum_{a_1\in\mathbb{F}_q}\mathrm{Teich}^{i_1}(a_1)\right)\ldots\left(\sum_{a_n\in\mathbb{F}_q}\mathrm{Teich}^{i_n}(a_n)\right)=0,
\end{align}

\noindent unless all the exponents $i_j$ are non-zero and
divisible by $q-1$. \qed
\end{proof}

Using Lemma \ref{lemma} on expressions of the form
$\sum_{x_i\in\mathbb{F}_{p^s}}\mathrm{Teich}^n(x_i)$, we obtain
the following simultaneous set of constraints on the values of the
$s_i$:
\begin{center}
\begin{math}
\begin{array}{ll}
q-1|\quad8s_{i}-4s_{\phi}-s_{\psi}& i=1,2\\
q-1|\quad4s_{i}-s_{\psi}& i=3,4,5\\
q-1|\quad4s_1+4s_2-4s_{\phi}-s_{\psi}& \\
q-1|\quad\sum_{i=1}^{5}s_i-s_{\psi}-s_{\phi}& \\
0\leq\; s_i,s_{\phi},s_{\psi}\; \leq (q-2)&i=1,2,3,4,5.
\end{array}
\end{math}
\end{center}

\noindent Hence define a vector of integers, $\textbf{v}^{\textbf{t}}=(v_1,v_2,v_3,v_4,v_5,v_{\phi},v_{\psi})$, such that:\\
\begin{center}
\begin{math}
\begin{array}{llll}
v_i(q-1)&=&8s_{i}-4s_{\phi}-s_{\psi}& i=1,2\\
v_i(q-1)&=&4s_{i}-s_{\psi}& i=3,4,5\\
v_{\phi}(q-1)&=&4s_1+4s_2-4s_{\phi}-s_{\psi}&\\
v_{\psi}(q-1)&=&\sum_{i=1}^{5}s_i-s_{\psi}-s_{\phi}\;.&
\end{array}
\end{math}
\end{center}
It is easy to see that we need to consider vectors $\textbf{v}$
such that:
\begin{eqnarray}
v_1+v_2+2(v_3+v_4+v_5)= 8v_{\psi}\nonumber\\
\frac{v_1+v_2}{2}=v_{\phi}\nonumber\\
0\leq v_i\leq7,\,i=1,2\nonumber\\
0\leq v_i\leq3,\,i=3,4,5.\nonumber\\
\label{cons} \label{vconditions}
\end{eqnarray}
Every vector of the form (\ref{vconditions}) comes from a set of
integers $(s_i,s_{\phi},s_{\psi})$ which satisfies the set of
constraints. From now on we shall only write down the first five
independent entries of $\textbf{v}$. Notice that given one
solution of (\ref{cons}), $(v_1,v_2,v_3,v_4,v_5)$, another
solution can be obtained by adding $(1,1,1,1,1)$ or $(4,4,0,0,0)$
and reducing $v_1, v_2$ mod 8; $v_3,v_4, v_5$ mod 4. In
(\ref{gencalc}), $\sum_{\textbf{v}}$, denotes summing over all
valid monomials, $\textbf{v}$. We can define an equivalence,
$\sim$, on the set of monomials: two monomials are equivalent if
they differ by a factor of $(1,1,1,1,1)$ $\mod 8$ in the first two
coordinates and $\mod 4$ last three coordinates. We can thus form
equivalence classes of $\sim$. Notice that permutations of the
form $P_2\times P_3$, where $P_2$, and $P_3$ are the permutations
of the first two and the last three coordinates, respectively,
leave products of Gauss sums unchanged, e.g. $\prod_{i=1,2}
G_{-\left(\frac{v_i(q-1)}{8}+\frac{s_{\phi}}{2}+\frac{s_{\psi}}{8}\right)}\prod_{i=3,4,5}
G_{-\left(\frac{v_i(q-1)}{4}+\frac{s_{\psi}}{4}\right)}$. We
define a concept of a `length of orbit' of classes, which is the
number of times the vector $(1,1,1,1,1)$ needs to be added to a
monomial until you return to a permutation of the monomial. For
the class of monomials labelled by $(4,0,1,1,0)$, for example, we
see that:
\begin{equation}
(4,0,1,1,0)\rightarrow(5,1,2,2,1)\rightarrow(6,2,3,3,2)\rightarrow(7,3,0,0,3)\rightarrow(0,4,1,1,0)\;.
\end{equation}
Since we return to (a permutation of) the same monomial after $4$
additions of $(1,1,1,1,1)$, we can regard this class has having an
orbit of length $4$.

Because permutations of monomials leave products of Gauss sums
unchanged, we can simply do the sum using one representative of
each class, and place a `permutation' factor
$\lambda_{\textbf{v}}$ in front, where it is clear that we need
\begin{equation}
\lambda_{\textbf{v}}=\;\#\emph{Permutations
of}\;\:\textbf{v}\times\emph{`Length of orbit'}/8.
\end{equation}
\pagebreak
Overall, there are $15$ classes of monomials:
\begin{center}
\begin{tabular}{|c|c|c|c|}\hline
Monomial $\textbf{v}$ & degree & Permutations $\lambda_{\textbf{v}}$ & Length of orbit\\
\hline\hline
(0,0,0,0,0) & 0 & 1 & 8\\
(7,1,0,0,0) & 8 & 2  & 8\\
(6,2,0,0,0) & 8 & 1  & 4\\ \hline
(0,0,0,2,2) & 8 & 3  & 8\\
(3,1,2,0,0) & 8 & 6 & 8\\
(4,0,2,0,0) & 8 & 3 & 4\\ \hline
(0,0,2,1,1) & 8 & 3  & 8\\
(6,0,1,0,0) & 8 & 6  & 8\\
(5,1,1,0,0) & 8 & 3  & 4\\ \hline
(4,0,1,1,0) & 8 & 3  & 4\\
(2,0,3,0,0) & 8 & 6  & 8\\
(1,1,0,0,3) & 8 & 3 & 8\\\hline
(0,0,3,1,0) & 8 & 6  & 8\\
(2,0,2,1,0) & 8 & 12  & 8\\
(7,3,2,1,0) & 16 & 6  & 4\\ \hline
\end{tabular}
\end{center}

In the above table we have used a representative of each monomial
class with minimal degree. The choice of representative is
arbitrary for the purposes of calculation, for the case $8|q-1$.
However, it shall become apparent that when $8 \nmid q-1$ the
divisibility properties of the $v_i$ become important, and
formulae will only take a representative of each class with
particular divisibility properties. This just reflects the way the
formulae are written, they can be rewritten, of course, to enable
the choice of a different representative. The choices here seem to
yield the simplest and most `beautiful' expressions. More
properties of these classes shall be discussed in Section
\ref{combin}. It can be shown that the table below of the $15$
monomial classes uses representatives that work whenever $4|q-1$,
for $4 \nmid q-1$ only four of these classes contribute, see Table
(\ref{fourclasses}).

\noindent
\begin{center}
\begin{tabular}{|c|c|c|c|}\hline
Monomial $\textbf{v}$ & degree & Permutations $\lambda_{\textbf{v}}$ & Length of orbit\\
\hline\hline
(0,0,0,0,0) & 0 & 1 & 8\\
(0,2,1,1,1) & 8 & 2  & 8\\
(0,4,2,2,2) & 16 & 1  & 4\\ \hline
(0,0,0,2,2) & 8 & 3  & 8\\
(2,0,1,3,3) & 16 & 6 & 8\\
(4,0,2,0,0) & 8 & 3 & 4\\ \hline
(0,0,2,1,1) & 8 & 3  & 8\\
(6,0,1,0,0) & 8 & 6  & 8\\
(0,4,0,3,3) & 16 & 3  & 4\\ \hline
(4,0,1,1,0) & 8 & 3  & 4\\
(2,0,3,0,0) & 8 & 6  & 8\\
(2,2,1,1,0) & 8 & 3 & 8\\\hline
(0,0,3,1,0) & 8 & 6  & 8\\
(2,0,2,1,0) & 8 & 12  & 8\\
(4,0,2,3,1) & 16 & 6  & 4\\ \hline
\end{tabular}
\end{center}

Recall that we remarked that $N_{\mathrm{exceptional}}$ is only non-zero when $4|q-1$, indeed $N_{\mathrm{exceptional}}$ could have been combined with  $N_{\mathrm{mon}}$ as part of the contributions from the monomials $(0,0,2,1,1)$ and $(1,1,0,0,3)$.

We shall continue the calculation of $N_{\mathrm{mon}}$, but for
the case $q=3 \mod 8$, as this is a typical non-trivial case. The
essential consideration at each step to ensure that the arguments,
$n$, of the Gauss sums, $G_n$, is integer. Thus the first step
would be to define $m=\frac{q-1}{2}\in\mathbb{Z}$, and
$t=s_{\phi}/4$. We need to ensure
$\frac{v_i}{4}m+\frac{s_{\phi}+t}{2}\in\mathbb{Z}$ and
$\frac{v_i}{2}m+t\in\mathbb{Z}$. Notice that for the case we are
considering, $\frac{3m+1}{4}=\frac{3q-1}{8}\in\mathbb{Z}$. First
we split the sums over $\textbf{v}$ and $t$, into two parts:
$(\textbf{v},t)$ and
$\left(\textbf{v}+(3,3,3,3,3),t+\frac{1}{2}\right)$. For the
second step, notice that the transformation on the integer
variable, $t$, of

\begin{equation}
t\rightarrow t+\frac{3m+1}{2}\Longleftrightarrow
\begin{cases}\textbf{v}\rightarrow\textbf{v}+(3,3,3,3,3)\;,\\t\rightarrow
t+\frac{1}{2}\;.\end{cases}
\end{equation}

So we replace the second sum over $\textbf{v}+(3,3,3,3,3)$, by
just extending the range of $t$ in the first sum by
$\frac{3m+1}{2}$:

\begin{eqnarray}
N_{\mathrm{mon}}(x_i\neq 0)
&=&
\frac{1}{q(q-1)^2}\sum_{\textbf{v}}\sum_{s_{\phi}=0}^{q-2}\sum_{0\leq t\leq\left\lfloor \frac{q-2}{4}\right\rfloor=\frac{q-3}{4}}\prod G_{-\left(\frac{v_i}{4}m+\frac{s_{\phi}+t}{2}\right)}\nonumber\\
& &
\times\prod G_{-\left(\frac{v_i}{2}m+t\right)} G_{4t}G_{s_{\phi}}\mathrm{Teich}^{-4t}(-8\psi)\mathrm{Teich}^{-s_{\phi}}(-2\phi)\nonumber\\
&&
+\frac{1}{q(q-1)^2}\sum_{\textbf{v}}\sum_{s_{\phi}=0}^{q-2}\sum_{0\leq t\leq\frac{q-7}{4}}\prod G_{-\left(\frac{(v_i+3)}{4}m+\frac{(s_{\phi}+t)}{2}+\frac{1}{4}\right)}\nonumber\\
& &
\times \prod G_{-\left(\frac{(v_i+3)}{2}m+t+\frac{1}{2}\right)}G_{4t+2}G_{s_{\phi}}\mathrm{Teich}^{-4t-2}(-8\psi)\mathrm{Teich}^{-s_{\phi}}(-2\phi)\nonumber\\
&=&
\frac{1}{q(q-1)^2}\sum_{s_{\phi}=0}^{q-2}\sum_{0\leq t\leq\left\lfloor \frac{q-2}{4}\right\rfloor=\frac{q-3}{4}\atop \frac{3q-1}{4}\leq t\leq(q-2)}\prod G_{-\left(\frac{v_i}{4}m+\frac{s_{\phi}+t}{2}\right)}\nonumber\\
& &
\times \prod G_{-\left(\frac{v_i}{2}m+t\right)}G_{4t}G_{s_{\phi}}\mathrm{Teich}^{-4t}(-8\psi)\mathrm{Teich}^{-s_{\phi}}(-2\phi)\;.\nonumber\\
\end{eqnarray}

Now we undertake an analogous step, this time introducing a new
variable, $u=\frac{s_{\phi}+t}{2}$, and allowing $u$ to be either
integer or half-integer. We again split the sum over the monomials
into two equal sized sets: $\textbf{v}$ and
$\textbf{v}+(2,2,2,2,2)$. For the second step, notice that
$\frac{m+1}{2}=\frac{q+1}{4}\in\mathbb{Z}$ and that the
transformation this time on both $t$ and $u$:

\begin{equation}
\begin{cases}t\rightarrow t+m\\u\rightarrow u+\frac{m+1}{2}\end{cases}\Longleftrightarrow \begin{cases}\textbf{v}\rightarrow\textbf{v}+(2,2,2,2,2)\\u\rightarrow u+\frac{1}{2}\;.\end{cases}
\end{equation}

However the transformation in $u$ follows from its dependence on
$t$. Hence again we just need to extend the sum over $t$:

\begin{eqnarray}
0\leq t\leq\frac{q-3}{4} \rightarrow \begin{cases} 0\leq t\leq\frac{q-3}{4},\\  \frac{q-1}{2}\leq t\leq\frac{3q-5}{4}\;.\end{cases}\nonumber\\
\frac{3q-1}{4}\leq t\leq q-2 \rightarrow \begin{cases} \frac{q+1}{4}\leq t\leq\frac{q-3}{2},\\  \frac{3q-1}{4}\leq t\leq q-2\;.\end{cases}\nonumber\\
\end{eqnarray}
which means having $0\leq t\leq q-2$, which is rather satisfying:

\begin{eqnarray}
N_{\mathrm{mon}}(x_i\neq 0)
&=&
\frac{1}{q(q-1)^2}\sum_{0\leq t\leq\frac{q-3}{4} \atop \frac{3q-1}{4}\leq t\leq(q-2)}\sum_{\left\lceil \frac{t}{2}\right\rceil\leq u\leq\left\lfloor \frac{q-2+t}{2}\right\rfloor}\prod G_{-\left(\frac{v_i}{4}m+u\right)}\nonumber\\
& &
\times \prod G_{-\left(\frac{v_i}{2}m+t\right)}G_{4t}G_{2u-t}\mathrm{Teich}^{-4t}(-8\psi)\mathrm{Teich}^{-s_{\phi}}(-2\phi)\nonumber\\
&&
+\frac{1}{q(q-1)^2}\sum_{0\leq t\leq\frac{q-3}{4} \atop \frac{3q-1}{4}\leq t\leq(q-2)}\sum_{\left\lceil \frac{t}{2}-1\right\rceil\leq u\leq\left\lfloor \frac{q-3+t}{2}\right\rfloor}\prod G_{-\left(\frac{(v_i+2)}{4}m+u+\frac{1}{2}\right)}\nonumber\\
& &
\times \prod G_{-\left(\frac{(v_i+2)}{2}m+t\right)}G_{4t}G_{2u+1-t}\mathrm{Teich}^{-4t-2}(-8\psi)\mathrm{Teich}^{-s_{\phi}}(-2\phi)\nonumber\\
&=&
\frac{1}{q(q-1)^2}\sum_{t=0}^{q-2}\sum_{\left\lceil \frac{t}{2}\right\rceil\leq u\leq\left\lfloor \frac{q-2+t}{2}\right\rfloor}\prod G_{-\left(\frac{v_i}{4}m+u\right)}\nonumber\\
& &
\times \prod G_{-\left(\frac{v_i}{2}m+t\right)}G_{4t}G_{2u-t}\mathrm{Teich}^{t}\left(\frac{-2\phi}{(8\psi)^4}\right)\mathrm{Teich}^{u}\left(\frac{1}{(2\phi)^2}\right)\;.\nonumber\\
\end{eqnarray}

Note that we have also carried out the following simple steps on
the Teichm\"{u}ller characters:
\begin{eqnarray}
\mathrm{Teich}^{-s_{\psi}}(-8\psi)\mathrm{Teich}^{-s_{\phi}}(-2\phi)&=&
\mathrm{Teich}^{-4t}(-8\psi)\mathrm{Teich}^{-2u+t}(-2\phi)\nonumber\\
&=&
\mathrm{Teich}^{t}\left(\frac{-2\phi}{(8\psi)^4}\right)\mathrm{Teich}^{-2u}({2\phi})\nonumber\\
&=&
\mathrm{Teich}^{u}\left(\frac{1}{(2\phi)^2}\right)\mathrm{Teich}^{t}\left(\frac{-2\phi}{(8\psi)^4}\right)\nonumber\\
&=& \mathrm{Teich}^{u}(\mu)\mathrm{Teich}^{t}(\Psi)\;,
\end{eqnarray}
where $\Psi$ and $\mu$ happen to be the large complex structure
coordinates:

\begin{equation}
\mu=\frac{1}{(2\phi)^2},\quad\Psi=\frac{-2\phi}{(8\psi)^4}\;.
\end{equation}

The final step involves extending the range of $u$. We can split
the sum over the $\textbf{v}$s into two:$\textbf{v}$ and
$\textbf{v}+(4,4,0,0,0)$, but:

\begin{equation}
\textbf{v}\rightarrow\textbf{v}+(2,2,2,2,2)\Longleftrightarrow
u\rightarrow u+m\;;
\end{equation}\\
hence we extend the range of $u$ from $\left\lceil
\frac{t}{2}\right\rceil\leq u\leq\left\lfloor
\frac{q-2+t}{2}\right\rfloor$ to $\left\lceil
\frac{t}{2}\right\rceil\leq u\leq\left\lfloor
\frac{2q-3+t}{2}\right\rfloor$ giving:

\begin{eqnarray}
N_{\mathrm{mon}}(x_i\neq 0)
&=&
\sum_{\textbf{v}}\lambda_{\textbf{v}}\frac{1}{q(q-1)^2}\sum_{t=0}^{q-2}\sum_{\left\lceil \frac{t}{2}\right\rceil\leq u\leq\left\lfloor \frac{2q-3+t}{2}\right\rfloor}\prod G_{-\left(\frac{v_i}{4}m+u\right)}\nonumber\\
& &
\times \prod G_{-\left(\frac{v_i}{2}m+t\right)}G_{4t}G_{2u-t}\mathrm{Teich}^{t}\left(\frac{-2\phi}{(8\psi)^4}\right)\mathrm{Teich}^{u}\left(\frac{1}{(2\phi)^2}\right)\;.\nonumber\\
\end{eqnarray}

A similar analysis for the case $q=7\, \mod \,8$ gives an
identical expression. Note that for these cases the allowed
monomials have to have the following properties: $4|v_i,\, i=1,2$,
and $2|v_i, \,i=3,4,5$. Hence, only four out of the fifteen
classes of monomial are allowed to contribute, these are:

\noindent \begin{center}
\begin{tabular}{|c|c|c|c|}\hline
Monomial $\textbf{v}$ & degree & Permutations $\lambda_{\textbf{v}}$ & Length of orbit \\
\hline\hline
(0,0,0,0,0) & 0 & 1 & 8\\
(0,4,2,2,2) & 8 & 1  & 4\\ \hline
(0,0,0,2,2) & 8 & 3  & 8\\
(4,0,2,0,0) & 8 & 3 & 4\\ \hline
\end{tabular}\label{fourclasses}
\end{center}

Now consider the case when only $x_1x_2\neq 0,\,x_3x_4x_5=0$:
\begin{eqnarray}
N_{\mathrm{mon}}(x_1x_2\neq 0,\,x_3x_4x_5=0)
&=&\frac{1}{q(q-1)}\sum_{y\in\mathbb{F}_p^\ast}g_{\phi}(y)g_8(y)^2[(g_4(y)+1)^3-g_4(y)^3]\nonumber\\
&=&
\frac{1}{q(q-1)}\sum_{y,x_6\in\mathbb{F}_p^\ast}\sum_{\textbf{x}\in(\mathbb{F}_p^\ast)^5}\Theta(yx_1^8x_6^4)\Theta(yx_2^8x_6^4)\Theta(yx_3^4)\nonumber\\
& &\times\Theta(yx_4^4)\Theta(yx_5^4)\Theta(-2y\phi x_1^4x_2^4x_6^4)\nonumber\\
&=&
\frac{1}{q(q-1)}\sum_{s_{\phi},s_i=0}^{q-2}\prod_{i=1}^{5} G_{-s_i}G_{-s_{\phi}}\mathrm{Teich}^{s_{\phi}}(-2\phi)\nonumber\\
&=&
\frac{1}{q(q-1)}\sum_{s_{\phi},s_i=0}^{q-2}\prod_{i=1}^{5} G_{-s_i} G_{s_{\phi}}\mathrm{Teich}^{-s_{\phi}}(-2\phi)\nonumber\\
&=&
\frac{-1}{q(q-1)}\sum_{s_{\phi}=0}^{q-2}\sum_{\textbf{v}s.t.\atop v_3v_4v_5=0}\prod G_{-\left(\frac{v_i(q-1)}{8}+\frac{s_{\phi}}{2}\right)}\prod G_{-\left(\frac{v_i(q-1)}{4}\right)}\nonumber\\
& &
\times G_{s_{\phi}}\mathrm{Teich}^{-s_{\phi}}(-2\phi)\;.\nonumber\\
\end{eqnarray}

Notice that we have again introduced monomials $\textbf{v}$. These
also have to satisfy (\ref{vconditions}), however in addition the
constraint, $v_3v_4v_5=0$ is required, as the expression
\linebreak
\hbox{$\sum_{y\in\mathbb{F}_p^\ast}[(g_4(y)+1)^3-g_4(y)^3]$} can
be considered as:
\begin{eqnarray*}
(\mbox{\emph{Non-zero arguments of Gauss sums}})-(\mbox{\emph{All
Gauss sums}})\\=-(\mbox{\emph{Gauss sums with at  least one zero
argument}}).
\end{eqnarray*}

Hence we need to restrict the allowable monomials with the
constraint: $\prod v_3v_4v_5=0$ and there is an overall minus
sign.

\noindent Now for $q=3\mod \,4$ we can write:
\begin{eqnarray}
N_{\mathrm{mon}}(x_1x_2\neq 0,\,x_3x_4x_5=0)
&=&
\frac{-1}{q(q-1)}\sum_{s_{\phi}=0}^{q-2}\sum_{\textbf{v}s.t.\atop \prod v_3v_4v_5=0}\prod G_{-\left(\frac{v_i}{4}m+\frac{s_{\phi}}{2}\right)}\prod G_{-\left(\frac{v_i}{2}m\right)}\nonumber\\
& &
\times G_{s_{\phi}}\mathrm{Teich}^{-s_{\phi}}(-2\phi)\nonumber\\
&=&
\frac{-1}{q(q-1)}\sum_{u=0}^{m-1}\sum_{\textbf{v}s.t.\atop v_3v_4v_5=0}\prod G_{-\left(\frac{v_i}{4}m+u\right)}\prod G_{-\left(\frac{v_i}{2}m\right)}\nonumber\\
& &
\times G_{2u}\mathrm{Teich}^{-u}(-2\phi)\nonumber\\
& &
+\frac{-1}{q(q-1)}\sum_{u=0}^{m-1}\sum_{\textbf{v}s.t.\atop v_3v_4v_5=0}\prod G_{-\left(\frac{(v_i+2)}{4}m+u+\frac{1}{2}\right)}\prod G_{-\left(\frac{v_i+2}{2}m\right)}\nonumber\\
& &
\times G_{2u+1}\mathrm{Teich}^{-(2u+1)}(-2\phi).\nonumber\\
\end{eqnarray}

Here we have defined $u=\frac{s_{\phi}}{2}$ and have split the sum
over the $\textbf{v}$s.(Notice that we could not have done this
had $8|q-1$ and in that case we only get the first term). Now we
can extend the sum over $u$ by noticing that

\begin{equation}
\textbf{v}\rightarrow\textbf{v}+(4,4,0,0,0)\Longleftrightarrow
u\rightarrow u+m;
\end{equation}\\
hence we can extend the range of $u$ from $0\leq u \leq m-1$ to $0
\leq u \leq q-2$:

\begin{eqnarray}
N_{\mathrm{mon}}(x_1x_2\neq 0,\,x_3x_4x_5=0)
&=&
-\frac{1}{q(q-1)}\sum_{u=0}^{q-2}\sum_{\textbf{v}s.t.\atop v_3v_4v_5=0}\prod G_{-\left(\frac{v_i}{4}m+u\right)}\prod G_{-\left(\frac{v_i}{2}m\right)}\nonumber\\
& &
\times G_{2u}\mathrm{Teich}^{u}\left(\frac{1}{(2\phi)^2}\right)\nonumber\\
& &
-\frac{1}{q(q-1)}\sum_{u=0}^{q-2}\sum_{\textbf{v}s.t.\atop v_3v_4v_5=0}\prod G_{-\left(\frac{(v_i+2)}{4}m+u+\frac{1}{2}\right)}\prod G_{-\left(\frac{v_i+2}{2}m\right)}\nonumber\\
& &
\times G_{2u+1}\mathrm{Teich}^{u}\left(\frac{1}{(2\phi)^2}\right)\mathrm{Teich}\left(-\frac{1}{2\phi}\right)\;.\nonumber\\
\end{eqnarray}

\noindent Finally the contribution from
$N_{\mathrm{mon}}(x_1x_2=0,x_3x_4x_5=0)$ needs to be found and
added:

\begin{eqnarray}
N_{\mathrm{mon}}(x_1x_2=0,x_3x_4x_5=0)
&=&
\sum_{y\in\mathbb{F}_p^\ast}[(g_8(y)+1)^2-g_8(y)^2](g_4(y)+1)^3\nonumber\\
&=&
-\frac{1}{q}\sum_{y\in\mathbb{F}_q^\ast}\sum_{b=0}^{1}\sum_{\textbf{v}s.t.\atop v_3v_4v_5\neq 0\, \&\,(v_i+4b)\neq 0}\lambda_{\textbf{v}}\prod_{i=1,2}G_{-\frac{(v_i+4b)}{4}m }\prod_{i=3,4,5}G_{-\frac{v_i}{2}m}\;.\nonumber\\
\end{eqnarray}

In total we obtain for $q=3\,\mod\,8$:\\
\begin{eqnarray}
N_{\mathrm{mon}} & = &
\frac{1}{q(q-1)^2}\sum_{t=0}^{\frac{q-3}{4}}\sum_{u=\left\lceil \frac{t}{2}\right\rceil}^{\left\lfloor \frac{2q-3+t}{2}\right\rfloor}\sum_{\textbf{v}}\lambda_{\textbf{v}}\prod_{i=1,2}G_{-(\frac{v_i}{4}m +u)}\prod_{i=3,4,5}G_{-(\frac{v_i}{2}m+t)}\nonumber\\
& &
\times G_{2u-t}G_{4t}\mathrm{Teich}^{u}\left(\frac{1}{(2\phi)^2}\right)\mathrm{Teich}^{t}\left(\frac{-2\phi}{(8\psi)^4}\right)\nonumber\\
& &
-\frac{1}{q(q-1)}\sum_{u=0}^{q-2}\sum_{\textbf{v}s.t.\atop v_3v_4v_5=0}\lambda_{\textbf{v}}\prod_{i=1,2}G_{-(\frac{v_i}{4}m +u)}\prod_{i=3,4,5}G_{-\frac{v_i}{2}m} G_{2u}\mathrm{Teich}^{u}\left(\frac{1}{(2\phi)^2}\right)\nonumber\\
& &
-\frac{1}{q(q-1)}\sum_{u=0}^{q-2}\sum_{\textbf{v}s.t.\atop v_3v_4v_5=0}\lambda_{\textbf{v}}\prod_{i=1,2}G_{-(\frac{v_i+2}{4}m +u+\frac{1}{2})}\prod_{i=3,4,5}G_{-\frac{v_i+2}{2}m} G_{2u+1}\mathrm{Teich}^{u}\frac{1}{(2\phi)^2}\nonumber\\
& &\times\mathrm{Teich}^{-1}\left(-2\phi\right)\nonumber\\
& &
-\frac{1}{q}\sum_{y\in\mathbb{F}_q^\ast}\sum_{b=0}^{1}\sum_{\textbf{v}\atop v_3v_4v_5\neq 0\, \&\,(v_i+4b)\neq 0}\lambda_{\textbf{v}}\prod_{i=1,2}G_{-\frac{(v_i+4b)}{4}m }\prod_{i=3,4,5}G_{-\frac{v_i}{2}m}\;.\nonumber\\
\end{eqnarray}

The computation for the other cases is similar and the results are
summarized in the next four sections.

\pagebreak

\section[Summary of Results]{Summary of
Results}\label{sectionsummaryofresults}
\subsection[Summary for $\phi=\psi= 0$]{Summary for $\phi=\psi= 0$}
\begin{enumerate}
\item When $8|q-1$:\\
\begin{eqnarray}
N_{\mathrm{mon}}
& = &
\frac{1}{q}\sum_{b=0}^{1}\sum_{a=0}^{3}\sum_{\textbf{v}}\lambda_{\textbf{v}}\prod_{i=1,2}G_{-(v_i+a+4b)k }\prod_{i=3,4,5}G_{-2(v_i+a)k}\;.\nonumber\\
\end{eqnarray}\\

\item When $4|q-1,\,8\nmid q-1$:\\
\begin{eqnarray}
N_{\mathrm{mon}}
& = &
\frac{1}{q}\sum_{a=0}^{3}\sum_{\textbf{v}}\lambda_{\textbf{v}}\prod_{i=1,2}G_{-\frac{(v_i+2a)}{2}l }\prod_{i=3,4,5}G_{-(v_i+2a)l}\;.\nonumber\\
\end{eqnarray}\\

\item When $4\nmid q-1$:\\
\begin{eqnarray}
N_{\mathrm{mon}}
& = &
\frac{1}{q}\sum_{b=0}^{1}\sum_{\textbf{v}}\lambda_{\textbf{v}}\prod_{i=1,2}G_{-\frac{(v_i+4b)}{4}m }\prod_{i=3,4,5}G_{-\frac{v_i}{2}m}\;.\nonumber\\
\end{eqnarray}
\end{enumerate}

\subsection[Summary for $\phi=0,\,\psi\neq 0$]{Summary for $\phi=0,\,\psi\neq 0$}
\begin{enumerate}
\item When $8|q-1$:\\
\begin{eqnarray}
N_{\mathrm{mon}}
& = &
\frac{1}{q(q-1)}\sum_{w=0}^{q-2}\sum_{\textbf{v}}\lambda_{\textbf{v}}\prod_{i=1,2}G_{-(v_ik +w)}\prod_{i=3,4,5}G_{-(2v_ik+2w)} G_{8w}\mathrm{Teich}^{w}\left(\frac{1}{(8\psi)^8}\right)\nonumber\\
&&-\frac{1}{q}\sum_{b=0}^{1}\sum_{a=0}^{3}\sum_{\textbf{v}}\lambda_{\textbf{v}}\prod_{i=1,2}G_{-(v_i+a+4b)k }\prod_{i=3,4,5}G_{-2(v_i+a)k}\;.\nonumber\\
\end{eqnarray}\\

\item When $4|q-1,\,8\nmid q-1$:\\
\begin{eqnarray}
N_{\mathrm{mon}}
& = &
\frac{1}{q(q-1)}\sum_{w=0}^{q-2}\sum_{\textbf{v}}\lambda_{\textbf{v}}
\prod_{i=1,2}G_{-(\frac{v_i}{2}l +w)}\prod_{i=3,4,5}G_{-(v_il+2w)} G_{8w}\mathrm{Teich}^{w}\left(\frac{1}{(8\phi)^8}\right)\nonumber\\
& & -\frac{1}{q}\sum_{a=0}^{3}\sum_{\textbf{v}}\lambda_{\textbf{v}}\prod_{i=1,2}G_{-\frac{(v_i+2a)}{2}l }\prod_{i=3,4,5}G_{-(v_i+2a)l}\;.\nonumber\\
\end{eqnarray}\\

\item When $4\nmid q-1$:\\
\begin{eqnarray}
N_{\mathrm{mon}}
& = &
\frac{1}{q(q-1)}\sum_{w=0}^{q-2}\sum_{\textbf{v}}\lambda_{\textbf{v}}
\prod_{i=1,2}G_{-(\frac{v_i}{4}m +w)}\prod_{i=3,4,5}G_{-(\frac{v_i}{2}m+2w)} G_{8w}\mathrm{Teich}^{w}\left(\frac{1}{(8\phi)^8}\right)\nonumber\\
& & -\frac{1}{q}\sum_{b=0}^{1}\sum_{\textbf{v}}\lambda_{\textbf{v}}\prod_{i=1,2}G_{-\frac{(v_i+4b)}{4}m }\prod_{i=3,4,5}G_{-\frac{v_i}{2}m}\;.\nonumber\\
\end{eqnarray}
\end{enumerate}

\pagebreak

\subsection[Summary for $\psi=0,\,\phi\neq 0$]{Summary for $\psi=0,\,\phi\neq 0$}
\begin{enumerate}
\item When $8|q-1$:\\
\begin{eqnarray}
N_{\mathrm{mon}}
& = &
\frac{1}{q(q-1)}\sum_{u=0}^{q-2}\sum_{\textbf{v}}\lambda_{\textbf{v}}\sum_{a=0}^{3}\prod_{i=1,2}G_{-((v_i+a)k +u)}\prod_{i=3,4,5}G_{-2(v_i+a)k} G_{2u}\mathrm{Teich}^{u}\left(\frac{1}{(2\phi)^2}\right)\nonumber\\
& & -\frac{1}{q}\sum_{b=0}^{1}\sum_{a=0}^{3}\sum_{\textbf{v}}\lambda_{\textbf{v}}\prod_{i=1,2}G_{-(v_i+a+4b)k }\prod_{i=3,4,5}G_{-2(v_i+a)k}\;.\nonumber\\
\end{eqnarray}

\item When $4|q-1,\,8\nmid q-1$:\\
\begin{eqnarray}
N_{\mathrm{mon}}
& = &
\frac{1}{q(q-1)}\sum_{u=0}^{q-2}\sum_{\textbf{v}}\lambda_{\textbf{v}}
\sum_{a=0}^{1}\prod_{i=1,2}G_{-(\frac{v_i+2a}{2}l +u)}\prod_{i=3,4,5}G_{-(v_i+2a)l} G_{2u}\mathrm{Teich}^{u}\left(\frac{1}{(2\phi)^2}\right)\nonumber\\
& &
+\frac{1}{q(q-1)}\sum_{u=0}^{q-2}\sum_{\textbf{v}}\lambda_{\textbf{v}}
\sum_{a=0}^{1}\prod_{i=1,2}G_{-(\frac{v_i+2a+1}{2}l +u+\frac{1}{2})}\prod_{i=3,4,5}G_{-(v_i+2a+1)l} G_{2u+1}\nonumber\\
& & \times\mathrm{Teich}^{u}\frac{1}{(2\phi)^2}\mathrm{Teich}^{-1}\left(-2\phi\right)\nonumber\\
& &
-\frac{1}{q}\sum_{b=0}^{1}\sum_{a=0}^{1}\sum_{\textbf{v}}\lambda_{\textbf{v}}\prod_{i=1,2}G_{-\frac{(v_i+2a+4b)}{2}l }\prod_{i=3,4,5}G_{-(v_i+2a)l}\;.\nonumber\\
\end{eqnarray}

\item When $4\nmid q-1$:\\
\begin{eqnarray}
N_{\mathrm{mon}} & = &
\frac{1}{q(q-1)}\sum_{u=0}^{q-2}\sum_{\textbf{v}}\lambda_{\textbf{v}}
\prod_{i=1,2}G_{-(\frac{v_i}{4}m +u)}\prod_{i=3,4,5}G_{-\frac{v_i}{2}m} G_{2u}\mathrm{Teich}^{u}\left(\frac{1}{(2\phi)^2}\right)\nonumber\\
& &
+\frac{1}{q(q-1)}\sum_{u=0}^{q-2}\sum_{\textbf{v}}\lambda_{\textbf{v}}
\prod_{i=1,2}G_{-(\frac{v_i+2}{4}m +u+\frac{1}{2})}\prod_{i=3,4,5}G_{-\frac{v_i+2}{2}m} G_{2u+1}\nonumber\\
& &\times\mathrm{Teich}^{u}\frac{1}{(2\phi)^2}\mathrm{Teich}^{-1}\left(-2\phi\right)\nonumber\\
& & -\frac{1}{q}\sum_{b=0}^{1}\sum_{\textbf{v}}\lambda_{\textbf{v}}\prod_{i=1,2}G_{-\frac{(v_i+4b)}{4}m }\prod_{i=3,4,5}G_{-\frac{v_i}{2}m}\;.\nonumber\\
\end{eqnarray}

\end{enumerate}

\pagebreak

\subsection[Summary for $\phi\psi\neq 0$]{Summary for $\phi\psi\neq 0$}
\label{generalexp}
\begin{enumerate}
\item When $8|q-1$:\\
\begin{eqnarray}
N_{\mathrm{mon}} & = &
\frac{1}{q(q-1)^2}\sum_{t=0}^{q-2}\sum_{u=\left\lceil \frac{t}{2}\right\rceil}^{\left\lfloor \frac{2q-3+t}{2}\right\rfloor}\sum_{\textbf{v}}\lambda_{\textbf{v}}\prod_{i=1,2}G_{-(v_ik +u)}\prod_{i=3,4,5}G_{-(2v_ik+t)}\nonumber\\
& &
\times G_{2u-t}G_{4t}\mathrm{Teich}^{u}\left(\frac{1}{(2\phi)^2}\right)\mathrm{Teich}^{t}\left(\frac{-2\phi}{(8\psi)^4}\right)\nonumber\\
& &
-\frac{1}{q(q-1)}\sum_{u=0}^{q-2}\sum_{a=0}^{3}\sum_{\textbf{v}s.t.\atop v_3v_4v_5=0}\lambda_{\textbf{v}}\prod_{i=1,2}G_{-((v_i+a)k +u)}\prod_{i=3,4,5}G_{-2(v_i+a)k} G_{2u}\nonumber\\
& &\times\mathrm{Teich}^{u}\left(\frac{1}{(2\phi)^2}\right)\nonumber\\
& &
-\frac{1}{q}\sum_{b=0}^{1}\sum_{a=0}^{3}\sum_{\textbf{v}\atop v_3v_4v_5\neq 0\, \&\,(v_i+a+4b)\neq 0}\lambda_{\textbf{v}}\prod_{i=1,2}G_{-(v_i+a+4b)k }\prod_{i=3,4,5}G_{-2(v_i+a)k}\;.\nonumber\\
\end{eqnarray}

\item When $4|q-1,\,8\nmid q-1$:\\
\begin{eqnarray}
N_{\mathrm{mon}} & = &
\frac{1}{q(q-1)^2}\sum_{t=0}^{q-2}\sum_{u=\left\lceil \frac{t}{2}\right\rceil}^{\left\lfloor \frac{2q-3+t}{2}\right\rfloor}\sum_{\textbf{v}}\lambda_{\textbf{v}}\prod_{i=1,2}G_{-(\frac{v_i}{2}l +u)}\prod_{i=3,4,5}G_{-(v_il+t)}\nonumber\\
& &
\times G_{2u-t}G_{4t}\mathrm{Teich}^{u}\left(\frac{1}{(2\phi)^2}\right)\mathrm{Teich}^{t}\left(\frac{-2\phi}{(8\psi)^4}\right)\nonumber\\
& &
-\frac{1}{q(q-1)}\sum_{u=0}^{q-2}\sum_{a=0}^{1}\sum_{\textbf{v}s.t.\atop v_3v_4v_5=0}\lambda_{\textbf{v}}\prod_{i=1,2}G_{-(\frac{(v_i+2a)}{2}l +u)}\prod_{i=3,4,5}G_{-(v_i+2a)l} G_{2u}\mathrm{Teich}^{u}\left(\frac{1}{(2\phi)^2}\right)\nonumber\\
& &
-\frac{1}{q(q-1)}\sum_{u=0}^{q-2}\sum_{a=0}^{1}\sum_{\textbf{v}s.t.\atop v_3v_4v_5=0}\lambda_{\textbf{v}}\prod_{i=1,2}G_{-(\frac{(v_i+2a+1)}{2}l+u+ \frac{1}{2})}\prod_{i=3,4,5}G_{-(v_i+2a+1)l} G_{2u+1}\nonumber\\
& &\times\mathrm{Teich}^{u}\frac{1}{(2\phi)^2}\mathrm{Teich}^{-1}\left(-2\phi\right)\nonumber\\
& &
-\frac{1}{q}\sum_{b=0}^{1}\sum_{a=0}^{1}\sum_{\textbf{v}\atop v_3v_4v_5\neq 0\, \&\,(v_i+2a+4b)\neq 0}\lambda_{\textbf{v}}\prod_{i=1,2}G_{-\frac{(v_i+2a+4b)}{2}l }\prod_{i=3,4,5}G_{-(v_i+2a)l}\;.\nonumber\\
\end{eqnarray}

\pagebreak

\item When $4\nmid q-1$:\\
\begin{eqnarray}
N_{\mathrm{mon}} & = &
\frac{1}{q(q-1)^2}\sum_{t=0}^{\frac{q-3}{4}}\sum_{u=\left\lceil \frac{t}{2}\right\rceil}^{\left\lfloor \frac{2q-3+t}{2}\right\rfloor}\sum_{\textbf{v}}\lambda_{\textbf{v}}\prod_{i=1,2}G_{-(\frac{v_i}{4}m +u)}\prod_{i=3,4,5}G_{-(\frac{v_i}{2}m+t)}\nonumber\\
& &
\times G_{2u-t}G_{4t}\mathrm{Teich}^{u}\left(\frac{1}{(2\phi)^2}\right)\mathrm{Teich}^{t}\left(\frac{-2\phi}{(8\psi)^4}\right)\nonumber\\
& &
-\frac{1}{q(q-1)}\sum_{u=0}^{q-2}\sum_{\textbf{v}s.t.\atop v_3v_4v_5=0}\lambda_{\textbf{v}}\prod_{i=1,2}G_{-(\frac{v_i}{4}m +u)}\prod_{i=3,4,5}G_{-\frac{v_i}{2}m} G_{2u}\mathrm{Teich}^{u}\left(\frac{1}{(2\phi)^2}\right)\nonumber\\
& &
-\frac{1}{q(q-1)}\sum_{u=0}^{q-2}\sum_{\textbf{v}s.t.\atop v_3v_4v_5=0}\lambda_{\textbf{v}}\prod_{i=1,2}G_{-(\frac{v_i+2}{4}m +u+\frac{1}{2})}\prod_{i=3,4,5}G_{-\frac{v_i+2}{2}m} G_{2u+1}\nonumber\\
& & \times\mathrm{Teich}^{u}\frac{1}{(2\phi)^2}\mathrm{Teich}^{-1}\left(-2\phi\right)\nonumber\\
& &
-\frac{1}{q}\sum_{y\in\mathbb{F}_q^\ast}\sum_{b=0}^{1}\sum_{\textbf{v}\atop v_3v_4v_5\neq 0\, \&\,(v_i+4b)\neq 0}\lambda_{\textbf{v}}\prod_{i=1,2}G_{-\frac{(v_i+4b)}{4}m }\prod_{i=3,4,5}G_{-\frac{v_i}{2}m}\;.\nonumber\\
\end{eqnarray}
\end{enumerate}

The number of points over $\mathbb{F}_q$ depends on the value of
$q\, \mod\, 8$. Note that $8|p^2-1$ for all primes $p$. Hence it
can be concluded that:

\begin{enumerate}
\item If $p-1=2\,\mod\,4$, then $p^{2r}-1=0\;\mod\,8$ and  $p^{2r-1}-1=2\, \mod\,4 \quad\forall r\in\mathbb{N}$,
\item If $p-1=0\,\mod\,4$, then $p^{2r}-1=0\;\mod\,8$ and  $p^{2r-1}-1=0\, \mod\,4 \quad\forall r\in\mathbb{N}$,
\item If $p-1=0\,\mod\,8$, then $p^{r}-1=0\;\mod\,8 \quad\forall r\in\mathbb{N}$.
\label{threecases}\end{enumerate}

\noindent Hence the zeta function can be split into odd and even
powers:

\begin{eqnarray}
\zeta(\phi,\psi,p,t)=&\equiv &\exp\left(\sum_{r \in
   \mathbb{N}} N_{r,p}(X)\frac{t^r}{r}\right)\nonumber\\
   &=&
   \exp\left(\sum_{r \in\mathbb{N}} N_{2r+1,p}(X)\frac{t^{2r+1}}{2r+1}\right)\exp\left(\sum_{r \in\mathbb{N}} N_{2r,p}(X)\frac{t^{2r}}{2r}\right)\;.
\end{eqnarray}

We conclude that for the three cases above the zeta function takes
the form:

\begin{enumerate}
\item$\zeta(\phi,\psi,p,t)=R_m(p,t)R_{\acute{k}}(p,t)$,
\item$\zeta(\phi,\psi,p,t)=R_l(p,t)R_{\acute{k}}(p,t)$,
\item$\zeta(\phi,\psi,p,t)=R_k(p,t)R_{\acute{k}}(p,t)$,
\end{enumerate}

\noindent where $R_{\acute{k}}(p,t)$ is the contribution from the
odd powers and $R_m(p,t)$, $R_l(p,t)$, $R_k(p,t)$ are the
contributions from the even powers.

\section[Rearrangement]{Rearrangement}\label{sectionrearrangement}
In this section we briefly mention ways to rewrite the previous
formulae. This rewriting is useful for the purpose of writing
computer code for counting the number of rational points. Notice
that (when $8|q-1$ and $\phi\psi\neq 0$) the contribution to the
sum for which $t=2ak$ ($a=0,1,\ldots,3$) (or for the other cases
$t=2al$ and $t=2am$) in the first term reduces to a multiple of
the second term:

\begin{eqnarray}
\mathrm{First}|_{t=ak,\,\forall a}&=&\frac{-1}{q(q-1)^2}\sum_{a=0}^{3}\sum_{\textbf{v}}\lambda_{\textbf{v}}\sum_{u=\left\lceil ak\right\rceil}^{\left\lfloor \frac{2q-3}{2}+ak\right\rfloor}\sum_{\textbf{v}}\lambda_{\textbf{v}}\prod_{i=1,2}G_{-(v_ik +u)}\prod_{i=3,4,5}G_{-2(v_i+a)k}\nonumber\\
& &
\times G_{2(u-ak)}\mathrm{Teich}^{u-ak}\left(\frac{1}{(2\phi)^2}\right)\nonumber\\
&=&
-\frac{1}{q(q-1)^2}\sum_{w=0}^{q-2}\sum_{a=0}^{3}\sum_{\textbf{v}}\lambda_{\textbf{v}}\prod_{i=1,2}G_{-((v_i+a)k +w)}\prod_{i=3,4,5}G_{-2(v_i+a)k} G_{2w}\mathrm{Teich}^{w}\left(\frac{1}{(2\phi)^2}\right)\nonumber\\
&=& \mathrm{Second}\;, \label{combine}
\end{eqnarray}\\

\noindent where the substitution $w=u-ak$ ($w=u-al$ or $w=u-am$)
was made. Hence the first two terms may be combined for suitable
$\textbf{v}$ such that we define the first term as:

\begin{eqnarray}
F(\textbf{v},t)\sum_{t=0}^{q-2}\sum_{u=\left\lceil \frac{t}{2}\right\rceil}^{\left\lfloor \frac{2q-3+t}{2}\right\rfloor}\sum_{\textbf{v}}\lambda_{\textbf{v}}\prod_{i=1,2}G_{-(v_ik +u)}\prod_{i=3,4,5}G_{-(2v_ik+t)}\nonumber\\
\times G_{2u-t}G_{4t}\mathrm{Teich}^{u}\left(\frac{1}{(2\phi)^2}\right)\mathrm{Teich}^{t}\left(\frac{-2\phi}{(8\psi)^4}\right)\;,\nonumber\\
\end{eqnarray}

\noindent where

\begin{equation}
F(\textbf{v},t)=\begin{cases}\frac{1}{q(q-1)^2}\quad
\mathrm{if}\quad 2k\nmid t\;,\\\frac{-1}{q(q-1)^2}\quad
\mathrm{if}\quad 2k|t\quad\mbox{and}\quad v_3v_4v_5\neq0\;,\\
\frac{-1}{(q-1)^2}\quad
 \mathrm{if}\quad 2k|t\quad\mbox{and}\quad v_3v_4v_5=0\;.\end{cases}
\end{equation}

Also, when both $t=2ak$ and $k|u$, i.e.
$u=(a+4b)k,\,a=0,1,\ldots,3,\,b=0,1$ the (combined) first term
(\ref{combine}) reduces to:

\begin{eqnarray}
\frac{1}{(q-1)^2}\sum_{b=0}^{1}\sum_{a=0}^{3}\sum_{\textbf{v}}\lambda_{\textbf{v}}\prod_{i=1,2}G_{-(v_i+a+4b)k }\prod_{i=3,4,5}G_{-2(v_i+a)k}\;,\nonumber\\
\end{eqnarray}
which is a multiple of the final term involving no Teichm\"{u}ller
characters; hence this term can also be combined.

Overall we can write an expression of the following form:

\begin{eqnarray}
N_{\mathrm{mon},s} & = &
\sum_{\textbf{v}}\sum_{t}\sum_{u}\beta_{t,u,\textbf{v},s}\mathrm{Teich}^{u}(\mu)\mathrm{Teich}^{t}(\Psi)+\delta(8\nmid
q-1)\sum_{u}\gamma_{u,\textbf{v},s}\mathrm{Teich}^{u}(\mu)\mathrm{Teich}^{-1}(-2\phi)\;;\nonumber\\
\end{eqnarray}

\noindent to do this we arrange for:

\begin{equation}
\beta_{t,u,\textbf{v},s} =
\frac{q(-1)^{s+t+1}\pi^{S(\textbf{v})}\prod_{l=0}^{s-1}\Gamma_p\left(1-\left<\frac{p^l4t}{q-1}\right>\right)\prod_{l=0}^{s-1}\Gamma_p\left(1-\left<\frac{p^l(2u-t)}{q-1}\right>\right)\mathrm{Teich}^{u}(\mu)\mathrm{Teich}^{t}(\Psi)}{(q-1)
\prod_{i=1,2}\prod_{l=0}^{s-1}\Gamma_p\left(1-\left<p^l(\frac{v_ik+u}{q-1})\right>\right)
\prod_{i=3,4,5}\prod_{l=0}^{s-1}\Gamma_p\left(1-\left<p^l(\frac{2v_ik+t}{4k})\right>\right)}\;,
\end{equation}

\noindent which is obtained by using the Gross--Koblitz formula.
Note that

\begin{equation}
S(\textbf{v})=\sum_{i=1,2} S(v_ik+u)+\sum_{i=3,4,5}
S(2v_ik+t)-S(4t)-S(2u-t)\;,
\end{equation}

\noindent where $S$ is the sum of $p$-adic digits and $\pi^{p-1}=-p$.\\

For the case $2k|t$:

\begin{eqnarray}
\beta_{2ak,u,\textbf{v},s} & = & \frac{-q^{-\delta(z_3)}}{q-1}\sum_{r_{\phi}=0}^{q-2}G_{s,2(u-ak)}
\prod_{i=1,2}G_{s,-(v_ik+u)}\prod_{i=3,4,5}G_{s,-2(v_i+a)k}\nonumber\\
& = &
\frac{(-1)^{t+1}q^{1-\delta(z_3)}\pi^{-S(2(u-ak))+\sum_{i=1,2}S(v_ik+u)+\sum_{i=3,4,5}S(2(v_i+a)k)}\Gamma_p\left(1-\left<\frac{p^l2(u-ak)}{q-1}\right>\right)}{(q-1)\prod_{i=1,2}\prod_{l=0}^{s-1}\Gamma_p\left(1-\left<\frac{p^l(v_ik+u)}{q-1}\right>\right)
\prod_{i=3,4,5}\prod_{l=0}^{s-1}\Gamma_p\left(1-\left<\frac{p^l2(v_i+a)k}{q-1}\right>\right)}\;.\nonumber\\
\end{eqnarray}

For the case $2k|t$ and $4k|(u-ak)$:

\begin{eqnarray}
\beta_{2ak,(a+4b)k,\textbf{v},s}& = & \frac{q^{-\delta_3}-(q-1)^2q^{-1}\delta_3(1-\delta_2)}{q-1}\prod_{i=1,2}G_{s,-(v_i+a+4b)k}\prod_{i=3,4,5}G_{s,-2(v_i+a)k}\nonumber\\
&=&
\frac{(q^{-\delta_3}-(q-1)^2q^{-1}\delta_3(1-\delta_2))(-1)^{s+t+1}\pi^{\sum_{i=1,2}S((v_i+a+4b)k)+\sum_{i=3,4,5}S(2(v_i+a)k)}}{(q-1)\prod_{i=1,2}\prod_{l=0}^{s-1}\Gamma_p\left(1-\left<\frac{p^l(v_i+a+4b)k}{q-1}\right>\right)
\prod_{i=3,4,5}\prod_{l=0}^{s-1}\Gamma_p\left(1-\left<\frac{p^l2(v_i+a)k}{q-1}\right>\right)}\;.\nonumber\\
\end{eqnarray}

Similarly for the simpler cases of when either or both of $\psi$
and $\phi$ are zero we obtain the following expressions. Where
$k=\frac{q-1}{8}$,$l=\frac{q-1}{4}$, and $m=\frac{q-1}{2}$.

Simple applications of the Gross--Koblitz formula for the Gauss
sums yield similar expressions for the other cases. We shall omit
reproduction of all these formulae, as they are obtained
analogously.

\pagebreak

\section[The Mirror Octics]{The Mirror Octics}
\label{mirroroctic}

\begin{figure}[htbp]
\centering
\includegraphics[scale=1]{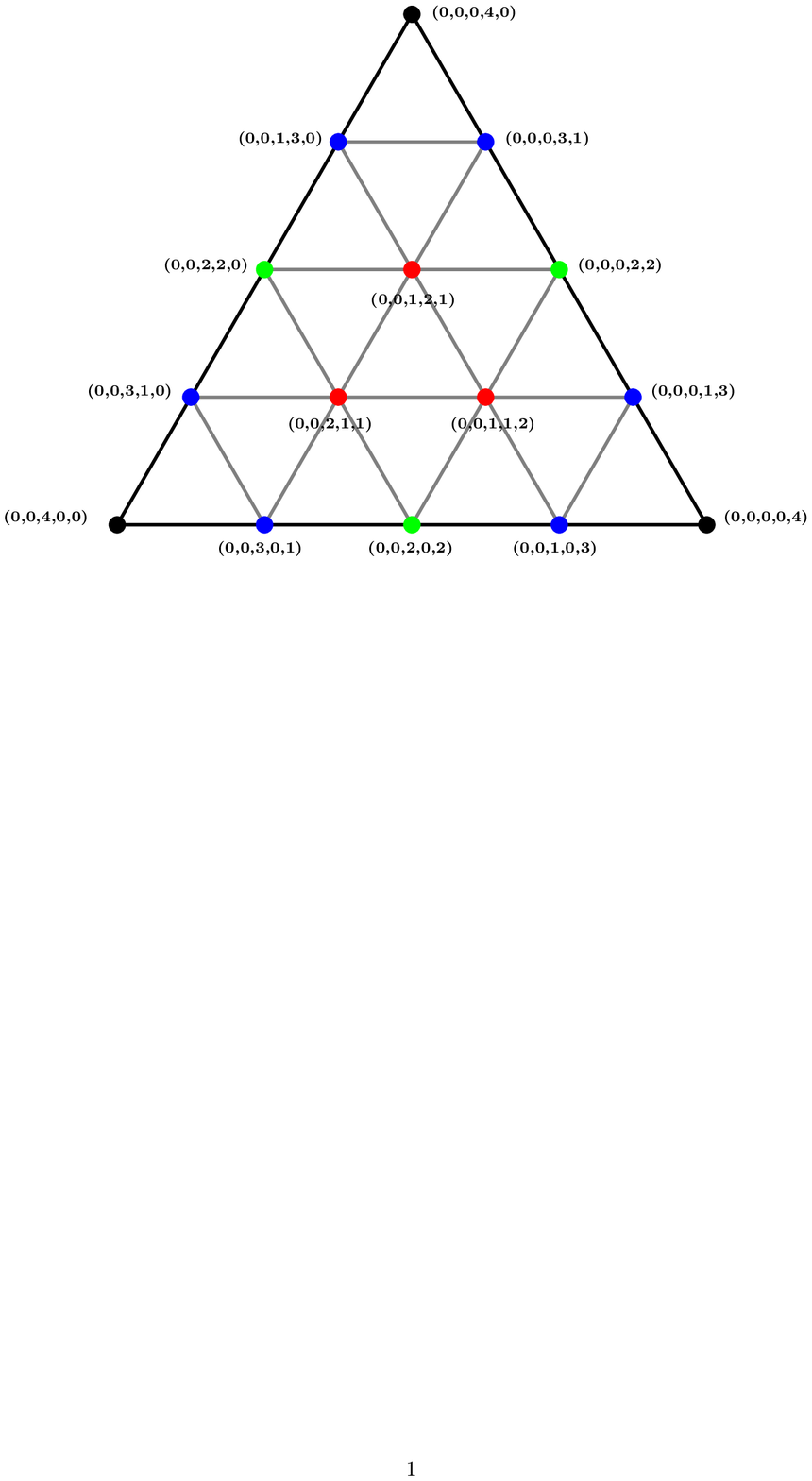}
\caption{The two-face with vertices $(0,0,4,0,0)$, $(0,0,0,4,0)$
and $(0,0,0,0,4)$.} \label{figtwoface1}
\end{figure}

\begin{figure}[htbp]
\centering
\includegraphics[scale=1]{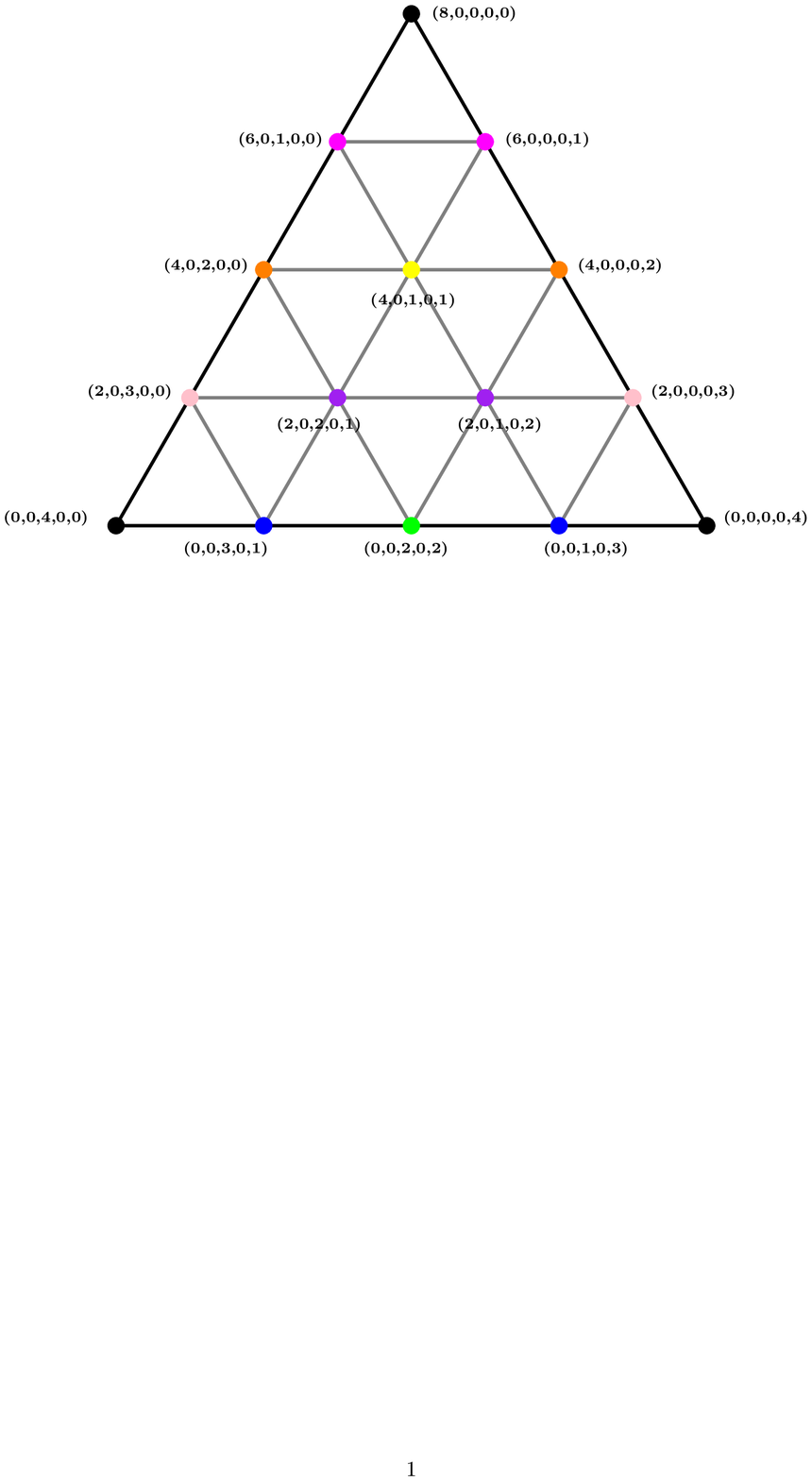}
\caption{A two-face with vertices $(8,0,0,0,0)$, $(0,0,4,0,0)$ and
$(0,0,0,0,4)$. } \label{figtwoface2}
\end{figure}

\begin{figure}[htbp]
\centering
\includegraphics[scale=1.2]{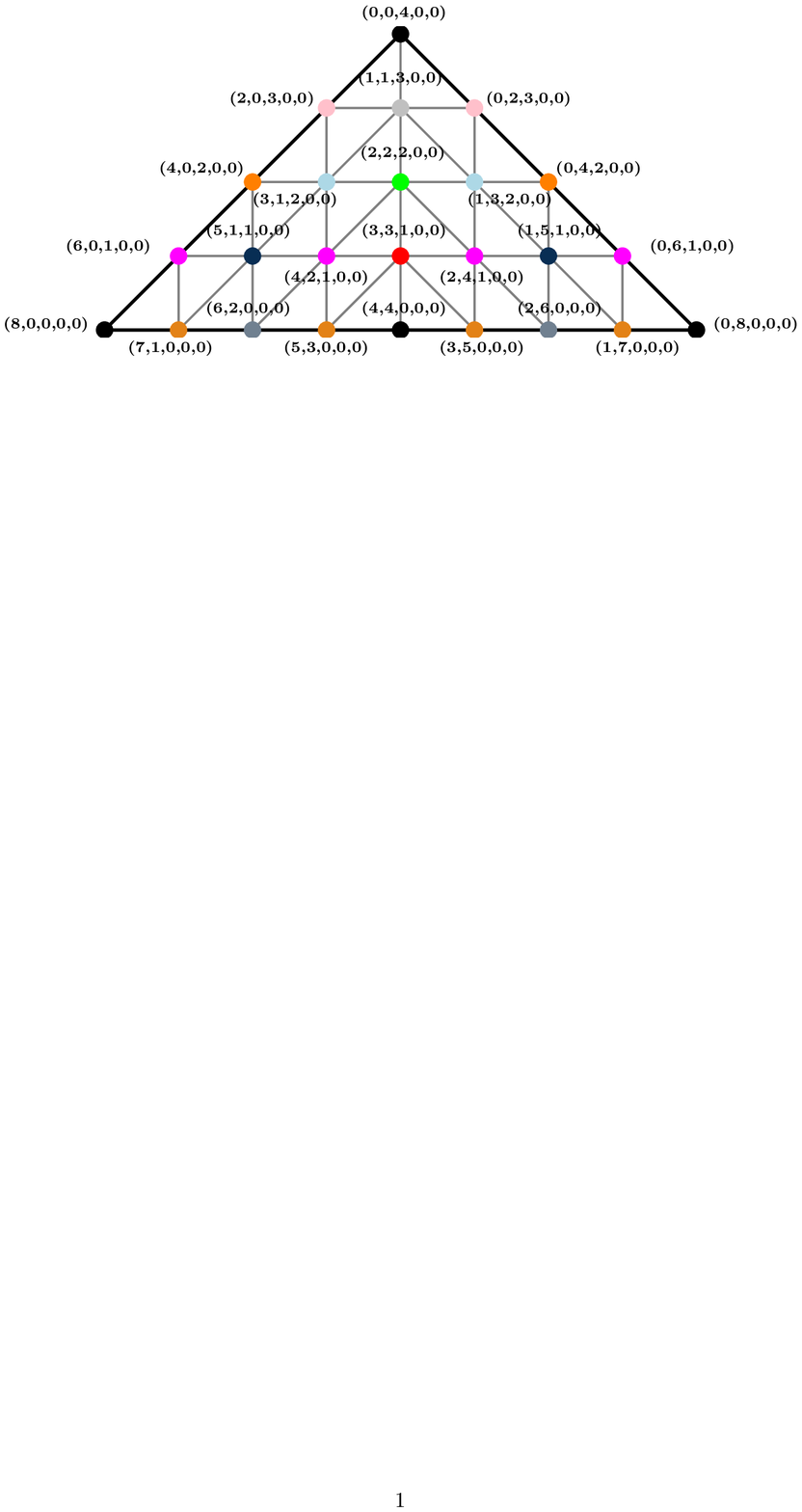}
\caption{A two-face with vertices $(8,0,0,0,0)$, $(0,8,0,0,0)$ and
$(0,0,4,0,0)$. } \label{figtwoface3}
\end{figure}

\begin{figure}[htbp]
\centering
\includegraphics[scale=1]{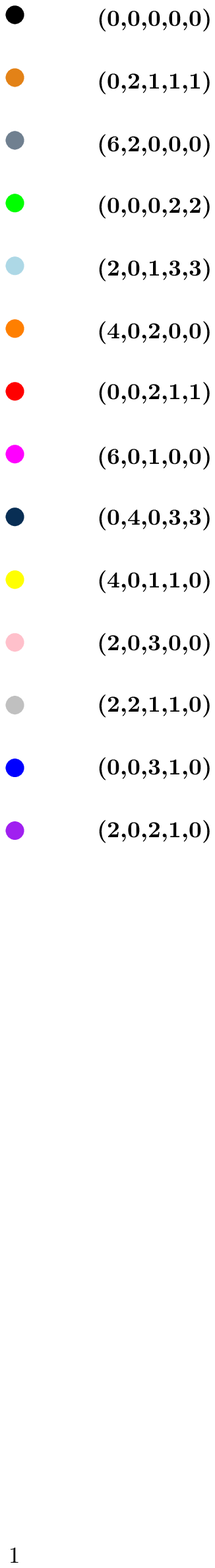}
\caption{Above is the key relating the $14$ colours in Figures
\ref{figtwoface1}, \ref{figtwoface2} and \ref{figtwoface3} to $14$
monomial classes. } \label{colourcode}
\end{figure}

The Batyrev mirror of the Octic is given by a hypersurface in the
toric variety,
$\frac{\mathbb{C}^{104}-F}{(\mathbb{C}^\ast)^{100}}$. We introduce
now $104$ Cox variables $\tilde{x}_\textbf{m}$ for the mirror, one
for each monomial $\textbf{m}$ in the Newton polyhedron (i.e. all
the degree $8$ monomials apart from the interior point
$(1,1,1,1,1)$). Let us write each monomial in the following way:
$(m_1, m_2,m_3,m_4,m_5,m_{\psi},m_{\phi})$, where
$m_{\phi}=\frac{m_1+m_2}{2}$ and
$m_{\psi}=\frac{m_1+m_2+2(m_3+m_4+m_5)}{8}$. Let us stack all
these monomials (with $(1,1,1,1,1,1,1)$ on the top) to form a
$105\times7$ matrix:

\begin{center}
$\begin{pmatrix}
1&1&1&1&1&1&1\\
8&0&0&0&0&1&4\\
0&8&0&0&0&1&4\\
0&0&4&0&0&1&0\\
0&0&0&4&0&1&0\\
0&0&0&0&4&1&0\\
4&4&0&0&0&1&4\\
7&1&0&0&0&1&4\\
\vdots&\vdots&\vdots&\vdots&\vdots&\vdots&\vdots\\
\end{pmatrix}$
\end{center}
Let the columns of this matrix be denoted
$(\textbf{w}_j,w_{\psi},w_{\phi})$, for $j=1,\ldots,5$, then the
mirror polynomial takes the form:

\begin{equation}
\tilde{P}(\tilde{x})=\tilde{x}^{w_1}+\tilde{x}^{w_2}+\tilde{x}^{w_3}+\tilde{x}^{w_4}+\tilde{x}^{w_5}-2\phi\tilde{x}^{w_{\phi}}-8\psi\tilde{x}^{w_{\psi}}\;.
\label{equationdefmirror}
\end{equation}
where $\tilde{x}^{w_j}:=\prod_{\textbf{m}}
\tilde{x}_\textbf{m}^{m_j}$. We can also define an extra
coordinate $\tilde{y}$ to be associated to the interior point.

\noindent We need to evaluate:
\begin{eqnarray}
q(q-1)^{100}\tilde{N} & = & \sum_{(\tilde{y},\tilde{\textbf{x}})\in(\mathbb{F}_q^{105}\diagdown F)}\Theta\left(\tilde{y}(\tilde{P}(\tilde{x})\right)\nonumber\\
&=&
q(q-1)^{100}(\tilde{N}(\tilde{y}=0)+\tilde{N}(\tilde{y}\neq 0))\;.\nonumber\\
\end{eqnarray}
A symmetric triangulation of the Newton polyhedron has been
chosen, as illustrated in \ref{figtwoface1}, \ref{figtwoface2},
and \ref{figtwoface3} (each colour represents one of the $14$
classes to which a monomial in the Newton polyhedron can belong).
The rule provided by toric geometry dictates that a subset of the
coordinates $\tilde{x}_{\textbf{m}}$ can be zero simultaneously if
and only if the corresponding \textbf{m}'s lie in the same cone.
Hence, as maximum of $3$ Cox variable are allowed to vanish at the
same time, and they must lie within the same `small triangle'.

First consider the case when none of the $105$ Cox variables
vanish:
\begin{eqnarray}
\sum_{(\tilde{y},\tilde{\textbf{x}})\in(\mathbb{F}_q^{\ast})^{105}}\Theta\left(\tilde{y}(\tilde{P}(\tilde{x})\right)
&=&
\sum_{(\tilde{y},\tilde{\textbf{x}})\in(\mathbb{F}_q^{\ast})^{105}}\Theta\left(\tilde{y}\left(\tilde{x}^{w_1}+\tilde{x}^{w_2}+\tilde{x}^{w_3}+\tilde{x}^{w_4}+\tilde{x}^{w_5}-2\tilde{\phi}\tilde{x}^{w_{\phi}}-8\tilde{\psi}\tilde{x}^{w_{\psi}}\right)\right)\nonumber\\
&=&
\sum_{(\tilde{y},\tilde{\textbf{x}})\in(\mathbb{F}_q^{\ast})^{105}}\Theta\left(\tilde{y}\tilde{x}^{w_1}\right)\Theta\left(\tilde{y}\tilde{x}^{w_2}\right)\Theta\left(\tilde{y}\tilde{x}^{w_3}\right)\Theta\left(\tilde{y}\tilde{x}^{w_4}\right)\Theta\left(\tilde{y}\tilde{x}^{w_5}\right)\nonumber\\
& &
\times\Theta\left(-2\phi\tilde{y}\tilde{x}^{w_{\phi}}\right)\Theta\left(-8\psi\tilde{y}\tilde{x}^{w_{\psi}}\right)\nonumber\\
&=&
\frac{(q-1)^{105}}{(q-1)^{7}}\sum_{0\leq s_i\leq q-2,\atop q-1|\sum_{i}s_im_i} \prod_{i=1}^{5}G_{-s_i}G_{s_{\phi}}G_{s_{\psi}}\mathrm{Teich}^{-s_{\psi}}(-8\psi)\mathrm{Teich}^{-s_{\phi}}(-2\phi)\;.\nonumber\\
\end{eqnarray}
Notice that here we are required to solve the constraints:
\begin{equation}
q-1|
(s_1m_1+s_2m_2+s_3m_3+s_4m_4+s_5m_5-s_{\phi}m_{\phi}-s_{\psi}m_{\psi})\quad
\forall \textbf{m}\;,
\end{equation}
where $\textbf{m}\equiv(m_1,m_2,m_3,m_4,m_5,m_{\psi},m_{\phi})$,
and $m_{\psi}=1$ and $m_{\phi}=\frac{m_1+m_2}{2}$.

\noindent Hence define a vector $(v)_{\textbf{m}}$ of length $105$
indexed by $\textbf{m}$, such that:
\begin{equation}
v_{\textbf{m}}(q-1)=\textbf{m}.\textbf{s}\;, \label{connie}
\end{equation}
where $\textit{s}\equiv(s_1,s_2,s_3,s_4,s_5,-s_{\psi},-s_{\phi})$.
We shall rename the $v$'s associated to the five vertex monomials
as follows:
\begin{eqnarray}
v_{(8,0,0,0,0,1,4)}\equiv v_1\;,\nonumber\\
v_{(0,8,0,0,0,1,4)}\equiv v_2\;,\nonumber\\
v_{(0,0,4,0,0,1,0)}\equiv v_3\;,\nonumber\\
v_{(0,0,0,4,0,1,0)}\equiv v_4\;,\nonumber\\
v_{(0,0,0,0,4,1,0)}\equiv v_5\;.\nonumber\\
\end{eqnarray}
It is clear that from applying (\ref{connie}) to the monomials,
$(8,0,0,0,0,1,4)$, $(0,8,0,0,0,1,4)$, $(0,0,4,0,0,1,0)$,
$(0,0,0,4,0,1,0)$, and $(0,0,0,4,0,1,0)$ we obtain:
\begin{eqnarray}
s_1=\frac{v_1(q-1)}{8}+\frac{s_{\phi}}{2}+\frac{s_{\psi}}{8}\;,\nonumber\\
s_2=\frac{v_2(q-1)}{8}+\frac{s_{\phi}}{2}+\frac{s_{\psi}}{8}\;,\nonumber\\
s_3=\frac{v_3(q-1)}{4}+\frac{s_{\psi}}{4}\;,\nonumber\\
s_4=\frac{v_4(q-1)}{4}+\frac{s_{\psi}}{4}\;,\nonumber\\
s_5=\frac{v_5(q-1)}{4}+\frac{s_{\psi}}{4}\;.\nonumber\\
\end{eqnarray}
By using that fact that, by the definition of the Cox variables
for this family of Calabi--Yau manifolds,
$m_1+m_2+2(m_3+m_4+m_5)=8$, and (\ref{connie}), it is easy to show
that:
\begin{equation}
v_{\textbf{m}}=\frac{(m_1v_1+m_2v_2+2(m_3v_3+m_4v_4+m_5v_5))}{8},\quad
\forall\, \textbf{m}\;.
\end{equation}
It is easy to check that the only solutions to these numerous
constraints is to have
\begin{equation}
(v_1,v_2,v_3,v_4,v_5)=(1,1,1,1,1)a,\quad a=0,1,\dots,7;
\end{equation}
where the last three digits are to be considered mod $4$;

\noindent hence,
\begin{eqnarray}
\sum_{(\tilde{y},\tilde{\textbf{x}})\in(\mathbb{F}_q^{\ast})^{105}}\Theta\left(\tilde{y}(\tilde{P}(\tilde{x})\right)
&=&
(q-1)^{98}\sum_{a=0}^{7} G_{-\left(\frac{a(q-1)}{8}+\frac{s_{\phi}}{2}+\frac{s_{\psi}}{8}\right)}^2G_{-\left(\frac{a(q-1)}{4}+\frac{s_{\psi}}{4}\right)}^3G_{s_{\phi}}G_{s_{\psi}}\nonumber\\
& &
\times\mathrm{Teich}^{-s_{\psi}}(-8\psi)\mathrm{Teich}^{-s_{\phi}}(-2\phi)\;.\nonumber\\
\end{eqnarray}
Notice that the above expression is almost equal to the
contribution to the number of points from the class of monomials,
$(0,0,0,0,0)$, in the original family of octic threefolds.

Now we need to consider the case when only monomials on the
two-face in Figure \ref{figtwoface1} with vertices, $(0,0,4,0,0)$,
$(0,0,0,4,0)$, and $(0,0,0,0,4)$, are allowed to vanish.

To begin with, consider the case $\tilde{x}_{(0,0,4,0,0)}=0$. Here
the defining equation reduces to:
\begin{equation}
\tilde{P}=\tilde{x}^{w_1}+\tilde{x}^{w_2}+\tilde{x}^{w_4}+\tilde{x}^{w_5}-2\phi\tilde{x}^{w_{\phi}}\;,
\end{equation}
and so:
\begin{eqnarray}
\sum_{(\tilde{y},\tilde{\textbf{x}})\in(\mathbb{F}_q^{104}\diagdown F),\atop\tilde{x}_{(0,0,4,0,0)}=0}\Theta\left(\tilde{y}(\tilde{P}(\tilde{x})\right)
&=&
(q-1)^{103}+(q-1)^{99}
\sum_{0\leq s_i\leq q-2,\atop q-1|\sum_{i}s_im_i} \prod_{i=1}^{5}G_{-s_i}G_{s_{\phi}}\mathrm{Teich}^{-s_{\phi}}(-2\phi)\;.\nonumber\\
\end{eqnarray}
The first term on the right hand side is the contribution to the
sum from when $\tilde{y}=0$. The second term was obtained from the
straightforward expansion of:
\begin{eqnarray}
\sum_{(\tilde{y},\tilde{\textbf{x}})\in(\mathbb{F}_q^{\ast})^{104}}\Theta\left(\tilde{y}\tilde{x}^{w_1}\right)\Theta\left(\tilde{y}\tilde{x}^{w_2}\right)\Theta\left(\tilde{y}\tilde{x}^{w_4}\right)\Theta\left(\tilde{y}\tilde{x}^{w_5}\right)
\Theta\left(-2\phi\tilde{y}\tilde{x}^{w_{\phi}}\right)\nonumber\\
\end{eqnarray}
in terms of Gauss sums.

We have constraints identical to (\ref{connie}), except that now
there is no $s_{\phi},s_3$ or $m_{\phi},m_3$:
\begin{eqnarray}
v_{\textbf{m}}(q-1)=\textbf{m}.\textbf{s}\quad \forall\,\textbf{m}\nonumber\\
q-1|\;(m_1s_1+m_2s_2+m_4s_4+m_5s_5-m_{\phi}s_{\phi})\;.\nonumber\\
\end{eqnarray}
This time the only solutions to the above constraints are
\begin{equation}
(v_1,v_2,v_3,v_4,v_5)=(0,0,0,0,0) \&\, (4,4,0,0,0).
\end{equation}
Hence we can rewrite the second term:
\begin{equation}
(q-1)^{99}
\sum_{a=0,4} G_{-\left(\frac{(q-1)}{8}a+\frac{s_{\phi}}{2}\right)}^2G_0^2G_{s_{\phi}}\mathrm{Teich}^{-s_{\phi}}(-2\phi);\\
\end{equation}
using the fact that $G_0=-1$, we obtain:
\begin{equation}
\sum_{(\tilde{y},\tilde{\textbf{x}})\in(\mathbb{F}_q^{104}\diagdown F),\atop\tilde{x}_{(0,0,4,0,0)}=0}\Theta\left(\tilde{y}(\tilde{P}(\tilde{x})\right)
=
(q-1)^{103}+
(q-1)^{99}
\sum_{a=0,1} G_{-\left(\frac{(q-1)}{2}a+\frac{s_{\phi}}{2}\right)}^2G_{s_{\phi}}\mathrm{Teich}^{-s_{\phi}}(-2\phi)\;.\\
\end{equation}

Now we can add up all the contributions from the various different
combinations of allowable vanishings on this two-face. In each
case, the contribution will have the form:
\begin{equation}
C(q-1)^{A}\pm
C(q-1)^{B}
\sum_{a=0,1} G_{-\left(\frac{(q-1)}{2}a+\frac{s_{\phi}}{2}\right)}^2G_{s_{\phi}}\mathrm{Teich}^{-s_{\phi}}(-2\phi)\;,\\
\end{equation}

\noindent where C can be thought of as the multiplicity of the
type of points under consideration, e.g. for the case when there
is only one variable vanishing and these are vertices, $C=3$, as
there are $3$ vertices.

\noindent The values of A,B,C are tabulated below:

\begin{center}
\begin{tabular}{|c|c|c|c|c|c|}\hline
Number of Points & Subsidiary condition & A& B &Sign& Multiplicity C \\\hline\hline
Point & Vertex & $103$ & $99$& $+$ &3\\ \hline
 & Interior to Edge & $103$ & $100$ &$-$&9\\ \hline
 & Interior to 2-Face & $103$ & $101$& $+$&3\\ \hline\hline
 Link & Edge Link & $102$ & $99$&$-$&12 \\ \hline
 &Interior Links &$102$ & $100$ &$+$&18\\ \hline\hline
Triangle & & $101$ & $99$&$+$ &16\\ \hline
\end{tabular}
\end{center}


\noindent We now have to consider cases where vanishing can take place on the other two-faces.\\

\noindent
\begin{tabular}{|c|c|c|}\hline
\# of Points & Subsidiary condition & Contribution
\\\hline\hline Point & Vertex, e.g. $(8,0,0,0,0)$&
$2(q-1)^{103}+2(q-1)^{100}\sum G_{-s_1}G_{-s_3}G_{-s_4}G_{-s_5}$\\
\hline
 & Interior to Edge & $18(q-1)^{103}+2(q-1)^{101}\sum G_{-s_1}G_{-s_3}G_{-s_4}$\\
 &e.g. $(7,1,0,0,0)$ & $7(q-1)^{103}+7(q-1)^{101}\sum G_{-s_3}G_{-s_4}G_{-s_5}$\\ \hline
 & Interior to 2-Face (Type 2)& $18(q-1)^{103}+2(q-1)^{102}\sum G_{-s_1}G_{-s_3}$\\
 &e.g. $(1,1,3,0,0)$ & $27(q-1)^{103}+27(q-1)^{102}\sum G_{-s_3}G_{-s_4}$\\ \hline\hline
 Link & Edge Link & $24(q-1)^{102}+24(q-1)^{100}\sum G_{-s_1}G_{-s_3}G_{-s_4}$\\ \hline
&Interior Link & $108(q-1)^{102}+108(q-1)^{101}\sum G_{-s_1}G_{-s_3}$ \\
& & $8(q-1)^{102}+108(q-1)^{100}\sum G_{-s_3}G_{-s_4}G_{-s_5}$ \\
& & $120(q-1)^{102}+120(q-1)^{101}\sum G_{-s_3}G_{-s_4}$ \\ \hline\hline
Triangle & & $96(q-1)^{101}+96(q-1)^{100}\sum G_{-s_1}G_{-s_3}$\\
& & $96(q-1)^{101}+96(q-1)^{100}\sum G_{-s_3}G_{-s_4}$ \\ \hline
\end{tabular} \\

It turns out that all the $s_i$'s in the above expressions have to
vanish (again due to a large number of constraints similar to
(\ref{connie})). Altogether we get the following contribution:

\begin{eqnarray}
72(q-1)^{103}+26(q-1)^{102}+192(q-1)^{101}+ 2(q-1)^{100}\sum G_{-s_1}G_{-s_3}G_{-s_4}G_{-s_5}\nonumber\\
+6(3q+1)(q-1)^{100}\sum G_{-s_1}G_{-s_3}G_{-s_4}+(7q+1)(q-1)^{100}\sum G_{-s_3}G_{-s_4}G_{-s_5}\nonumber\\
+3(9q^2+22q+1)(q-1)^{100}\sum
G_{-s_3}G_{-s_4}+6(1+12q+3q^2)(q-1)^{100}\sum
G_{-s_1}G_{-s_3}\;.\nonumber\\
\end{eqnarray}
Hence, using the fact that $G_0=-1$ we obtain:

\begin{eqnarray}
\tilde{N} &=&
\frac{1}{(q-1)^{2}}\sum_{a=0}^{7} G_{-\left(\frac{a(q-1)}{8}+\frac{s_{\phi}}{2}+\frac{s_{\psi}}{8}\right)}^2G_{-\left(\frac{a(q-1)}{4}+\frac{s_{\psi}}{4}\right)}^3G_{s_{\phi}}G_{s_{\psi}}\nonumber\\
& &
\times\mathrm{Teich}^{-s_{\psi}}(-8\psi)\mathrm{Teich}^{-s_{\phi}}(-2\phi)\nonumber\\
&&
+\frac{1}{(q-1)}(1+3q+3q^2)\sum_{a=0,1} G_{-\left(\frac{(q-1)}{2}a+\frac{s_{\phi}}{2}\right)}^2G_{s_{\phi}}\mathrm{Teich}^{-s_{\phi}}(-2\phi)\nonumber\\
&& -2+80q+83q^2+q^3\;.
\end{eqnarray}

Perusal of Section (\ref{generalexp}) shows that the contribution
to the number of points from $(0,0,0,0,0)$ to the expression for
the original family of octics $N_{(0,0,0,0,0)}$ is given by:

\begin{eqnarray}
N_{(0,0,0,0,0)} &=&
    \frac{1}{(q-1)^{2}}\sum_{a=0}^{7} G_{-\left(\frac{a(q-1)}{8}+\frac{s_{\phi}}{2}+\frac{s_{\psi}}{8}\right)}^2G_{-\left(\frac{a(q-1)}{4}+\frac{s_{\psi}}{4}\right)}^3G_{s_{\phi}}G_{s_{\psi}}\nonumber\\
& &
\times\mathrm{Teich}^{-s_{\psi}}(-8\psi)\mathrm{Teich}^{-s_{\phi}}(-2\phi)\nonumber\\
&&
+\frac{1}{(q-1)}\sum_{a=0,1} G_{-\left(\frac{(q-1)}{2}a+\frac{s_{\phi}}{2}\right)}^2G_{s_{\phi}}\mathrm{Teich}^{-s_{\phi}}(-2\phi)\;;\nonumber\\
\end{eqnarray}
hence we can rewrite the total in the form:
\begin{eqnarray}
\tilde{N}&=&    -2+80q+83q^2+q^3 + N_{(0,0,0,0,0)} +\frac{3(q+1)}{(q-1)}\sum_{a=0,1}\sum_{s_{\phi}=0}^{q-2} G_{-\left(\frac{(q-1)}{2}a+\frac{s_{\phi}}{2}\right)}^2G_{s_{\phi}}\mathrm{Teich}^{-s_{\phi}}(-2\phi)\nonumber\\
&=&
-2+80q+83q^2+q^3 + N_{(0,0,0,0,0)} +\frac{3(q+1)}{(q-1)}\sum_{u=0}^{q-2} G_{-u}^2G_{2u}\mathrm{Teich}^{u}\left(\frac{1}{(2\phi)^2}\right)\;.\nonumber\\
\label{almostthere}
\end{eqnarray}
It turns out that the final term in (\ref{almostthere}) is related
to the number of points on a Calabi--Yau manifold of dimension
zero, which is described in \cite{COV1}. This Calabi--Yau is
defined as the points which satisfy the following equation in
$\mathbb{F}_p\mathbb{P}^1$ (projective space with coordinates in
$\mathbb{F}_p$):
\begin{eqnarray}
P(x)&=&x_1^2+x_2^2-2\phi x_1x_2\nonumber\\
&=&(x_1-\phi x_2)^2-(\phi^2-1)x_2^2\;.\nonumber\\
\end{eqnarray}
The rewriting of the equation in the second line makes it clear
that the number of points over $\mathbb{F}_p\mathbb{P}^1$ is given
by:
\begin{align}
N(\phi)=1+\left(\frac{\phi^2-1}{p}\right)=\begin{cases}1,\quad
\text{if}\quad \phi^2-1 \equiv 0\, \mod\,p\;,\\2,\quad
\text{if}\quad \phi^2-1\quad \text{is a non-zero sqaure residue of
p}\;,\\0,\quad \text{if}\quad \phi^2-1\quad \text{is a non-sqaure
residue of p}\;,\\\end{cases} \label{legendrezeroCY}
\end{align}
where $\left(\frac{a}{p}\right)$ is the Legendre symbol.

Now computing $N$ in the usual way gives:
\begin{equation}
N=1-\frac{1}{q}+\frac{1}{q(q-1)}\sum_{u=0}^{q-2}G_u^2G_{2u}\mathrm{Teich}^u\left(\frac{1}{(2\phi)^2}\right)\;.
\label{gausszeroCY}
\end{equation}
Equating (\ref{legendrezeroCY}) and (\ref{gausszeroCY}) gives us:
\begin{equation}
\sum_{u=0}^{q-2}G_u^2G_{2u}\mathrm{Teich}^u\left(\frac{1}{(2\phi)^2}\right)=(q-1)\left(1+q\left(\frac{\phi^2-1}{p}\right)\right)\;,
\end{equation}
Generalizing to $\mathbb{F}_{p^s}$  we obtain:
\begin{eqnarray}
\tilde{N} &=& 1+83q+83q^2+q^3 +
N_{(0,0,0,0,0)}+3q(q+1)\left(\frac{\phi^2-1}{p}\right)^s\;.\nonumber\\
\end{eqnarray}
%
This leads to a zeta function of the form:
\begin{equation}
\tilde{\zeta}(\phi,\psi)=\frac{R_{\textbf{(0,0,0,0,0)}}}{(1-t)(1-pt)^{83}(1-p^2t)^{83}\left(1-\left(\frac{\phi^2-1}{p}\right)pt\right)^3\left(1-\left(\frac{\phi^2-1}{p}\right)p^2t\right)^3(1-p^3t)}\;.
\end{equation}

The appearance of the number $83$ almost certainly has a deeper
significance, because for the mirror octics,
$H^{1,1}_{\mathrm{toric}}=83$, whereas $H^{1,1}=86$, as explained
in Chapter \ref{chapmirrorsymmetry}. This phenomenon shall be
discussed in more detail in Section \ref{summarymirrorzeta}.


\section[Some Combinatorics]{Some Combinatorics}
\label{combin} From the previous computations it is clear that
combinatorics plays an important r\^{o}le in the decomposition of
the zeta function into parts associated to a particular monomials.
To examine this carefully we study the Calabi--Yau $(n-3)$-fold, a
hypersurface in $\mathbb{P}^n$ defined by the following equation:

\begin{equation}
P(\textbf{x})=\sum_{i=1}^{n}x_i^n-n\psi\prod_{i=1}^{n}x_i\;.
\end{equation}
An analysis very similar to that for the quintic three-folds
\cite{COV1} and the K3's, leads to seeking monomials of length n,
$\textbf{v}=(v_1,v_2,v_3,...,v_n), \quad v_i\leq(n-1)$, which add
up to a multiple of $n$, $\sum_{i=1}^{n}v_i=rn$, where
$r=0,1,2,...,(n-1)$. However we wish to group these monomials into
equivalence classes that are related by addition of the monomial,
$(1,1,1,...,1)$, and taking the result mod $n$ and permutations.

We would like to find all the different classes of monomials. We
can associate to each monomial $\textbf{v}=(v_1,v_2,v_3,...,v_n)$,
a partition of $rn$, $0^{a_0}1^{a_1}2^{a_2}...(n-1)^{a_{n-1}}$,
where exactly $a_k$ of the $v_i$ take the value $i$: e.g. for
$n=7$ we have a class:

\begin{center}
\begin{tabular}{|c|ccccccc|}\hline
&\multicolumn{7}{|c|}{$a_i$ for each $i$}\\
 Monomial $\textbf{v}$     & 0&1&2&3&4&5&6\\ \hline
(5,2,0,0,0,0,0)&5&1&0&0&0&1&0\\
(6,3,1,1,1,1,1)&0&5&1&0&0&0&1\\
(0,4,2,2,2,2,2)&1&0&5&1&0&0&0\\
(1,5,3,3,3,3,3)&0&1&0&5&1&0&0\\
(2,6,4,4,4,4,4)&0&0&1&0&5&1&0\\
(3,0,5,5,5,5,5)&0&0&0&1&0&5&1\\
(4,1,6,6,6,6,6)&1&0&0&0&1&0&5\\\hline
\end{tabular}
\end{center}
It is easy to see that cyclic interchange of the $a_i$ lead to
monomials in the same class.

For the two-parameter family of octics we needed to find
$\textbf{v}$, such that:

\begin{eqnarray}
v_1+v_2+2(v_3+v_4+v_5)&=& 8v_{\psi}\nonumber\\
\frac{v_1+v_2}{2}&=&v_{\phi}\nonumber\\
&0\leq v_i\leq7,\,&i=1,2\nonumber\\
&0\leq v_i\leq3,\,&i=3,4,5\nonumber\\
\end{eqnarray}

These are precisely the following monomial classes:
\begin{center}
\begin{tabular}{|c|cccccccc|cccc|}\hline
   &\multicolumn{8}{|c|}{$a_i$ for $(v_1,v_2)$}& \multicolumn{4}{|c|}{$a_i$ for $(v_3,v_4,v_5)$}\\
Monomial $\textbf{v}$   &0&1&2&3&4&5&6&7& 0&1&2&3\\ \hline
(0,0,0,0,0)&2&0&0&0&0&0&0&0& 3&0&0&0\\
(7,1,0,0,0)&0&1&0&0&0&0&0&1& 3&0&0&0\\
(6,2,0,0,0)&0&0&1&0&0&0&1&0& 3&0&0&0\\\hline
(0,0,0,2,2)&2&0&0&0&0&0&0&0& 1&0&2&0\\
(3,1,2,0,0)&0&1&0&1&0&0&0&0& 2&0&1&0\\
(4,0,2,0,0)&1&0&0&0&1&0&0&0& 2&0&1&0\\
(0,0,2,1,1)&2&0&0&0&0&0&0&0& 0&2&1&0\\
(6,0,1,0,0)&1&0&0&0&0&0&1&0& 2&1&0&0\\
(5,1,1,0,0)&0&1&0&0&0&1&0&0& 2&1&0&0\\
(4,0,1,1,0)&1&0&0&0&1&0&0&0& 1&2&0&0\\
(2,0,3,0,0)&1&0&1&0&0&0&0&0& 2&0&1&0\\
(1,1,3,0,0)&0&2&0&0&0&0&0&0& 2&0&0&1\\\hline
(0,0,3,1,0)&2&0&0&0&0&0&0&0& 1&1&0&1\\
(2,0,2,1,0)&1&0&1&0&0&0&0&0& 1&1&1&0\\
(7,3,2,1,0)&0&0&0&1&0&0&0&1& 1&1&1&0\\\hline
\end{tabular}
\end{center}

The integers $a_i$ are given for each class of monomials up to
cyclic interchange. The integers $(v_1,v_2)$ and $(v_3,v_4,v_5)$
can take values $0,1,\ldots, 7$ and $0,1,\ldots,3$ respectively.


\chapter[Zeta Functions]{Zeta Functions}\label{chapzeta}
\label{zeta} In this chapter we shall present and discuss the zeta
functions obtained from the formulae derived in the previous
chapter. The Gross--Koblitz formula (\ref{grosskoblitz}) was used
repeatedly to expand the Gauss sums in terms of $p$-adic Gamma
functions. The computation being extremely heavy, this task was
undertaken by writing extensive Mathematica code.

For the octic threefolds results for the number of points over
$\mathbb{F}_{p^r}$ are known for up to $r=2$ for the primes up to
$73$. For the primes up to $23$ the number of points could be
computed up to $r=3$. For smaller primes higher values of $r$ were
computable, e.g. for $p=3$, we computed up to $r=8$. To compute
the zeta function fully for larger primes would require
considerably more computing power and memory than currently
typically available top of the range desktop machines. It turns
out that calculation up to $r=3$ is usually sufficient for
determining the full zeta function. In the Appendix we record the
zeta functions found for primes from $3$ to $17$. The results for
$19$ and $23$ are not included as they would take up far too much
space, and inclusion would not be significantly more illuminating.
In both cases below, the K3 surfaces and the octic threefolds, $2$
is the only prime of bad reduction.

An outline of this chapter is as follows: firstly, in Section
\ref{sectionk3zeta} we present the zeta functions obtained for the
much simpler case of the quartic K3 surfaces. Discussion of the
zeta functions for octic threefolds begins in Section
\ref{octicdiscussion}. Speculations concerning relations to Siegel
modular forms are given in Section \ref{siegel}. A summary of the
results for the mirror is given in Section
\ref{summarymirrorzeta}.

\section[Zeta Function of a Family of K3 Surfaces]{Zeta Function of a Family of K3
Surfaces}\label{sectionk3zeta} For a K3 surface the zeta function
(\ref{generalzeta}) takes the form:

\begin{equation}
Z(X/\mathbb{F}_p,t)=\frac{1}{(1-t)P_2^{(p)}(t)(1-p^2t)}\;,
\end{equation}
where $P_2^{(p)}(t)$ is of degree $22$ because the Hodge diamond is of the form:\\

\begin{center}
\begin{tabular}{ccccc}
 & & 1 &     &\\
 &0&   &0    &\\
1& &$20$& &1\\
 &0&   &0    &\\
 & & 1 &     &\\
\end{tabular}
\end{center}

The one-parameter family of K3 surfaces in consideration is
defined by a quartic in $\mathbb{P}_3$:

\begin{equation}
P(x)=x_1^4+x_2^4+x_3^4+x_4^4-4\psi x_1x_2x_3x_4\;,
\end{equation}

\noindent with singularities at $\psi^4=1$.

The expression for the number of rational points for this variety
is entirely analogous with that of the one-parameter family of
Quintic threefolds \cite{COV1} (Indeed, this family of K3's was
considered by one of the authors of \cite{COV1}, Xenia de la Ossa,
\cite{delaossa}):

\begin{equation}
N=1+q+q^2+\frac{1}{q(q-1)}\sum_{\textbf{v}}\lambda_{\textbf{v}}
\sum_{n=0}^{q-2}\beta_{q,n,\textbf{v}}\mathrm{Teich}^{n}\left(\frac{1}{(4\psi)^4}\right)\;,\\
\label{k3points}
\end{equation}

\noindent where, when $4|q-1$ with $l=\frac{q-1}{4}$ and $l \nmid
n$:
\begin{equation}
\beta_{q,n,\textbf{v}}=\begin{cases}G_{4n}\prod_{i=1}^{4}G_{-(n+v_il)},\quad
4|q-1\;,\\G_{4n}\prod_{i=1}^{4}G_{-(n+\frac{v_i}{2}m)}, 4\nmid
q-1\;,\\\end{cases}
\end{equation}

\noindent for $l|n$:

\begin{equation}
\beta_{q,al,\textbf{v}}=\begin{cases}-q\prod_{i=1}^{4}G_{-(n+v_il)}\;,
\\-\prod_{i=1}^{4}G_{-(n+v_il)}\;,\end{cases}
\end{equation}

\noindent and for $m|n$:

\begin{equation}
\beta_{q,al,\textbf{v}}=\begin{cases}-q\prod_{i=1}^{4}G_{-(n+\frac{v_i}{2}m)},\quad
z(\textbf{v})\neq0\;,
\\-\prod_{i=1}^{4}G_{-(n+\frac{v_i}{2}m)},\quad
z(\textbf{v})=0\;.\end{cases}
\end{equation}

\noindent Here we have defined $z(\textbf{u})$ as the function
which counts the number of components of a vector $\textbf{u}$
that are zero $\mod \,4$. The monomials that contribute are:

\begin{center}
\begin{tabular}{|c|c|}\hline
Monomial $\textbf{v}$& $\lambda_{\textbf{v}}$\\\hline
$(0,0,0,0)$&1\\
$(2,2,0,0)$&3\\
$(3,1,0,0)$&12\\\hline
\end{tabular}
\end{center}




\noindent Below are the forms of the zeta function computed for
many primes. In all cases it is clear from (\ref{k3points}) that
the zeta function is given by an expression of the form:

\begin{equation}
\frac{1}{(1-t)(1-pt)(1-p^2t)R_{\textbf{(0,0,0,0)}}(p,\psi,t)R_{\textbf{(2,2,0,0)}}^3(p,\psi,t)R_{\textbf{(3,1,0,0)}}^{12}(p,\psi,t)}\;,
\end{equation}

\noindent where $R_{\textbf{v}}(p,\psi,t)$ is the contribution to
the zeta function from each monomial.

The zeta function for the mirror is given by:
\begin{eqnarray}
\frac{1}{(1-t)(1-pt)^{19}(1-p^2t)R_{\textbf{(0,0,0,0)}}(p,\psi,t)}\;.
\end{eqnarray}

\noindent The number $19$ is significant as this is the Picard
number of the varieties.

It shall be seen that at the singularity $\psi^4=1$, the $R_{\textbf{v}}$ degenerate:\\

\begin{center}
\begin{tabular}{|c|c|c|}\hline
\multicolumn{3}{|c|}{Degree of Contribution $R_{\textbf{v}}(t)$} \\ \hline
Monomial $\textbf{v}$ & Smooth & $\psi^4=1$ \\
\hline\hline
$(0,0,0,0)$&$3$&$2$\\
$(2,2,0,0)$&$2$&$1$\\
$(3,1,0,0)$&$1$&$0$\\
\hline
\end{tabular}
\end{center}

The degeneration at $\phi^4=1$ is analogous to that of conifold
points in threefolds (as we shall see later with the octic
threefolds), although geometrically this singularity is not a
conifold, it is a curve of $A_1$ singularities.

\noindent In fact the monomials obtained are as follows:
\begin{itemize}
\item Smooth cases:
\begin{eqnarray}
R_{(0,0,0,0)}&=&(1\pm pt)(1-a(p)t+p^2t)\;,\nonumber\\
R_{(2,2,0,0)}&=&(1\pm pt)(1\pm pt)\;,\nonumber\\
R_{(3,1,0,0)}&=&\begin{cases}[(1-pt)(1+pt)]^{\frac{1}{2}}\quad
\mathrm{when} p=3\,\mod\,4\;,\\(1\pm pt)\,
\quad\mathrm{otherwise}\;.\end{cases}\nonumber\\
\end{eqnarray}

\item $\psi^4=1$:
\begin{eqnarray}
R_{(0,0,0,0)}&=&(1-a(p)t+p^2t)\;,\nonumber\\
R_{(2,2,0,0)}&=&(1\pm pt)\;,\nonumber\\
R_{(3,1,0,0)}&=&1\;.\nonumber\\
\end{eqnarray}
\end{itemize}

The coefficient $a(p)$ appears to always take the form
$a(p)=2\,\mod\,4$. Of course, $2$ is special in that it is the bad
prime.


It may be observed that all the quadratic factors used to build
$R_{\textbf{v}}$ satisfy the functional equation:

\begin{eqnarray}
R_{\textbf{v}}\left(\frac{1}{p^2t}\right)=\frac{1}{p^2t^2}R_{\textbf{v}}\left(t\right).
\end{eqnarray}

\noindent This can be considered as a `decomposition' of the
functional equation of the whole zeta function (\ref{functional}).
See the end of Section \ref{octicdiscussion} below for a more
detailed discussion.

Two incidences of potential modularity were observed for the cases
$\psi=0$ and $\psi=1$. The $R_{(0,0,0,0,0)}$ factor is of the
form: $(1\pm pt)(1+a_pt+p^2t^2)$. Hence, we except them to be
related to the $L$-functions of weight $3$ cusp forms. Perusal of
William Stein's tables \cite{Stein}, yields the following two
candidates (the $r$th integer denotes the coefficient
corresponding to the $r$th prime):

\noindent For $\psi=0$:
\begin{eqnarray*}
 E[16,2] &=& [0,0,-6,0,0,10,-30,0,0,42,0,-70,18,0,0,90,0,-22,0,0,-110,0,\\
&&0,-78,130,-198,0,0,-182,-30,0,0,210,0,-102,0,170,0,0,330,0,\\
&&-38,0,-190,-390,0,0];
\end{eqnarray*}

\noindent For $\psi=1$:
\begin{eqnarray*}
E[32,2] &=& [0,2,0,0,-14,0,2,34,0,0,0,0,-46,-14,0,0,82,0,-62,0,-142,0,\\
&&-158,146,-94,0,0,178,0,98,0,-62,-238,-206,0,0,0,322,0,0,34\\
&&,0,0,98,0,0,226].
\end{eqnarray*}

\noindent In each case the level $N$ of the congruence subgroup
$\Gamma_1(N)$ was a power of the bad prime $2$. It should be noted
that in all cases the coefficient, $b$, in the quadratic
$1+bt+p^3t^2$ is always $2\mod 4$.

\subsection[Zeta Functions for the K3 Surfaces]{Zeta Functions for the K3 Surfaces}
In this section we record the $R_{\textbf{v}}$ factors of the zeta
function for primes $p=3,\ldots,41$.

\noindent
\begin{tabular}{|c|c|c|c|}\hline
\multicolumn{4}{|c|}{$p=3$}\\\hline
$\psi$ & (0,0,0,0) &(2,2,0,0) & (3,1,0,0)\\
\hline\hline
0 & $(1-3t)^2(1+3t)$ & $(1-3t)(1+3t)$ & $[(1-3t)(1+3t)]^{\frac{1}{2}}$\\
1,2& $(1+2t+3^2t^2)$ & $(1-3t)$ & 1\\ \hline
\end{tabular}
\\

\noindent
\begin{tabular}{|c|c|c|c|}\hline
\multicolumn{4}{|c|}{$p=5$}\\\hline
$\psi$ & (0,0,0,0) &(2,2,0,0) & (3,1,0,0)\\
\hline\hline
0 & $(1-5t)(1+6t+5^2t^2)$ & $(1-5t)^2$ & $(1+5t)$\\
1,2,3,4& $(1-5t)(1+5t)$ & $(1+5t)$ & 1\\ \hline
\end{tabular}
\\

\noindent
\begin{tabular}{|c|c|c|c|}\hline
\multicolumn{4}{|c|}{$p=7$}\\\hline
$\psi$ & (0,0,0,0) &(2,2,0,0) & (3,1,0,0)\\
\hline\hline
0 & $(1-7t)^2(1+7t)$ & $(1-7t)(1+7t)$ & $[(1-7t)(1+7t)]^{\frac{1}{2}}$\\
1,6& $(1-7t)(1+7t)$ & $(1+7t)$ & 1\\
2,5& $(1-7t)(1+10t+7^2t^2)$ & $(1-7t)^2$ & $[(1-7t)(1+7t)]^{\frac{1}{2}}$\\
3,4& $(1+7t)(1-6t+7^2t^2)$&$(1-7t)(1+7t)$&$[(1-7t)(1+7t)]^{\frac{1}{2}}$\\\hline
\end{tabular}
\\

\noindent
\begin{tabular}{|c|c|c|c|}\hline
\multicolumn{4}{|c|}{$p=11$}\\\hline
$\psi$ & (0,0,0,0) &(2,2,0,0) & (3,1,0,0)\\
\hline\hline
0 & $(1-11t)^2(1+11t)$ & $(1-11t)(1+11t)$ & $[(1-11t)(1+11t)]^{\frac{1}{2}}$\\
1,10& $(1-14t+11t^2)$ & $(1-11t)$ & 1\\
2,9& $(1-11t)(1+6t+11^2t^2)$ & $(1+11t)^2$ & $[(1-11t)(1+11t)]^{\frac{1}{2}}$\\
3,8& $(1-11t)(1+18t+11^2t^2)$&$(1-11t)^2$&$[(1-11t)(1+11t)]^{\frac{1}{2}}$\\
4,7& $(1+11t)(1-14t+11^2t^2)$&$(1-11t)(1+11t)$&$[(1-11t)(1+11t)]^{\frac{1}{2}}$\\
5,6& $(1+11t)(1+10t+11^2t^2)$&$(1-11t)(1+11t)$&$[(1-11t)(1+11t)]^{\frac{1}{2}}$\\\hline
\end{tabular}
\\

\noindent
\begin{tabular}{|c|c|c|c|}\hline
\multicolumn{4}{|c|}{$p=13$}\\\hline
$\psi$ & (0,0,0,0) &(2,2,0,0) & (3,1,0,0)\\
\hline\hline
0 & $(1-13t)(1-10t+13^2t^2)$ & $(1-13t)^2$ & $(1+13t)$\\
1,12,5,8& $(1-13t)(1+13t)$ & $(1+13t)$ & 1\\
2,11,3,10 & $(1+13t)(1+6t+13^2t^2)$ & $(1-13t)(1+13t)$ & $(1-13t)$\\
4,9,6,7 & $(1+13t)(1-13t)^2$ & $(1-13t)(1+13t)$ & $(1-13t)$\\\hline
\end{tabular}
\\

\noindent
\begin{tabular}{|c|c|c|c|}\hline
\multicolumn{4}{|c|}{$p=17$}\\\hline
$\psi$ & (0,0,0,0) &(2,2,0,0) & (3,1,0,0)\\
\hline\hline
0 & $(1-17t)(1+30t+17^2t^2)$ & $(1-17t)^2$ & $(1-17t)$\\
1,16,4,13& $(1-2t+17^2t^2)$ & $(1-17t)$ & 1\\
2,15,8,9 & $(1-17t)(1+18t+17^2t^2)$ & $(1+17t)^2$ & $(1-17t)$\\
3,14,5,12 & $(1-17t)^2(1+17t)$ & $(1-17t)(1+17t)$ & $(1+17t)$\\
6,11,7,10 & $(1+17t)(1-2t+17^2t^2)$ & $(1-17t)(1+17t)$ & $(1+17t)$\\\hline
\end{tabular}
\\

\noindent
\begin{tabular}{|c|c|c|c|}\hline
\multicolumn{4}{|c|}{$p=19$}\\\hline
$\psi$ & (0,0,0,0) &(2,2,0,0) & (3,1,0,0)\\
\hline\hline
0 & $(1-19t)^2(1+19t)$ & $(1-19t)(1+19t)$ & $[(1-19t)(1+19t)]^{\frac{1}{2}}$\\
1,18& $(1+34t+19t^2)$ & $(1-19t)$ & 1\\
2,17& $(1+19t)(1-6t+19^2t^2)$ & $(1-19t)(1+19t)$ & $[(1-19t)(1+19t)]^{\frac{1}{2}}$\\
3,16& $(1-19t)(1+2t+19^2t^2)$&$(1-19t)^2$&$[(1-19t)(1+19t)]^{\frac{1}{2}}$\\
4,15& $(1+19t)(1-19t)^2$&$(1-19t)(1+19t)$&$[(1-19t)(1+19t)]^{\frac{1}{2}}$\\
5,14,9,10& $(1-19t)(1+22t+19^2t^2)$&$(1+19t)^2$&$[(1-19t)(1+19t)]^{\frac{1}{2}}$\\
6,13,8,11& $(1+19t)(1-30t+19^2t^2)$&$(1-19t)(1+19t)$&$[(1-19t)(1+19t)]^{\frac{1}{2}}$\\
7,12& $(1-19t)(1+34t+19^2t^2)$&$(1-19t)^2$&$[(1-19t)(1+19t)]^{\frac{1}{2}}$\\\hline
\end{tabular}
\\

\noindent
\begin{tabular}{|c|c|c|c|}\hline
\multicolumn{4}{|c|}{$p=23$}\\\hline
$\psi$ & (0,0,0,0) &(2,2,0,0) & (3,1,0,0)\\
\hline\hline
0 & $(1-23t)^2(1+23t)$ & $(1-23t)(1+23t)$ & $[(1-23t)(1+23t)]^{\frac{1}{2}}$\\
1,22& $(1-23t)(1+23t)$ & $(1+23t)$ & 1\\
2,21& $(1+23t)(1-38t+23^2t^2)$ & $(1-23t)(1+23t)$ & $[(1-23t)(1+23t)]^{\frac{1}{2}}$\\
3,20& $(1+23t)(1+26t+23^2t^2)$&$(1-23t)(1+23t)$&$[(1-23t)(1+23t)]^{\frac{1}{2}}$\\
4,19& $(1-23t)(1+42t+23t^2)$&$(1-23t)^2$&$[(1-23t)(1+23t)]^{\frac{1}{2}}$\\
5,18& $(1-23t)(1-18t+23^2t^2)$&$(1+23t)^2$&$[(1-23t)(1+23t)]^{\frac{1}{2}}$\\
6,17& $(1+23t)(1-38t+23^2t^2)$&$(1-23t)(1+23t)$&$[(1-23t)(1+23t)]^{\frac{1}{2}}$\\
7,16& $(1-23t)(1+23t)^2$&$(1+23t)^2$&$[(1-23t)(1+23t)]^{\frac{1}{2}}$\\
8,15& $(1-23t)(1+10t+23^2t^2)$&$(1-23t)^2$&$[(1-23t)(1+23t)]^{\frac{1}{2}}$\\
9,14& $(1+23t)(1-14t+23^2t^2)$&$(1+23t)(1-23t)$&$[(1-23t)(1+23t)]^{\frac{1}{2}}$\\
10,13& $(1+23t)(1-14t+23^2t^2)$&$(1+23t)(1-23t)$&$[(1-23t)(1+23t)]^{\frac{1}{2}}$\\
11,12& $(1-23t)(1+10t+23^2t^2)$&$(1-23t)^2$&$[(1-23t)(1+23t)]^{\frac{1}{2}}$\\
\hline
\end{tabular}
\\

\noindent
\begin{tabular}{|c|c|c|c|}\hline
\multicolumn{4}{|c|}{$p=29$}\\\hline
$\psi$ & (0,0,0,0) &(2,2,0,0) & (3,1,0,0)\\
\hline\hline
0 & $(1-29t)(1-42t+29^2t^2)$ & $(1-29t)^2$ & $(1+29t)$\\
1,28,12,17& $(1-29t)(1+29t)$ & $(1+29t)$ & 1\\
2,27,5,24 & $(1+29t)(1-26t+29^2t^2)$ & $(1-29t)(1+29t)$ & $(1-29t)$\\
3,26,7,22 & $(1-29t)(1+29t)^2$ & $(1+29t)^2$ & $(1+29t)$\\
4,25,10,19 & $(1-29t)(1-6t+29^2t^2)$ & $(1+29t)^2$ & $(1+29t)$\\
6,23,14,15 & $(1+29t)(1-26t+29^2t^2)$ & $(1-29t)(1+29t)$ & $(1-29t)$\\
8,21,9,20 & $(1-29t)(1+54t+29^2t^2)$ & $(1-29t)^2$ & $(1+29t)$\\
11,18,13,16 & $(1-29t)(1+22t+29^2t^2)$ & $(1-29t)^2$ & $(1+29t)$\\\hline
\end{tabular}
\\

\noindent
\begin{tabular}{|c|c|c|c|}\hline
\multicolumn{4}{|c|}{$p=31$}\\\hline
$\psi$ & (0,0,0,0) &(2,2,0,0) & (3,1,0,0)\\
\hline\hline
0 & $(1-31t)^2(1+31t)$ & $(1-31t)(1+31t)$ & $[(1-31t)(1+31t)]^{\frac{1}{2}}$\\
1,30& $(1-31t)(1+31t)$ & $(1+31t)$ & 1\\
2,29& $(1+31t)(1-30t+31^2t^2)$ & $(1-31t)(1+31t)$ & $[(1-31t)(1+31t)]^{\frac{1}{2}}$\\
3,28& $(1-31t)(1+31t)^2$&$(1+31t)^2$&$[(1-31t)(1+31t)]^{\frac{1}{2}}$\\
4,27& $(1-31t)(1+58t+31t^2)$&$(1-31t)^2$&$[(1-31t)(1+31t)]^{\frac{1}{2}}$\\
5,26& $(1-31t)(1+26t+31^2t^2)$&$(1-31t)^2$&$[(1-31t)(1+31t)]^{\frac{1}{2}}$\\
6,25& $(1+31t)(1-54t+23^2t^2)$&$(1-31t)(1+31t)$&$[(1-31t)(1+31t)]^{\frac{1}{2}}$\\
7,24& $(1+31t)(1-31t)^2$&$(1-31t)(1+31t)$&$[(1-31t)(1+31t)]^{\frac{1}{2}}$\\
8,23& $(1+31t)(1+10t+31^2t^2)$&$(1-31t)(1+31t)$&$[(1-31t)(1+31t)]^{\frac{1}{2}}$\\
9,22& $(1-31t)(1-2t+31^2t^2)$&$(1+31t)^2$&$[(1-31t)(1+31t)]^{\frac{1}{2}}$\\
10,21& $(1+31t)(1-30t+31^2t^2)$&$(1+31t)(1-31t)$&$[(1-31t)(1+31t)]^{\frac{1}{2}}$\\
11,20& $(1-31t)(1-38t+31^2t^2)$&$(1+31t)^2$&$[(1-31t)(1+31t)]^{\frac{1}{2}}$\\
12,19& $(1+31t)(1+10t+31^2t^2)$&$(1-31t)(1+31t)$&$[(1-31t)(1+31t)]^{\frac{1}{2}}$\\
13,18& $(1-31t)(1+58t+31^2t^2)$&$(1-31t)^2$&$[(1-31t)(1+31t)]^{\frac{1}{2}}$\\
14,17& $(1+31t)(1+10t+31^2t^2)$&$(1-31t)(1+31t)$&$[(1-31t)(1+31t)]^{\frac{1}{2}}$\\
15,16& $(1-31t)(1-2t+31^2t^2)$&$(1+31t)^2$&$[(1-31t)(1+31t)]^{\frac{1}{2}}$\\
\hline
\end{tabular}
\\

\noindent
\begin{tabular}{|c|c|c|c|}\hline
\multicolumn{4}{|c|}{$p=37$}\\\hline
$\psi$ & (0,0,0,0) &(2,2,0,0) & (3,1,0,0)\\
\hline\hline
0 & $(1-37t)(1+70t+37^2t^2)$ & $(1-37t)^2$ & $(1+37t)$\\
1,36,6,31& $(1-37t)(1+37t)$ & $(1+37t)$ & 1\\
2,35,12,25 & $(1+37t)(1-42t+37^2t^2)$ & $(1-37t)(1+37t)$ & $(1-37t)$\\
3,34,18,19 & $(1+37t)(1+54t+37^2t^2)$ & $(1-37t)(1+37t)$ & $(1-37t)$\\
4,33,13,24 & $(1-37t)(1+10t+37^2t^2)$ & $(1+37t)^2$ & $(1+37t)$\\
5,32,7,30 & $(1+37t)(1-37t)^2$ & $(1-37t)(1+37t)$ & $(1-37t)$\\
8,29,11,26 & $(1-37t)(1+70t+37^2t^2)$ & $(1-37t)^2$ & $(1+37t)$\\
9,28,17,20 & $(1-37t)(1+10t+37^2t^2)$ & $(1+37t)^2$ & $(1+37t)$\\
10,27,14,23 & $(1-37t)(1+38t+37^2t^2)$ & $(1-37t)^2$ & $(1+37t)$\\
15,22,16,21 & $(1+37t)(1-37t)^2$ & $(1-37t)(1+37t)$ & $(1-37t)$\\
\hline
\end{tabular}
\\

\noindent
\begin{tabular}{|c|c|c|c|}\hline
\multicolumn{4}{|c|}{$p=41$}\\\hline
$\psi$ & (0,0,0,0) &(2,2,0,0) & (3,1,0,0)\\
\hline\hline
0 & $(1-41t)(1-18t+41^2t^2)$ & $(1-41t)^2$ & $(1-41t)$\\
1,40,9,32& $(1+46t+41^2t^2)$ & $(1-41t)$ & 1\\
2,39,18,23 & $(1+41t)(1-50t+41^2t^2)$ & $(1-41t)(1+41t)$ & $(1+41t)$\\
3,38,14,27 & $(1-41t)(1+78t+41^2t^2)$ & $(1-41t)^2$ & $(1-41t)$\\
4,37,5,36 & $(1-41t)(1+66t+41^2t^2)$ & $(1+41t)^2$ & $(1-41t)$\\
6,35,13,28 & $(1+41t)(1-50t+41^2t^2)$ & $(1-41t)(1+41t)$ & $(1+41t)$\\
7,34,19,22 & $(1+41t)(1+46t+41^2t^2)$ & $(1-41t)(1+41t)$ & $(1+41t)$\\
8,33,10,31 & $(1-41t)(1-62t+41^2t^2)$ & $(1+41t)^2$ & $(1-41t)$\\
11,30,17,24 & $(1+41t)(1-41t)^2$ & $(1-41t)(1+41t)$ & $(1+41t)$\\
12,29,15,26 & $(1+41t)(1-50t+41^2t^2)$ & $(1-41t)(1+41t)$ & $(1+41t)$\\
16,25,20,21 & $(1+41t)(1-50t+41^2t^2)$ & $(1-41t)(1+41t)$ & $(1+41t)$\\
\hline
\end{tabular}
\\

\pagebreak

\section[Zeta Function of the Octic Calabi--Yau threefolds]{Zeta Function of the Octic Calabi--Yau threefolds}
\label{octicdiscussion}

\subsection[Basic Form of Zeta Function]{Basic Form of the Zeta
Function} It was found that the number of points can be written in
the following way:
\begin{equation}
N_{r}(\psi)=N_{\textbf{(0,0,0,0,0)}}+\sum_{\textbf{v}}\lambda_{\textbf{v}}N_{\textbf{v},r}(\psi)\;,
\end{equation}

\noindent where the $N_{\textbf{v},r}$ are not in general either
positive or integral. Hence, for each $(\phi,\psi)$ the zeta
function is given by:

\begin{equation}
\zeta(\phi,\psi,p,t)=\frac{R_{\text{excep}}(t)\prod_{\textbf{v}}R_{\textbf{v}}(t)^{\lambda_{\textbf{v}}}}{(1-t)(1-pt)^2(1-p^2t)^2(1-p^3t)}\;,
\end{equation}

\noindent where $R_{\textbf{v}}(t)$ is the contribution to the
zeta function associated to the monomial $\textbf{v}$.

$R_{\text{excep}}(t)$, a sextic, which always factorizes into the
cube of a quadratic, can be thought of as a contribution from the
exceptional divisor, which as explained in (\ref{singularity}) was
a ruled surface over a curve of genus $3$. A curve of genus $3$
has a Hodge diamond of the form:

\begin{center}
\begin{tabular}{ccc}
  & 1 &     \\
 3&   &3    \\
  & 1 &     \\
\end{tabular}\\
\end{center}

\noindent thus, its zeta function would also have a sextic in the
numerator. The factor $R_{\text{excep}}(t)$ is independent of the
parameters $(\phi, \psi)$. It could, however, be included as part
of the contribution from $(0,0,2,1,1)$ and $(2,2,1,1,0)$ combined.
The presence of $R_{\text{excep}}(t)$ arises from having
singularities in the ambient space, in this case
$\mathbb{P}^4_{(1,1,2,2,2)}$, which need to be resolved. It also
reflects the fact that not all complex structure deformations can
be described as monomial deformations, since for this case
$h^{2,1}_{\mathrm{poly}}=83$, $h^{2,1}=86$. This prompts the
following conjecture:

\begin{conjecture}
For a family of Calabi--Yau hypersurfaces, if
$h^{2,1}>h^{2,1}_{\mathrm{poly}}$, the numerator of the zeta
function has a factor that is independent of the parameters
labelling each member of the family. This factor, denoted
$R_{\text{excep}}(t)$, will have degree $2\acute{h}$, where
$\acute{h}=h^{2,1}-h^{2,1}_{\mathrm{poly}}$.
\end{conjecture}

For the family of quintic threefolds\cite{COV2} no such parameter
independent factor was found in the zeta function, in keeping with
the above conjecture, as in that case,
\hbox{$h^{2,1}_{\mathrm{poly}}=h^{2,1}=101$}. \pagebreak

For various values of $p$, $R_{\text{excep}}(t)$ takes the form:
\begin{center}
\begin{tabular}{|c|c|}\hline
$p$& $R_{\text{excep}}(t)$\\\hline
$3$&$(1+3^3t^2)^3$\\
$5$&$(1-2.5t+5^3t^2)^3$\\
$7$&$(1+7^3t^2)^3$\\
$11$&$(1+11^3t^2)^3$\\
$13$&$(1+6.13t+13^3t^2)^3$\\
$17$&$(1-2.17t+17^3t^2)^3$\\
$19$&$(1+19^3t^2)^3$\\
$23$&$(1+23^3t^2)^3$\\
$29$&$(1-10.29t+29^3t^2)^3$\\
$31$&$(1+31^3t^2)^3$\\
$37$&$(1-2.37t+37^3t^2)^3$\\
$41$&$(1-10.41t+41^3t^2)^3$\\
$43$&$(1+43^3t^2)^3$\\
$47$&$(1+47^3t^2)^3$\\
$53$&$(1+14.53t+53^3t^2)^3$\\
$59$&$(1+59^3t^2)^3$\\
$61$&$(1-10.61t+61^3t^2)^3$\\
$67$&$(1+67^3t^2)^3$\\
$71$&$(1+71^3t^2)^3$\\
$73$&$(1+6.73t+73^3t^2)^3$\\
\hline\end{tabular}
\end{center}
When $p\neq1\,\mod\,4$ it seems that
$R_{\text{excep}}(t)=(1+p^3t^2)^3$.\\
When $p=1\,\mod\,4$, $R_{\text{excep}}(t)=(1+b(p)pt+p^3t^2)^3$.

\subsection[Behaviour at Singularities]{Behaviour at Singularities}
The degree of the rational function, $R_{\textbf{v}}(t)$ ,
obtained from the contribution to the rational points associated
to each monomial class is dependent on the point in complex
structure moduli space, which specified by the values of $\phi$
and $\psi$. By `degree' of a rational function,
$\left(f/g\right)$, we mean
$\deg\left(f/g\right)=\deg(f)-\deg(g)$.

It is apparent that not only does the zeta function change
dramatically at the discriminant locus, but the change appears to
be characterized by the type of singularity. Hence we propose the
following conjecture:

\begin{conjecture}[Singularity Conjecture 1]
The degeneration of the degree of each piece of the zeta function
$\deg(R_{\textbf{v}})$ of a Calabi--Yau manifold at a singular
point in the moduli space is determined by the nature of the
singularity.
\end{conjecture}

\noindent The following table summarizes this degeneration at
various types of singularity:

\begin{center}
\begin{tabular}{|c|c|c|c|c|}\hline
\multicolumn{5}{|c|}{Degree of Contribution $R_{\textbf{v}}(t)$ According to Singularity} \\ \hline
Monomial $\textbf{v}$ & Smooth & Conifold &$\phi^2=1$& Conifold and $\phi^2=1$ \\
\hline\hline
(0,0,0,0,0) & 6 & 5& 4& 3 \\
(0,2,1,1,1) & 4 & 3& 2& 1  \\
(6,2,0,0,0) & 4 & 3& 2& 1  \\ \hline
(0,0,0,2,2) & 4 & 3& 3& 2 \\
(2,0,1,3,3) & 2 & 1&1&0 \\
(4,0,2,0,0) & 4 & 3& 3& 2 \\ \hline
(0,0,2,1,1) & 3& 2&2&1  \\
(6,0,1,0,0) & 3& 2&2&1  \\
(0,4,0,3,3) & 4 &3&3& 2  \\
(4,0,1,1,0) & 4&3&3& 2  \\
(2,0,3,0,0) & 3& 2&2&1  \\
(2,2,1,1,0) & 3& 2&2&1 \\\hline
(0,0,3,1,0) & 2&1&2&1   \\
(2,0,2,1,0) & 2&1&2& 1  \\
(4,0,2,3,1) & 0&-1& 0& -1  \\ \hline
\end{tabular}\label{degeneration}
\end{center}

It can be shown that the contribution from $(4,0,3,2,1)$ is only
non-trivial (i.e. not $1$) at conifold points:

\begin{equation}
R_{(4,0,3,2,1)}(t)=\delta((8\psi^4+\phi)^2-1)
\begin{cases}\frac{1}{[(1-p^2t)(1+p^2t)]^{\frac{1}{2}}};\quad 4 \nmid
(p-1)\\ \\ \frac{1}{1+p^2t};\quad 4|(p-1),\;8\nmid(p-1)\\
\\\frac{1}{1-p^2t};\quad 8|(p-1)\end{cases}
\end{equation}

\begin{conjecture}[Degenerations due to Isolated $A_1$ Singularities]
For isolated $A_1$ singularities \textup{(}ordinary double points
or conifold points\textup{)}, the degree of each contribution $\deg(R_{\textbf{v}})$ goes
down by exactly $1$.
\end{conjecture}

\begin{conjecture}[Singularity Conjecture 2]
Where two types of singularity coincide, the total degeneration in
degree of each contribution $\deg(R_{\textbf{v}})$ is the sum of the degenerations due to
each type of singularity. \label{sing2}
\end{conjecture}

For instance, Table (\ref{degeneration}) shows that when there is
both a conifold singularity and a $\phi^2=1$ singularity, the
total degeneration is indeed as specified by Conjecture
(\ref{sing2}).

The monomials in the table have been categorized according to
their combinatorial properties, as described in Section
\ref{combin}. In addition, the pairs of monomials
\hbox{$(0,0,1,1,2)\, \&\, (2,2,1,1,0)$}, \hbox{$(6,0,1,0,0)\, \&\,
(2,0,3,0,0)$} and \hbox{$(0,4,3,3,0)\, \&\, (4,0,1,1,0)$} have
been separated because together they display some very interesting
properties which will be discussed in Section
\ref{pairsofmonomials}. The degeneration due to the $\phi^2=1$
singular locus seems to depend on the combinatorial properties of
the monomial, since the degree goes down by $2$ for all the
monomials except for the last three monomials $(0,0,3,1,0)$,
$(2,0,2,1,0)$ and $(4,0,3,2,1)$ whose corresponding degrees remain
unchanged.

\subsection[Functional Equation]{Functional
Equation}\label{sectfunctional} It shall be seen that the rational
functions obtained for the various pieces of the zeta function all
satisfy a functional equation. In fact it is the functional
equation desired if it were possible to `break up' $\chi$ in
(\ref{functional}) into its `components', based on the formula for
the Euler characteristic in terms of the Betti numbers, i.e. up to
sign the degree of the polynomial can replace $\chi$ in
(\ref{functional}):

\begin{equation}
\chi=\sum_{i=0}^{2d}(-1)^ib_i\;.
\end{equation}
So for sextics, it would be desirable for the following functional
equation to hold:

\begin{equation}
R_{\textbf{v}}\left(\frac{1}{p^3t}\right)=\frac{1}{p^9t^6}R_{\textbf{v}}\left(t\right)\;,
\end{equation}

\noindent for all values of $(\phi,\psi)$. Hence, if
$R_{\textbf{v}}$ is a sextic one would expect it to have the
following form:

\begin{eqnarray}
1+a_1t+a_2t^2+a_3t^3+a_2p^3t^4+a_1p^6t^5+p^9t^6\;.
\end{eqnarray}

This is indeed the form of the sextics observed. It can be
observed that the $R_{\textbf{v}}$ always decomposes into
(fractional powers of) even degree polynomials satisfying an
analogous functional equation, and some linear terms which are
$(1\pm p^ut),\,u=1,2$.

Similarly all quartics encountered are of the form:

\begin{eqnarray}
1+a_1t+a_2t^2+a_1p^3t^3+p^6t^4\;.
\end{eqnarray}

\subsection[Interesting Pairings]{Interesting Pairings}\label{pairsofmonomials}
\begin{figure}[htbp]
\centering
\includegraphics[scale=1]{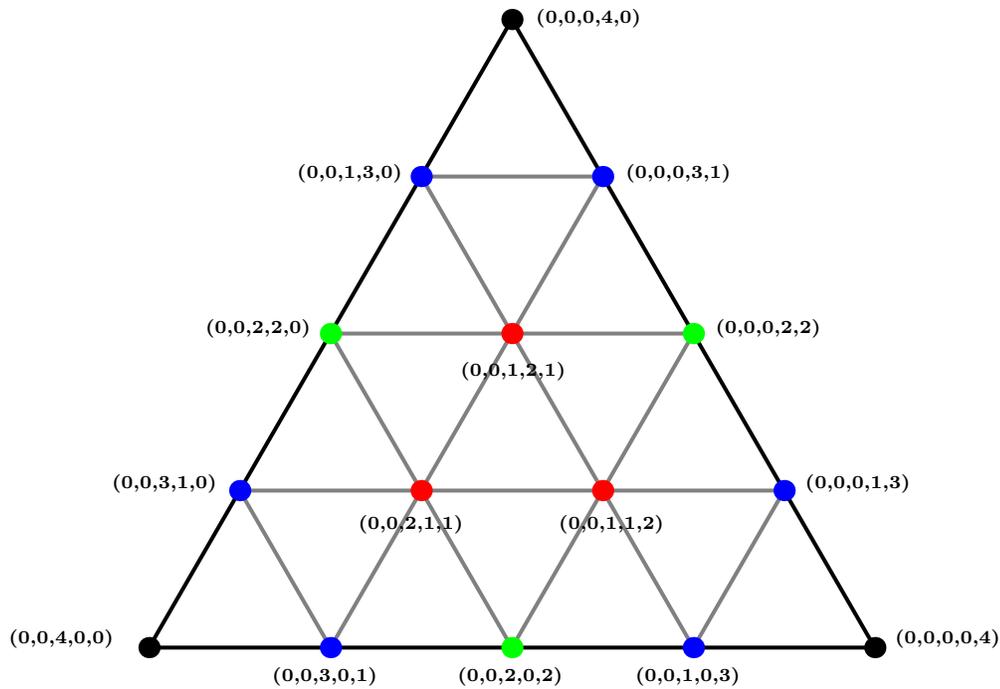}
\caption{The two-face with vertices $(0,0,4,0,0)$, $(0,0,0,4,0)$
and $(0,0,0,0,4)$.} \label{figtwoface1diff}
\end{figure}

\begin{figure}[htbp]
\centering
\includegraphics[scale=1]{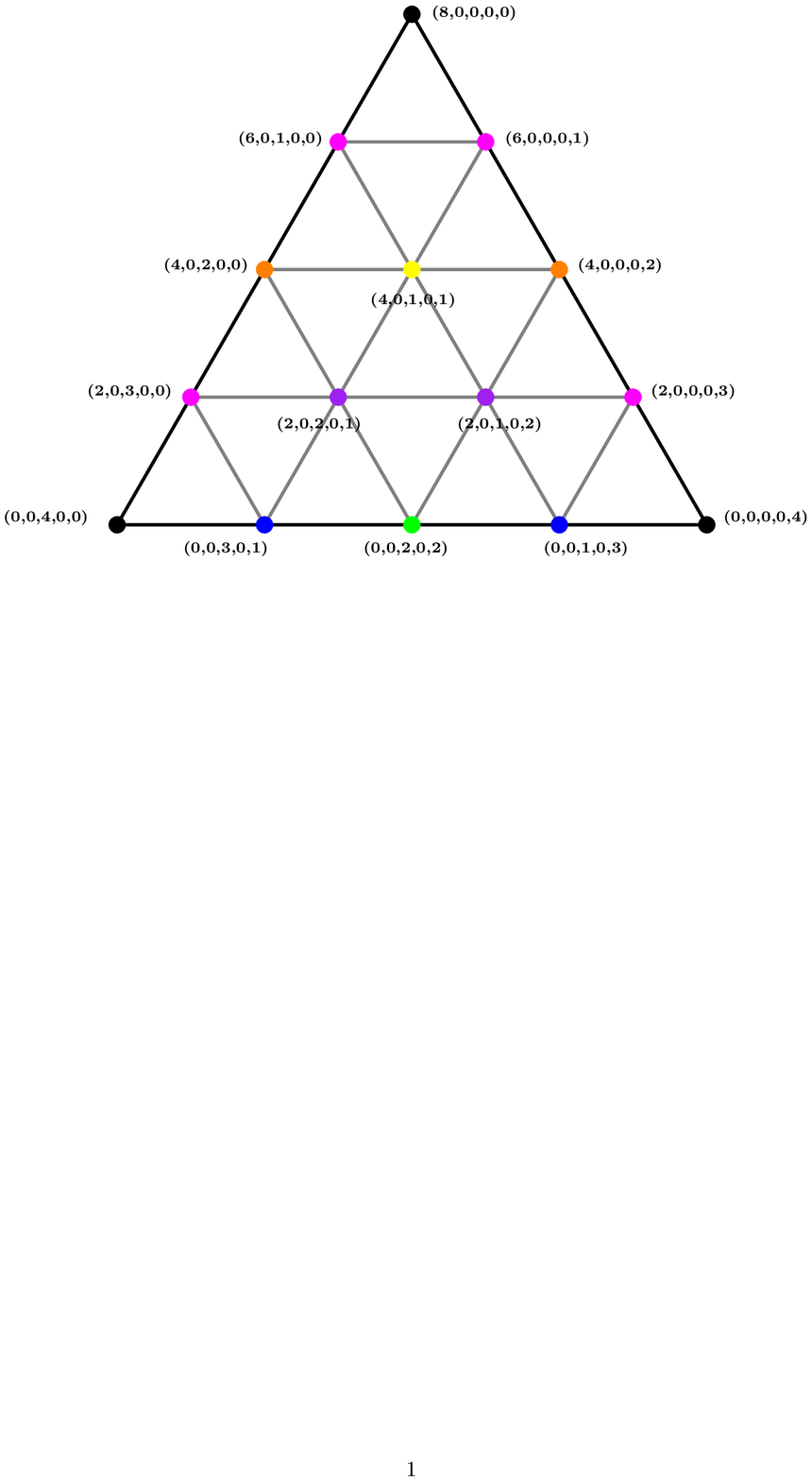}
\caption{A two-face with vertices $(8,0,0,0,0)$, $(0,0,4,0,0)$ and
$(0,0,0,0,4)$. } \label{figtwoface2diff}
\end{figure}

\begin{figure}[htbp]
\centering
\includegraphics[scale=1.2]{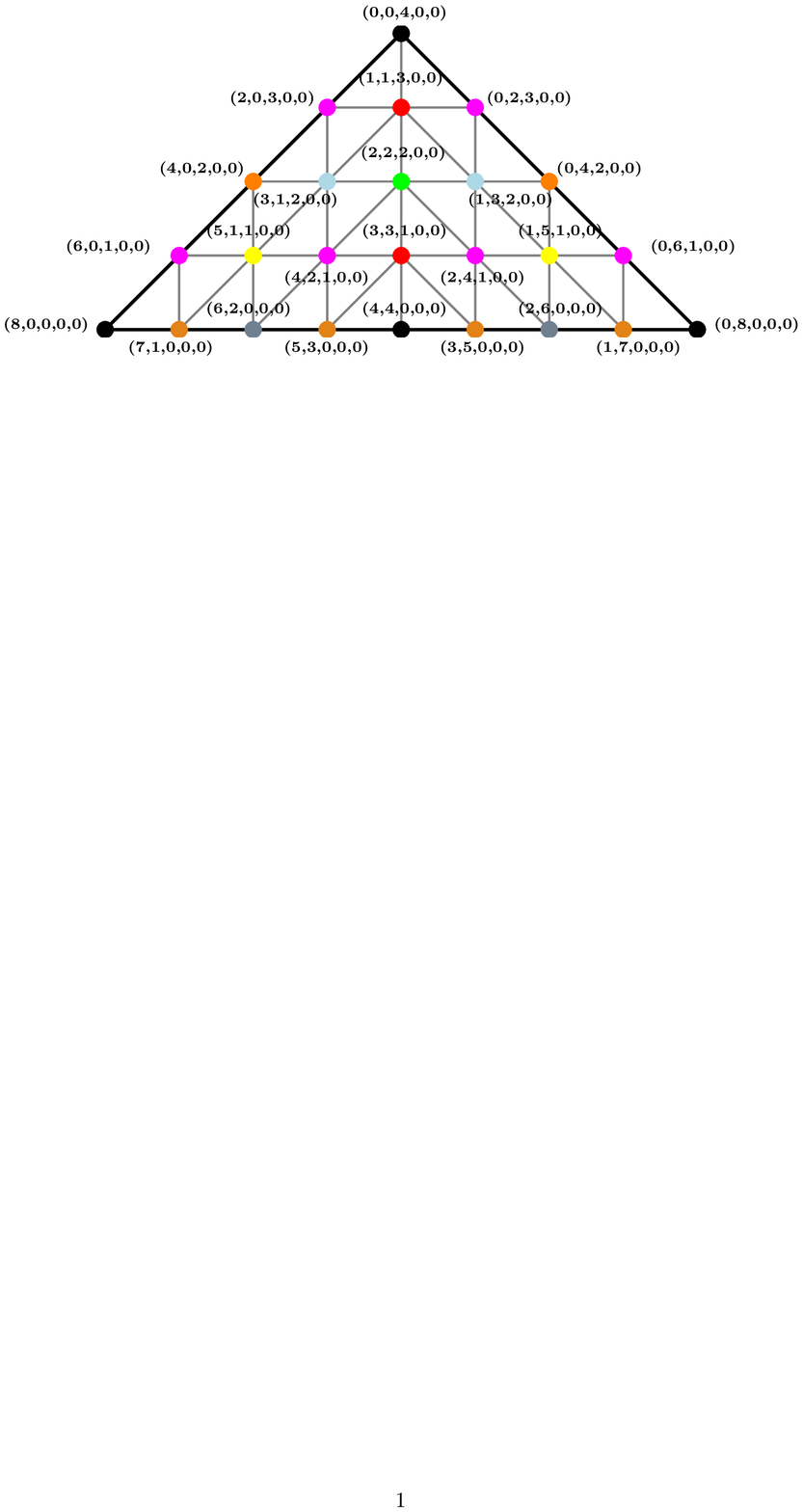}
\caption{A two-face with vertices $(8,0,0,0,0)$, $(0,8,0,0,0)$ and
$(0,0,4,0,0)$. } \label{figtwoface3diff}
\end{figure}

\begin{figure}[htbp]
\centering
\includegraphics[scale=1]{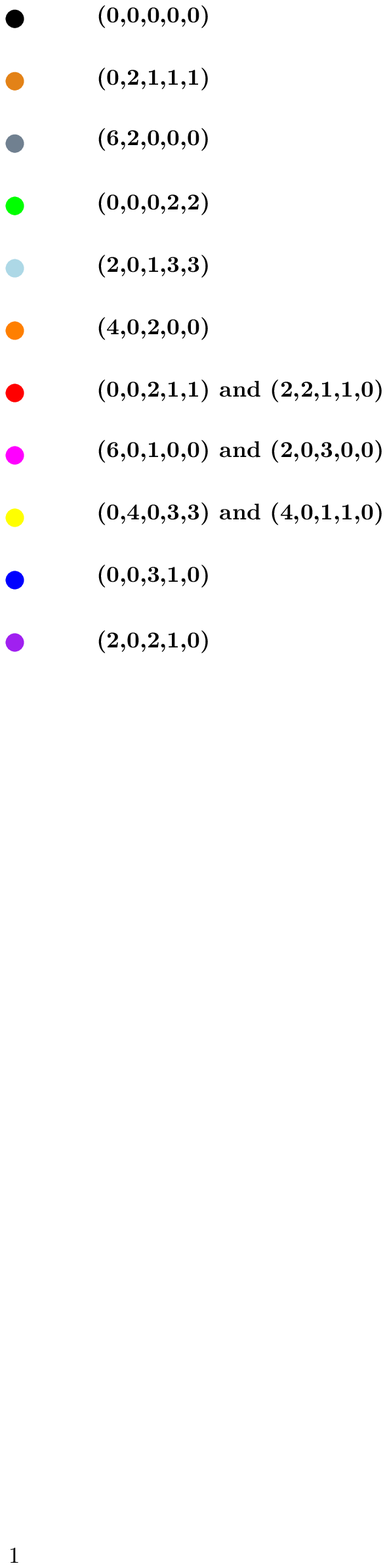}
\caption{Above is the key relating the $11$ colours in Figures
\ref{figtwoface1diff}, \ref{figtwoface2diff} and
\ref{figtwoface3diff} to monomial classes. } \label{colourcode2}
\end{figure}

So far our calculations indicate that for $4|p-1$, (i.e. for the
primes for which we have calculated this is $p=5, 13, 17$) the
contributions from the pairs of monomials \linebreak
\hbox{$(0,0,1,1,2)\, \&\, (2,2,1,1,0)$}, \hbox{$(6,0,1,0,0)\, \&\,
(2,0,3,0,0)$} and \hbox{$(0,4,3,3,0)\, \&\, (4,0,1,1,0)$} multiply
together give polynomials or fractional powers of polynomials
(individually they gives infinite series). Notice that the value
of $\lambda_{\textbf{v}}$ of the monomials in each pair are the
same. In addition it can be observed that, if you add up the
various pairs (mod $8(4)$ for the first two (last three) digits)
you always obtain monomials that are equivalent to $(0,0,0,0,0)$:
\begin{eqnarray}
(0,0,1,1,2)+(2,2,1,1,0)=(2,2,2,2,2)\;,\nonumber\\
(6,0,1,0,0)+(2,0,3,0,0)=(0,0,0,0,0)\;,\nonumber\\
(0,4,3,3,0)+(4,0,1,1,0)=(4,4,0,0,0)\;.\nonumber\\ \label{inverses}
\end{eqnarray}
It has been `experimentally' observed that for the cases where
$4\nmid p-1$, these monomials individually contribute polynomials.
This hints of analogues of the $\mathcal{A}$ and $\mathcal{B}$
Euler curves found in \cite{COV2}, these curves could be
associated to certain pairs of monomials (namely,
\hbox{$(4,1,0,0,0)\, \&\, (2,3,0,0,0)$} and \linebreak
\hbox{$(3,1,1,0,0)\, \&\, (2,2,1,0,0)$}) whose contributions
multiplied together to give fractional powers of polynomials. In
that case, the relatively simple form of the solutions to the
Picard--Fuchs equations, namely hypergeometric functions, enabled
them to be associated to Euler's integral. This led directly to
the presence of genus $4$ curves. In our case, there appears to be
no such fortuitous correspondence, as the associated Picard--Fuchs
equations are complicated partial differential equations, as can
be seen from the diagrams in Chapter \ref{chappicardfuchs}.

In Figures \ref{figtwoface1diff}. \ref{figtwoface2diff} and
\ref{figtwoface3diff}, we depict the points (in the Newton
polyhedron for the mirror) associated to these pairs of monomials
in the same colour (The colour-coding is shown in Figure \ref{colourcode2}) . As well as combinatorial
properties and perhaps as a consequence of (\ref{inverses}), these
pairs also occupy similar positions on the two-faces, e.g.
\hbox{$(6,0,1,0,0)\, \&\, (2,0,3,0,0)$} lie on sides of the
two-faces, whereas the other pairs are all interior to (not
necessarily the same) two-faces. This is in contrast to the case
in \cite{COV2}, where \hbox{$(4,1,0,0,0)\, \&\, (2,3,0,0,0)$} lie
entirely on edges of two-faces and \hbox{$(3,1,1,0,0)\,
\&\, (2,2,1,0,0)$} lie exclusively in the interior of each two
face.

Hence the largest polynomials that need to be found in order to
fully determine the zeta function are octics, namely those
corresponding to $R_{(0,4,0,3,3)}\times R_{(0,4,0,3,3)}$. However
it was found that in all cases, this octic factorises. An octic
with the appropriate reflection formula is fully defined by only
$4$ parameters, $(a_1,a_2,a_3,a_4)$:

\begin{equation}
1+a_1t+a_2t^2+a_3t^3+a_4t^4+a_3p^3t^5+a_2p^6t^6+a_1p^9t^7+p^{12}t^8\;.
\end{equation}
The Weil Conjectures alone only indicate a zeta function of the
form (\ref{Weil174}), with a polynomial of degree $174$ in the
numerator of the form $P_j^{(p)}(t)= \prod_{l=1}^{b_j} (1-
\alpha_l^{(j)}(p)t)$, with $\|\alpha_j^{(i)}(p)\| = p^{i/2}$. This
mean that we need to compute $\mathcal{N}_{p,r}$ up to $r=174$,
however given the decomposition, the zeta function can be
determined by computing up to $r=4$ and in practise $r=3$.

It is likely that the quartics and sextics encountered are related
to Siegel modular forms. This shall be explained in more detail in
Section \ref{siegel}.

\subsection[The Locus $\phi^2=1$ and Modularity]{The Locus $\phi^2=1$ and
Modularity}\label{modularity}
 As expected from the fact that on
the singular locus $\phi^2=1$ the family is birationally
equivalent to a one-parameter family of Calabi--Yau threefolds
after resolution of singularities. Indeed when $\phi^2=1$ and
there is no conifold singularity, we always obtain a zeta function
of the form:

\begin{equation}
\frac{R_{128}}{(1-t)(1-p^2t)^2(1-p^3t)}\;,
\end{equation}

\noindent where $R_{128}$ is a polynomial of degree $128$. The
quartic denominator is thus in keeping with the fact that we are
now dealing with a family birationally equivalent to a
one-parameter family, i.e. $h^{1,1}=1$.

Recall that the Hodge diamond of this one-parameter birational
model (after smoothing was):

\begin{center}
\begin{tabular}{ccccccc}
 & & & 1 &    & &\\
  &&0&   &0 &   &\\
 &0& &$1$& &0&\\
1& &$89$& & $89$& & 1\\
 &0& &$1$& &0&\\
   &&0&   &0 &   &\\
 & & & 1 &    & &\\
\end{tabular}\\
\end{center}

Hence after smoothing we would expect the numerator to be a
polynomial of degree $180$, and the denominator to be of the form
$(1-t)(1-pt)(1-p^2t)(1-p^3t)$.

When there is also a conifold singularity, the form of the zeta
function agrees with the results of \cite{Villegas}. In that paper
it is found that at the conifold points, the factor $R_0(t)$ in
the numerator of the zeta function (which corresponds to our
$R_{(0,0,0,0,0)}$) takes the following form:

\begin{equation}
R_0(t)=\left(1-\left(\frac{D}{p}\right)pt\right)(1-a_pt+p^3t^2)\;.
\end{equation}

Generically for $\phi^2=1$, $R_0(t)$ is a of degree four, but at
conifold points, it degenerates to the above cubic. One can verify
that $a_p$ is the $p$-th coefficient of a modular newform of
weight $4$ for a congruence subgroup for the form $\Gamma_0(N)$of
$SL_2(\mathbb{Z})$. The modular forms were identified by computing
$a_p$ using $p$ the $p$-adic formulae employed here and in
\cite{COV1,COV2}, and by comparison with the tables of W.Stein
\cite{Stein}. For this case, $D=2^3, N=2^4$; the value of $a_p$ is
listed below for the first few primes:

\begin{tabular}{|c||c|c|c|c|c|c|c|c|c|c|c|}\hline
 $p$& 2 & 3& 5& 7& 11&13&17&19&23&29&31\\
\hline
$a_p$ &0&4&-2&-24&44&22&50&-44&56&198&160\\
\hline
\end{tabular}
\\

\begin{tabular}{|c||c|c|c|c|c|c|c|c|c|c|}\hline
 $p$&37&41&43&47&53&59&61&67&71&73\\
\hline$a_p$ & -162&-198&-52&-528&-242&668&550&-188&-728&154\\
\hline
\end{tabular}
\\

\noindent The values so far have been confirmed up to $p=73$.
It should be noted that the coefficients of these forms are always
$0\mod 2$.

The cusp form above, denoted $V_{2,4}(q)$, admits an
$\eta$-product expression:

\begin{equation}
V_{2,4}(q)=\eta(q^2)^{-4}\eta(q^4)^{16}\eta(q^8)^{-4}\;.
\end{equation}
Our calculations verify the results of \cite{Villegas}. The cases
of ours which correspond to those in \cite{Villegas} are when both
singularities coincide and $\psi\neq0$. When $\psi=0$, we also
observed some factorization reminiscent of modularity, a possible
cusp form is of level $N=32=2^5$ (a higher power of the bad prime,
$2$):

\begin{tabular}{|c||c|c|c|c|c|c|c|c|c|c|c|}\hline
 $p$& 2 & 3& 5& 7& 11&13&17&19&23&29&31\\
\hline
$a_p$ &0&0&22&0&0&-18&-94&0&0&-130&0\\
\hline
\end{tabular}

\begin{tabular}{|c||c|c|c|c|c|c|c|c|c|c|}\hline
 $p$&37&41&43&47&53&59&61&67&71&73\\
\hline$a_p$ &214&-230&0&0&518&0&830&0&0&1098\\
\hline
\end{tabular}

\noindent Notice that our results predict that this modular form
will only have coefficients $a_p$ for $p$ such that
\hbox{$4|p-1$}.

There was found to be agreement so far, for all the primes
computed, that is, up to $p=73$. We are using the notation of
Section \ref{notation}, later employed extensively in the tables.
Here, $(a)_1\equiv(1+at)$ and $(a)_2\equiv(1+at+p^3t^2)$.

\noindent
\begin{tabular}{|c|c|c|c|c|c|}\hline
Prime &3&5&7&11&13 \\\hline
$\psi\neq0$&$(3)_1(-4)_2$&$(5)_1(2)_2$&$(-7)_1(24)_2$&$(11)_1(-44)_2$&$(13)_1(-22)_2$\\
$\psi=0$&$(-3)_1(0)_2$&$(-5)_1(-22)_2$&$(-7)_1(0)_2$&$(-11)_1(0)_2$&$(-13)_1(18)_2$\\\hline
\end{tabular}

\noindent
\begin{tabular}{|c|c|c|c|c|c|}\hline
Prime &17&19&23&29&31 \\\hline
$\psi\neq0$&$(-17)_1(-50)_2$&$(19)_1(44)_2$&$(-23)_1(-56)_2$&$(29)_1(-198)_2$&$(-31)_1(-160)_2$\\
$\psi=0$&$(-17)_1(94)_2$&$(-19)_1(0)_2$&$(-23)_1(0)_2$&$(-29)_1(130)_2$&$(-31)_1(0)_2$\\\hline
\end{tabular}

\noindent
\begin{tabular}{|c|c|c|c|c|c|}\hline
Prime &37&41&43&47&53 \\\hline
$\psi\neq0$&$(37)_1(162)_2$&$(-41)_1(198)_2$&$(43)_1(52)_2$&$(-47)_1(528)_2$&$(53)_1(242)_2$\\
$\psi=0$&$(-37)_1(-214)_2$&$(-41)_1(230)_2$&$(-43)_1(0)_2$&$(-47)_1(0)_2$&$(-53)_1(-518)_2$\\\hline
\end{tabular}

\noindent
\begin{tabular}{|c|c|c|c|c|c|}\hline
Prime &59&61&67&71&73 \\\hline
$\psi\neq0$&$(59)_1(-668)_2$&$(61)_1(-550)_2$&$(67)_1(188)_2$&$(-71)_1(728)_2$&$(-73)_1(-154)_2$\\
$\psi=0$&$(-59)_1(0)_2$&$(-61)_1(-830)_2$&$(-67)_1(0)_2$&$(-71)_1(0)_2$&$(-73)_1(-1098)_2$\\\hline
\end{tabular}

The level $N$ of the modular form is always a product of powers of
the bad primes, this satisfies a well known `folk' conjecture:

\begin{conjecture}[Modularity Folklore]
For a $d$-dimensional Calabi--Yau variety, the powers, $\alpha_p$,
 that occur in the expansion in the bad primes, $N=\prod_{p\;\text{bad}}p^{\alpha_p}$, of  the level $N$ of the weight $d+1$ modular
form for some $\Gamma_{0}(N)$, increases with the `badness' of the
singularity.
\end{conjecture}
Hence in the case above, when the manifolds had a singularity of
the type $C_{\mathrm{con}}\bigcap C_{1}$ (see Section
\ref{singularity}), $N=2^4$, however when the singularity was at
the intersection of the loci $C_{\mathrm{con}}\bigcap C_{1}\bigcap
C_{0}$, $N=2^5$.

This is generalization of the case for modular forms related to
elliptic curves \hbox{($d=1$)}: An elliptic curve $E$ has bad
reduction when it has a singularity modulo $p$. The type of
singularity determines the power of $p$ that occurs in the
conductor. If the singularity is a `node' corresponding to a
double root of the polynomial, the curve is said to have
`multiplicative reduction' and $p$ occurs to the first power in
the conductor. If the singularity is a `cusp' corresponding to a
triple root, $E$ is said to have `additive reduction', and $p$
occurs in the conductor with a power of $2$ or more. The conductor
is equal to the level of the associated elliptic modular form.

\subsection[$\psi=0$]{$\psi=0$}
 Generally when $\psi=0$ (a special locus in the moduli space),
 there seemed to be more factorization of the $R_{(0,0,0,0,0)}$'s,
 in fact it can be factorized into a product of quadratic terms with coefficients in
  $\mathbb{Q}(\sqrt{2})$ if $4 \nmid (p-1)$ and  $\mathbb{Q}(\sqrt{p})$, when
  $4|p-1$ but $8\nmid p-1$:

\begin{center}
\begin{tabular}{|c|c|}\hline
Prime $p$ & $\phi=0$ \\\hline
 $3$ &      $(0)_2(-3.2\sqrt{2})_2(3.2\sqrt{2})_2$ \\
 $5$ & $(-10)_2(-2^3\sqrt{5})_2(2^3\sqrt{5})_2$\\
 $7$ & $(0)_2^3$\\
 $11$ & $(0)_2(-11.2\sqrt{2})_2(11.2\sqrt{2})_2$\\
  $13$
 &$(6.13)_2(-2^3.3\sqrt{13}t)_2(2^3.3\sqrt{13})_2$\\
$17$&$(2(45-16\sqrt{2}))_2(2(45+16\sqrt{2}))_2$\\\hline
\end{tabular}
\end{center}
\noindent In general, factorization may indicate decomposition of
Galois representations into those of lower dimension.


\section[Siegel Modular Forms]{Siegel Modular Forms}\label{siegel}
The sextics and quartics which constitute parts of the zeta
function take the same form as the denominator of either the
standard or the spinor $L$-function for a Siegel modular forms of
the symplectic groups $Sp(4,\mathbb{Z})$ and $Sp(6,\mathbb{Z})$.
As remarked previously in \ref{modularitylanglands}, a connection
to Siegel modular forms is expected according to the Langlands
program. At the moment this identification is preliminary, owing
to the fact that there are currently no readily available
published tables akin to those for elliptic modular forms
\cite{Stein}, consisting of eigenvalues of Siegel modular forms
(although there is an online modular forms database under
construction by Nils-Peter Skoruppa \cite{Skoruppa}, which may be
of use in the future). This is because, unlike for the elliptic
modular case, there is no obvious connection between the known
invariants of the Hecke algebra (the Satake $p$-parameters, which
shall be defined in the next section) and the Fourier coefficients
of a Hecke eigenform. The `problem of multiplicity one'(the
eigenvalues determining a modular form) is still a hard problem.
The author is grateful to Valery Gritsenko for clarifying this
point.

A natural question to ask would be how would Siegel modular forms
be related to the variety in question. In fact, they arise very
naturally when studying modular group which arises from
considering changes of basis in the Special Geometry description
of the space of complex structures. A brief review is given in
Appendix \ref{sectionsymplectic}; see \cite{CdO},
\cite{Strominger} for more details. The arithmetic properties of
this modular group warrant future study. Unfortunately, as
stressed above, identification of any particular Siegel modular
form with the arithmetic has not been possible yet.

\begin{definition}[Siegel upper half-plane]
The Siegel upper half plane, $\mathcal{H}_n$, of genus $n\in\mathbb{Z}^{+}$ is the
set of symmetric $n\times n$ complex matrices having positive
definite imaginary part:
\begin{equation}
\mathcal{H}_n=\{Z=X+iY \in M_n(\mathbb{C}; Z^t=Z, Y>0\}\;.
\end{equation}
\end{definition}
The space $\mathcal{H}_n$ is a complex manifold of complex
dimension $n(n+1)/2$.

\begin{definition}[Siegel Modular Form]
A Siegel modular form $F$ of genus $n\in\mathbb{Z}^{+}$ and weight
$k\in\mathbb{Z}^{+}$ is a holomorphic function on the Siegel upper
half plane $\mathcal{H}_n$ such that
\begin{equation}
F((AZ+B)(CZ+D)^{-1})=(det(CZ+D))^kF(Z)\quad \textrm{for}\quad
{A\,B\choose C\,D}\in Sp(n,\mathbb{Z})\;.
\end{equation}
\end{definition}

Siegel modular forms of genus $1$ are classical modular forms on
the upper half plane which transform under $SL(2,\mathbb{Z})$.

The space of Siegel modular form of weight $k$ and genus $n$,
denoted $\mathcal{M}_k(\Gamma^n)$, is a finite dimensional vector
space over $\mathbb{C}$ upon which various linear operators act.
The Hecke operators are an important family of linear operators,
useful for characterizing Siegel modular forms.
$\mathcal{M}_k(\Gamma^n)$ has a basis of forms which are
simultaneous eigenforms for all the Hecke operators. For $n=1$,
the Hecke algebra consists of operators $T(m) \forall
m\in\mathbb{Z}^+$. For arbitrary genus there are analogues of
$T(m)$, but also additional operators \cite{Andrianov}.

\noindent Some properties of Hecke operators:
\begin{enumerate}
\item $H_n$ can be broken into $p$th components as $H_n=\otimes_p
H_{n,p}$ over all primes $p$.

\item There is an isomorphism
$\mathrm{Hom}_{\mathbb{C}}(H_{n,p},\mathbb{C})\cong(\mathbb{C}^{\times})^{n+1}/W$
where $W$ is the Weyl group.

\item The Weyl group $W$ has an action on $(n+1)$-tuples
$(\beta_0,\ldots,\beta_n)\in (\mathbb{C}^{\times})^{n+1}$ which is
generated by all permutations of the elements
$(\beta_1,\ldots,\beta_n)$ and the maps \linebreak
$(\beta_0,\beta_1,\ldots,\beta_i,\ldots,\beta_n)\rightarrow(\beta_0\beta_i,\beta_1,\ldots,\beta_i^{-1},\ldots,\beta_n)$
for $i=1,\ldots, n$.

\item For $f$ a given simultaneous eigenform of all Hecke
operators $T\in H_n$ with respective eigenvalues $\lambda_f(T)$,
the map $T\mapsto \gamma_f(T)$ is an element of
$\mathrm{Hom}_{\mathbb{C}}(H_{n,p},\mathbb{C})$.
\end{enumerate}

\begin{definition}
The Satake $p$-parameters associated to the eigenform
$f\in\mathcal{M}_k(\Gamma^n)$ are the elements of the
$(n+1)$-tuple, $(a_{0,p},a_{1,p},\ldots,a_{n,p})$ in
$(\mathbb{C}^{\times})^{n+1}/W$, which is the image of the map
$T\rightarrow\lambda_f(T)$ under the isomorphism
$\mathrm{Hom}_{\mathbb{C}}(H_{n,p},\mathbb{C})\cong(\mathbb{C}^{\times})^{n+1}/W$.
\end{definition}

Let $f$ be a Siegel modular form of degree $n$ for the full
modular group $\Gamma_n=Sp(2n,\mathbb{Z})$. If $f$ is an
eigenfunction for the action of the Hecke algebra, then there are
two $L$-function associated with $f$. Let $a_0,a_1,\ldots,a_n$ be
the Satake parameters of $f$, and define:

\begin{definition}[Standard $L$-function]
\begin{equation}
D_{f,p}(T)=(1-T)\prod_{i=1}^{n}(1-a_{i,p}T)(1-a_{i,p}^{-1}T)\;,
\end{equation}
then the standard zeta function is

\begin{equation}
D_f(s)=\prod_p[D_{f,p}(p^{-s})]^{-1}\;.
\end{equation}
\end{definition}

\begin{definition}[Spinor Zeta Function]
Let $F\in\mathcal{M}_k(\Gamma^n)$ be the simultaneous eigenform
for all the Hecke operators in $H_n$. Define
\begin{equation}
Z_{f,p}(T)=(1-a_{0,p}T)\prod_{r=1}^{n}\prod_{1\leq
i_1<i_2\ldots<i_r\leq n}(1-a_{0,p}a_{i_1,p}\ldots a_{i_r,p}T)\;,
\end{equation}
The spinor zeta function is
$Z_f(s)=\prod_{p}[Z_{f,p}(p^{-s})]^{-1}$\;.
\end{definition}

For genus $1$, the spinor zeta function is equal to the usual
$L$-function associated to a weight $k$ eigenform, up to
normalization:

\begin{equation}
a(1)Z_F(s)=\sum_{n=1}^{\infty}a(n)n^{-s}=\prod_p(1-\gamma_f(p)p^{-s}+p^{k-1}(p^{-s})^2)^{-1}\;.
\end{equation}
Thus we may think of the spinor zeta function as a generalization
of the usual genus $1$ $L$-functions.

For $n=2$ the spinor $L$-function reads:
\begin{eqnarray*}
Z_{f,p}(T)&=&(1-a_{0,p}T)(1-a_{0,p}a_{1,p}T)(1-a_{0,p}a_{2,p}T)(1-a_{0,p}a_{1,p}a_{2,p}T)\\
&=&1-\lambda_f(p)T+(\lambda_f(p)^2-\lambda(p^2)-p^{2k-4})T^2-\lambda_f(p)p^{2k-3}T^3+p^{4k-6}T^4\;.
\end{eqnarray*}

\noindent Therefore comparing with $1+aT+bT^2+ap^3T^3+p^6T^4$, we
see that we are searching modular forms of weight $3$ i.e. $k=3$.

For $n=3$ part of the Standard $L$-function with $T=p^{3/2}t$ is
precisely of the form of the sextic factors in the zeta functions,
namely:
\begin{equation}
\frac{D_{f,p}(T)}{(1-T)}\Bigg|_{T=p^{3/2}t}=
1+at+bt^2+ct^3+bp^3t^4+cp^6t^5+p^9t^6\;,
\end{equation}
where,
\begin{equation}
a=\left(a_1+a_2+a_3+\frac{1}{a_1}+\frac{1}{a_2}+\frac{1}{a_3}\right)p^{3/2}\;,
\end{equation}

\begin{equation}
b=\left(3+\frac{a_1}{a_2}+\frac{a_2}{a_3}+\frac{a_1}{a_3}+\frac{a_3}{a_1}+\frac{a_2}{a_3}+\frac{a_3}{a_2}
+a_1a_2+a_1a_3+a_2a_3+\frac{1}{a_1a_2}+\frac{1}{a_2a_3}+\frac{1}{a_3a_1}\right)p^3\;,
\end{equation}
and
\begin{equation}\footnotesize{
c=\left(a_1a_2a_3+2a_1+2a_2+2a_3+\frac{2}{a_1}+\frac{2}{a_2}+\frac{2}{a_3}+\frac{a_1a_2}{a_3}+\frac{a_1a_3}{a_2}+\frac{a_2a_3}{a_1}
+\frac{a_3}{a_1a_2}+\frac{a_2}{a_1a_3}+\frac{a_1}{a_2a_3}\right)p^{9/2}}\;.
\end{equation}

\noindent Similarly we could have set $T=p^{3/2}t$ for $n=2$ to
obtain:
\begin{eqnarray}
\frac{D_{f,p}(T)}{(1-T)}\Bigg|_{T=p^{3/2}t}&=&1-\left(a_1+a_2+\frac{1}{a_1}+\frac{1}{a_2}\right)p^{3/2}t\nonumber\\
&&+\left(2+\frac{a_1}{a_2}+\frac{a_2}{a_1}+a_1a_2+\frac{1}{a_1a_2}\right)p^3t^2\nonumber\\
&&-\left(a_1+a_2+\frac{1}{a_1}+\frac{1}{a_2}\right)p^{9/2}t^3+t^4\;,\nonumber\\
\end{eqnarray}
which has the form of the quartic factors discovered in the zeta
function, $1+at+bt^2+ap^3t^3+p^6t^4$. Here:

\begin{equation}
a=-\left(a_1+a_2+\frac{1}{a_1}+\frac{1}{a_2}\right)p^{3/2}\;,
\end{equation}
and
\begin{equation}
b=\left(2+\frac{a_2}{a_1}+\frac{a_1}{a_2}+a_1a_2+\frac{1}{a_1a_2}\right)p^{3}\;.
\end{equation}

Hence, we able, in principal, to derive the Satake parameters of
any potential associated modular form from our data, by solving
for the $a_i$ given the coefficients of the quartic or sextic
factors using the formulae above.

\subsection[List of Eigenvalues $\lambda(p)$ and $\lambda(p^2)$]{Lists of Eigenvalues $\lambda(p)$ and
$\lambda(p^2)$}\label{sectioneigenvalues} In this section, we
record potential values of $\lambda(p)$ and $\lambda(p^2)$ that
are obtained from quartics for the primes $3,5$ and $7$. The
associated values of $\phi$ and $\psi$ are given, and also the
monomial, $\textbf{v}$, whose associated contribution to the zeta
function, $R_{\textbf{v}}$, contained the quartic as a factor (up
to fractional powers).

\begin{tabular}{|c|c|c|c|c|}\hline
\multicolumn{5}{|c|}{p=3}\\\hline $\phi$&$\psi$&Monomial
$\textbf{v}$&Quartic&$\lambda(p^2)$\\\hline
$0$&$0$&$(0,0,0,0,0)$&$(0,-18)_4$&$9$\\
&&$(0,0,0,2,2)$& & \\
&&$(0,0,2,1,1)$& & \\
&&$(2,2,1,1,0)$& & \\\hline
$0$&$1,2$&$(0,0,0,0,0)$&$(-6)_2(8)_2$&$-11$\\\hline
$1$&$1,2$&$(0,0,0,0,0)$&$(2,-18)_4$&$13$\\\hline
\end{tabular}

\begin{tabular}{|c|c|c|c|c|}\hline
\multicolumn{5}{|c|}{p=5}\\\hline $\phi$&$\psi$&Monomial
$\textbf{v}$&Quartic&$\lambda(p^2)$\\\hline
$0$&$0$&$(0,0,0,0,0)$&$(0,-70)_4$&$45$\\\hline
$0$&$1,2,3,4$&$(0,0,0,0,0)$&$(-4,-90)_4$&$81$\\\hline
$2$&$1,2,3,4$&$(0,0,0,0,0)$&$(-5,-1200)_4$&$1200$\\\hline
$3$&$1,2,3,4$&$(0,0,0,0,0)$&$(-12,190)_4$&$-71$\\\hline
$4$&$1,2,3,4$&$(0,0,0,0,0)$&$(-14)_2(10)_2$&$-119$\\\hline
\end{tabular}

\begin{tabular}
{|c|c|c|c|c|}\hline \multicolumn{5}{|c|}{p=7}\\\hline
$\phi$&$\psi$&Monomial $\textbf{v}$&Quartic&$\lambda(p^2)$\\\hline
$0$&$2,5$&$(0,0,0,0,0)$&$(-8,-210)_4$&$225$\\\hline
$0$&$0$&$(0,0,0,0,0)$&$(-14,-98)_4$&$245$\\\hline
$2$&$0$&$(0,0,0,0,0)$&$(0,2.3.7^2)_4$&$-343$\\\hline
$2$&$1,6$&$(0,2,1,1,1)$&$(0,98)_4$&$-147$\\\hline
$0$&$3,4$&$(0,0,2,1,1)\&(2,2,1,1,0)$&$(0,-490)_4$&$441$\\
&&$(6,0,1,0,0)\&(2,0,3,0,0)$& &\\
&&$(0,4,0,3,3)\&(4,0,1,1,0)$& &\\
&&$(2,2,1,1,0)$& & \\\hline
$2$&$2,5$&$(0,2,1,1,1)$& &\\\hline
$3$&$1,6$&$(0,0,0,0,0)$&$(-28,2.5.7^2)_4$&$245$\\\hline
$0$&$1,6$&$(0,0,0,0,0)$&$(20,34.7)_4$&$113$\\\hline
$2$&$3,4$&$(0,0,0,0,0)$&$(-26,2.29.7)_4$&$221$\\\hline
$4$&$2,5$&$(0,0,0,0,0)$&$(4,2.3.11.7)_4$&$-495$\\\hline
$4$&$3,4$&$(0,0,0,0,0)$&$(-16,2.7)_4$&$193$\\\hline
$5$&$1,6$&$(0,0,0,0,0)$&$(-6,-2.3.7)_4$&$29$\\\hline
$1$&$1,6$&$(0,0,0,0,0)$&$(-2,-98)_4$&$53$\\\hline
$1$&$2,5$&$(0,0,0,0,0)$&$(-12,-2.3.11)_4$&$161$\\\hline
$1$&$3,4$&$(0,0,0,0,0)$&$(18,2.41.7)_4$&$-299$\\\hline
$6$&$1,6$&$(0,0,0,0,0)$&$(-28,2.3.11.7)_4$&$273$\\\hline
$6$&$3,4$&$(0,0,0,0,0)$&$(8,2.5.7)_4$&$-55$\\\hline
\end{tabular}

\pagebreak
\section[Summary for the Mirror Octics]{Summary for the Mirror
Octics}\label{summarymirrorzeta} In Section \ref{mirroroctic} it
was found that:
\begin{eqnarray*}
\tilde{N} &=& 1+83q+83q^2+q^3 +
N_{(0,0,0,0,0)}+3q(q+1)\left(\frac{\phi^2-1}{p}\right)^s.
\end{eqnarray*}
where $q=p^s$ and $\left(\frac{a}{p}\right)$ is the Legendre
symbol.

\noindent It is clear that the form of the zeta function for the
mirror is one of the following:

\begin{enumerate}
\item When $\phi^2-1 \neq 0\,\mod\, p$, we get either:

\[
\tilde{\zeta}(\phi,\psi)=\left\{\begin{array}{l@{\quad
\quad}l}\frac{R_{\textbf{(0,0,0,0,0)}}}{(1-t)(1-pt)^{86}(1-p^2t)^{86}(1-p^3t)}
& \mathrm{if}\quad\left(\frac{\phi^2-1}{p}\right)=1,\\\\
\frac{R_{\textbf{(0,0,0,0,0)}}}{(1-t)(1-pt)^{83}(1-p^2t)^{83}(1+pt)^{3}(1+p^2t)^{3}(1-p^3t)}&\mathrm{if}\quad\left(\frac{\phi^2-1}{p}\right)=-1.
\end{array}\right.
\]\\

In this case $R_{\textbf{(0,0,0,0,0)}}$ has degree 6 or 5 (if in
addition $(8\psi^4+\phi)^2-1\,\mod\,p$).

\item When $\phi^2-1 = 0\,\mod\, p$ we get:

\begin{equation}
\tilde{\zeta}(\phi,\psi)=\frac{R_{\textbf{(0,0,0,0,0)}}}{(1-t)(1-pt)^{83}(1-p^2t)^{83}(1-p^3t)}\;;
\end{equation}

\noindent in this case $R_{\textbf{(0,0,0,0,0)}}$ either has
degree $4$ or (if in addition $(8\psi^4+\phi)^2-1=0\,\mod\,p$)
$3$.

\end{enumerate}

The zeta functions of the original family and the mirror both have
a sextic, $R_{\textbf{(0,0,0,0,0)}}$, in common, this is due to
the $6$ periods that they share. This confirms the equivalence
between the periods and the monomials first observed for the
quintic in \cite{COV1}, where $4$ periods were shared, and the
factor $R_{\textbf{(0,0,0,0,0)}}$ was generically a quartic. This
observation shall be explained in Section \ref{sectionlast}.

The splitting of the denominator according to the value of
$\phi^2$ may have a deeper significance. It was noted in Chapter
\ref{chapmirrorsymmetry} $h^{1,1}_{\mathrm{toric}}=83$, whereas
$h^{1,1}=86$ for the mirror octics. For the original family,
$h^{1,1}=h^{1,1}_{\mathrm{toric}}=2$, and correspondingly, no
splitting was observed in the denominator. This prompts the following conjecture:

\begin{conjecture}[Toric Splitting of Denominator]
For Calabi--Yau threefolds for which $h^{1,1}(X)\neq
h^{1,1}_{\mathrm{toric}}(X)$. Let
$\tilde{h}=h^{1,1}(X)-h^{1,1}_{\mathrm{toric}}(X)$ the denominator
of the zeta function will take the form:

\begin{equation}
(1-t)(1-pt)^{h^{1,1}_{\mathrm{toric}}}(1-p^2t)^{h^{1,1}_{\mathrm{toric}}}(1-\chi
pt)^{\tilde{h}}(1-\chi p^2t)^{\tilde{h}}(1-p^3t)\;,
\end{equation}

\noindent where $\chi$ is a character that depends on the defining
equation of a singular locus. At this same singular locus, the
denominator will take the form:

\begin{equation}
(1-t)(1-pt)^{h^{1,1}_{\mathrm{toric}}}(1-p^2t)^{h^{1,1}_{\mathrm{toric}}}(1-p^3t)\;.
\end{equation}

\end{conjecture}


\chapter[Summary of Results, Conjectures, and Open Questions]{Summary of Results, Conjectures, and Open Questions}\label{chaplast}
\normalsize

\section[Results and Conjectures on the Form of the Zeta Function]{Results and Conjectures on the Form of the Zeta Function}\label{sectionlast}

%
To begin, we recall the result of our main calculation. A priori,
for arbitrary octic hypersurfaces in $\mathbb{P}_4{}^{\,(1, 1, 2,
2, 2)}$, one would expect a zeta function of the form:

\begin{equation}
Z(X/\mathbb{F}_p,t)=\frac{R_{174}(t)}{(1-t)R_2(t)\acute{R}_2(t)(1-p^3t)}\;,
\end{equation}\\
however, restricting to a two-parameter family containing only
monomials which are invariant under the group of automorphisms,
$G$, we find a factorization of the zeta function of the form:

\begin{equation}
\zeta(\phi,\psi,p,t)=\frac{R_{\text{excep}}(t)\prod_{\textbf{v}}R_{\textbf{v}}(t)^{\lambda_{\textbf{v}}}}{(1-t)(1-pt)^2(1-p^2t)^2(1-p^3t)}\;,
\end{equation}\\

\noindent where each $\textbf{v}$ labels a classes of
representations of $G$. We would not expect such a beautiful
decomposition in general. The mirror is obtained by quotienting
out this special sub-family by the group of automorphisms, $G$
\cite{GP}. The zeta function of the mirror shares a factor
$R_{(0,0,0,0,0)}$which has degree equal to the number of shared
periods. Hence, in the case of the quintic threefolds,
$R_{(0,0,0,0,0)}$ was a quartic, and for the octic threefolds
$R_{(0,0,0,0,0)}$ was a sextic. This gives rise to the following
theorem:
\pagebreak
\begin{theorem}[Mirror Theorem for Hypersurfaces of Fermat Type]
For mirror pairs of Calabi--Yau threefolds which can be described
using the Greene--Plesser construction \textup{(}i.e.
hypersurfaces of Fermat type in weighted projective
space\textup{)}, the $G$-invariant part, i.e. $R_{(0,0,0,0,0)}$ of
the zeta function of the Calabi--Yau threefold is preserved. The
degree of this factor is equal to the number of shared periods.
\end{theorem}

This was observed for the quintic in \cite{COV2}, and for the
octic here. It would be interesting to compute the zeta functions
for mirror symmetry pairs of Calabi--Yau manifolds that are not of
Fermat type. This should be possible using existing methods, i.e.
using Cox variables and the Batyrev procedure.

\begin{proof}
For Fermat hypersurfaces the dual of (the larger) Newton
polyhedron consists of a subset of the points of the original,
this is because the mirror is formed by taking the quotient by the
Greene--Plesser group of automorphisms. All monomials in this
`smaller' polyhedron are invariant under the Greene--Plesser group
of automorphisms. Each point corresponds to a period, as was shown
in Chapter \ref{chappicardfuchs}.

The number of shared periods for Fermat hypersurfaces is equal to
the number of monomials in the smaller of the Newton polyhedra
excluding the unique interior point. For the family of manifolds
that have defining equations given by the monomials in the smaller
polyhedron (in our example, for instance, this is the mirror octic
threefolds defined by (\ref{equationdefmirror})), the shared
periods are all the periods, and the number of periods is equal to
the dimension of the third cohomology group. Let us denote this
family, the `mirror family', for convenience.

The method of calculation of the number of rational points by
using Gauss sums consisting of Dwork's character, gives the number
of points in terms of the monomials or equivalently, the periods.
In both families there is a contribution to the number of rational
points coming from the $G$-invariant monomials. However applying
the Weil conjectures for the `mirror family' gives us that the
degree of the numerator of the zeta function has to be equal to
the third Betti number, $b_3$. The contribution to the number of
rational points that gives the denominator is not dependent on the
monomials/periods, and hence the numerator of the zeta function
comes from the contribution to the number of points from the
$G$-invariant monomials. Let this numerator be denoted
$R_{(0,0,0,0,0)}$.

The special $G$-invariant sub-family of the original family that
we choose also has a contribution to the zeta function in terms of
the $G$-invariant monomials/periods. Hence, the numerator of the
zeta function of this family contains the factor
$R_{(0,0,0,0,0)}$. \qed
\end{proof}

The calculation of the number of points on the original family led
to a parameter invariant part, $R_{\text{excep}}(t)$, and prompted
the following conjecture:

\begin{conjecture}[Non-monomial splitting of the Numerator]
For a family of three-dimensional Calabi--Yau hypersurfaces, if
$h^{2,1}>h^{2,1}_{\mathrm{poly}}$, the numerator of the zeta
function has a factor that is independent of the parameters
labelling each member of the family. This factor, denoted
$R_{\text{excep}}(t)$, will have degree $2\acute{h}$, where
$\acute{h}=h^{2,1}-h^{2,1}_{\mathrm{poly}}$.\label{nonmonomialsplit}
\end{conjecture}

As remarked before, the findings for the family of quintic
threefolds\cite{COV2}, where no such parameter independent factor
was found in the zeta function, is in keeping with the above
conjecture, as in that case,
$h^{2,1}_{\mathrm{poly}}=h^{2,1}=101$.

The calculation of the mirror octic in Section \ref{mirroroctic}
hinted that the arithmetic may `know' about the monomial-divisor
map, i.e. the monomial and non-monomial deformations of the
defining polynomial for the original family and equivalently the
toric and non-toric divisors of the mirror.

\begin{conjecture}[Toric Splitting of Denominator]
For Calabi--Yau threefolds for which $h^{1,1}(X)\neq
h^{1,1}_{\mathrm{toric}}(X)$. Let
$\tilde{h}=h^{1,1}(X)-h^{1,1}_{\mathrm{toric}}(X)$ the denominator
of the zeta function takes the form:

\begin{equation}
(1-t)(1-pt)^{h^{1,1}_{\mathrm{toric}}}(1-p^2t)^{h^{1,1}_{\mathrm{toric}}}(1-\chi
pt)^{\tilde{h}}(1-\chi p^2t)^{\tilde{h}}(1-p^3t)\;,
\end{equation}

\noindent where $\chi$ is a character that depends on the defining
equation of a singular locus. At this singular locus, the
denominator will take the form:

\begin{equation}
(1-t)(1-pt)^{h^{1,1}_{\mathrm{toric}}}(1-p^2t)^{h^{1,1}_{\mathrm{toric}}}(1-p^3t)\;.
\end{equation}

\label{toricsplit}
\end{conjecture}

\begin{remark}Indeed the above two conjectures may really be viewed as a
single mirror symmetric conjecture viewed from different sides,
because from Batyrev's formulae, $\acute{h}=\tilde{h}$, i.e.
$h^{2,1}(X)-h^{2,1}_{\mathrm{poly}}(X)=h^{1,1}(Y)-h^{1,1}_{\mathrm{toric}}(Y)$,
for a mirror symmetric pair of manifolds $(X,Y)$.
\end{remark}

\begin{conjecture}[Singularity Conjectures]\hfill
\begin{enumerate}
\item Degenerations due to Singularities:\\
The degeneration of the degree of each piece of the zeta function
$\deg(R_{\textbf{v}})$ of a Calabi--Yau manifold at a singular
point in the moduli space is determined by the nature of the
singularity.

\item Degenerations due to Isolated $A_1$ Singularities:\\
For isolated $A_1$ singularities \textup{(}ordinary double points
or conifold points\textup{)}, the degree of each contribution $\deg(R_{\textbf{v}})$ goes
down by exactly $1$.

\item `Degeneration Conservation':\\
 Where two types of singularity
coincide, the total degeneration in degree of each contribution $\deg(R_{\textbf{v}})$ is
the sum of the degenerations due to each type of singularity.
\end{enumerate}
\end{conjecture}

It would be interesting to classify the types of degeneration that
occur at singular points in the moduli space. This `arithmetic
classification' of singularities may differ from geometric
classifications; comparisons would be illuminating. The
establishment of a procedure for relating the zeta function of a
singular variety to the zeta functions of its resolutions would
also be desirable.

One way to proceed which is likely to yield quick results would be
to study K3 surfaces. Mirror symmetry of K3 surfaces has a
slightly different flavour: the Hodge diamond remains unchanged.
Here the significant objects are the Picard lattices. The Picard
number of a K3 and its mirror add up to $20$, by Dolgachev's
construction of Mirror symmetry \cite{Dol}. K3 manifolds often
admit an elliptic Weierstrass fibrations. Given such a fibration,
the Picard number can be calculated by using the Shioda--Tate
formula. For K3 manifolds with both a toric description and a
Weierstrass description, the behaviour of the zeta function at
singularities can be explored, in a more systematic manner due to
the Kodaira classification of singular fibres.


The following conjecture (not by the author) is widely believed:
\begin{conjecture}[Modularity Folklore]
For a $d$-dimensional Calabi--Yau variety, the powers $\alpha_p$
 that occur in the expansion in the bad primes, $N=\prod_{p\;\text{bad}}p^{\alpha_p}$, of  the level $N$ of the weight $d+1$ modular
form for some $\Gamma_{0}(N)$, increases with the `badness' of the
singularity.
\end{conjecture}

The level $N$ of the modular form is a product of powers of the
bad primes. The powers are related to the type of singularity.

As remarked before in Section \ref{modularity}, this is similar to
the case for modular forms. $E$ has bad reduction when it has a
singularity modulo $p$ with the type of singularity determining
the power of $p$ that occurs in the conductor which equal to the
level of the associated weight two elliptic modular form.

\section[Open Questions]{Open Questions}
\begin{enumerate}

\item
Finding analogues to the Euler curves of \cite{COV2}. It is
apparent that this may be found by deeper analysis of the
Picard--Fuchs equations corresponding to the pairs of
monomials, namely, $(0,0,2,1,1)\, \&\, (2,2,1,1,0)$, $(6,0,1,0,0)\, \&\,
(2,0,3,0,0)$ and $(0,4,0,3,3)\, \&\, (4,0,1,1,0)$. In \cite{COV2}
the geometric origin of Euler curves $\mathcal{A}$ and $\mathcal{B}$ curves is
not clear, and they only arise through studying the periods of the
Calabi--Yau manifold. Also it is known that the Jacobian of the
Calabi--Yau manifold contains the Jacobians of the curves,
$\mathcal{A}$ and $\mathcal{B}$. However Jacobians of an
arithmetic variety often contain the Jacobians of lower
dimensional varieties, so studying these particular Picard--Fuchs
equations in more detail would be a first step to shedding light
on this problem.

\item
Both the K3 surfaces and the octic threefolds, exhibiting
interesting behaviour $2$-adically, ($2$ being the only prime of
bad reduction). The $5$-adic analysis of the zeta functions of the
quintic threefolds in \cite{COV2} ($5$ was the only bad prime for
this case), suggested arithmetic analogues of the large complex
structure limit. These computations may also aid the formulation
of a `quantum corrected' zeta function. With this aim in mind, it
would also be illuminating to study other families of Calabi--Yau
manifolds with other bad primes, particularly models with more
than one bad~prime.

\item
Determination of the Siegel modular forms connected with these
local zeta functions are a key step in figuring out the full
(global) Hasse--Weil zeta functions. However this is likely to be
difficult due to the reasons described in Section \ref{siegel}.

\item
The fact that the number of points may be written in terms of the
periods of the Calabi--Yau manifold, hints at a possible
arithmetic analogue of Special K\"{a}hler geometry. Also further
investigation of the vector-valued Siegel modular form, which
arises from Special Geometry, from an arithmetic perspective could
be undertaken.



\item
Arithmetic Mirror Symmetry may, as is hoped, have some physical
significance. Indeed the original conformal field theoretic
constructions of Mirror symmetry remains an important tool.
\cite{Schimmrigk} shows it may be possible to derive certain
conformal field theoretic quantities from the algebraic
Calabi--Yau variety by reducing the variety over finite fields. In
particular, the fusion rules of the underlying conformal field
theories can be derived from the number theoretical structure of
the cohomological Hasse--Weil $L$-function. The Hasse--Weil
$L$-series can be interpreted as Hecke $L$-series of an algebraic
number field. In the simplest case is that of Gepner's
construction of tensor models of minimal $N=2$ superconformal
field theories.

\item
Another connection to physics comes from the `attractor mechanism'
for dyonic supersymmetric black holes \cite{FKS}. The mechanism
gives rise to equations on the Hodge structure of Calabi--Yau
manifolds, providing a framework which naturally isolates certain
arithmetic varieties. Several conjectures about arithmetic
properties of these `attractor' varieties were made in a long
paper by Gregory Moore \cite{Moore1} (a summary of which was
provided in \cite{Moore2}). We shall not give details here, but
shall mention that any rigid Calabi--Yau is automatically an
attractor variety and modular K3 surfaces over $\mathbb{Q}$ turn
out to be attractor varieties. All known examples of (elliptic)
modular Calabi--Yau varieties are rigid and hence automatically
attractors. A relationship between attractors and modularity would
be quite fascinating.
\end{enumerate}

\appendix
\chapter[Symplectic Transformations in the Space of Complex
Structures]{Symplectic Transformations in the Space of Complex
Structures}\label{sectionsymplectic}
Here we recall some results concerning the symplectic modular
group which arise from Special Geometry, we follow the conventions
and notation of \cite{CdO}.

Let $\mathcal{M}_z$ be the Calabi--Yau manifold at the point $z^a$
in the moduli space. The space of complex structures can be
described in terms of the periods of the holomorphic three-form,
$\Omega$. Let $(A^a,B_b),\;a,b,=0\ldots,b_{21}$ be a canonical
homology basis for $H_3(\mathcal{M},\mathbb{Z})$ and let
$(\alpha_a,\beta^b)$ be the dual cohomology basis such that:

\begin{eqnarray}
\int_{A_b}\alpha_a=\int_{\mathcal{M}}\alpha_a\wedge\beta^b=\delta_a^{\,b},\quad\int_{B_a}\beta_b=\int_{\mathcal{M}}\beta^b\wedge\alpha_a=-\delta_a^{\,b}\;.
\end{eqnarray}

Also let:
\begin{eqnarray}
z^a=\int_{A^a}\Omega,\quad\mathcal{G}_a=\int_{B_a}\Omega\;.
\end{eqnarray}
The $(z^a, \mathcal{G}_a)$ are periods of $\Omega$.

The three-form $\Omega$ can be written as:
$\Omega=z^a\alpha_a-\mathcal{G}_a\beta^a$. This decomposition
rests on an arbitrary choice of basis $(\alpha_a,\beta^b)$ for
$H^3(\mathcal{M},\mathbb{Z})$. If
$(\tilde{\alpha}_a,\tilde{\beta}^b)$ is another symplectic basis
for $H^3(\mathcal{M},\mathbb{Z})$, then it can be related to
$(\alpha_a,\beta^b)$ by a the following transformation:

\begin{eqnarray}
\begin{pmatrix}\tilde{\alpha}\\\tilde{\beta}\end{pmatrix}=\begin{pmatrix}A&B\\C&D\\\end{pmatrix}\begin{pmatrix}\alpha\\\beta\end{pmatrix},\quad\begin{pmatrix}A&B\\C&D\\\end{pmatrix}\in\Sp(2b_{21}+2;\mathbb{Z}).
\end{eqnarray}

\noindent this transformation leads to a transformation rule for
the complex structure coordinates and the prepotential as follows:

\begin{eqnarray}
\begin{pmatrix}\tilde{\partial\mathcal{G}}\\\tilde{z}\end{pmatrix}=\begin{pmatrix}A&B\\C&D\\\end{pmatrix}\begin{pmatrix}\partial\mathcal{G}\\z\end{pmatrix}.
\end{eqnarray}

The action on the period matrix turns out to be:
\begin{equation}
\mathbb{G}\rightarrow\tilde{\mathbb{G}}=(A\mathbb{G}+B)(C\mathbb{G}+D)^{-1},
\end{equation}

\noindent this is analogous to the action of the symplectic group
on the period matrices of Riemann surfaces. We can rewrite the
transformation law for $z$ in the form:

\begin{equation}
z(\mathbb{G})\rightarrow
z(\tilde{\mathbb{G}})=(C\mathbb{G}+D)z(\mathbb{G}).
\end{equation}

Hence $z(\mathbb{G})$ is a vector valued modular form of
$\Sp(2b_{21}+2;\mathbb{Z})$ of weight $1$. Unfortunately, however,
we remarked in the Section \ref{siegel} that we need to find forms
of weight $3$.

 \pagebreak

\chapter[Results for the Octic threefolds]{Results for the Octic
threefolds}\label{appendix}
\subsection[New Notation]{New Notation}\label{notation}
Occasionally for clarity and conciseness, the following notation
shall be used in the tables:
\begin{center}
\begin{tabular}{|c|c|}\hline
Notation&Polynomial, for prime $p$\\\hline
$(a)_1$&$(1+at)$\\
$(a)_2$&$(1+at+p^3t^2)$\\
$(a,b)_4$&$(1+at+bt^2+ap^3t^3+p^6t^4)$\\
$(a,b,c)_6$&$(1+at+bt^2+ct^3+bp^3t^4+ap^6t^5+p^9t^6)$\\
$(a,b,c,d)_8$&$(1+at+bt^2+ct^3+dt^4+cp^3t^5+bp^6t^6+ap^9t^7+p^{12}t^8)$\\\hline
\end{tabular}
\end{center}

\subsection[Numerical Data Obtained]{Numerical Data Obtained}
In the table below we give the maximum $r$ for which the number of rational points over $\mathbb{F}_{p^r}$ was computed for each prime~$p$.
\begin{center}
\begin{tabular}{|c|c|}\hline
Prime $p$& Maximum $r$\\\hline
$3$&$8$\\
$5$&$5$\\
$7$&$4$\\
$11$&$3$\\
$13$&$3$\\
$17$&$3$\\\hline
\end{tabular}
\end{center}
The zeta functions for $p=19,23$ were also computed (using data up to $r=3$), but the results were too large to display in this thesis. For primes up to $73$, the numerical data up to $r=2$ has been obtained and this data was used to check modularity, but again there is insufficient space to display the corresponding zeta functions.

\pagebreak
\section[Summary for p=3]{Summary for p=3}
\subsubsection[Original Family]{Original Family}
$\phi=0, \psi=0$:

\begin{equation}
\frac{(0,-2p^2)_4^7(-2p)_2^7(2p)_2^7(0,-2p^3)_4^{6}(0)_2^{47}}{(-1)_1(-p)_1^2(-p^2)_1^2(-p^3)_1}
\end{equation}

\noindent$\phi=0, \psi=1,2$:

\begin{equation}
\frac{(2p)_2^5(8)_2(-p)_1^{27}(p)_1^{29}(0)_2^{12}(0,-2p^2)_4^3(0,-2p)_2^5}{(-1)_1(-p)_1^2(-p^2)_1^2(-p^3)_1}\\
\end{equation}

\noindent$\phi=1, \psi=0$:

\begin{equation}
\frac{(-p)_1^{9}(p)_1^{10}(0)_2^{26}}{(-1)_1(-p^2)_1^6(p^2)_1^3(-p^3)_1}
\end{equation}

\noindent$\phi=1, \psi=1,2$:

\begin{equation}
\frac{(-p)_1(p)_1^9(0)_2^{14}(2,-2p^2)_4(-2p)_2^{18}(2p)_2^{13}(0,-2p^2)_4^6}{(-1)_1(-p^2)_1^2(-p^3)_1}\\
\end{equation}

\noindent$\phi=2, \psi=0$:

\begin{equation}
\frac{(-p)_1^{8}(p)_1^{11}(0)_2^{26}}{(-1)_1(-p^2)_1^6(p^2)_1^3(-p^3)_1}
\end{equation}

\noindent$\phi=2, \psi=1,2$:

\begin{equation}
\frac{(-p)_1^{26}(p)_1^{36}(0)_2^6(-4)_2}{(-1)_1(-
p^2)_1^8(p^2)_1^6(-p^3t)_1}
\end{equation}

\pagebreak

\subsubsection[Mirror]{Mirror}
\noindent For $p=3$:\\
\begin{eqnarray}
\left(\frac{\phi^2-1}{p}\right)=\begin{cases}-1\quad \phi=0,\\0\quad \mathrm{otherwise}\end{cases}
\end{eqnarray}

\noindent $\phi=0,\psi=0$ (Smooth):\\
\begin{eqnarray}
\frac{(1+27t^2)(1-18t^2+729t^4)}{(1-t)(1-3t)^{83}(1-9t)^{83}(1+3t)^{3}(1+9t)^{3}(1-27t)}
\end{eqnarray}

\noindent $\phi=0,\psi=1,2$ (Conifold):\\
\begin{eqnarray}
\frac{(1-3t)(1-6t+27t^2)(1+8t+27t^2)}{(1-t)(1-3t)^{83}(1-9t)^{83}(1+3t)^{3}(1+9t)^{3}(1-27t)}
\end{eqnarray}

\noindent $\phi=1,\psi=1,2$ ($\phi^2=1$):\\
\begin{eqnarray}
    \frac{(1+2t-18t^2+54t^3+729t^4)}{(1-t)(1-3t)^{83}(1-9t)^{83}(1-27t)}
\end{eqnarray}

\noindent $\phi=1,\psi=0$ (Both):\\
\begin{eqnarray}
    \frac{(1-3t)(1+27t^2)}{(1-t)(1-3t)^{83}(1-9t)^{83}(1-27t)}
\end{eqnarray}

\noindent $\phi=2,\psi=0$ (Both):\\
\begin{eqnarray}
    \frac{(1+3t)(1+27t^2)}{(1-t)(1-3t)^{83}(1-9t)^{83}(1-27t)}
\end{eqnarray}

\noindent $\phi=2,\psi=1,2$ (Both):\\
\begin{eqnarray}
    \frac{(1+3t)(1-4t+27t^2)}{(1-t)(1-3t)^{83}(1-9t)^{83}(1-27t)}
\end{eqnarray}

\noindent Notice that $a_3$ of $V_{2,4}(q)$ is $4$, hence
agreement with \cite{Villegas}. \pagebreak
\subsection[Summary for p=5]{Summary for p=5}
\subsubsection[Original Family]{Original Family}
The contribution from the exceptional divisor is: $(1-10t+125t^2)^3$.

\noindent $\phi=0, \psi=0$:
\begin{equation}
\frac{(0,-7.5.2)_4(-2.5)_2^{13}(2.5)_2^{20}(-4.5)_2^{13}(4.5)_2^{13}(0)_2^{20}}{(-1)_1(-5)_1^2(-5^2)_1^2(-5^3)_1}
\end{equation}


\noindent $\phi=0, \psi=1,2,3,4$:
\begin{equation}
\frac{(0,-2p^2)_4^3(-2p)_2^5(2p)_2^5(8)_2(-p)_1^{27}(p)_1^{29}(0)_2^{12}}{(-1)_1(-p^2)_1^5(p^2)_1^3(-p^3)_1}
\end{equation}

\noindent $\phi=1, \psi=0$:
\begin{equation}
\frac{(-5)_1(5)_1^{18}(-22)_2(-2.5)_2^{18}(2.5)_2^{7}}{(-1)_1(-5^2)_1^3(5^2)_1^6(-5^3)_1}
\end{equation}


\noindent $\phi=1, \psi=1,2,3,4$:
\begin{eqnarray}
\frac{(-5)_1(5)_1^9(0)_2^{14}(-2.5p)_2^{18}(2.5)_2^{13}(0,-2.5^2)_4^6(2,-2.5^2)_4}{(-1)_1(-5^2)_1^2(-5^3)_1}\nonumber\\
\end{eqnarray}


\noindent $\phi=2, \psi=0$:
\begin{equation}
\frac{(-5)_1^{8}(5)_1^{11}(0)_2^{26}}{(-1)_1(-5^2)_1^6(5^2)_1^3(-5^3)_1}
\end{equation}


\noindent $\phi=2, \psi=1,2,3,4$:
\begin{equation}
\frac{(-5)_1^{26}(5)_1^{36}(0)_2^6(-4)_2}{(-1)(-5^2)_1^8(5^2)_1^6(-5^3)_1}
\end{equation}


\noindent $\phi=3, \psi=0$:
\begin{eqnarray}
\frac{(4)_2(10)_2^{26}(-10)_2^{33}(20)_2^{9}(-20)_2^{18}}{(-1)_1(-5)_1^2(-5^2)_1^2(-5^3)_1}
\end{eqnarray}

\noindent $\phi=3, \psi=1,2,3,4$:
\begin{eqnarray}
\frac{(5)_1^{58}(-12,190)_4(10)_2(-10)_2^{14}(20)_2^{3}(-20)_2^{9}}{(-1)_1(-5)_1^2(-5^2)_1^2(-5^3)_1(5^2)_1^6}
\end{eqnarray}

\noindent $\phi=4, \psi=0$:
\begin{equation}
\frac{(-5)_1^{13}(5)_1^{8}(10)_2^{12}(-10)_2^{13}(22)_2}{(-1)_1(-5)_1^2(-5^2)_1^3(-5^3)_1(5^2)_1^6}
\end{equation}

\noindent $\phi=4, \psi=1,2,3,4$:
\begin{equation}
\frac{(0,-2.5^3)_4^{9}(-14)_2(10)_2^{20}(-10)_2^{26}(5)_1^{6}(-5)_1^{6}}{(-1)_1(-5)_1^2(-5^2)_1^2(-5^3)_1}
\end{equation}

\pagebreak
\subsubsection[Mirror]{Mirror}
\noindent For $p=5$:\\
\begin{eqnarray}
\left(\frac{\phi^2-1}{p}\right)=\begin{cases}-1\quad \phi=2,3,\\0\quad \phi=1,4\\1\quad\phi=0\end{cases}
\end{eqnarray}

\noindent $\phi=0,\psi=0$ (Smooth):\\
\begin{eqnarray}
\frac{(1-10t+5^3t^2)(1-70t^2+5^6t^4)}{(1-t)(1-5t)^{86}(1-25t)^{86}(1-125t)}
\end{eqnarray}

\noindent $\phi=0,\psi=1,2,3,4$ (Smooth):\\
\begin{eqnarray}
\frac{(1+5^3t^2)(1-4t-90t^2-4.5^3t^3+5^6t^4)}{(1-t)(1-5t)^{86}(1-25t)^{86}(1-125t)}
\end{eqnarray}

\noindent $\phi=2,\psi=0$ (Smooth):\\
\begin{eqnarray}
\frac{(1-10t+5^3t^2)(1+10t+5^3t^2)(1-4t+5^3t^2)}{(1-t)(1-5t)^{83}(1-25t)^{83}(1+5t)^{3}(1+25t)^{3}(1-125t)}
\end{eqnarray}

\noindent $\phi=2,\psi=1,2,3,4$ (Smooth):\\
\begin{eqnarray}
\frac{(1-5t^2-1200t^3-5.5^3t^4+5^9t^6)}{(1-t)(1-5t)^{83}(1-25t)^{83}(1+5t)^{3}(1+25t)^{3}(1-125t)}
\end{eqnarray}

\noindent $\phi=3,\psi=0$ (Smooth):\\
\begin{eqnarray}
\frac{(1+4t+5^3t^2)(1+10t+5^3t^2)^2}{(1-t)(1-5t)^{83}(1-25t)^{83}(1+5t)^{3}(1+25t)^{3}(1-125t)}
\end{eqnarray}

\noindent $\phi=3,\psi=1,2,3,4$ (Conifold):\\
\begin{eqnarray}
\frac{(1+5t)(1-12t+190t^2-12.5^3t^3+5^6t^4)}{(1-t)(1-5t)^{83}(1-25t)^{83}(1+5t)^{3}(1+25t)^{3}(1-125t)}
\end{eqnarray}

\noindent $\phi=4,\psi=1,2,3,4$ ($\phi^2=1$):\\
\begin{eqnarray}
\frac{(1-14t+5^3t^2)(1+10t+5^3t^2)}{(1-t)(1-5t)^{83}(1-25t)^{83}(1-125t)}
\end{eqnarray}

\noindent $\phi=1,\psi=0$ (Both):\\
\begin{eqnarray}
\frac{(1-5t)(1-22t+5^3t^2)}{(1-t)(1-5t)^{83}(1-25t)^{83}(1-125t)}
\end{eqnarray}

\noindent $\phi=4,\psi=0$ (Both):\\
\begin{eqnarray}
\frac{(1-5t)(1+22t+5^3t^2)}{(1-t)(1-5t)^{83}(1-25t)^{83}(1-125t)}
\end{eqnarray}

\noindent $\phi=1,\psi=1,2,3,4$ (Both):\\
\begin{eqnarray}
\frac{(1+5t)(1+2t+5^3t^2)}{(1-t)(1-5t)^{83}(1-25t)^{83}(1-125t)}
\end{eqnarray}

Again the case $\phi=4,\psi=1,2,3,4$ confirms \cite{Villegas}.

\pagebreak
\subsection[Summary for p=7]{Summary for p=7}
We shall only record the summary for the mirror:
\subsubsection[Mirror]{Mirror}
\noindent For $p=7$:\\
\begin{eqnarray}
\left(\frac{\phi^2-1}{p}\right)=\begin{cases}-1\quad \phi=0,2,5,\\0\quad \phi=1,6\\1\quad\phi=3,4\end{cases}
\end{eqnarray}

\noindent $\phi=0,\psi=0$ (Smooth):\\
\begin{eqnarray}
\frac{(1+7^3t^2)^3}{(1-t)(1-7t)^{83}(1-49t)^{83}(1+7t)^{3}(1+49t)^{3}(1-343t)}
\end{eqnarray}

\noindent $\phi=0,\psi=2,5$ (Smooth):\\
\begin{eqnarray}
\frac{(1-14t+7^3t^2)(1-8t-30.7t^2+8.7^3t^3+7^6t^4)}{(1-t)(1-7t)^{83}(1-49t)^{83}(1+7t)^{3}(1+49t)^{3}(1-343t)}
\end{eqnarray}

\noindent $\phi=0,\psi=3,4$ (Smooth):\\
\begin{eqnarray}
\frac{(1-6t-147t^2+44.7^2t^3-147.7^3t^4+6.7^6t^5+7^9t^6)}{(1-t)(1-7t)^{83}(1-49t)^{83}(1+7t)^{3}(1+49t)^{3}(1-343t)}
\end{eqnarray}

\noindent $\phi=2,\psi=0$ (Smooth):\\
\begin{eqnarray}
\frac{(1+28t+7^3t^2)(1+2.3.7^2t^2+7^6t^4)}{(1-t)(1-7t)^{83}(1-49t)^{83}(1+7t)^{3}(1+49t)^{3}(1-343t)}
\end{eqnarray}

\noindent $\phi=2,\psi=1,6$ (Smooth):\\
\begin{eqnarray}
\frac{(1+4t+19.7t^2-8.7^2t^3+19.7.7^3t^4+4.7^6t^5+7^9t^6)}{(1-t)(1-7t)^{83}(1-49t)^{83}(1+7t)^{3}(1+49t)^{3}(1-343t)}
\end{eqnarray}

\noindent $\phi=2,\psi=2,5$ (Smooth):\\
\begin{eqnarray}
\frac{(1-2t+3.7t^2-2^2.3.5.7^2t^3+3.7.7^3t^4-2.7^6t^5+7^9t^6)}{(1-t)(1-7t)^{83}(1-49t)^{83}(1+7t)^{3}(1+49t)^{3}(1-343t)}
\end{eqnarray}

\noindent $\phi=3,\psi=0$ (Smooth):\\
\begin{eqnarray}
\frac{(1+7^3t^2)(1+28t+7^3t^2)(1-28t+7^3t^2)}{(1-t)(1-7t)^{86}(1-49t)^{86}(1-343t)}
\end{eqnarray}

\noindent $\phi=3,\psi=1,6$ (Smooth):\\
\begin{eqnarray}
\frac{(1-4t+7^2t^2+2^3.7^2t^3+7^2.7^3t^4-4.7^6t^5+7^9t^6)}{(1-t)(1-7t)^{86}(1-49t)^{86}(1-343t)}
\end{eqnarray}

\noindent $\phi=3,\psi=2,5$ (Smooth):\\
\begin{eqnarray}
\frac{(1-6t+59.7t^2+2^2.7^2.11t^3+59.7.7^3t^4-6.7^6t^5+7^9t^6)}{(1-t)(1-7t)^{86}(1-49t)^{86}(1-343t)}
\end{eqnarray}

\noindent $\phi=3,\psi=3,4$ (Smooth):\\
\begin{eqnarray}
\frac{(1+14t-37.7t^2+47.7^2.2^2t^3-37.7.7^3t^4+14.7^6t^5+7^9t^6)}{(1-t)(1-7t)^{86}(1-49t)^{86}(1-343t)}
\end{eqnarray}

\noindent $\phi=4,\psi=1,6$ (Smooth):\\
\begin{eqnarray}
\frac{(1+16t+343t^2)(1+28t+343t^2)(1-28t+343t^2)}{(1-t)(1-7t)^{86}(1-49t)^{86}(1-343t)}
\end{eqnarray}

\noindent $\phi=5,\psi=0$ (Smooth):\\
\begin{eqnarray}
\frac{(1-28t+343t^2)(1+2.3.7^2t^2+7^6t^4)}{(1-t)(1-7t)^{83}(1-49t)^{83}(1+7t)^{3}(1+49t)^{3}(1-343t)}
\end{eqnarray}

\noindent $\phi=5,\psi=2,5$ (Smooth):\\
\begin{eqnarray}
\frac{(1+14t+43.7t^2+2^2.7^2.17t^3+43.7.7^3t^4+14.7^6t^5+7^9t^6)}{(1-t)(1-7t)^{83}(1-49t)^{83}(1+7t)^{3}(1+49t)^{3}(1-343t)}
\end{eqnarray}

\noindent $\phi=5,\psi=3,4$ (Smooth):\\
\begin{eqnarray}
\frac{(1-4t+3^3.7t^2+2^3.7^2.11t^3+3^3.7.7^3t^4-4.7^6t^5+7^9t^6)}{(1-t)(1-7t)^{83}(1-49t)^{83}(1+7t)^{3}(1+49t)^{3}(1-343t)}
\end{eqnarray}

\noindent $\phi=0,\psi=1,6$ (Conifold):\\
\begin{eqnarray}
\frac{(1-7t)(1+20t+34.7t^2+20.7^3t^3+7^6t^4)}{(1-t)(1-7t)^{83}(1-49t)^{83}(1+7t)^{3}(1+49t)^{3}(1-343t)}
\end{eqnarray}

\noindent $\phi=2,\psi=3,4$ (Conifold):\\
\begin{eqnarray}
\frac{(1+7t)(1-26t+2.29.7t^2+26.7^3t^3+7^6t^4)}{(1-t)(1-7t)^{83}(1-49t)^{83}(1+7t)^{3}(1+49t)^{3}(1-343t)}
\end{eqnarray}

\noindent $\phi=4,\psi=2,5$ (Conifold):\\
\begin{eqnarray}
\frac{(1+7t)(1+4t+2.3.11.7t^2+4.7^3t^3+7^6t^4)}{(1-t)(1-7t)^{86}(1-49t)^{86}(1-343t)}
\end{eqnarray}

\noindent $\phi=4,\psi=3,4$ (Conifold):\\
\begin{eqnarray}
\frac{(1-7t)(1-16t+2.7t^2-16.7^3t^3+7^6t^4)}{(1-t)(1-7t)^{86}(1-49t)^{86}(1-343t)}
\end{eqnarray}

\noindent $\phi=5,\psi=1,6$ (Conifold):\\
\begin{eqnarray}
\frac{(1+7t)(1-6t-2.3.7t^2-6.7^3t^3+7^6t^4)}{(1-t)(1-7t)^{83}(1-49t)^{83}(1+7t)^{3}(1+49t)^{3}(1-343t)}
\end{eqnarray}

\noindent $\phi=1,\psi=1,6$ ($\phi^2=1$):\\
\begin{eqnarray}
\frac{(1-2t-98t^2-2.7^3t^3+7^6t^4)}{(1-t)(1-7t)^{83}(1-49t)^{83}(1-343t)}
\end{eqnarray}

\noindent $\phi=1,\psi=2,5$ ($\phi^2=1$):\\
\begin{eqnarray}
\frac{(1-12t-2.3.11.7t^2-12.7^3t^3+7^6t^4)}{(1-t)(1-7t)^{83}(1-49t)^{83}(1-343t)}
\end{eqnarray}

\noindent $\phi=1,\psi=3,4$ ($\phi^2=1$):\\
\begin{eqnarray}
\frac{(1+18t+2.41.7t^2+18.7^3t^3+7^6t^4)}{(1-t)(1-7t)^{83}(1-49t)^{83}(1-343t)}
\end{eqnarray}

\noindent $\phi=6,\psi=1,6$ ($\phi^2=1$):\\
\begin{eqnarray}
\frac{(1-28t+2.3.11.7t^2-28.7^3t^3+7^6t^4)}{(1-t)(1-7t)^{83}(1-49t)^{83}(1-343t)}
\end{eqnarray}

\noindent $\phi=6,\psi=3,4$ ($\phi^2=1$):\\
\begin{eqnarray}
\frac{(1+8t+2.5.7t^2+8.7^3t^3+7^6t^4)}{(1-t)(1-7t)^{83}(1-49t)^{83}(1-343t)}
\end{eqnarray}

\noindent $\phi=1,\psi=0$ (Both):\\
\begin{eqnarray}
\frac{(1-7t)(1+343t^2)}{(1-t)(1-7t)^{83}(1-49t)^{83}(1-343t)}
\end{eqnarray}

\noindent $\phi=6,\psi=0$ (Both):\\
\begin{eqnarray}
\frac{(1+7t)(1+343t^2)}{(1-t)(1-7t)^{83}(1-49t)^{83}(1-343t)}
\end{eqnarray}

\noindent $\phi=6,\psi=2,5$ (Both):\\
\begin{eqnarray}
\frac{(1-7t)(1+24t+343t^2)}{(1-t)(1-7t)^{83}(1-49t)^{83}(1-343t)}
\end{eqnarray}

\pagebreak

\subsection[p=3 Complete Tables]{p=3 Complete Tables}
For $p=3$, $R_{\text{excep}}(t)=(1+27t^2)^3$.\\

\noindent For $\phi=0,\psi=0$:\\

 \noindent
\begin{tabular}{|c|c|c|}\hline
Monomial $\textbf{v}$ & Contribution & Power $\lambda_{\textbf{v}}$\\
\hline\hline
(0,0,0,0,0) & $(1+27t^2)(1-18t^2+729t^4)$ & 1 \\
(0,2,1,1,1) & $(1+27t^2)[(1-6t+27t^2)(1+6t+27t^2)]^{\frac{1}{2}}$ & 2  \\
(6,2,0,0,0) & $(1+27t^2)^2$ & 1  \\ \hline
(0,0,0,2,2) & $(1-18t^2+729t^4)$ & 3  \\
(2,0,1,3,3) & $(1+27t^2)$ & 6 \\
(4,0,2,0,0) & $(1+27t^2)^2$ & 3 \\ \hline
(0,0,2,1,1) & $(1+27t^2)^{\frac{1}{2}}(1-18t^2+729t^4)^{\frac{1}{2}}$ & 3  \\
(6,0,1,0,0) & $(1+27t^2)^{\frac{1}{2}}(1-27t^2)$ & 6  \\
(0,4,0,3,3) & $(1+27t^2)^2$ & 3  \\
(4,0,1,1,0) & $(1+27t^2)^2$ & 3  \\
(2,0,3,0,0) & $(1+27t^2)^{\frac{1}{2}}(1-27t^2)$& 6  \\
(2,2,1,1,0) & $(1+27t^2)^{\frac{1}{2}}(1-18t^2+729t^4)^{\frac{1}{2}}$ & 3 \\\hline
(0,0,3,1,0) & $(1+27t^2)$ & 6  \\
(2,0,2,1,0) & $[(1-6t+27t^2)(1+6t+27t^2)]^{\frac{1}{2}}$ & 12  \\
(4,0,2,3,1) & $1$ & 6  \\ \hline
\end{tabular}
\\[2cm]

\noindent The corresponding zeta function is:

\begin{equation}
\frac{(1-2p^2t^2+p^6t^4)^7(1-2pt+p^3t^2)^7(1+2pt+p^3t^2)^7(1-p^3t^2)^{12}(1+p^3t^2)^{47}}{(1-t)(1-pt)^2(1-p^2t)^2(1-p^3t)}.
\end{equation}

\pagebreak
\noindent For $\phi=0,\psi=1,2$ which is a conifold point:\\

\noindent
\begin{tabular}{|c|c|c|}\hline
Monomial $\textbf{v}$ & Contribution & Power $\lambda_{\textbf{v}}$ \\
\hline\hline
(0,0,0,0,0) & $(1-3t)(1-6t+27t^2)(1+8t+27t^2)$ & 1 \\
(0,2,1,1,1) & $[(1-3t)(1+3t)(1-6t+27t^2)(1+6t+27t^2)]^{\frac{1}{2}}$ & 2  \\
(6,2,0,0,0) & $(1+3t)(1+6t+27t^2)$ & 1  \\ \hline
(0,0,0,2,2) & $(1-3t)(1+6t+27t^2)$ & 3  \\
(2,0,1,3,3) & $[(1-3t)(1+3t)]^{\frac{1}{2}}$ & 6 \\
(4,0,2,0,0) & $(1+3t)(1-6t+27t^2)$ & 3 \\ \hline
(0,0,2,1,1) & $[(1-3t)(1+3t)(1+27t^2)]^{\frac{1}{2}}$ & 3  \\
(6,0,1,0,0) & $[(1-3t)(1+3t)(1+27t^2)]^{\frac{1}{2}}$ & 6  \\
(0,4,0,3,3) & $[(1-3t)(1+3t)(1-18t^2+729t^4)]^{\frac{1}{2}}$ & 3  \\
(4,0,1,1,0) & $[(1-3t)(1+3t)(1-18t^2+729t^4)]^{\frac{1}{2}}$ & 3  \\
(2,0,3,0,0) & $[(1-3t)(1+3t)(1+27t^2)]^{\frac{1}{2}}$& 6  \\
(2,2,1,1,0) & $[(1-3t)(1+3t)(1+27t^2)]^{\frac{1}{2}}$ & 3 \\\hline
(0,0,3,1,0) & $[(1-3t)(1+3t)]^{\frac{1}{2}}$ & 6  \\
(2,0,2,1,0) & $[(1-3t)(1+3t)]^{\frac{1}{2}}$ & 12  \\
(4,0,2,3,1) & $\frac{1}{((1-9t)(1+9t))^{\frac{1}{2}}}$ & 6  \\ \hline
\end{tabular}
\\[2cm]

\noindent The corresponding zeta function is: {\footnotesize
\begin{equation}
\frac{(1-2p^2t^2+p^6t^4)^3(1-2pt+p^3t^2)^5(1+2pt+p^3t^2)^5(1+8t+p^3t^2)(1-pt)^{27}(1+pt)^{29}(1+p^3t^2)^{12}}{(1-t)(1-p^2t)^5(1-p^3t)(1+p^2t)^3}.
\end{equation}}

\pagebreak
\noindent For $\phi=1,\psi=0$ which is both a conifold point and $\phi^2=1$:\\

\noindent
\begin{tabular}{|c|c|c|}\hline
Monomial $\textbf{v}$ & Contribution & Power $\lambda_{\textbf{v}}$ \\
\hline\hline
(0,0,0,0,0) & $(1-3t)(1+27t^2)$ & 1 \\
(0,2,1,1,1) & $[(1-3t)(1+3t)]^{\frac{1}{2}}$ & 2  \\
(6,2,0,0,0) & $\frac{(1+27t^2)}{(1-9t)}$ & 1  \\ \hline
(0,0,0,2,2) & $(1+27t^2)$ & 3  \\
(2,0,1,3,3) & $1$ & 6 \\
(4,0,2,0,0) & $(1+27t^2)$ & 3 \\ \hline
(0,0,2,1,1) & $(1+27t^2)^{\frac{1}{2}}$ & 3  \\
(6,0,1,0,0) & $(1+27t^2)^{\frac{1}{2}}$ & 6  \\
(0,4,0,3,3) & $(1+27t^2)$ & 3  \\
(4,0,1,1,0) & $(1+27t^2)$ & 3  \\
(2,0,3,0,0) & $(1+27t^2)^{\frac{1}{2}}$& 6  \\
(2,2,1,1,0) & $(1+27t^2)^{\frac{1}{2}}$ & 3 \\\hline
(0,0,3,1,0) & $[(1-3t)(1+3t)]^{\frac{1}{2}}$ & 6  \\
(2,0,2,1,0) & $[(1-3t)(1+3t)]^{\frac{1}{2}}$ & 12  \\
(4,0,2,3,1) & $\frac{1}{((1-9t)(1+9t))^{\frac{1}{2}}}$ & 6  \\ \hline
\end{tabular}
\\[2cm]

\noindent The corresponding zeta function is:

\begin{equation}
\frac{(1-pt)^{9}(1+pt)^{10}(1+p^3t^2)^{26}}{(1-t)(1-p^2t)^6(1-p^3t)(1+p^2t)^3}.
\end{equation}

\pagebreak
\noindent For $\phi=1,\psi=1,2$, here $\phi^2=1$:\\

\noindent
\begin{tabular}{|c|c|c|}\hline
Monomial $\textbf{v}$ & Contribution & Power $\lambda_{\textbf{v}}$ \\
\hline\hline
(0,0,0,0,0) & $(1+2t-18t^2+54t^3+729t^4)$ & 1 \\
(0,2,1,1,1) & $(1+27t^2)$ & 2  \\
(6,2,0,0,0) & $(1+6t+27t^2)$ & 1  \\ \hline
(0,0,0,2,2) & $(1+3t)(1-6t+27t^2)$ & 3  \\
(2,0,1,3,3) & $[(1-3t)(1+3t)]^{\frac{1}{2}}$ & 6 \\
(4,0,2,0,0) & $(1+3t)(1-6t+27t^2)$ & 3 \\ \hline
(0,0,2,1,1) & $(1-18t^2+729t^4)^{\frac{1}{2}}$ & 3  \\
(6,0,1,0,0) & $[(1-6t+27t^2)(1+6t+27t^2)]^{\frac{1}{2}}$ &6  \\
(0,4,0,3,3) & $(1+27t^2)^{\frac{1}{2}}(1-18t^2+729t^4)^{\frac{1}{2}}$ & 3  \\
(4,0,1,1,0) & $(1+27t^2)^{\frac{1}{2}}(1-18t^2+729t^4)^{\frac{1}{2}}$ & 3  \\
(2,0,3,0,0) & $[(1-6t+27t^2)(1+6t+27t^2)]^{\frac{1}{2}}$& 6  \\
(2,2,1,1,0) & $(1-18t^2+729t^4)^{\frac{1}{2}}$ & 3 \\\hline
(0,0,3,1,0) & $(1+27t^2)$ & 6  \\
(2,0,2,1,0) & $[(1-6t+27t^2)(1+6t+27t^2)]^{\frac{1}{2}}$ & 12  \\
(4,0,2,3,1) & $1$ & 6  \\ \hline
\end{tabular}
\\[2cm]

\noindent The corresponding zeta function is:

{\footnotesize
\begin{equation}
\frac{(1-pt)(1+pt)^9(1+p^3t^2)^{14}(1+2t-2p^2t^2+2p^3t^3+p^6t^4)(1-2pt+p^3t^2)^{18}(1+2pt+p^3t^2)^{13}(1-2p^2t^2+p^6t^4)^6}{(1-t)(1-p^2t)^2(1-p^3t)}.
\end{equation}}

\pagebreak
\noindent For $\phi=2,\psi=0$ which is both a conifold point and $\phi^2=1$:\\

\noindent
\begin{tabular}{|c|c|c|}\hline
Monomial $\textbf{v}$ & Contribution & Power $\lambda_{\textbf{v}}$ \\
\hline\hline
(0,0,0,0,0) & $(1+3t)(1+27t^2)$ & 1 \\
(0,2,1,1,1) & $[(1-3t)(1+3t)]^{\frac{1}{2}}$ & 2  \\
(6,2,0,0,0) & $\frac{(1+27t^2)}{(1-9t)}$& 1  \\ \hline
(0,0,0,2,2) & $(1+27t^2)$ & 3  \\
(2,0,1,3,3) & $1$ & 6 \\
(4,0,2,0,0) & $(1+27t^2)$ & 3 \\ \hline
(0,0,2,1,1) & $(1+27t^2)^{\frac{1}{2}}$ & 3  \\
(6,0,1,0,0) & $(1+27t^2)^{\frac{1}{2}}$ & 6  \\
(0,4,0,3,3) & $(1+27t^2)$ & 3  \\
(4,0,1,1,0) & $(1+27t^2)$ & 3  \\
(2,0,3,0,0) & $(1+27t^2)^{\frac{1}{2}}$& 6  \\
(2,2,1,1,0) & $(1+27t^2)^{\frac{1}{2}}$ & 3 \\\hline
(0,0,3,1,0) & $[(1-3t)(1+3t)]^{\frac{1}{2}}$ & 6  \\
(2,0,2,1,0) & $[(1-3t)(1+3t)]^{\frac{1}{2}}$ & 12  \\
(4,0,2,3,1) & $\frac{1}{((1-9t)(1+9t))^{\frac{1}{2}}}$ & 6  \\ \hline
\end{tabular}
\\[2cm]

\noindent The corresponding zeta function is:

\begin{equation}
\frac{(1-pt)^{8}(1+pt)^{11}(1+p^3t^2)^{26}}{(1-t)(1-p^2t)^6(1-p^3t)(1+p^2t)^3}.
\end{equation}

\pagebreak
\noindent For $\phi=2,\psi=1,2$ which is both a conifold point and $\phi^2=1$:\\

\noindent
\begin{tabular}{|c|c|c|}\hline
Monomial $\textbf{v}$ & Contribution & Power $\lambda_{\textbf{v}}$ \\
\hline\hline
(0,0,0,0,0) & $(1+3t)(1-4t+27t^2)$ & 1 \\
(0,2,1,1,1) & $[(1-3t)(1+3t)]^{\frac{1}{2}}$ & 2  \\
(6,2,0,0,0) & $(1+3t)$ & 1 \\ \hline
(0,0,0,2,2) & $(1-3t)(1+3t)$ & 3  \\
(2,0,1,3,3) & $\left(\frac{(1-3t)(1+3t)}{(1-9t)(1+9t)}\right)^{\frac{1}{2}}$ & 6 \\
(4,0,2,0,0) & $(1+3t)^2$ & 3 \\ \hline
(0,0,2,1,1) & $[(1-3t)(1+3t)]^{\frac{1}{2}}$ & 3  \\
(6,0,1,0,0) & $[(1-3t)(1+3t)]^{\frac{1}{2}}$ & 6  \\
(0,4,0,3,3) & $[(1+3t)(1-3t)(1+27t^2)]^{\frac{1}{2}}$ & 3  \\
(4,0,1,1,0) & $[(1+3t)(1-3t)(1+27t^2)]^{\frac{1}{2}}$ & 3  \\
(2,0,3,0,0) & $[(1-3t)(1+3t)]^{\frac{1}{2}}$& 6  \\
(2,2,1,1,0) & $[(1-3t)(1+3t)]^{\frac{1}{2}}$ & 3 \\\hline
(0,0,3,1,0) & $[(1-3t)(1+3t)]^{\frac{1}{2}}$ & 6  \\
(2,0,2,1,0) & $[(1-3t)(1+3t)]^{\frac{1}{2}}$ & 12  \\
(4,0,2,3,1) & $\frac{1}{((1-9t)(1+9t))^{\frac{1}{2}}}$ & 6  \\ \hline
\end{tabular}
\\[2cm]

\noindent The corresponding zeta function is:

\begin{equation}
\frac{(1-pt)^{26}(1+pt)^{36}(1+p^3t^2)^6(1-4t+p^3t^2)}{(1 - t)(1 -
p^2t)^8(1 + p^2t)^6(1 - p^3t)}.
\end{equation}
\pagebreak
\subsection[p=5 Complete Tables]{p=5 Complete Tables}
For $p=5$, $R_{\text{excep}}(t)=(1-2.5t+125t^2)^3$.

\noindent
\begin{tabular}{|c|c|c|}\hline
\multicolumn{3}{|c|}{$\phi=0,\psi=0$}\\\hline
Monomial $\textbf{v}$ & Contribution & Power $\lambda_{\textbf{v}}$ \\
\hline\hline
(0,0,0,0,0) & $(1-10t+125t^2)(1-70t^2+15625t^4)$ & 1 \\
(0,2,1,1,1) & $(1+125t^2)(1+10t+125t^2)$ & 2  \\
(4,0,2,2,2) & $(1-20t+125t^2)(1+20t+125t^2)$ & 1  \\ \hline
(0,0,0,2,2) & $(1+125t^2)^2$ & 3  \\
(2,0,1,3,3) & $(1+10t+125t^2)$ & 6 \\
(4,0,2,0,0) & $(1-20t+125t^2)(1+20t+125t^2)$ & 3 \\ \hline
$(0,0,2,1,1)\times(2,2,1,1,0)$  & $(1+125t^2)^2(1-10t+125t^2)$ & 3  \\\hline
$(6,0,1,0,0)\times(2,0,3,0,0)$ & $(1-20t+125t^2)(1-10t+125t^2)(1+20t+125t^2)$ & 6  \\\hline
$(0,4,0,3,3)\times (4,0,1,1,0)$& $(1-20t+125t^2)(1+10t+125t^2)^2(1+20t+125t^2)$ & 3  \\\hline
(0,0,3,1,0) & $(1+10t+125t^2)$ & 6  \\
(2,0,2,1,0) & $(1+125t^2)$ & 12 \\
(4,0,2,3,1) & $1$ & 6  \\ \hline
\end{tabular}
\\[1cm]

\noindent
\begin{tabular}{|c|c|}\hline
\multicolumn{2}{|c|}{$\phi=0,\psi=1,2,3,4$}\\\hline
Monomial $\textbf{v}$ & Contribution  \\
\hline\hline
(0,0,0,0,0) & $(1+125t^2)(1-4t-90t^2-500t^3+15625t^4)$  \\
(0,2,1,1,1) & $(1-20t+125t^2)(1+10t+125t^2)$  \\
(4,0,2,2,2) & $(1+125t^2)(1+10t+125t^2)$  \\ \hline
(0,0,0,2,2) & $(1+20t+125t^2)(1-10t+125t^2)$  \\
(2,0,1,3,3) & $(1+10t+125t^2)$  \\
(4,0,2,0,0) & $(1+125t^2)(1-10t+125t^2)$  \\ \hline
$(0,0,2,1,1)\times(2,2,1,1,0)$ & $(1-20t+125t^2)(1+10t+125t^2)(1+20t+125t^2)$ \\\hline
$(6,0,1,0,0)\times(2,0,3,0,0)$ & $(1-10t+75t^2-1500t^3+75.5^3t^4-10.5^6t^5+5^9t^6)$ \\\hline
$(0,4,0,3,3)\times (4,0,1,1,0)$& $(1-20t+125t^2)(1-10t+125t^2)(1+10t+125t^2)(1+20t+125t^2)$ \\\hline
(0,0,3,1,0) & $(1-10t+125t^2)$  \\
(2,0,2,1,0) & $(1-10t+125t^2)$   \\
(4,0,2,3,1) & $1$ \\ \hline
\end{tabular}
\\

\pagebreak
\noindent
\begin{tabular}{|c|c|c|}\hline
\multicolumn{3}{|c|}{$\phi=1,\psi=0$}\\\hline
Monomial $\textbf{v}$ & Contribution & Power $\lambda_{\textbf{v}}$\\
\hline\hline
(0,0,0,0,0) & $(1-5t)(1-22t+125t^2)$ & 1 \\
(0,2,1,1,1) & $(1-5t)$ & 2  \\
(6,2,0,0,0) & $\frac{(1+10t+125t^2)}{(1-25t)}$ & 1  \\ \hline
(0,0,0,2,2) & $(1+10t+125t^2)$ & 3  \\
(2,0,1,3,3) & $1$ & 6 \\
(4,0,2,0,0) & $(1+10t+125t^2)$ & 3 \\ \hline
$(0,0,2,1,1)\times(2,2,1,1,0)$ & $(1-10t+125t^2)$ & 3  \\\hline
$(6,0,1,0,0)\times(2,0,3,0,0)$  & $(1-10t+125t^2)$ & 6  \\\hline
(0,4,0,3,3) & $(1-10t+125t^2)$ & 3  \\
(4,0,1,1,0) & $(1-10t+125t^2)$ & 3  \\\hline
(0,0,3,1,0) & $(1+5t)$ & 6  \\
(2,0,2,1,0) & $(1+5t)$ & 12  \\
(4,0,2,3,1) & $\frac{1}{(1+25t)}$ & 6  \\ \hline
\end{tabular}
\\[1cm]

\noindent
\begin{tabular}{|c|c|c|}\hline
\multicolumn{3}{|c|}{$\phi=1,\psi=1,2,3,4$}\\\hline
Monomial $\textbf{v}$ & Contribution & Power $\lambda_{\textbf{v}}$\\
\hline\hline
(0,0,0,0,0) & $(1+5t)(1+2t+125t^2)$ & 1 \\
(0,2,1,1,1) & $(1-5t)$ & 2  \\
(6,2,0,0,0) & $(1-5t)$ & 1  \\ \hline
(0,0,0,2,2) & $(1-5t)(1+5t)$ & 3  \\
(2,0,1,3,3) & $1$ & 6 \\
(4,0,2,0,0) & $(1-5t)(1+5t)$ & 3 \\ \hline
(0,0,2,1,1) & $(1+5t)$ & 3  \\
(6,0,1,0,0) & $(1-5t)$ & 6  \\\hline
$(0,4,0,3,3)\times (4,0,1,1,0)$ & $(1+5t)^2(1-10t+125t^2)$ & 3  \\\hline
(2,0,3,0,0) & $(1-5t)$& 6  \\
(2,2,1,1,0) & $(1+5t)$ & 3 \\\hline
(0,0,3,1,0) & $(1+5t)$ & 6  \\
(2,0,2,1,0) & $(1-5t)$ & 12  \\
(4,0,2,3,1) & $\frac{1}{(1+25t)}$ & 6  \\ \hline
\end{tabular}

\pagebreak \noindent
\begin{tabular}{|c|c|c|}\hline
\multicolumn{3}{|c|}{$\phi=2,\psi=0$}\\\hline
Monomial $\textbf{v}$ & Contribution & Power $\lambda_{\textbf{v}}$\\
\hline\hline
(0,0,0,0,0) & $(1-10t+125t^2)(1+10t+125t^2)(1-4t+125t^2)$ & 1 \\
(0,2,1,1,1) & $(1-10t+125t^2)(1-20t+125t^2)$ & 2  \\
(6,2,0,0,0) & $(1-10t+125t^2)(1+20t+125t^2)$ & 1  \\ \hline
(0,0,0,2,2) & $(1-10t+125t^2)(1+20t+125t^2)$ & 3  \\
(2,0,1,3,3) & $(1+20t+125t^2)$ & 6 \\
(4,0,2,0,0) & $(1-10t+125t^2)(1-20t+125t^2)$ & 3 \\ \hline
$(0,0,2,1,1)\times(2,2,1,1,0)$  & $(1+10t+125t^2)^2(1-20t+125t^2)$ & 3  \\\hline
$(6,0,1,0,0)\times(2,0,3,0,0)$ & $(1+10t+125t^2)(1-10t+125t^2)(1-20t+125t^2)$ & 6  \\\hline
$(0,4,0,3,3)\times (4,0,1,1,0)$ & $(1-10t+125t^2)(1+10t+125t^2)^2(1+20t+125t^2)$ & 3  \\\hline
(0,0,3,1,0) & $(1-10t+125t^2)$ & 6  \\
(2,0,2,1,0) & $(1+10t+125t^2)$ & 12  \\
(4,0,2,3,1) & $1$ & 6  \\ \hline
\end{tabular}
\\[2cm]
\noindent
\begin{tabular}{|c|c|c|}\hline
\multicolumn{3}{|c|}{$\phi=2,\psi=1,2,3,4$}\\\hline
Monomial $\textbf{v}$ & Contribution & Power $\lambda_{\textbf{v}}$\\
\hline\hline
(0,0,0,0,0) & $(1-5t^2-1200t^3-5.5^3t^4+5^9t^6)$ & 1 \\
(0,2,1,1,1) & $(1-50t^2+15625t^4)$ & 2  \\
(6,2,0,0,0) & $(1+125t^2)(1+10t+125t^2)$ & 1  \\ \hline
(0,0,0,2,2) & $(1+10t+125t^2)(1-20t+125t^2)$ & 3  \\
(2,0,1,3,3) & $(1+10t+125t^2)$ & 6 \\
(4,0,2,0,0) & $(1+125t^2)(1+10t+125t^2)$ & 3 \\ \hline
$(0,0,2,1,1)\times(2,2,1,1,0)$  & $(1+10t+125t^2)^2(1-20t+125t^2)$ & 3  \\\hline
$(6,0,1,0,0)\times(2,0,3,0,0)$ & $(1-25t^2-2000t^3-25.5^3t^4+5^9t^6)$ & 6  \\\hline
$(0,4,0,3,3)\times (4,0,1,1,0)$ & $(1-125t^2)^2(1-20t+125t^2)(1+10t+125t^2)$ & 3  \\\hline
(0,0,3,1,0) & $(1+10t+125t^2)$ & 6  \\
(2,0,2,1,0) & $(1+125t^2)$ & 12  \\
(4,0,2,3,1) & $1$ & 6  \\ \hline
\end{tabular}
\\
\pagebreak

\noindent
\begin{tabular}{|c|c|c|}\hline
\multicolumn{3}{|c|}{$\phi=3,\psi=0$}\\\hline
Monomial $\textbf{v}$ & Contribution & Power $\lambda_{\textbf{v}}$\\
\hline\hline
(0,0,0,0,0) & $(1+4t+125t^2)(1+10t+125t^2)^2$ & 1 \\
(0,2,1,1,1) & $(1-20t+125t^2)(1+10t+125t^2)$ & 2  \\
(6,2,0,0,0) & $(1-20t+125t^2)(1+10t+125t^2)$ & 1  \\ \hline
(0,0,0,2,2) & $(1-20t+125t^2)(1+10t+125t^2)$ & 3  \\
(2,0,1,3,3) & $(1+20t+125t^2)$ & 6 \\
(4,0,2,0,0) & $(1+20t+125t^2)(1+10t+125t^2)$ & 3 \\ \hline
$(0,0,2,1,1)\times(2,2,1,1,0)$& $(1-10t+125t^2)(1+10t+125t^2)(1-20t+125t^2)$ & 3  \\\hline
$(6,0,1,0,0)\times(2,0,3,0,0)$& $(1-20t+125t^2)(1-10t+125t^2)(1+10t+125t^2)$ & 6  \\\hline
$(0,4,0,3,3)\times (4,0,1,1,0)$& $(1-20t+125t^2)(1-10t+125t^2)^2(1+10t+125t^2)$ & 3  \\\hline
(0,0,3,1,0) & $(1-10t+125t^2)$ & 6  \\
(2,0,2,1,0) & $(1-10t+125t^2)$ & 12  \\
(4,0,2,3,1) & $1$ & 6  \\ \hline
\end{tabular}
\\[1cm]

\noindent
\begin{tabular}{|c|c|c|}\hline
\multicolumn{3}{|c|}{$\phi=3,\psi=1,2,3,4$}\\\hline
Monomial $\textbf{v}$ & Contribution & Power $\lambda_{\textbf{v}}$\\
\hline\hline
(0,0,0,0,0) & $(1+5t)(1-12t+190t^2-12.5^3t^3+15625t^4)$ & 1 \\
(0,2,1,1,1) & $(1+5t)(1-10t+125t^2)$ & 2  \\
(6,2,0,0,0) & $(1+5t)(1+10t+125t^2)$ & 1  \\ \hline
(0,0,0,2,2) & $(1+5t)(1-10t+125t^2)$ & 3  \\
(2,0,1,3,3) & $(1+5t)$ & 6 \\
(4,0,2,0,0) & $(1+5t)(1-10t+125t^2)$ & 3 \\ \hline
$(0,0,2,1,1)\times(2,2,1,1,0)$& $(1+5t)^2(1-20t+125t^2)$ & 3  \\\hline
$(6,0,1,0,0)\times(2,0,3,0,0)$ & $(1+5t)^2(1-20t+125t^2)$ & 6  \\\hline
$(0,4,0,3,3)\times (4,0,1,1,0)$ & $(1+5t)^2(1+20t+125t^2)(1-10t+125t^2)$ & 3  \\\hline
(0,0,3,1,0) & $(1+5t)$ & 6  \\
(2,0,2,1,0) & $(1+5t)$ & 12  \\
(4,0,2,3,1) & $\frac{1}{(1+25t)}$ & 6  \\ \hline
\end{tabular}
\\
\noindent
\begin{tabular}{|c|c|c|}\hline
\multicolumn{3}{|c|}{$\phi=4,\psi=0$}\\\hline
Monomial $\textbf{v}$ & Contribution & Power $\lambda_{\textbf{v}}$\\
\hline\hline
(0,0,0,0,0) & $(1-5t)(1+22t+125t^2)$ & 1 \\
(0,2,1,1,1) & $(1+5t)$ & 2  \\
(6,2,0,0,0) & $\frac{(1-10t+125t^2)}{(1-25t)}$ & 1  \\ \hline
(0,0,0,2,2) & $(1-10t+125t^2)$ & 3  \\
(2,0,1,3,3) & $1$ & 6 \\
(4,0,2,0,0) & $(1-10t+125t^2)$ & 3 \\ \hline
$(0,0,2,1,1)\times(2,2,1,1,0)$ & $(1+10t+125t^2)$ & 3  \\\hline
$(6,0,1,0,0)\times(2,0,3,0,0)$ & $(1+10t+125t^2)$ & 6  \\\hline
$(0,4,0,3,3)\times (4,0,1,1,0)$ & $(1-10t+125t^2)(1+10t+125t^2)$ & 3  \\\hline
(0,0,3,1,0) & $(1+5t)$ & 6  \\
(2,0,2,1,0) & $(1-5t)$ & 12  \\
(4,0,2,3,1) & $\frac{1}{(1+25t)}$ & 6  \\ \hline
\end{tabular}\\[1cm]

\noindent
\begin{tabular}{|c|c|c|}\hline
\multicolumn{3}{|c|}{$\phi=4,\psi=1,2,3,4$}\\\hline
Monomial $\textbf{v}$ & Contribution & Power $\lambda_{\textbf{v}}$\\
\hline\hline
(0,0,0,0,0) & $(1-14t+125t^2)(1+10t+125t^2)$ & 1 \\
(0,2,1,1,1) & $(1-10t+125t^2)$ & 2  \\
(6,2,0,0,0) & $(1+10t+125t^2)$ & 1  \\ \hline
(0,0,0,2,2) & $(1-5t)(1+10t+125t^2)$ & 3  \\
(2,0,1,3,3) & $(1+5t)$ & 6 \\
(4,0,2,0,0) & $(1-5t)(1+10t+125t^2)$ & 3 \\ \hline
(0,0,2,1,1) & $(1-125t^2)$ & 3  \\
(6,0,1,0,0) & $(1-10t+125t^2)$ & 6  \\\hline
$(0,4,0,3,3)\times (4,0,1,1,0)$ & $(1-125t^2)^2(1-10t+125t^2)$ & 3  \\ \hline
(2,0,3,0,0) & $(1-10t+125t^2)$& 6  \\
(2,2,1,1,0) & $(1-125t^2)$ & 3 \\\hline
(0,0,3,1,0) & $(1-10t+125t^2)$ & 6  \\
(2,0,2,1,0) & $(1+10t+125t^2)$ & 12  \\
(4,0,2,3,1) & $1$ & 6  \\ \hline
\end{tabular}
\\
\pagebreak
\subsection[p=7 Complete Tables]{p=7 Complete Tables}
For $p=7$, $R_{\text{excep}}(t)=(1+343t^2)^3$. \nopagebreak
{\footnotesize\subsubsection[Smooth Cases]{Smooth Cases} \noindent
\begin{tabular}{|c|c|c|}\hline
\multicolumn{3}{|c|}{$\phi=0, \psi=0$}\\\hline
Monomial $\textbf{v}$ & Contribution & Power $\lambda_{\textbf{v}}$ \\
\hline\hline
(0,0,0,0,0) & $(1+343t^2)^3$ & 1 \\
(0,2,1,1,1) & $(1+343t^2)^2$ & 2  \\
(6,2,0,0,0) & $(1+343t^2)^2$ & 1  \\ \hline
(0,0,0,2,2) & $(1+343t^2)^2$ & 3  \\
(2,0,1,3,3) & $(1+343t^2)$ & 6 \\
(4,0,2,0,0) & $(1+343t^2)^2$ & 3 \\ \hline
(0,0,2,1,1) & $(1+343t^2)^{3/2}$ & 3  \\
(6,0,1,0,0) & $(1+343t^2)^{3/2}$ & 6  \\
(0,4,0,3,3) & $(1+343t^2)^2$ & 3  \\
(4,0,1,1,0) & $(1+343t^2)^2$ & 3  \\
(2,0,3,0,0) & $(1+343t^2)^{3/2}$& 6  \\
(2,2,1,1,0) & $(1+343t^2)^{3/2}$ & 3 \\\hline
(0,0,3,1,0) & $(1+343t^2)$ & 6  \\
(2,0,2,1,0) & $(1+343t^2)$ & 12  \\
(4,0,2,3,1) & $1$ & 6  \\ \hline
\end{tabular}

\noindent
\begin{tabular}{|c|c|c|}\hline
\multicolumn{3}{|c|}{\pagebreak $\phi=0, \psi=2,5$}\\\hline
Monomial $\textbf{v}$ & Contribution & Power $\lambda_{\textbf{v}}$ \\
\hline\hline
(0,0,0,0,0) & $(1-14t+343t^2)(1-8t-30.7t^2-8.7^3t^3+7^6t^4)$ & 1 \\
(0,2,1,1,1) & $(1-28t+343t^2)(1+28t+343t^2)$ & 2  \\
(6,2,0,0,0) & $(1+343t^2)(1-14t+343t^2)$ & 1  \\ \hline
(0,0,0,2,2) & $(1-14t-98t+14.7t^3+7^6t^4)$ & 3  \\
(2,0,1,3,3) & $(1-7t)(1+7t)$ & 6 \\
(4,0,2,0,0) & $(1+343t^2)(1+14t+343t^2)$ & 3 \\ \hline
(0,0,2,1,1) & $[(1-28t+343t^2)(1+28t+343t^2)(1-7t)(1+7t)]^{\frac{1}{2}}$ & 3  \\
(6,0,1,0,0) & $[(1+98t^2+7^6t^4)(1-7t)(1+7t)]^{\frac{1}{2}}$ & 6  \\
(0,4,0,3,3) & $(1+98t^2+7^6t^4)$ & 3  \\
(4,0,1,1,0) & $(1+98t^2+7^6t^4)$ & 3  \\
(2,0,3,0,0) & $[(1+98t^2+7^6t^4)(1-7t)(1+7t)]^{\frac{1}{2}}$& 6  \\
(2,2,1,1,0) & $[(1-28t+343t^2)(1+28t+343t^2)(1-7t)(1+7t)]^{\frac{1}{2}}$ & 3 \\\hline
(0,0,3,1,0) & $(1-7t)(1+7t)$ & 6  \\
(2,0,2,1,0) & $[(1-14t+343t^2)(1+14t+343t^2)]^{\frac{1}{2}}$ & 12  \\
(4,0,2,3,1) & $1$ & 6  \\ \hline
\end{tabular}
\\

\noindent
\begin{tabular}{|c|c|c|}\hline
\multicolumn{3}{|c|}{$\phi=0, \psi=3,4$}\\\hline
Monomial $\textbf{v}$ & Contribution & Power $\lambda_{\textbf{v}}$ \\
\hline\hline
(0,0,0,0,0) & $(1+6t-147t^2-44.7^2t^3-147.7^3t^4+6.7^6t^5+7^9t^6)$ & 1 \\
(0,2,1,1,1) & $(1+343t^2)(1-7t)(1+7t)(1-490t^2+7^6t^4)$ & 2  \\
(6,2,0,0,0) & $(1+28t+343t^2)(1+14t+343t^2)$ & 1  \\ \hline
(0,0,0,2,2) & $(1-28t+343t^2)(1+14t+343t^2)$ & 3  \\
(2,0,1,3,3) & $[(1-14t+343t^2)(1+14t+343t^2)]^{\frac{1}{2}}$ & 6 \\
(4,0,2,0,0) & $(1+28t+343t^2)(1-14t+343t^2)$ & 3 \\ \hline
(0,0,2,1,1) & $[(1+343t^2)(1-490t^2+7^6t^4)]^{\frac{1}{2}}$ & 3  \\
(6,0,1,0,0) & $[(1+343t^2)(1-490t^2+7^6t^4)]^{\frac{1}{2}}$ & 6  \\
(0,4,0,3,3) & $(1+343t^2)[(1-490t^2+7^6t^4)]^{\frac{1}{2}}$ & 3  \\
(4,0,1,1,0) & $(1+343t^2)[(1-490t^2+7^6t^4)]^{\frac{1}{2}}$ & 3  \\
(2,0,3,0,0) & $[(1+343t^2)(1-490t^2+7^6t^4)]^{\frac{1}{2}}$& 6  \\
(2,2,1,1,0) & $[(1+343t^2)(1-490t^2+7^6t^4)]^{\frac{1}{2}}$ & 3 \\\hline
(0,0,3,1,0) & $(1+343t^2)$ & 6  \\
(2,0,2,1,0) & $(1-7t)(1+7t)$ & 12  \\
(4,0,2,3,1) & $1$ & 6  \\ \hline
\end{tabular}
\\[1cm]

\noindent
\begin{tabular}{|c|c|c|}\hline
\multicolumn{3}{|c|}{$\phi=2, \psi=0$}\\\hline
Monomial $\textbf{v}$ & Contribution & Power $\lambda_{\textbf{v}}$ \\
\hline\hline
(0,0,0,0,0) & $(1+28t+343t^2)(1+2.3.7^2t^2+7^6t^4)$ & 1 \\
(0,2,1,1,1) & $[(1+343t^2)(1-14t+343t^2)(1+14t+343t^2)]^{\frac{1}{2}}$ & 2  \\
(6,2,0,0,0) & $(1+343t^2)$ & 1  \\ \hline
(0,0,0,2,2) & $(1+343t^2)(1-7t)(1+7t)$ & 3  \\
(2,0,1,3,3) & $(1+343t^2)$ & 6 \\
(4,0,2,0,0) & $(1+343t^2)^2$ & 3 \\ \hline
(0,0,2,1,1) & $(1+343t^2)[(1-7t)(1+7t)]^{\frac{1}{2}}$ & 3  \\
(6,0,1,0,0) & $(1-343t^2)(1+343t^2)^{1/2}$ & 6  \\
(0,4,0,3,3) & $(1+343t^2)^2$ & 3  \\
(4,0,1,1,0) & $(1+343t^2)^2$ & 3  \\
(2,0,3,0,0) & $(1-343t^2)(1+343t^2)^{1/2}$& 6  \\
(2,2,1,1,0) & $(1+343t^2)[(1-7t)(1+7t)]^{\frac{1}{2}}$ & 3 \\\hline
(0,0,3,1,0) & $(1-7t)(1+7t)$ & 6  \\
(2,0,2,1,0) & $[(1-14t+343t^2)(1+14t+343t^2)]^{\frac{1}{2}}$ & 12  \\
(4,0,2,3,1) & $1$ & 6  \\ \hline
\end{tabular}
\\

\noindent
\begin{tabular}{|c|c|c|}\hline
\multicolumn{3}{|c|}{$\phi=2, \psi=1,6$}\\ \hline
Monomial $\textbf{v}$ & Contribution & Power $\lambda_{\textbf{v}}$ \\
\hline\hline
(0,0,0,0,0) & $(1+4t+19.7t^2-8.7^2t^3+19.7.7^3t^4+4.7^6t^5+7^9t^6)$ & 1 \\
(0,2,1,1,1) & $(1+98t^2+7^6t^4)$ & 2  \\
(6,2,0,0,0) & $(1+343t^2)(1+28t+343t^2)$ & 1  \\ \hline
(0,0,0,2,2) & $(1-28t+343t^2)(1+28t+343t^2)$ & 3  \\
(2,0,1,3,3) & $(1-14t+343t^2)(1+14t+343t^2)$ & 6 \\
(4,0,2,0,0) & $(1+343t^2)(1-28t+343t^2)$ & 3 \\ \hline
(0,0,2,1,1) & $[(1+343t^2)(1-490t^2+7^6t^4)]^{\frac{1}{2}}$ & 3  \\
(6,0,1,0,0) & $[(1+343t^2)(1-28t+343t^2)(1+28t+343t^2)]^{1/2}$ & 6  \\
(0,4,0,3,3) & $(1+2.3.7^2t^2+7^6t^4)$ & 3  \\
(4,0,1,1,0) & $(1+2.3.7^2t^2+7^6t^4)$ & 3  \\
(2,0,3,0,0) & $[(1+343t^2)(1-28t+343t^2)(1+28t+343t^2)]^{1/2}$& 6  \\
(2,2,1,1,0) & $[(1+343t^2)(1-490t^2+7^6t^4)]^{\frac{1}{2}}$ & 3 \\\hline
(0,0,3,1,0) & $(1+343t^2)$ & 6  \\
(2,0,2,1,0) & $(1-7t)(1+7t)$ & 12  \\
(4,0,2,3,1) & $1$ & 6  \\ \hline
\end{tabular}
\\[1cm]

\noindent
\begin{tabular}{|c|c|c|}\hline
\multicolumn{3}{|c|}{ $\phi=2, \psi=2,5$}\\\hline
Monomial $\textbf{v}$ & Contribution & Power $\lambda_{\textbf{v}}$ \\
\hline\hline
(0,0,0,0,0) & $(1-2t+3.7t^2-2^2.3.5.7^2t^3+3.7.7^3t^4-2.7^6t^5+7^9t^6)$ & 1 \\
(0,2,1,1,1) & $[(1+343t^2)(1-7t)(1+7t)(1-490t^2+7^6t^4)]^{1/2}$ & 2  \\
(6,2,0,0,0) & $(1-14t+343t^2)(1+28t+343t^2)$ & 1  \\ \hline
(0,0,0,2,2) & $(1-14t+343t^2)(1+28t+343t^2)$ & 3  \\
(2,0,1,3,3) & $(1+343t^2)$ & 6 \\
(4,0,2,0,0) & $(1-14t+343t^2)(1+28t+343t^2)$ & 3 \\ \hline
(0,0,2,1,1) & $[(1+343t^2)(1-28t+343t^2)(1+28t+343t^2)]^{\frac{1}{2}}$ & 3  \\
(6,0,1,0,0) & $[(1+343t^2)(1-28t+343t^2)(1+28t+343t^2)]^{1/2}$ & 6  \\
(0,4,0,3,3) & $[(1-7t)(1+7t)(1+343t^2)(1-28t+343t^2)(1+28t+343t^2)]^{1/2}$ & 3  \\
(4,0,1,1,0) & $[(1-7t)(1+7t)(1+343t^2)(1-28t+343t^2)(1+28t+343t^2)]^{1/2}$ & 3  \\
(2,0,3,0,0) & $[(1+343t^2)(1-28t+343t^2)(1+28t+343t^2)]^{1/2}$& 6  \\
(2,2,1,1,0) & $[(1+343t^2)(1-28t+343t^2)(1+28t+343t^2)]^{\frac{1}{2}}$ & 3 \\\hline
(0,0,3,1,0) & $(1+343t^2)$ & 6  \\
(2,0,2,1,0) & $(1-7t)(1+7t)$ & 12  \\
(4,0,2,3,1) & $1$ & 6  \\ \hline
\end{tabular}
\\

\noindent
\begin{tabular}{|c|c|c|}\hline
\multicolumn{3}{|c|}{$\phi=3,4, \psi=0$}\\\hline
Monomial $\textbf{v}$ & Contribution & Power $\lambda_{\textbf{v}}$ \\
\hline\hline
(0,0,0,0,0) & $(1+343t^2)(1+28t+343t^2)(1-28t+343t^2)$ & 1 \\
(0,2,1,1,1) & $(1+343t^2)(1-7t)(1+7t)$ & 2  \\
(6,2,0,0,0) & $(1+343t^2)^2$ & 1  \\ \hline
(0,0,0,2,2) & $[(1-14t+343t^2)(1+14t+343t^2)]^{1/2}(1-343t^2)$ & 3  \\
(2,0,1,3,3) & $(1+343t^2)$ & 6 \\
(4,0,2,0,0) & $(1+343t^2)^2$ & 3 \\ \hline
(0,0,2,1,1) & $[(1+343t^2)(1-28t+343t^2)(1+28t+343t^2)]^{\frac{1}{2}}$ & 3  \\
(6,0,1,0,0) & $(1+343t^2)^{3/2}$ & 6  \\
(0,4,0,3,3) & $(1+343t^2)^2$ & 3  \\
(4,0,1,1,0) & $(1+343t^2)^2$ & 3  \\
(2,0,3,0,0) & $(1+343t^2)^{3/2}$& 6  \\
(2,2,1,1,0) & $[(1+343t^2)(1-28t+343t^2)(1+28t+343t^2)]^{\frac{1}{2}}$ & 3 \\\hline
(0,0,3,1,0) & $(1+343t^2)$ & 6  \\
(2,0,2,1,0) & $(1-7t)(1+7t)$ & 12  \\
(4,0,2,3,1) & $1$ & 6  \\ \hline
\end{tabular}
\\[1cm]

\noindent
\begin{tabular}{|c|c|c|}\hline
\multicolumn{3}{|c|}{$\phi=3, \psi=1,6$}\\\hline
Monomial $\textbf{v}$ & Contribution & Power $\lambda_{\textbf{v}}$ \\
\hline\hline
(0,0,0,0,0) & $(1-4t+7^2t^2+2^3.7^2t^3+7^2.7^3t^4-4.7^6t^5+7^9t^6)$ & 1 \\
(0,2,1,1,1) & $(1+343t^2)(1-7t)(1+7t)$ & 2  \\
(6,2,0,0,0) & $(1-14t+343t^2)(1+14t+343t^2)$ & 1  \\ \hline
(0,0,0,2,2) & $(1-28t+2.5.7^2t^2-28.7^3t^3+7^6t^4)$ & 3  \\
(2,0,1,3,3) & $(1+7t)(1-7t)$ & 6 \\
(4,0,2,0,0) & $(1-14t+343t^2)(1+14t+343t^2)$ & 3 \\ \hline
(0,0,2,1,1) & $[(1+343t^2)(1-28t+343t^2)(1+28t+343t^2)]^{\frac{1}{2}}$ & 3  \\
(6,0,1,0,0) & $[(1-7t)(1+7t)(1-28t+343t^2)(1+28t+343t^2)]^{\frac{1}{2}}$ & 6  \\
(0,4,0,3,3) & $(1+343t^2)(1-7t)(1+7t)$ & 3  \\
(4,0,1,1,0) & $(1+343t^2)(1-7t)(1+7t)$ & 3  \\
(2,0,3,0,0) & $[(1-7t)(1+7t)(1-28t+343t^2)(1+28t+343t^2)]^{\frac{1}{2}}$& 6  \\
(2,2,1,1,0) & $[(1+343t^2)(1-28t+343t^2)(1+28t+343t^2)]^{\frac{1}{2}}$ & 3 \\\hline
(0,0,3,1,0) & $(1+343t^2)$ & 6  \\
(2,0,2,1,0) & $(1-7t)(1+7t)$ & 12  \\
(4,0,2,3,1) & $1$ & 6  \\ \hline
\end{tabular}
\\

\noindent
\begin{tabular}{|c|c|c|}\hline
\multicolumn{3}{|c|}{$\phi=3, \psi=2,5$}\\\hline
Monomial $\textbf{v}$ & Contribution & Power $\lambda_{\textbf{v}}$ \\
\hline\hline
(0,0,0,0,0) & $(1-6t+59.7t^2+2^2.7^2.11t^3+59.7.7^3t^4-6.7^6t^5+7^9t^6)$ & 1 \\
(0,2,1,1,1) & $[(1-14t+343t^2)(1+14t+343t^2)]^{\frac{1}{2}}(1-343t^2)$ & 2  \\
(6,2,0,0,0) & $(1+14t+343t^2)(1+28t+343t^2)$ & 1  \\ \hline
(0,0,0,2,2) & $(1-14t+343t^2)(1+28t+343t^2)$ & 3  \\
(2,0,1,3,3) & $(1+343t^2)$ & 6 \\
(4,0,2,0,0) & $(1-28t+343t^2)(1-14t+343t^2)$ & 3 \\ \hline
(0,0,2,1,1) & $[(1-7t)(1+7t)(1-28t+343t^2)(1+28t+343t^2)]^{1/2}$ & 3  \\
(6,0,1,0,0) & $[(1-7t)(1+7t)(1+98t^2+7^6t^4)]^{1/2}$ & 6  \\
(0,4,0,3,3) & $(1-14t+343t^2)(1+14t+343t^2)$ & 3  \\
(4,0,1,1,0) & $(1-14t+343t^2)(1+14t+343t^2)$ & 3  \\
(2,0,3,0,0) & $[(1-7t)(1+7t)(1+98t^2+7^6t^4)]^{1/2}$& 6  \\
(2,2,1,1,0) & $[(1-7t)(1+7t)(1-28t+343t^2)(1+28t+343t^2)]^{1/2}$ & 3 \\\hline
(0,0,3,1,0) & $(1-7t)(1+7t)$ & 6  \\
(2,0,2,1,0) & $[(1-14t+343t^2)(1+14t+343t^2)]^{1/2}$ & 12  \\
(4,0,2,3,1) & $1$ & 6  \\ \hline
\end{tabular}
\\[1cm]

\noindent
\begin{tabular}{|c|c|c|}\hline
\multicolumn{3}{|c|}{$\phi=3, \psi=3,4$}\\\hline
Monomial $\textbf{v}$ & Contribution & Power $\lambda_{\textbf{v}}$ \\
\hline\hline
(0,0,0,0,0) & $(1+14t-37.7t^2+47.7^2.2^2t^3-37.7.7^3t^4+14.7^6t^5+7^9t^6)$ & 1 \\
(0,2,1,1,1) & $[(1-28t+343t^2)(1+28t+343t^2)(1-14t+343t^2)(1+14t+343t^2)]^{\frac{1}{2}}$ & 2  \\
(6,2,0,0,0) & $(1-14t+343t^2)(1+28t+343t^2)$ & 1  \\ \hline
(0,0,0,2,2) & $(1+14t+343t^2)(1-28t+343t^2)$ & 3  \\
(2,0,1,3,3) & $[(1-14t+343t^2)(1+14t+343t^2)]^{\frac{1}{2}}$ & 6 \\
(4,0,2,0,0) & $(1-28t+343t^2)(1+14t+343t^2)$ & 3 \\ \hline
(0,0,2,1,1) & $[(1-7t)(1+7t)(1-28t+343t^2)(1+28t+343t^2)]^{1/2}$ & 3  \\
(6,0,1,0,0) & $[(1-7t)(1+7t)(1-28t+343t^2)(1+28t+343t^2)]^{1/2}$ & 6  \\
(0,4,0,3,3) & $(1-14t+343t^2)(1+14t+343t^2)$ & 3  \\
(4,0,1,1,0) & $(1-14t+343t^2)(1+14t+343t^2)$ & 3  \\
(2,0,3,0,0) & $[(1-7t)(1+7t)(1-28t+343t^2)(1+28t+343t^2)]^{1/2}$& 6  \\
(2,2,1,1,0) & $[(1-7t)(1+7t)(1-28t+343t^2)(1+28t+343t^2)]^{1/2}$ & 3 \\\hline
(0,0,3,1,0) & $(1+343t^2)$ & 6  \\
(2,0,2,1,0) & $(1+343t^2)$ & 12  \\
(4,0,2,3,1) & $1$ & 6  \\ \hline
\end{tabular}
\\

\noindent
\begin{tabular}{|c|c|c|}\hline
\multicolumn{3}{|c|}{$\phi=4, \psi=1,6$}\\\hline
Monomial $\textbf{v}$ & Contribution & Power $\lambda_{\textbf{v}}$ \\
\hline\hline
(0,0,0,0,0) & $(1+16t+343t^2)(1+28t+343t^2)(1-28t+343t^2)$ & 1 \\
(0,2,1,1,1) & $[(1+98t^2+7^6t^4)(1+343t^2)(1-7t)(1+7t)]^{\frac{1}{2}}$ & 2  \\
(6,2,0,0,0) & $(1+343t^2)(1+28t+343t^2)$ & 1  \\ \hline
(0,0,0,2,2) & $(1-343t^2)[(1-14t+343t^2)(1+14t+343t^2)]^{1/2}$ & 3  \\
(2,0,1,3,3) & $[(1-14t+343t^2)(1+14t+343t^2)]^{1/2}$ & 6 \\
(4,0,2,0,0) & $(1+343t^2)(1+28t+343t^2)$ & 3 \\ \hline
(0,0,2,1,1) & $(1+343t^2)^{3/2}$ & 3  \\
(6,0,1,0,0) & $[(1-7t)(1+7t)(1+98t^2+7^6t^4)]^{1/2}$ & 6  \\
(0,4,0,3,3) & $[(1+28t+343t^2)(1-28t+343t^2)]^{1/2}$ & 3  \\
(4,0,1,1,0) & $[(1+28t+343t^2)(1-28t+343t^2)]^{1/2}$ & 3  \\
(2,0,3,0,0) & $[(1-7t)(1+7t)(1+98t^2+7^6t^4)]^{1/2}$& 6  \\
(2,2,1,1,0) & $(1+343t^2)^{3/2}$ & 3 \\\hline
(0,0,3,1,0) & $(1-7t)(1+7t)$ & 6  \\
(2,0,2,1,0) & $[(1-14t+343t^2)(1+14t+343t^2)]^{1/2}$ & 12  \\
(4,0,2,3,1) & $1$ & 6  \\ \hline
\end{tabular}
\\[1cm]

\noindent
\begin{tabular}{|c|c|c|}\hline
\multicolumn{3}{|c|}{$\phi=5, \psi=0$}\\\hline
Monomial $\textbf{v}$ & Contribution & Power $\lambda_{\textbf{v}}$ \\
\hline\hline
(0,0,0,0,0) & $(1-28t+343t^2)(1+2.3.7^2t^2+7^6t^4)$ & 1 \\
(0,2,1,1,1) & $[(1+343t^2)(1-14t+343t^2)(1+14t+343t^2)]^{1/2}$ & 2  \\
(6,2,0,0,0) & $(1+343t^2)^2$ & 1  \\ \hline
(0,0,0,2,2) & $[(1-7t)(1+7t)(1+343t^2)$ & 3  \\
(2,0,1,3,3) & $(1+343t^2)$ & 6 \\
(4,0,2,0,0) & $(1+343t^2)^2$ & 3 \\ \hline
(0,0,2,1,1) & $(1+343t^2)[(1-7t)(1+7t)]^{\frac{1}{2}}$ & 3  \\
(6,0,1,0,0) & $(1+343t^2)^{\frac{1}{2}}(1-343t^2)$ & 6  \\
(0,4,0,3,3) & $(1+343t^2)^2$ & 3  \\
(4,0,1,1,0) & $(1+343t^2)^2$ & 3  \\
(2,0,3,0,0) & $(1+343t^2)^{\frac{1}{2}}(1-343t^2)$& 6  \\
(2,2,1,1,0) & $(1+343t^2)[(1-7t)(1+7t)]^{\frac{1}{2}}$ & 3 \\\hline
(0,0,3,1,0) & $(1-7t)(1+7t)$ & 6  \\
(2,0,2,1,0) & $[(1-14t+343t^2)(1+14t+343t^2)]^{1/2}$ & 12  \\
(4,0,2,3,1) & $1$ & 6  \\ \hline
\end{tabular}
\\

\noindent
\begin{tabular}{|c|c|c|}\hline
\multicolumn{3}{|c|}{$\phi=5, \psi=2,5$}\\\hline
Monomial $\textbf{v}$ & Contribution & Power $\lambda_{\textbf{v}}$ \\
\hline\hline
(0,0,0,0,0) & $(1+14t+43.7t^2+2^2.7^2.17t^3+43.7.7^3t^4+14.7^6t^5+7^9t^6)$ & 1 \\
(0,2,1,1,1) & $[(1+343t^2)(1-7t)(1+7t)(1+98t^2+7^6t^4)]^{\frac{1}{2}}$ & 2  \\
(6,2,0,0,0) & $(1+343t^2)(1+14t+343t^2)$ & 1  \\ \hline
(0,0,0,2,2) & $(1+14t-98t^2+14.7^3t^3+7^6t^4)$ & 3  \\
(2,0,1,3,3) & $(1-7t)(1+7t)$ & 6 \\
(4,0,2,0,0) & $(1+343t^2)(1-14t+343t^2)$ & 3 \\ \hline
(0,0,2,1,1) & $[(1+343t^2)(1-28t+343t^2)(1+28t+343t^2)]^{1/2}$ & 3  \\
(6,0,1,0,0) & $[(1+343t^2)(1-28t+343t^2)(1+28t+343t^2)]^{1/2}$ & 6  \\
(0,4,0,3,3) & $(1+98t^2+7^6t^4)$ & 3  \\
(4,0,1,1,0) & $(1+98t^2+7^6t^4)$ & 3  \\
(2,0,3,0,0) & $[(1+343t^2)(1-28t+343t^2)(1+28t+343t^2)]^{1/2}$& 6  \\
(2,2,1,1,0) & $[(1+343t^2)(1-28t+343t^2)(1+28t+343t^2)]^{1/2}$ & 3 \\\hline
(0,0,3,1,0) & $(1+343t^2)$ & 6  \\
(2,0,2,1,0) & $(1+343t^2)$ & 12  \\
(4,0,2,3,1) & $1$ & 6  \\ \hline
\end{tabular}
\\[1cm]

\noindent
\begin{tabular}{|c|c|c|}\hline
\multicolumn{3}{|c|}{$\phi=5, \psi=3,4$}\\\hline
Monomial $\textbf{v}$ & Contribution & Power $\lambda_{\textbf{v}}$ \\
\hline\hline
(0,0,0,0,0) & $(1-4t+3^3.7t^2+2^3.7^2.11t^3+3^3.7.7^3t^4-4.7^6t^5+7^9t^6)$ & 1 \\
(0,2,1,1,1) & $[(1+343t^2)(1-7t)(1+7t)(1+98t^2+7^6t^4)]^{\frac{1}{2}}$ & 2  \\
(6,2,0,0,0) & $(1+28t+343t^2)^2$ & 1  \\ \hline
(0,0,0,2,2) & $(1-7t)(1+7t)(1+343t^2)$ & 3  \\
(2,0,1,3,3) & $(1-7t)(1+7t)$ & 6 \\
(4,0,2,0,0) & $(1-28t+343t^2)(1+28t+343t^2)$ & 3 \\ \hline
(0,0,2,1,1) & $[(1-7t)(1+7t)(1-28t+343t^2)(1+28t+343t^2)]^{\frac{1}{2}}$ & 3  \\
(6,0,1,0,0) & $[(1-7t)(1+7t)(1-14t+343t^2)(1+14t+343t^2)]^{\frac{1}{2}}$ & 6  \\
(0,4,0,3,3) & $(1+343t^2)(1-7t)(1+7t)$ & 3  \\
(4,0,1,1,0) & $(1+343t^2)(1-7t)(1+7t)$ & 3  \\
(2,0,3,0,0) & $[(1-7t)(1+7t)(1-14t+343t^2)(1+14t+343t^2)]^{\frac{1}{2}}$& 6  \\
(2,2,1,1,0) & $[(1-7t)(1+7t)(1-28t+343t^2)(1+28t+343t^2)]^{\frac{1}{2}}$ & 3 \\\hline
(0,0,3,1,0) & $(1-7t)(1+7t)$ & 6  \\
(2,0,2,1,0) & $[(1-14t+343t^2)(1+14t+343t^2)]^{1/2}$ & 12  \\
(4,0,2,3,1) & $1$ & 6  \\ \hline
\end{tabular}
\\

\subsubsection[Pure Conifold cases]{Pure Conifold cases}
\noindent
\begin{tabular}{|c|c|c|}\hline
\multicolumn{3}{|c|}{$\phi=0, \psi=1,6$}\\\hline
Monomial $\textbf{v}$ & Contribution & Power $\lambda_{\textbf{v}}$ \\
\hline\hline
(0,0,0,0,0) & $(1-7t)(1+20t+34.7t^2+20.7^3t^3+7^6t^4)$ & 1 \\
(0,2,1,1,1) & $(1+343t^2)[(1-7t)(1+7t)]^{\frac{1}{2}}$ & 2  \\
(6,2,0,0,0) & $(1-7t)(1+28t+343t^2)$ & 1  \\ \hline
(0,0,0,2,2) & $(1-7t)(1+28t+343t^2)$ & 3  \\
(2,0,1,3,3) & $[(1-7t)(1+7t)]^{\frac{1}{2}}$ & 6 \\
(4,0,2,0,0) & $(1-7t)(1-28t+343t^2)$ & 3 \\ \hline
(0,0,2,1,1) & $[(1+343t^2)(1-7t)(1+7t)]^{\frac{1}{2}}$ & 3  \\
(6,0,1,0,0) & $[(1+343t^2)(1-7t)(1+7t)]^{\frac{1}{2}}$ & 6  \\
(0,4,0,3,3) & $(1+343t^2)[(1-7t)(1+7t)]^{\frac{1}{2}}$ & 3  \\
(4,0,1,1,0) & $(1+343t^2)[(1-7t)(1+7t)]^{\frac{1}{2}}$ & 3  \\
(2,0,3,0,0) & $[(1+343t^2)(1-7t)(1+7t)]^{\frac{1}{2}}$& 6  \\
(2,2,1,1,0) & $[(1+343t^2)(1-7t)(1+7t)]^{\frac{1}{2}}$ & 3 \\\hline
(0,0,3,1,0) & $[(1-7t)(1+7t)]^{\frac{1}{2}}$ & 6  \\
(2,0,2,1,0) & $[(1-7t)(1+7t)]^{\frac{1}{2}}$ & 12  \\
(4,0,2,3,1) & $\frac{1}{((1-49t)(1+49t))^{\frac{1}{2}}}$ & 6  \\ \hline
\end{tabular}
\\[1cm]

\noindent
\begin{tabular}{|c|c|c|}\hline
\multicolumn{3}{|c|}{$\phi=2, \psi=3,4$}\\\hline
Monomial $\textbf{v}$ & Contribution & Power $\lambda_{\textbf{v}}$ \\
\hline\hline
(0,0,0,0,0) & $(1+7t)(1-26t+2.29.7t^2+26.7^3t^3+7^6t^4)$ & 1 \\
(0,2,1,1,1) & $[(1-14t+343t^2)(1+14t+343t^2)(1-7t)(1+7t)]^{\frac{1}{2}}$ & 2  \\
(6,2,0,0,0) & $(1-7t)(1+14t+343t^2)$ & 1  \\ \hline
(0,0,0,2,2) & $(1+7t)(1+14t+343t^2)$ & 3  \\
(2,0,1,3,3) & $[(1-7t)(1+7t)]^{\frac{1}{2}}$ & 6 \\
(4,0,2,0,0) & $(1-7t)(1-14t+343t^2)$ & 3 \\ \hline
(0,0,2,1,1) & $[(1+343t^2)(1-7t)(1+7t)]^{\frac{1}{2}}$ & 3  \\
(6,0,1,0,0) & $[(1+343t^2)(1-7t)(1+7t)]^{\frac{1}{2}}$ & 6  \\
(0,4,0,3,3) & $[(1-28t+343t^2)(1+28t+343t^2)(1-7t)(1+7t)]^{\frac{1}{2}}$ & 3  \\
(4,0,1,1,0) & $[(1-28t+343t^2)(1+28t+343t^2)(1-7t)(1+7t)]^{\frac{1}{2}}$ & 3  \\
(2,0,3,0,0) & $[(1+343t^2)(1-7t)(1+7t)]^{\frac{1}{2}}$& 6  \\
(2,2,1,1,0) & $[(1+343t^2)(1-7t)(1+7t)]^{\frac{1}{2}}$ & 3 \\\hline
(0,0,3,1,0) & $[(1-7t)(1+7t)]^{\frac{1}{2}}$ & 6  \\
(2,0,2,1,0) & $[(1-7t)(1+7t)]^{\frac{1}{2}}$ & 12  \\
(4,0,2,3,1) & $\frac{1}{((1-49t)(1+49t))^{\frac{1}{2}}}$ & 6  \\ \hline
\end{tabular}
\\

\noindent
\begin{tabular}{|c|c|c|}\hline
\multicolumn{3}{|c|}{$\phi=4, \psi=2,5$}\\\hline
Monomial $\textbf{v}$ & Contribution & Power $\lambda_{\textbf{v}}$ \\
\hline\hline
(0,0,0,0,0) & $(1+7t)(1+4t+2.3.11.7t^2+4.7^3t^3+7^6t^4)$ & 1 \\
(0,2,1,1,1) & $[(1-28t+343t^2)(1+28t+343t^2)(1-7t)(1+7t)]^{\frac{1}{2}}$ & 2  \\
(6,2,0,0,0) & $(1-7t)(1+28t+343t^2)$ & 1  \\ \hline
(0,0,0,2,2) & $(1+7t)(1+28t+343t^2)$ & 3  \\
(2,0,1,3,3) & $[(1-7t)(1+7t)]^{\frac{1}{2}}$ & 6 \\
(4,0,2,0,0) & $(1-7t)(1+28t+343t^2)$ & 3 \\ \hline
(0,0,2,1,1) & $[(1+343t^2)(1-7t)(1+7t)]^{\frac{1}{2}}$ & 3  \\
(6,0,1,0,0) & $[(1+343t^2)(1-7t)(1+7t)]^{\frac{1}{2}}$ & 6  \\
(0,4,0,3,3) & $[(1-28t+343t^2)(1+28t+343t^2)(1-7t)(1+7t)]^{\frac{1}{2}}$ & 3  \\
(4,0,1,1,0) & $[(1-28t+343t^2)(1+28t+343t^2)(1-7t)(1+7t)]^{\frac{1}{2}}$ & 3  \\
(2,0,3,0,0) & $[(1+343t^2)(1-7t)(1+7t)]^{\frac{1}{2}}$& 6  \\
(2,2,1,1,0) & $[(1+343t^2)(1-7t)(1+7t)]^{\frac{1}{2}}$ & 3 \\\hline
(0,0,3,1,0) & $[(1-7t)(1+7t)]^{\frac{1}{2}}$ & 6  \\
(2,0,2,1,0) & $[(1-7t)(1+7t)]^{\frac{1}{2}}$ & 12  \\
(4,0,2,3,1) & $\frac{1}{((1-49t)(1+49t))^{\frac{1}{2}}}$ & 6  \\ \hline
\end{tabular}
\\[1cm]

\noindent
\begin{tabular}{|c|c|c|}\hline
\multicolumn{3}{|c|}{$\phi=4, \psi=3,4$}\\\hline
Monomial $\textbf{v}$ & Contribution & Power $\lambda_{\textbf{v}}$ \\
\hline\hline
(0,0,0,0,0) & $(1-7t)(1-16t+2.7t^2-16.7^3t^3+7^6t^4)$ & 1 \\
(0,2,1,1,1) & $[(1-28t+343t^2)(1+28t+343t^2)(1-7t)(1+7t)]^{\frac{1}{2}}$ & 2  \\
(6,2,0,0,0) & $(1-7t)(1+343t^2)$ & 1  \\ \hline
(0,0,0,2,2) & $(1-7t)(1+343t^2)$ & 3  \\
(2,0,1,3,3) & $[(1-7t)(1+7t)]^{\frac{1}{2}}$ & 6 \\
(4,0,2,0,0) & $(1-7t)(1+343t^2)$ & 3 \\ \hline
(0,0,2,1,1) & $[(1+343t^2)(1-7t)(1+7t)]^{\frac{1}{2}}$ & 3  \\
(6,0,1,0,0) & $[(1+343t^2)(1-7t)(1+7t)]^{\frac{1}{2}}$ & 6  \\
(0,4,0,3,3) & $[(1-28t+343t^2)(1+28t+343t^2)(1-7t)(1+7t)]^{\frac{1}{2}}$ & 3  \\
(4,0,1,1,0) & $[(1-28t+343t^2)(1+28t+343t^2)(1-7t)(1+7t)]^{\frac{1}{2}}$ & 3  \\
(2,0,3,0,0) & $[(1+343t^2)(1-7t)(1+7t)]^{\frac{1}{2}}$& 6  \\
(2,2,1,1,0) & $[(1+343t^2)(1-7t)(1+7t)]^{\frac{1}{2}}$ & 3 \\\hline
(0,0,3,1,0) & $[(1-7t)(1+7t)]^{\frac{1}{2}}$ & 6  \\
(2,0,2,1,0) & $[(1-7t)(1+7t)]^{\frac{1}{2}}$ & 12  \\
(4,0,2,3,1) & $\frac{1}{((1-49t)(1+49t))^{\frac{1}{2}}}$ & 6  \\ \hline
\end{tabular}
\\
\pagebreak

\noindent
\begin{tabular}{|c|c|c|}\hline
\multicolumn{3}{|c|}{$\phi=5, \psi=1,6$}\\\hline
Monomial $\textbf{v}$ & Contribution & Power $\lambda_{\textbf{v}}$ \\
\hline\hline
(0,0,0,0,0) & $(1+7t)(1-6t-2.3.7t^2-6.7^3t^3+7^6t^4)$ & 1 \\
(0,2,1,1,1) & $[(1-14t+343t^2)(1+14t+343t^2)(1-7t)(1+7t)]^{\frac{1}{2}}$ & 2  \\
(6,2,0,0,0) & $(1-7t)(1-14t+343t^2)$ & 1  \\ \hline
(0,0,0,2,2) & $(1+7t)(1-14t+343t^2)$ & 3  \\
(2,0,1,3,3) & $[(1-7t)(1+7t)]^{\frac{1}{2}}$ & 6 \\
(4,0,2,0,0) & $(1-7t)(1+14t+343t^2)$ & 3 \\ \hline
(0,0,2,1,1) & $[(1+343t^2)(1-7t)(1+7t)]^{\frac{1}{2}}$ & 3  \\
(6,0,1,0,0) & $[(1+343t^2)(1-7t)(1+7t)]^{\frac{1}{2}}$ & 6  \\
(0,4,0,3,3) & $[(1-28t+343t^2)(1+28t+343t^2)(1-7t)(1+7t)]^{\frac{1}{2}}$ & 3  \\
(4,0,1,1,0) & $[(1-28t+343t^2)(1+28t+343t^2)(1-7t)(1+7t)]^{\frac{1}{2}}$ & 3  \\
(2,0,3,0,0) & $[(1+343t^2)(1-7t)(1+7t)]^{\frac{1}{2}}$& 6  \\
(2,2,1,1,0) & $[(1+343t^2)(1-7t)(1+7t)]^{\frac{1}{2}}$ & 3 \\\hline
(0,0,3,1,0) & $[(1-7t)(1+7t)]^{\frac{1}{2}}$ & 6  \\
(2,0,2,1,0) & $[(1-7t)(1+7t)]^{\frac{1}{2}}$ & 12  \\
(4,0,2,3,1) & $\frac{1}{((1-49t)(1+49t))^{\frac{1}{2}}}$ & 6  \\ \hline
\end{tabular}
\\[1cm]

\subsubsection[$\phi^2=1$ only]{$\phi^2=1$ only}

\noindent
\begin{tabular}{|c|c|c|}\hline
\multicolumn{3}{|c|}{$\phi=1, \psi=1,6$}\\\hline
Monomial $\textbf{v}$ & Contribution & Power $\lambda_{\textbf{v}}$ \\
\hline\hline
(0,0,0,0,0) & $(1-2t-98t^2-2.7^3t^3+7^6t^4)$ & 1 \\
(0,2,1,1,1) & $(1-7t)(1+7t)$ & 2  \\
(6,2,0,0,0) & $(1-14t+343t^2)$ & 1  \\ \hline
(0,0,0,2,2) & $(1+7t)(1-14t+343t^2)$ & 3  \\
(2,0,1,3,3) & $[(1-7t)(1+7t)]^{\frac{1}{2}}$ & 6 \\
(4,0,2,0,0) & $(1+7t)(1-14t+343t^2)$ & 3 \\ \hline
(0,0,2,1,1) & $[(1+343t^2)(1-7t)(1+7t)]^{\frac{1}{2}}$ & 3  \\
(6,0,1,0,0) & $[(1-14t+343t^2)(1+14t+343t^2)]^{\frac{1}{2}}$ & 6  \\
(0,4,0,3,3) & $[(1+343t^2)^2(1-7t)(1+7t)]^{\frac{1}{2}}$ & 3  \\
(4,0,1,1,0) & $[(1+343t^2)^2(1-7t)(1+7t)]^{\frac{1}{2}}$ & 3  \\
(2,0,3,0,0) & $[(1-14t+343t^2)(1+14t+343t^2)]^{\frac{1}{2}}$& 6  \\
(2,2,1,1,0) & $[(1+343t^2)(1-7t)(1+7t)]^{\frac{1}{2}}$ & 3 \\\hline
(0,0,3,1,0) & $(1-7t)(1+7t)$ & 6  \\
(2,0,2,1,0) & $[(1-14t+343t^2)(1+14t+343t^2)]^{\frac{1}{2}}$ & 12  \\
(4,0,2,3,1) & $1$ & 6  \\ \hline
\end{tabular}
\\

 \noindent
\begin{tabular}{|c|c|c|}\hline
\multicolumn{3}{|c|}{$\phi=1, \psi=2,5$}\\\hline
Monomial $\textbf{v}$ & Contribution & Power $\lambda_{\textbf{v}}$ \\
\hline\hline
(0,0,0,0,0) & $(1-12t-2.3.11.7t^2-12.7^3t^3+7^6t^4)$ & 1 \\
(0,2,1,1,1) & $(1+343t^2)$ & 2  \\
(6,2,0,0,0) & $(1+28t+343t^2)$ & 1  \\ \hline
(0,0,0,2,2) & $(1+7t)(1-28t+343t^2)$ & 3  \\
(2,0,1,3,3) & $[(1-7t)(1+7t)]^{\frac{1}{2}}$ & 6 \\
(4,0,2,0,0) & $(1+7t)(1-28t+343t^2)$ & 3 \\ \hline
(0,0,2,1,1) & $(1+343t^2)$ & 3  \\
(6,0,1,0,0) & $(1-7t)(1+7t)$ & 6  \\
(0,4,0,3,3) & $(1+343t^2)^{\frac{3}{2}}$ & 3  \\
(4,0,1,1,0) & $(1+343t^2)^{\frac{3}{2}}$ & 3  \\
(2,0,3,0,0) & $(1-7t)(1+7t)$& 6  \\
(2,2,1,1,0) & $(1+343t^2)$ & 3 \\\hline
(0,0,3,1,0) & $(1+343t^2)$ & 6  \\
(2,0,2,1,0) & $(1-7t)(1+7t)$ & 12  \\
(4,0,2,3,1) & $1$ & 6  \\ \hline
\end{tabular}
\\[1cm]

 \noindent
\begin{tabular}{|c|c|c|}\hline
\multicolumn{3}{|c|}{$\phi=1, \psi=3,4$}\\\hline
Monomial $\textbf{v}$ & Contribution & Power $\lambda_{\textbf{v}}$ \\
\hline\hline
(0,0,0,0,0) & $(1+18t+2.41.7t^2+18.7^3t^3+7^6t^4)$ & 1 \\
(0,2,1,1,1) & $(1-7t)(1+7t)$ & 2  \\
(6,2,0,0,0) & $(1-14t+343t^2)$ & 1  \\ \hline
(0,0,0,2,2) & $(1+7t)(1+14t+343t^2)$ & 3  \\
(2,0,1,3,3) & $[(1-7t)(1+7t)]^{\frac{1}{2}}$ & 6 \\
(4,0,2,0,0) & $(1+7t)(1+14t+343t^2)$ & 3 \\ \hline
(0,0,2,1,1) & $[(1+343t^2)(1-7t)(1+7t)]^{\frac{1}{2}}$ & 3  \\
(6,0,1,0,0) & $[(1-14t+343t^2)(1+14t+343t^2)]^{\frac{1}{2}}$ & 6  \\
(0,4,0,3,3) & $[(1+343t^2)^2(1-7t)(1+7t)]^{\frac{1}{2}}$ & 3  \\
(4,0,1,1,0) & $[(1+343t^2)^2(1-7t)(1+7t)]^{\frac{1}{2}}$ & 3  \\
(2,0,3,0,0) & $[(1-14t+343t^2)(1+14t+343t^2)]^{\frac{1}{2}}$& 6  \\
(2,2,1,1,0) & $[(1+343t^2)(1-7t)(1+7t)]^{\frac{1}{2}}$ & 3 \\\hline
(0,0,3,1,0) & $(1-7t)(1+7t)$ & 6  \\
(2,0,2,1,0) & $[(1-14t+343t^2)(1+14t+343t^2)]^{\frac{1}{2}}$ & 12  \\
(4,0,2,3,1) & $1$ & 6  \\ \hline
\end{tabular}
\\

 \noindent
\begin{tabular}{|c|c|c|}\hline
\multicolumn{3}{|c|}{$\phi=6, \psi=1,6$}\\\hline
Monomial $\textbf{v}$ & Contribution & Power $\lambda_{\textbf{v}}$ \\
\hline\hline
(0,0,0,0,0) & $(1-28t+2.3.11.7t^2-28.7^3t^3+7^6t^4)$ & 1 \\
(0,2,1,1,1) & $(1+343t^2)$ & 2  \\
(6,2,0,0,0) & $(1+28t+343t^2)$ & 1  \\ \hline
(0,0,0,2,2) & $(1-7t)(1-14t+343t^2)$ & 3  \\
(2,0,1,3,3) & $[(1-7t)(1+7t)]^{\frac{1}{2}}$ & 6 \\
(4,0,2,0,0) & $(1+7t)(1+28t+343t^2)$ & 3 \\ \hline
(0,0,2,1,1) & $(1-7t)(1+7t)$ & 3  \\
(6,0,1,0,0) & $(1-7t)(1+7t)$ & 6  \\
(0,4,0,3,3) & $[(1+343t^2)(1-28t+343t^2)(1+28t+343t^2)]^{\frac{1}{2}}$ & 3  \\
(4,0,1,1,0) & $[(1+343t^2)(1-28t+343t^2)(1+28t+343t^2)]^{\frac{1}{2}}$ & 3  \\
(2,0,3,0,0) & $(1-7t)(1+7t)$& 6  \\
(2,2,1,1,0) & $(1-7t)(1+7t)$ & 3 \\\hline
(0,0,3,1,0) & $(1+343t^2)$ & 6  \\
(2,0,2,1,0) & $(1+343t^2)$ & 12  \\
(4,0,2,3,1) & $1$ & 6  \\ \hline
\end{tabular}
\\[1cm]

 \noindent
\begin{tabular}{|c|c|c|}\hline
\multicolumn{3}{|c|}{$\phi=6, \psi=3,4$}\\\hline
Monomial $\textbf{v}$ & Contribution & Power $\lambda_{\textbf{v}}$ \\
\hline\hline
(0,0,0,0,0) & $(1+8t+2.5.7t^2+8.7^3t^3+7^6t^4)$ & 1 \\
(0,2,1,1,1) & $(1+343t^2)$ & 2  \\
(6,2,0,0,0) & $(1+343t^2)$ & 1  \\ \hline
(0,0,0,2,2) & $(1-7t)(1+343t^2)$ & 3  \\
(2,0,1,3,3) & $\left[(1-7t)(1+7t)\right]^{\frac{1}{2}}$ & 6 \\
(4,0,2,0,0) & $(1+7t)(1+343t^2)$ & 3 \\ \hline
(0,0,2,1,1) & $(1-7t)(1+7t)$ & 3  \\
(6,0,1,0,0) & $(1+343t^2)$ & 6  \\
(0,4,0,3,3) & $[(1+343t^2)(1-28t+343t^2)(1+28t+343t^2)]^{\frac{1}{2}}$ & 3  \\
(4,0,1,1,0) & $[(1+343t^2)(1-28t+343t^2)(1+28t+343t^2)]^{\frac{1}{2}}$ & 3  \\
(2,0,3,0,0) & $(1+343t^2)$& 6  \\
(2,2,1,1,0) & $(1-7t)(1+7t)$ & 3 \\\hline
(0,0,3,1,0) & $(1+343t^2)$ & 6  \\
(2,0,2,1,0) & $(1-7t)(1+7t)$ & 12  \\
(4,0,2,3,1) & $1$ & 6  \\ \hline
\end{tabular}
\\

\subsubsection[Both Singularities]{Both Singularities}
Cases where both singularities coincide; when $\psi \neq0$ these
display modularity, confirming the observations of Villegas:\\
\noindent
\begin{tabular}{|c|c|c|}\hline
\multicolumn{3}{|c|}{$\phi=1, \psi=0$}\\\hline
Monomial $\textbf{v}$ & Contribution & Power $\lambda_{\textbf{v}}$ \\
\hline\hline
(0,0,0,0,0) & $(1-7t)(1+343t^2)$ & 1 \\
(0,2,1,1,1) & $[(1-7t)(1+7t)]^{\frac{1}{2}}$ & 2  \\
(6,2,0,0,0) & $\frac{(1+343t^2)}{(1-49t)}$ & 1  \\ \hline
(0,0,0,2,2) & $(1+343t^2)$ & 3  \\
(2,0,1,3,3) & $1$ & 6 \\
(4,0,2,0,0) & $(1+343t^2)$ & 3 \\ \hline
(0,0,2,1,1) & $(1+343t^2)^{1/2}$ & 3  \\
(6,0,1,0,0) & $(1+343t^2)^{1/2}$ & 6  \\
(0,4,0,3,3) & $(1+343t^2)$ & 3  \\
(4,0,1,1,0) & $(1+343t^2)$ & 3  \\
(2,0,3,0,0) & $(1+343t^2)^{1/2}$& 6  \\
(2,2,1,1,0) & $(1+343t^2)^{1/2}$ & 3 \\\hline
(0,0,3,1,0) & $[(1-7t)(1+7t)]^{\frac{1}{2}}$ & 6  \\
(2,0,2,1,0) & $[(1-7t)(1+7t)]^{\frac{1}{2}}$ & 12  \\
(4,0,2,3,1) & $\frac{1}{((1-49t)(1+49t))^{\frac{1}{2}}}$ & 6  \\ \hline
\end{tabular}
\\[1cm]

\noindent
\begin{tabular}{|c|c|c|}\hline
\multicolumn{3}{|c|}{$\phi=6, \psi=0$}\\\hline
Monomial $\textbf{v}$ & Contribution & Power $\lambda_{\textbf{v}}$ \\
\hline\hline
(0,0,0,0,0) & $(1+7t)(1+343t^2)$ & 1 \\
(0,2,1,1,1) & $[(1-7t)(1+7t)]^{\frac{1}{2}}$ & 2  \\
(6,2,0,0,0) & $\frac{(1+343t^2)}{(1-49t)}$ & 1  \\ \hline
(0,0,0,2,2) & $(1+343t^2)$ & 3  \\
(2,0,1,3,3) & $1$ & 6 \\
(4,0,2,0,0) & $(1+343t^2)$ & 3 \\ \hline
(0,0,2,1,1) & $(1+343t^2)^{1/2}$ & 3  \\
(6,0,1,0,0) & $(1+343t^2)^{1/2}$ & 6  \\
(0,4,0,3,3) & $(1+343t^2)$ & 3  \\
(4,0,1,1,0) & $(1+343t^2)$ & 3  \\
(2,0,3,0,0) & $(1+343t^2)^{1/2}$& 6  \\
(2,2,1,1,0) & $(1+343t^2)^{1/2}$ & 3 \\\hline
(0,0,3,1,0) & $[(1-7t)(1+7t)]^{\frac{1}{2}}$ & 6  \\
(2,0,2,1,0) & $[(1-7t)(1+7t)]^{\frac{1}{2}}$ & 12  \\
(4,0,2,3,1) & $\frac{1}{((1-49t)(1+49t))^{\frac{1}{2}}}$ & 6  \\ \hline
\end{tabular}
\\

 \noindent
\begin{tabular}{|c|c|c|}\hline
\multicolumn{3}{|c|}{$\phi=6, \psi=2,5$ (these show
modularity)}\\\hline
Monomial $\textbf{v}$ & Contribution & Power $\lambda_{\textbf{v}}$ \\
\hline\hline
(0,0,0,0,0) & $(1-7t)(1+24t+343t^2)$ & 1 \\
(0,2,1,1,1) & $[(1-7t)(1+7t)]^{\frac{1}{2}}$ & 2  \\
(6,2,0,0,0) & $(1-7t)$ & 1  \\ \hline
(0,0,0,2,2) & $(1-7t)^2$ & 3  \\
(2,0,1,3,3) & $[\frac{(1-7t)(1+7t)}{(1-49t)(1+49t)}]^{\frac{1}{2}}$ & 6 \\
(4,0,2,0,0) & $(1-7t)(1+7t)$ & 3 \\ \hline
(0,0,2,1,1) & $[(1-7t)(1+7t)]^{1/2}$ & 3  \\
(6,0,1,0,0) & $[(1-7t)(1+7t)]^{1/2}$ & 6  \\
(0,4,0,3,3) & $[(1+343t^2)(1-7t)(1+7t)]^{1/2}$ & 3  \\
(4,0,1,1,0) & $[(1+343t^2)(1-7t)(1+7t)]^{1/2}$ & 3  \\
(2,0,3,0,0) & $[(1-7t)(1+7t)]^{1/2}$& 6  \\
(2,2,1,1,0) & $[(1-7t)(1+7t)]^{1/2}$ & 3 \\\hline
(0,0,3,1,0) & $[(1-7t)(1+7t)]^{\frac{1}{2}}$ & 6  \\
(2,0,2,1,0) & $[(1-7t)(1+7t)]^{\frac{1}{2}}$ & 12  \\
(4,0,2,3,1) & $\frac{1}{((1-49t)(1+49t))^{\frac{1}{2}}}$ & 6  \\ \hline
\end{tabular}
\\[1cm]}



\pagebreak
\subsection[p=11 Complete Tables]{p=11 Complete Tables}
For $p=11$, $R_{\text{excep}}(t)=(1+11^3t^2)^3$.\\
 \small \noindent
\begin{tabular}{|c|c|c|c|}\hline
Monomial $\textbf{v}$ & $\phi=0, \psi=0$ & $\phi=1,10, \psi=0$ &$\phi=2,9, \psi=0$\\
\hline\hline
(0,0,0,0,0) & $(0)_2(0,1694)_4$ & $(-11)_1(0)_2$ & $(-44)_2(44)_2^2$\\
(0,2,1,1,1) & $(0)_2[(-66)_2(66)_2]^{1/2}$ &$[(-11)_1(11)_1]^{1/2}$ & $(0)_2^2$\\
(6,2,0,0,0) & $(0)_2^2$ & $\frac{(0)_2}{(-11^2)_1}$ & $(0)_2^2$
\\ \hline
(0,0,0,2,2) & $(0,1694)_4$& $(0)_2$ & $(-44)_2(44)_2$  \\
(2,0,1,3,3) & $(0)_2$ & $1$ & $(0)_2$ \\
(4,0,2,0,0) & $(0)_2^2$& $(0)_2$& $(0)_2^2$
\\ \hline
$(0,0,2,1,1)$
&$(0)_2^{1/2}(0,1694)_4^{1/2}$&$(0)_2^{1/2}$&$[(0)_2(-44)_2(44)_2]^{1/2}$\\
$(6,0,1,0,0)$&$(0)_2^{1/2}(11\sqrt{11})_1(-11\sqrt{11})_1$&$(0)_2^{1/2}$&$(0)_2^{3/2}$\\
$(0,4,0,3,3)$&$(0)_2^2$&$(0)_2$&$(0)_2^2$\\
$(4,0,1,1,0)$&$(0)_2^2$&$(0)_2$&$(0)_2^2$\\
$(2,0,3,0,0)$&$(0)_2^{1/2}(11\sqrt{11})_1(-11\sqrt{11})_1$&$(0)_2^{1/2}$&$(0)_2^{3/2}$\\
$(2,2,1,1,0)$&$(0)_2^{1/2}(0,1694)_4^{1/2}$&$(0)_2^{1/2}$&$[(0)_2(-44)_2(44)_2]^{1/2}$
\\\hline
(0,0,3,1,0) & $(0)_2$  & $[(-11)_1(11)_1]^{1/2}$ & $[(-44)_2(44)_2]^{1/2}$\\
(2,0,2,1,0) & $[(-66)_2(66)_2]^{1/2}$&$[(-11)_1(11)_1]^{1/2}$ & $(0)_2$ \\
(4,0,2,3,1) & $1$  & $[(-11^2)_1(11^2)_1]^{-1/2}$&$1$ \\ \hline
\end{tabular}
\\[1cm]
\noindent
\begin{tabular}{|c|c|c|c|}\hline
Monomial $\textbf{v}$ & $\phi=3,8 \psi=0$ & $\phi=4,7 \psi=0$ &$\phi=5,6 \psi=0$\\
\hline\hline
$(0,0,0,0,0)$&$(0)_2(0,2.7.11^2)_4$ & $(0)_2^2(-44)_2$ & $(44)_2(0,-1210)_4$\\
$(0,2,1,1,1)$&$(0,2420)_4$ &$(0,1694)_4$ & $(0,1694)_4$\\
$(6,2,0,0,0)$&$(0)_2^2$ & $(0)_2^2$ & $(0)_2^2$\\ \hline
$(0,0,0,2,2)$&$(0,2.7.11^2)_4$& $(0)_2^2$ & $[(0,-2^2.5.11^2)_4]^{1/2}$  \\
$(2,0,1,3,3)$&$(0)_2$ & $(0)_2$ & $(0)_2$ \\
$(4,0,2,0,0)$&$(0)_2^2$& $(0)_2^2$& $(0)_2^2$\\\hline
$(0,0,2,1,1)$&$[(0)_2(0,1694)_4]^{1/2}$&$(0)_2^{3/2}$&$[(0)_2(0,-1210)_4]^{1/2}$\\
$(6,0,1,0,0)$&$(0)_2^{1/2}(11\sqrt{11})_1(-11\sqrt{11})_1$&$(0)_2^{3/2}$&$(0)_2^{3/2}$\\
$(0,4,0,3,3)$&$(0)_2^2$&$(0)_2^2$&$(0)_2^2$\\
$(4,0,1,1,0)$&$(0)_2^2$&$(0)_2^2$&$(0)_2^2$\\
$(2,0,3,0,0)$&$(0)_2^{1/2}(11\sqrt{11})_1(-11\sqrt{11})_1$&$(0)_2^{3/2}$&$(0)_2^{3/2}$\\
$(2,2,1,1,0)$&$[(0)_2(0,1694)_4]^{1/2}$&$(0)_2^{3/2}$&$[(0)_2(0,-1210)_4]^{1/2}$
\\\hline
$(0,0,3,1,0)$&$(0)_2$  & $[(-44)_2(44)_2]^{1/2}$ & $[(-44)_2(44)_2]^{1/2}$\\
$(2,0,2,1,0)$&$[(-22)_2(22)_2]^{1/2}$&$[(-44)_2(44)_2]^{1/2}$ & $[(-44)_2(44)_2]^{1/2}$ \\
$(4,0,2,3,1)$&$1$&$1$&$1$ \\ \hline
\end{tabular}
\\

\noindent
\begin{tabular}{|c|c|c|c|}\hline
Monomial $\textbf{v}$ & $\phi=0 \psi=1,10$ & $\phi=0 \psi=2,9$ &$\phi=0 \psi=3,8$\\
\hline\hline
$(0,0,0,0,0)$&$(-52,1969,-60984)_6$ & $(4,-231,18392)_6$ & $(-11)_1(46,1782)_4$\\
$(0,2,1,1,1)$&$(0)_2[(-33)_2(33)_2]^{1/2}$ &$(0)_2(-11)_1(11)_2$ & $[(0)_2(-22)_2(22)_2]^{1/2}$\\
$(6,2,0,0,0)$&$(0)_2(-44)^2$ & $(0)_2^2$ & $(-11)_1(-22)_2$\\
\hline
$(0,0,0,2,2)$&$(0,2.7.11^2)_4$& $(0)_2^2$ & $[(-11)_1(11)_1]^{1/2}$  \\
$(2,0,1,3,3)$&$(0)_2$ & $(0)_2$ & $(11)_1(22)_2$ \\
$(4,0,2,0,0)$&$(0)_2^2$& $(0)_2^2$& $(0,1210)_4^{1/2}$\\\hline
$(0,0,2,1,1)$&$[(0)_2^{1/2}(-11)_1(11)_1$&$(0)_2^{1/2}(-11)_1(11)_1$&$(0,1210)_4^{1/2}$
\\
$(6,0,1,0,0)$&$[(-11)_1(11)_1]^{1/2}(0)_2$&$[(0)_2(-44)_2(44)_2]^{1/2}$&$(0,1210)_4^{1/2}$\\
$(0,4,0,3,3)$&$[(-11)_1(11)_1(-55)_2(55)_2]^{1/2}$&$[(-11)_1(11)_1(-55)_2(55)_2]^{1/2}$&$[(0)_2(0,242)_4]^{1/2}$\\
$(4,0,1,1,0)$&$[(-11)_1(11)_1(-55)_2(55)_2]^{1/2}$&$[(-11)_1(11)_1(-55)_2(55)_2]^{1/2}$&$[(0)_2(0,242)_4]^{1/2}$\\
$(2,0,3,0,0)$&$[(-11)_1(11)_1]^{1/2}(0)_2$&$[(0)_2(-44)_2(44)_2]^{1/2}$&$(0,1210)_4^{1/2}$\\
$(2,2,1,1,0)$&$[(0)_2^{1/2}(-11)_1(11)_1$&$(0)_2^{1/2}(-11)_1(11)_1$&$(0,1210)_4^{1/2}$
\\\hline
$(0,0,3,1,0)$&$(0)_2$  & $[(-44)_2(44)_2]^{1/2}$ & $[(-11)_1(11)_1]^{1/2}$\\
$(2,0,2,1,0)$&$[(-66)_2(66)_2]^{1/2}$&$[(-44)_2(44)_2]^{1/2}$ & $[(-11)_1(11)_1]^{1/2}$ \\
$(4,0,2,3,1)$&$1$&$1$&$[(-11^2)_1(11^2)_1]^{-1/2}$ \\ \hline
\end{tabular}
\\[1cm]

\noindent
\begin{tabular}{|c|c|c|}\hline
Monomial $\textbf{v}$ & $\phi=0, \psi=4,7$ & $\phi=0, \psi=5,6$ \\
\hline\hline
$(0,0,0,0,0)$&$(30,-231,-5324)_6$ & $(22,99.11,18876)_6$ \\
$(0,2,1,1,1)$&$[(0_2(-11)_1(11)_1(0,1694)_4]^{1/2}$ &$(0)_2[(-66)_2(66)_2]^{1/2}$ \\
$(6,2,0,0,0)$&$(0)_2(66)_2$ & $(-22)_2(44)_2$ \\
\hline
$(0,0,0,2,2)$&$(-22,2.3.11^2)_4$& $(-22)_2(66)_2$ \\
$(2,0,1,3,3)$&$[(-44)_2(44)_2]^{1/2}$ & $(0)_2$ \\
$(4,0,2,0,0)$&$(0)_2(-66)_2$& $(22)_2(44)_2$\\\hline
$(0,0,2,1,1)$&$[(0)_2(0,-1210)_4]^{1/2}$&$(0)_2^{1/2}(-11)_1(11)_1$\\
$(6,0,1,0,0)$&$[(0)_2(0,-1210)_4]^{1/2}$&$[(0)_2(-44)_2(44)_2]^{1/2}$\\
$(0,4,0,3,3)$&$(-22)_2(22)_2$&$[(-11)_1(11)_1(-55)_2(22)_2$\\
$(4,0,1,1,0)$&$(-22)_2(22)_2$&$[(-11)_1(11)_1(-55)_2(22)_2$\\
$(2,0,3,0,0)$&$[(0)_2(0,-1210)_4]^{1/2}$&$[(0)_2(-44)_2(44)_2]^{1/2}$\\
$(2,2,1,1,0)$&$[(0)_2(0,-1210)_4]^{1/2}$&$(0)_2^{1/2}(-11)_1(11)_1$\\\hline
$(0,0,3,1,0)$&$[(-44)_2(44)_2]^{1/2}$  & $[(-44)_2(44)_2]^{1/2}$ \\
$(2,0,2,1,0)$&$(0)_2$&$[(-44)_2(44)_2]^{1/2}$\\
$(4,0,2,3,1)$&$1$&$1$ \\ \hline
\end{tabular}
\\
\noindent
\begin{tabular}{|c|c|c|c|}\hline
Monomial $\textbf{v}$ & $\phi=1 \psi=1,10$ & $\phi=1 \psi=2,9$ &$\phi=1 \psi=3,8$\\
\hline\hline
$(0,0,0,0,0)$&$(56,2486)_4$ & $(-10,-418)_4$ & $(-66,2750)_4$\\
$(0,2,1,1,1)$&$[(-44)_2(44)_2]^{1/2}$ &$(0)_2$ & $(0)_2$\\
$(6,2,0,0,0)$&$(44)_2$ & $(66)_2$ & $(-22)_2$\\
\hline
$(0,0,0,2,2)$&$(11)_1(-44)_2$& $(11)_1(-66)_2$ & $(11)_1(22)_2$  \\
$(2,0,1,3,3)$&$[(-11)_1(11)_1]^{1/2}$ & $[(-11)_1(11)_1]^{1/2}$ & $[(-11)_1(11)_1]^{1/2}$ \\
$(4,0,2,0,0)$&$(11)_1(-44)_2$&$(11)_1(-66)_2$&$(11)_1(22)_2$\\\hline
$(0,0,2,1,1)$&$(0,-1210)_4^{1/2}$&$(0,1694)_4^{1/2}$&$(0,1694)_4^{1/2}$
\\
$(6,0,1,0,0)$&$[(-44)_2(44)_2]^{1/2}$&$[(-66)_2(66)_2]^{1/2}$&$[(-22)_2(22)_2]^{1/2}$\\
$(0,4,0,3,3)$&$[(0)_2(0,-1210)_4]^{1/2}$&$[(0)_2(0,1694)_4]^{1/2}$&$[(0)_2(0,1694)_4]^{1/2}$\\
$(4,0,1,1,0)$&$[(0)_2(0,-1210)_4]^{1/2}$&$[(0)_2(0,1694)_4]^{1/2}$&$[(0)_2(0,1694)_4]^{1/2}$\\
$(2,0,3,0,0)$&$[(-44)_2(44)_2]^{1/2}$&$[(-66)_2(66)_2]^{1/2}$&$[(-22)_2(22)_2]^{1/2}$\\
$(2,2,1,1,0)$&$(0,-1210)_4^{1/2}$&$(0,1694)_4^{1/2}$&$(0,1694)_4^{1/2}$\\\hline
$(0,0,3,1,0)$&$[(-44)_2(44)_2]^{1/2}$  & $(0)_2$ & $(0)_2$\\
$(2,0,2,1,0)$&$(0)_2$&$[(-22)_2(22)_2]^{1/2}$ & $[(-66)_2(66)_2]^{1/2}$ \\
$(4,0,2,3,1)$&$1$&$1$&$1$ \\ \hline
\end{tabular}
\\[1cm]
\noindent
\begin{tabular}{|c|c|c|}\hline
Monomial $\textbf{v}$ & $\phi=1, \psi=4,7$ & $\phi=1, \psi=5,6$ \\
\hline\hline
$(0,0,0,0,0)$&$(30,1342)_4$ & $(-4,-858)_4$ \\
$(0,2,1,1,1)$&$(0_2$ &$[(-44)_2(44)_2]^{1/2}$ \\
$(6,2,0,0,0)$&$(-22)^2$ & $(0)_2$ \\
\hline
$(0,0,0,2,2)$&$(11)_1(22)_2$& $(11)_1(0)_2$ \\
$(2,0,1,3,3)$&$[(-11)_1(11)_1]^{1/2}$ & $[(-11)_1(11)_1]^{1/2}$ \\
$(4,0,2,0,0)$&$(11)_1(22)_2$& $(11)_1(0)_2$\\\hline
$(0,0,2,1,1)$&$(0,1694)_4^{1/2}$&$(0,-1210)_4^{1/2}$\\
$(6,0,1,0,0)$&$[(-22)_2(22)_2]^{1/2}$&$(0)_2$\\
$(0,4,0,3,3)$&$[(0)_2(0,1694)_4]^{1/2}$&$[(0)_2(0,-1210)_4]^{1/2}$\\
$(4,0,1,1,0)$&$[(0)_2(0,1694)_4]^{1/2}$&$[(0)_2(0,-1210)_4]^{1/2}$\\
$(2,0,3,0,0)$&$[(-22)_2(22)_2]^{1/2}$&$(0)_2$\\
$(2,2,1,1,0)$&$(0,1694)_4^{1/2}$&$(0,-1210)_4^{1/2}$\\\hline
$(0,0,3,1,0)$&$(0)_2$  & $[(-44)_2(44)_2]^{1/2}$ \\
$(2,0,2,1,0)$&$[(-22)_2(22)_2]^{1/2}$&$[(-44)_2(44)_2]^{1/2}$\\
$(4,0,2,3,1)$&$1$&$1$ \\ \hline
\end{tabular}%
\\
\pagebreak
\subsection[p=13 Complete Tables]{p=13 Complete Tables}
For $p=13$, $R_{\text{excep}}(t)=(1+6.13t+13^3t^2)^3$.\\
\noindent
%

Note that for $\phi=7,\,\psi=1,5,8,12$, the contribution from
$(0,4,0,3,3)$ and $(4,0,1,1,0)$ was
$(1-2.13t+2.13.13^3t^3+13^6t^4)$ each (which is not of the
`canonical' form), which squares to the octic in the table.

\noindent
\begin{tabular}{|c|c|}\hline
\multicolumn{2}{|c|}{$\phi=7$}\\\hline
Monomial $\textbf{v}$ & $\psi=4,6,7,9$\\
\hline\hline
(0,0,0,0,0) &  $(13)_1(0,2.5.3.13)_4$ \\
(0,2,1,1,1) &  $(13)_1(-2.13)_2$\\
(6,2,0,0,0)&$(13)_1(-2.13)_2$\\
\hline
(0,0,0,2,2)  & $(13)_1(-2.13)_2$\\
(2,0,1,3,3)  & $(13)_1$ \\
(4,0,2,0,0)&$(13)_1(-2.13)_2$
\\ \hline
$(0,0,2,1,1)\times (2,2,1,1,0)$& $(-4.13,2.13^2,2.5.13^3)_6$ \\
$(6,0,1,0,0)\times (2,0,3,0,0)$& $(-4.13,2.13^2,2.5.13^3)_6$\\
$(0,4,0,3,3)\times(4,0,1,1,0)$&$(2.13)_2(2^3.13)_2(-2^3.13,29.13^2)_4$\\\hline
(0,0,3,1,0)  & $(13)_1$ \\
(2,0,2,1,0)  & $(13)_1$ \\
(4,0,2,3,1)&$(13^2)_1^{-1}$ \\ \hline
\end{tabular}%
\\




\pagebreak
\subsection[p=17 Complete Tables]{p=17 Complete Tables}
For $p=17$, $R_{\text{excep}}(t)=(1-2.17t+17^3t^2)^3$.\\

\footnotesize \noindent
%

In the above tables the contributions from $(0,4,0,3,3)\times
(4,0,1,1,0) $ could not be determined uniquely without computing
the number of points over $\mathbb{F}_{17^r}$ for $r=4$.

\noindent
%

In the above tables the contributions from $(0,4,0,3,3)\times
(4,0,1,1,0) $ could not be determined uniquely without computing
the number of points over $\mathbb{F}_{17^r}$ for $r=4$.
\\

\end{document}